# Possible high-temperature superconductors predicted from electronic structure and data-filtering algorithms

M. Klintenberg[1, *] and O. Eriksson[1, †]

[1] Department of Physics and Astronomy, Uppsala University, Box 516, SE-75120, Uppsala, Sweden

(Dated: September 30, 2011)

We report here the completion of the electronic structure of the majority of the known stoichiometric inorganic compounds, as listed in the International Crystal Structure Data-base (ICSD). We make a detailed comparison of the electronic structure, crystal geometry and chemical bonding of cuprate high temperature superconductors, with the calculated over sixty thousand electronic structures. Based on compelling similarities of the electronic structures in the normal state and a data-filtering technique, we propose that high temperature superconductivity is possible for electron- or hole-doping in a much larger group of materials than previously considered. The indentified materials are composed of over one hundred layered compounds, most which hitherto are untested with respect to their super conducting properties. Of particular interest are the following materials; $Ca_2(CuBr_2O_2)$, $K_2CoF_4$, $Sr_2(MoO_4)$ and $Sr_4V_3O_{10}$, which are discussed in detail.

Electronic structure, and the chemical binding which is the direct consequence of it, is responsible for all materials properties. This applies to the equation of state, hardness, elasticity, catalytic activity, surface tension, work function, magnetism, conductivity, lattice dynamics and superconductivity. Theory based on density functional theory[1], so called *ab-initio* theory, has recently matured to reliably reproduce the electronic structure of materials, and almost all properties associated with it. Hence, ab-initio theory has become an indispensable tool for analyzing experimental results, and even for predicting novel properties[2]. We report here on the completion of the calculation of the electronic structures of the majority of the synthesized stochimetric compounds reported in ICSD[3], which amounts to over sixty thousand compounds. Our study may be viewed as the materials scientists counterpart to the Human Genome mapping[4, 5] in bio-medical science, with a similar potential for impact, albeit with the ambition to identify new functional materials. In this comparison between materials science and life-science, the electronic structure of a material corresponds to the genome of a biological system. It is the electronic structure which ultimately governs the materials property. Having established an electronic structure data-base we argue that it is possible to use what is best described as a data-filtering approach,[6] so that several new materials with selected properties can be identified. For example, the methodology was recently used to successfully predict several new topological insulators[7]. In this report it is high temperature superconductors we focus on.

The electronic structure data-base was first generated by extracting structural information from the Inorganic Crystal Structure Database (ICSD)[3], and using this structural information we performed first principles calculations of the electronic structure. The electronic structure was calculated within the local density approximation (LDA) in combination with a full potential linear muffin-tin orbital (FP-LMTO) method[8]. Although large data-bases of any kind may be used in a variety of ways, we aim to illustrate one particular application. The method may best be described as data-filtering. The basic philosophy is to identify a known class of materials which has been well characterized with respect to a certain property (e.g. superconductivity). If these materials have conspicuous and unique similarities in the underlying electronic structure (the 'code'), one may make a comparison of the electronic structure of other materials, which may not have been subjected to a detailed experimental investigation of the relevant materials property. If a large similarity is found in the electronic structure and possibly also other materials properties like the crystal structure, one may expect a similarity in the materials properties in general.

One hundred years ago superconductivity was discovered[9]. Seventy five years later the field of high temperature superconductivity started, with the discovery of doped $La_2CuO_4$, that had an ordering temperature of 30 K[10]. This discovery of a so called cuprate (copper oxide) superconductor, was quickly followed by several other cuprates with high superconducting temperature. Among the most studied cuprate superconductors is $YBa_2Cu_3O_{6+x}$ (YBCO)[11], which has a critical temperature of 93 K. The critical temperature of $Bi_2Sr_2CaCu_2O_8$ (BISCO)[12] is similar. Currently, the highest critical temperature is found in $HgBa_2Ca_2Cu_3O_{8+x}$ at $\sim$ 135-160 K[13, 14]. Unfortunately a well established theory is lacking for the origin of the d-wave superconductivity in these materials, and several reviews outlining different mechanisms may be found, e.g. in Refs.15–19. The lack of a firm theoretical understanding of the microscopic mechanism behind the pairing in these materials has made it impossible to pre-dict new compounds with higher critical superconducting temperatures, critical fields or critical currents. This unfortunately makes this sub-branch of materials science unique, since most other properties of the materials, e.g. magnetism, optical conductivity, elasticity, phase- and structural stability, in general can now be predicted from *ab-initio* theory[20] with good accuracy.

The resonant valence bond model[21] was suggested early on as a possible mechanisms for superconductivity in the cuprates. In addition, spin-bag theories, antiferromagnetic Fermi liquid theory, nested Fermi liquid theory, and a van Hove scenario have been reviewed in Ref.19 and references therein. Here mechanisms based on spinon, holon and anyons are also described. Furthermore, one has considered microscopic mechanisms based on the Hubbard model,[22] antiferromagnetic paramagnons[23] as well as the so called t-J model[24]. In addition a skyrmion like electronic state has been suggested, where a hole is suggested to be moving clock-wise or counter clock-wise on the oxygen plaquette, and this movement couples to transversal magnetic excitations[25].

The basic hypothesis of the present study is that whatever the mechanism or combination of mechanisms that cause the pairing of charge carriers, there is a crucial aspect in that this takes place in a unique electronic structure and crystal geometry. Namely, that of a quasi two-dimensional/layered crystal structure, in which the d-shell of a transition metal atom hybridizes strongly with p-orbitals of ligand atoms. In the cuprates this is manifested in a band of primary $d_{x^2-y^2}$ character that hybridizes with oxygen p-orbitals. We also suggest that it is important that in the normal state, only one single hybridized band cuts through the Fermi level ($E_F$), for each CuO$_2$-plane. This is exactly the situation for the cuprate superconductors, and angular resolved photoemision of the over-doped compounds result in a Fermi surface which is in agreement with that of electronic structure calculations (e.g. as noted in Ref.26 and references therein). One should of course bare in mind that undoped cuprates are Mott insulators (see e.g.Ref.19), where the direct incorporation of electron-electron interaction must be taken into account in order to get the correct electronic structure. Once this is done, the hole doping that drives superconductivity appears in a single band of Cu $d_{x^2-y^2}$ and O p-character. It is this feature we focus on.

A natural question is if there are other compounds that may be promising candidates for superconductivity, due to a similar electronic structure and crystal geometry as those of the cuprates. We suggest that this is the case and as we shall see below we propose over one hundred compounds which have a crystal geometry and electronic structure which is similar to that of the normal state of the cuprates. Amongst previous studies of superconductivity and the electronic structure we note in particular the work of Pavarini et al.[27] who correlated parameters which determine the electronic structure of cuprates to the critical temperature of these materials, and the works of Refs.28 and 29, where the electronic structure of a nicelate material was tuned by confinement and correlation effects to give an electronic structure that is similar to that of the cuprates.

We have identified potential superconducting compounds by the following data-filtering approach. We have let almost all known stoichiometric compounds as listed by ICSD pass though this filter which has resulted in a list of 139 materials. There are three criteria used in this filtering process: i) A layered crystal structure, with characteristics of the electronic structure reflecting this layered geometry. This can be achieved by inspection of the crystal structure itself, but it is sufficient to investigate the size of the Brillouin-zone, which for two-dimensional materials is compressed in one direction and in addition has a very weak dispersion of the energy bands along this direction. ii) The electronic structure of the cuprates have in the normal state a characteristic feature in that only one band (per CuO$_2$-layer) cuts through $E_F$, and that this band has a dispersion that roughly follows a cosine like dispersion (which is easily understood from tight-binding analysis). iii) The bands the cut through $E_F$ are the result of strongly hybridization between ligand state $p$ and transition metal $d$ orbitals. This last criteria can be fulfilled by analyzing the character of the eigenstates close to $E_F$, i.e the orbital character for each atom (and site) in the unit cell.

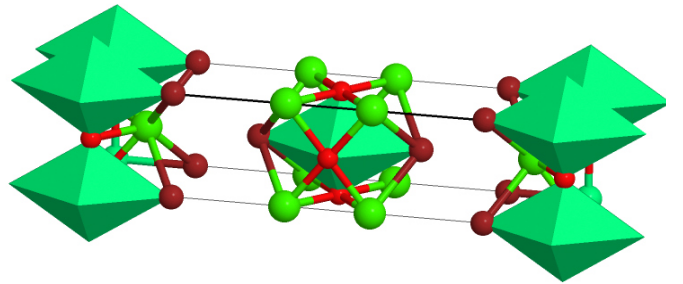

FIG. 1: (Color) Crystal structure of Ca$_2$(CuBr$_2$O$_2$). The Br, Ca, Cu and O atoms are shown in brown, light green, dark green, and red, respectively [3]. The Cu atom sits in an octahedron (marked in green) built up of four oxygen atoms and two Br atoms in apical position.

In the appendix (presented in the Supplementary Information) a list of the suggested 139 superconducting compounds is given, together with figures of these materials electronic- and crystal structures. The 139 compounds are materials that result from the data-filtering criteria outlined above. Slightly different criteria would result in a slightly modified list of possible superconductors. We suggest that electron/hole doping of this list of materials should also be tried in order to find possible new high-temperature superconductors. It is also possible that compounds where a substitution of iso-valent elements to the materials given in the appendix are strong candidates for superconductivity.

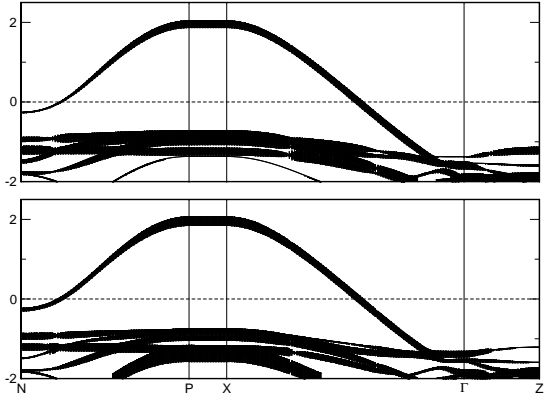

FIG. 2: Electronic structure of the normal state of $Ca_2(CuBr_2O_2)$. The upper panel shows the d-projected bands and the lower panel the p-projected bands, in a so called fatband representation where the amount of d- or p wave character is represented by the thickness of the bands. The Fermi level is at zero and is marked by a horizontal dashed line.

Note that for each compound the electronic structure is given in the so called fat-band representation, to highlight the angular momentum character of the bands. This means that the fatter a band is, the higher degree of a certain angular momentum does it have. In particular we show the amount of transition metal d-character and the ligand p-character of the energy bands.

Before entering a discussion about possible new superconductors we note that in the list of suggested materials one finds all established high-temperature superconducting cuprates. This is not surprising since the datafiltering algorithms were set to capture materials with a cuprate like electronic and crystalline structure. Nevertheless it is rewarding that these materials are found, and this may be viewed as an internal test of the filtering algorithms. Also, other materials with strong electron-phonon coupling are found, e.g. $NbSe_2$ which is known to have a charge density wave state.

The compounds listed in the Appendix form in tetragonal, hexagonal, orthorhombic, monoclinic and cubic crystal structures. Oxides form a large group of compounds in this list, but many non-oxygen based compounds are also found. A detailed analysis of all the materials suggested to be superconducting that are listed in the appendix, is too lengthy for this communication, but we point out that all identified materials have a layered crystal structure as well as few bands (per chemical building block) that cut $E_F$. In addition, these bands are for the majority of compounds, composed of strongly hybridized transition metal d-orbitals and ligand p-orbitals. To give an example of the information that is available in our list of suggested materials, the nature of the chemical bonding, geometry and the electronic structure of four of these materials have been selected, and are analyzed in full detail below.

## $Ca_2(CuBr_2O_2)$

The compound $Ca_2(CuBr_2O_2)$ crystallizes in the space group I 4/mmm (139), in a tetragonal body centered structure (see Fig.1). The Cu atoms are in a 2+ state and are located in a layered crystal structure, with O atoms in the same plane as the Cu atoms forming $CuO_2$-plaquettes. Unlike the cuprates there is in this compound no apical oxygen atom. Instead there is in the out-of-plane position a Br atom that has a Ca atom as a nearest neighbor (see Fig.1). The crystal structure is clearly a layered one with a Cu-O network in a planar geometry that is similar to that of the cuprates.

The electronic structure of the normal state is shown in Fig.2 and is seen to be very similar to that of the cuprates, with a single hybridized Cu d - O p band that cuts $E_F$ and with an energy dispersion that roughly follows what is expected from a tight-binding analysis, where $\epsilon(\mathbf{k}) = \epsilon_0 - 2t(cos(k_xa) + cos(k_ya))$ (here $t$ is the hopping parameter and $a$ the in-plane lattice constant). A symmetry analysis of the orbitals that build up this band is that it is composed of Cu $d_{x^2-y^2}$ character that hybridizes with oxygen $p_x$ and $p_y$ orbitals. In the fatband representation shown in Fig.2, it is seen that this band is essentially an equal mixture of O p and Cu d orbitals. This feature of the electronic structure is similar as that of the known cuprate superconductors.

Electron or hole-doping of this material may be an possible route towards finding new high-temperature superconductors. In this search it is of-course necessary to consider other halides than Br and other alkali earths than Ca. Oxygen doping on the Br site should also be tried.

## $K_2CoF_4$

The compound $K_2CoF_4$ crystallizes in the space group I 4/mmm (139), also in a tetragonal body centered structure (see Fig.3). The Co atoms are positioned in layers together with a square network of F atoms. There is in addition F atoms in apical positions, in a way that is very similar to that of the oxygen positions in the cuprate

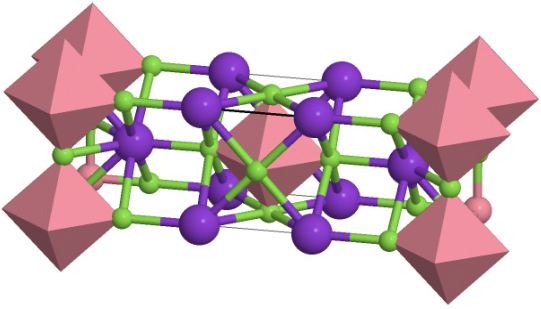

FIG. 3: (Color) Crystal structure of $K_2CoF_4$. The K, Co and F atoms are shown in purple, pink and green, respectively [3]. The F atoms form octahedral cages (marked in pink) which surround the Co atoms.

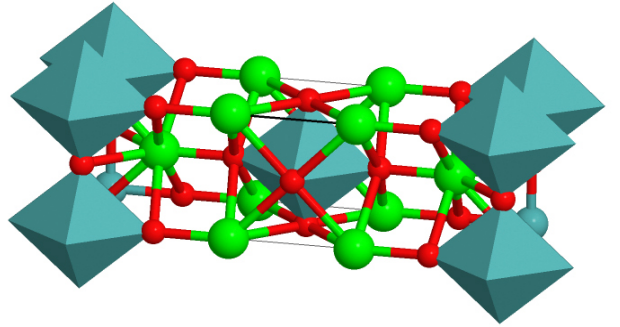

FIG. 5: (Color) Crystal structure of $Sr_2(MoO_4)$. The Sr, Mo and O atoms are shown in green, grayish blue and red, respectively [3]. The O atoms form octahedral cages (marked in grayish blue) which surround the Cu atoms.

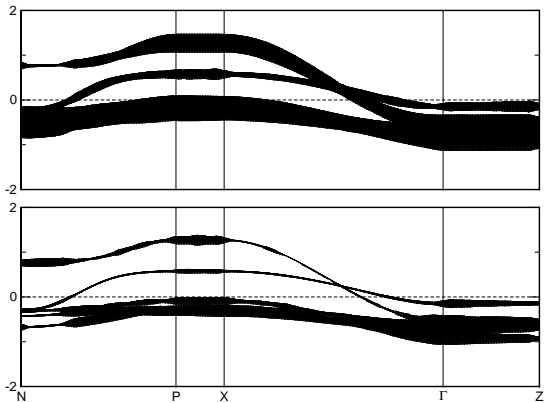

FIG. 4: Electronic structure of the normal state of $K_2CoF_4$. The upper panel shows the d-projected bands and the lower panel the p-projected bands, in a fat-band representation where the amount of d- or p wave character is represented by the thickness of the bands. The Fermi level is at zero and is marked by a horizontal dashed line.

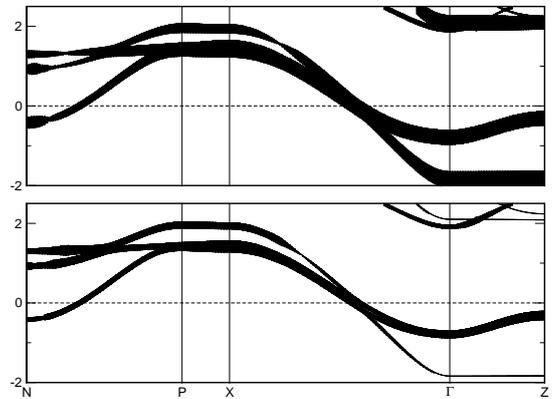

FIG. 6: Electronic structure of the normal state of $Sr_2(MoO_4)$. The upper panel shows the d-projected bands and the lower panel the p-projected bands, in a fat-band representation where the amount of d- or p wave character is represented by the thickness of the bands. The Fermi level is at zero and is marked by a horizontal dashed line.

superconductors. The Co atom is in a +2 state in this compound.

The electronic structure is shown in Fig.4 and is found to be very similar to that of the cuprates. The wavefunction characters of the two bands that cut $E_F$ and have cosine like dispersion, are mainly composed of Co d ($e_g$) and F $p_x$ and $p_y$ orbitals. The lowest band also has some admixture of $d_{xz}$ and $d_{yz}$ orbitals. Electron or hole-doping of this material may be a possible route towards finding new high-temperature superconductors. In this search it is of-course necessary to consider other halides than F and other Group IX elements than Co. Oxygen doping for F should also be tried.

### $Sr_2(MoO_4)$

The compound $Sr_2(MoO_4)$ crystallizes in the space group I 4/mmm (139), also in a tetragonal body centered structure (see Fig.5). The Mo atoms are positioned in layers together with a square network of O atoms. There is in addition O atoms in apical positions, and the crystal structure is indeed the same as that of $La_2CuO_4$, with Mo taking the place of the Cu atoms. The Mo atom is in a +4 state in this compound.

The electronic structure is show in Fig.6 and we note that the three bands that cross $E_F$ have a cosine like dispersion, where the upper band has mainly Mo $d_{x^2-y^2}$ character hybridized with O $p_x$ and $p_y$ orbitals. The two lower bands have mainly Mo $d_{xz}$ and Mo $d_{yz}$ character that hybridize with O $p_z$ states. The hybridization between the Mo d-states and ligand p-states is stronger than for $K_2CoF_4$, as is clear when comparing Fig.4 and Fig.6. Electron or hole-doping of this material may be a possible route towards finding new high-temperature superconductors. In this search it is of-course necessary to consider other alkali-earths than Sr and other Group VI elements than Mo.

### Sr$_4$V$_3$O$_{10}$

The compound $Sr_4V_3O_{10}$ crystallizes in the space group I 4/mmm (139), also in a tetragonal body centered structure (see Fig.7). The V atoms are located in layers with each V atom in the center of a octahedron which is built up of O atoms. The V atom is in a 4+ state in this compound which means that it is the electron-hole symmetric counterpart of a Cu atom in a 2+ configuration.

The electronic structure is similar to that of the cuprates with a single hybridized band along the P-N direction of the Brillouin-zone, that crosses $E_F$. Along the $\Gamma$-X direction, three bands cross $E_F$. The bands around $E_F$ are composed of V d orbitals that hybridize with O p orbitals. Electron or hole-doping of this material may be a possible route towards finding new high-temperature superconductors. In this search it is of-course necessary to consider other alkali-earths than Sr and other Group V elements than vanadium.

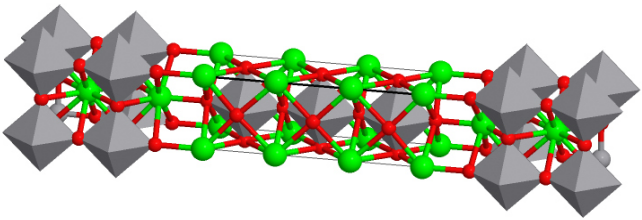

FIG. 7: (Color) Crystal structure of $Sr_4V_3O_{10}$. The Sr, V and O atoms are shown in green, gray and red, respectively [3]. Octahedra (marked in gray) built up of O atoms surround each V atom.

We report the completion of the calculation of almost all known inorganic compounds, which we combine with a data-filtering approach to identify new candidate materials for high-temperature superconductivity. Criteria involving known properties of the normal state of the cuprate superconductors was used in the filtering process. There is one more conspicuous property of the known cuprate superconductors, namely the strong antiferromagnetic inter-atomic exchange interaction between Cu atoms, which is caused by a superexchange mechanism. Since it is unknown if or how this influences the superconductivity we did not use this property as a filtering criterion. Several of the discussed materials are, however, likely to be spin-polarized, but to allow for a larger group of possible new superconductors we did not use spin-polarization as a criterion in the data-filtering process. Instead we used electronic structures from spin-degenerate calculations. If spin-polarization would occur for some of the discussed materials, it is likely to not change the characteristic electronic structure, with the cosine like band that cuts $E_F$. Experience with the cuprates show this.

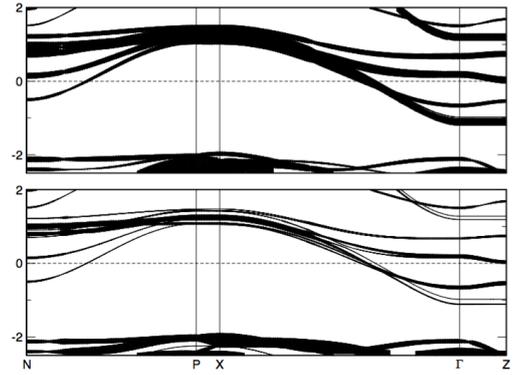

FIG. 8: Electronic structure of the normal state of $Sr_4V_3O_{10}$. The upper panel shows the d-projected bands and the lower panel the p-projected bands, in a fat-band representation where the amount of d- or p wave character is represented by the thickness of the bands. The Fermi level is at zero and is marked by a horizontal dashed line.

It is not unlikely that the materials presented here as candidates for high temperature superconductivity have an electronic structure that is best described from a theory that includes electron-correlation effects somewhat more accurately than that given by the LDA. A natural extension of our work is hence to use dynamical mean field theory to address this issue, and such studies are indeed underway. It is however important to point out that the characteristics of the electronic structure we have used in our filtering approach is a feature which for the cuprate superconductors is in some sense independent of the level of electron correlation. Namely, the high $T_C$ cuprates are all doped in such a way that they in the normal state have an electronic structure which coincides with the filtering criterion used here.

Combining an accurate data-base of the electronic structure with efficient data-filtering protocols to identify new materials, can of course be adapted also for the new class of FeAs based superconductors, albeit, with different filtering criteria than used above. In addition, other properties like thermo electricity, solid state lubrication, new permanent magnets and materials relevant for renewable energy technology are other examples where the described methodology for finding new materials may be applied.

We acknowledge support from the Swedish Research Council (VR), Göran Gustafssons Stiftelse, Swedish Foundation for Strategic Research (SSF), eSSENCE, and the Swedish National Allocations Committee (SNIC/SNAC). O.E. also acknowledges support from ERC (project 247062 - ASD) and the KAW foundation. We thank Oscar Granäs for helpful discussion regarding the implementation of fat-bands. The authors have contributed equally to this work. Critical reading of Dr.L.Kurland is highly appreciated.

---

[*] Electronic address: mattias@physics.uu.se
[†] Electronic address: olle.eriksson@physics.uu.se

# Supplementary information

The table below shows compound name, space group, space group number, bravais lattice and ICSD reference number. Using the ICSD web-site [1], and the ICSD reference number, allows the reader to generate pictures of the crystal structure.

In the fat-band representation the thickness of the band corresponds to the amount of $s$-, $p$- or $d$-character of the band. This is calculated as follows. Let $n(\mathbf{k})$ be the occupation at $\mathbf{k}$ given by

$$n(\mathbf{k}) = Tr(O(\mathbf{k})\rho(\mathbf{k}))$$

where $O(\mathbf{k})$ and $\rho(\mathbf{k})$ is the overlap and density, respectively. The fatness $f_l$ for eigenvalue $\nu$ can then be calculated using

$$f_l = \frac{Tr(O(\mathbf{k})\rho^{\nu_l}(\mathbf{k}))}{Tr(O(\mathbf{k})\rho(\mathbf{k}))}$$

where element $ij$ of the density matrix is given in terms of weights $(w)$ and eigenvectors $(Z)$ by:

$$\rho_{ij}(\mathbf{k}) = \sum_{\nu} w_{\nu,\mathbf{k}} Z_i(\mathbf{k},\nu) Z_j^{\dagger}(\mathbf{k},\nu)$$

$$\rho_{ij}^{\nu_l}(\mathbf{k}) = w_{\nu_l,\mathbf{k}} Z_i(\mathbf{k},\nu_l) Z_j^{\dagger}(\mathbf{k},\nu_l)$$

$\sum_l f_l$ sums up to one electron and $l$ runs over the complete basis. Note that in the figures below the Fermi level is at zero. Also, in the band plots we make a projection on atomic and l-resolved fat bands.

―――――――

| Material | Space group | (#) | Bravais lattice | ICSD # |
|---|---|---|---|---|
| AuCuZn$_2$ | F m -3 m | (225) | cubic face-centred | 150571 |
| AgAuZn$_2$ | F m -3 m | (225) | cubic face-centred | 604792 |
| CuNi$_2$Sb | F m -3 m | (225) | cubic face-centred | 53320 |
| CuNi$_2$Sn | F m -3 m | (225) | cubic face-centred | 103068 |
| ErPt$_2$ | F d -3 m S | (227) | cubic face-centred | 103287 |
| EuPt$_2$ | F d -3 m S | (227) | cubic face-centred | 103430 |
| HoPt$_2$ | F d -3 m S | (227) | cubic face-centred | 104441 |
| NaPt$_2$ | F d -3 m S | (227) | cubic face-centred | 644945 |
| BaPt$_2$ | F d -3 m S | (227) | cubic face-centred | 616039 |
| CuSe | P 63/m m c | (194) | hexagonal primitive | 240 |
| KAuTe | P 63/m m c | (194) | hexagonal primitive | 40165 |
| RbAuTe | P 63/m m c | (194) | hexagonal primitive | 75026 |
| CdInGaS$_4$ | P -3 m 1 | (164) | hexagonal primitive | 20785 |
| ZrNCl | P -3 m 1 | (164) | hexagonal primitive | 25506 |
| KCuSe | P 63/m m c | (194) | hexagonal primitive | 12157 |
| KCuTe | P 63/m m c | (194) | hexagonal primitive | 12158 |
| Li$_2$ZnGe | P -3 m 1 | (164) | hexagonal primitive | 53678 |
| Li$_2$ZnSi | P -3 m 1 | (164) | hexagonal primitive | 16221 |
| AuYO$_2$ | P 63/m m c | (194) | hexagonal primitive | 95675 |
| AgAlO$_2$ | P 63/m m c | (194) | hexagonal primitive | 300020 |
| CuBr | P 63 m c | (186) | hexagonal primitive | 30092 |
| Ca$_2$CuZn$_2$P$_3$ | P 63/m m c | (194) | hexagonal primitive | 89517 |
| Al$_5$C$_3$N | P 63 m c | (186) | hexagonal primitive | 26859 |
| Ca$_3$Cu$_2$Zn$_2$P$_4$ | P -3 m 1 | (164) | hexagonal primitive | 89515 |
| Eu$_3$Cu$_2$Zn$_2$P$_4$ | P -3 m 1 | (164) | hexagonal primitive | 89516 |
| Cu$_4$(S$_2$)$_2$(CuS)$_2$ | P 63/m m c | (194) | hexagonal primitive | 26968 |

| Material | Space group | (#) | Bravais lattice | ICSD # |
|---|---|---|---|---|
| LaKPdO$_3$ | C 1 2/m 1 | (12) | monoclinic base-centred | 417108 |
| BaY$_2$F$_8$ | C 1 2/m 1 | (12) | monoclinic base-centred | 74359 |
| AgCuS | C m c m | (63) | orthorhombic base-centred | 30233 |
| LaSeTe$_2$ | C m c m | (63) | orthorhombic base-centred | 413171 |
| NbS$_2$ | C m 2 m | (38) | orthorhombic base-centred | 67443 |
| BaNiY$_2$O$_5$ | I m m m | (71) | orthorhombic body-centred | 68795 |
| RuOCl$_2$ | I m m m | (71) | orthorhombic body-centred | 83883 |
| Bi$_2$(CO$_3$)O$_2$ | I m m 2 | (44) | orthorhombic body-centred | 94740 |
| Al$_2$Ba$_3$Ge$_2$ | I m m m | (71) | orthorhombic body-centred | 52612 |
| Ba$_3$Al$_2$Si$_2$ | I m m m | (71) | orthorhombic body-centred | 100128 |
| Ba$_3$Al$_2$Sn$_2$ | I m m m | (71) | orthorhombic body-centred | 9565 |
| NbSe$_2$ | F m 2 m | (42) | orthorhombic face-centred | 71339 |
| TaS$_2$ | F m 2 m | (42) | orthorhombic face-centred | 280988 |
| TaS$_2$ | F 2 m m | (42) | orthorhombic face-centred | 67651 |
| TaSe$_2$ | F m 2 m | (42) | orthorhombic face-centred | 72198 |
| Tl$_2$Ba$_2$CuO$_6$ | F m m m | (69) | orthorhombic face-centred | 41569 |
| TiNCl | P m m n S | (59) | orthorhombic primitive | 27396 |
| Pb$_2$Ba$_2$YCuCu$_2$O$_8$ | P 2 21 2 | (17) | orthorhombic primitive | 66088 |
| Pb$_2$Sr$_2$YCu$_3$O$_8$ | P 2 21 2 | (17) | orthorhombic primitive | 66587 |
| YBa$_2$Cu$_3$O$_{6.5}$ | P m m m | (47) | orthorhombic primitive | 75697 |
| YBa$_2$Cu$_3$O$_{6.5}$ | P m m m | (47) | orthorhombic primitive | 96016 |
| EuBa$_2$Cu$_3$O$_7$ | P m m m | (47) | orthorhombic primitive | 81171 |
| Ba$_2$GdCu$_3$O$_7$ | P m m m | (47) | orthorhombic primitive | 56514 |
| HoBa$_2$Cu$_3$O$_7$ | P m m m | (47) | orthorhombic primitive | 68044 |
| LaBa$_2$Cu$_3$O$_7$ | P m m m | (47) | orthorhombic primitive | 81167 |
| NdBa$_2$Cu$_3$O$_7$ | P m m m | (47) | orthorhombic primitive | 81169 |
| PrBa$_2$Cu$_3$O$_7$ | P m m m | (47) | orthorhombic primitive | 81168 |
| SmBa$_2$Cu$_3$O$_7$ | P m m m | (47) | orthorhombic primitive | 71705 |
| Ba$_2$YCu$_3$O$_7$ | P m m m | (47) | orthorhombic primitive | 202770 |
| Ba$_2$YCu$_3$O$_7$ | P m m m | (47) | orthorhombic primitive | 77737 |
| LaBa$_2$Cu$_3$O$_8$ | P m m m | (47) | orthorhombic primitive | 85291 |
| Na$_3$Cu$_4$S$_4$ | P b a m | (55) | orthorhombic primitive | 10004 |
| Sr$_4$V$_3$O$_{10}$ | I 4/m m m | (139) | tetragonal body-centred | 73698 |
| MoB | I 41/a m d S | (141) | tetragonal body-centred | 24280 |
| WB | I 41/a m d S | (141) | tetragonal body-centred | 24281 |
| Yb(AgS$_2$) | I 41 m d | (109) | tetragonal body-centred | 27091 |
| LaI$_2$ | I 4/m m m | (139) | tetragonal body-centred | 202452 |
| SmCu$_2$Si$_2$ | I 4/m m m | (139) | tetragonal body-centred | 106843 |
| TbCu$_2$Si$_2$ | I 4/m m m | (139) | tetragonal body-centred | 106844 |
| Cu$_2$TmSi$_2$ | I 4/m m m | (139) | tetragonal body-centred | 53349 |
| Cu$_2$YSi$_2$ | I 4/m m m | (139) | tetragonal body-centred | 23551 |
| Li$_2$PdH$_2$ | I 4/m m m | (139) | tetragonal body-centred | 108534 |
| Na$_2$PdH$_2$ | I 4/m m m | (139) | tetragonal body-centred | 68071 |
| (Cu$_2$S$_2$)(Sr$_2$NiO$_2$) | I 4/m m m | (139) | tetragonal body-centred | 88424 |
| Ca$_2$(CuBr$_2$O$_2$) | I 4/m m m | (139) | tetragonal body-centred | 1028 |
| Sr$_2$CoO$_2$Br$_2$ | I 4/m m m | (139) | tetragonal body-centred | 151789 |
| CuSr$_2$Br$_2$O$_2$ | I 4/m m m | (139) | tetragonal body-centred | 1178 |
| Ca$_2$(CuCl$_2$O$_2$) | I 4/m m m | (139) | tetragonal body-centred | 1027 |
| Ca$_2$CuO$_2$Cl$_2$ | I 4/m m m | (139) | tetragonal body-centred | 83117 |
| Sr$_2$CuO$_2$Cl$_2$ | I 4/m m m | (139) | tetragonal body-centred | 4087 |
| Tl$_2$Ba$_2$CaCu$_2$O$_8$ | I 4/m m m | (139) | tetragonal body-centred | 78592 |
| Bi$_2$Sr$_2$CaCu$_2$O$_8$ | I 4/m m m | (139) | tetragonal body-centred | 68188 |
| Ce$_2$BiO$_2$ | I 4/m m m | (139) | tetragonal body-centred | 9099 |
| Ce$_2$SbO$_2$ | I 4/m m m | (139) | tetragonal body-centred | 9100 |
| CePd$_2$Si$_2$ | I 4/m m m | (139) | tetragonal body-centred | 621852 |
| CePt$_2$Si$_2$ | I 4/m m m | (139) | tetragonal body-centred | 52895 |
| Cu$_2$ErGe$_2$ | I 4/m m m | (139) | tetragonal body-centred | 53251 |
| ErCu$_2$Si$_2$ | I 4/m m m | (139) | tetragonal body-centred | 106845 |
| Cu$_2$GdSi$_2$ | I 4/m m m | (139) | tetragonal body-centred | 64825 |

| Material | Space group | (#) | Bravais lattice | ICSD # |
|---|---|---|---|---|
| $Cu_2HoGe_2$ | I 4/m m m | (139) | tetragonal body-centred | 53270 |
| $YCu_2Ge_2$ | I 4/m m m | (139) | tetragonal body-centred | 52764 |
| $Cu_2HoSi_2$ | I 4/m m m | (139) | tetragonal body-centred | 53289 |
| $NdCu_2Si_2$ | I 4/m m m | (139) | tetragonal body-centred | 106842 |
| $Eu_2(VO_4)$ | I 4/m m m | (139) | tetragonal body-centred | 89000 |
| $K_2(NiF_4)$ | I 4/m m m | (139) | tetragonal body-centred | 15576 |
| $K_2(NiF_4)$ | I 4/m m m | (139) | tetragonal body-centred | 631720 |
| $Rb_2(NiF_4)$ | I 4/m m m | (139) | tetragonal body-centred | 69682 |
| $La_2(NiO_4)$ | I 4/m m m | (139) | tetragonal body-centred | 1179 |
| $La_2(NiO_4)$ | I 4/m m m | (139) | tetragonal body-centred | 33536 |
| $La_2PdO_4$ | I 4/m m m | (139) | tetragonal body-centred | 40262 |
| $Sr_2(MoO_4)$ | I 4/m m m | (139) | tetragonal body-centred | 152123 |
| $Sr_2(RuO_4)$ | I 4/m m m | (139) | tetragonal body-centred | 157401 |
| $Sr_2VO_4$ | I 4/m m m | (139) | tetragonal body-centred | 72219 |
| $Cs_2AgF_4$ | I 4/m m m | (139) | tetragonal body-centred | 16254 |
| $K_2CoF_4$ | I 4/m m m | (139) | tetragonal body-centred | 33522 |
| $Rb_2CoF_4$ | I 4/m m m | (139) | tetragonal body-centred | 69683 |
| $Gd_2(CuO_4)$ | I 4/m m m | (139) | tetragonal body-centred | 41844 |
| $In_2CuO_4$ | I 4/m m m | (139) | tetragonal body-centred | 39475 |
| $La_2(CuO_4)$ | I 4/m m m | (139) | tetragonal body-centred | 41643 |
| $Ba_2CoF_6$ | I 4/m m m | (139) | tetragonal body-centred | 21057 |
| $Ba_2NiF_6$ | I 4/m m m | (139) | tetragonal body-centred | 21056 |
| $Ba_2(ZnF_6)$ | I 4/m m m | (139) | tetragonal body-centred | 21054 |
| $(Cu_2S_2)(Sr_2CuO_2)$ | I 4/m m m | (139) | tetragonal body-centred | 88423 |
| $Ba_2Cu_3O_4Br_2$ | I 4/m m m | (139) | tetragonal body-centred | 36128 |
| $Ba_2Cu_3O_4Cl_2$ | I 4/m m m | (139) | tetragonal body-centred | 355 |
| $Ca_3Cu_2O_4Br_2$ | I 4/m m m | (139) | tetragonal body-centred | 69182 |
| $Ca_3Cu_2O_4Cl_2$ | I 4/m m m | (139) | tetragonal body-centred | 69181 |
| $La_3Ni_2O_6$ | I 4/m m m | (139) | tetragonal body-centred | 249209 |
| $K_3Ni_2F_7$ | I 4/m m m | (139) | tetragonal body-centred | 33523 |
| $Sr_3V_2O_7$ | I 4/m m m | (139) | tetragonal body-centred | 71320 |
| $Sr_3(V_2O_7)$ | I 4/m m m | (139) | tetragonal body-centred | 71451 |
| $K_3Co_2F_7$ | I 4/m m m | (139) | tetragonal body-centred | 33524 |
| $K_3Cu_2F_7$ | I 4/m m m | (139) | tetragonal body-centred | 15373 |
| $La_4Ni_3O_8$ | I 4/m m m | (139) | tetragonal body-centred | 173372 |
| $K_5Te_3$ | I 4/m | (87) | tetragonal body-centred | 96743 |
| $CaSmCuO_3Cl$ | P 4/n m m Z | (129) | tetragonal primitive | 86428 |
| $HgBa_2CaCu_2O_6$ | P 4/m m m | (123) | tetragonal primitive | 75725 |
| $HgBa_2CaCu_2O_6$ | P 4/m m m | (123) | tetragonal primitive | 83087 |
| $TlYBa_2Cu_2O_7$ | P 4/m m m | (123) | tetragonal primitive | 74163 |
| $TlCaSr_2Cu_2O_7$ | P 4/m m m | (123) | tetragonal primitive | 74165 |
| $NdBa_2Cu_2NbO_8$ | P 4/m m m | (123) | tetragonal primitive | 44255 |
| $Sr_2CoO_3Cl$ | P 4/n m m Z | (129) | tetragonal primitive | 91750 |
| $HgBa_2CuO_4$ | P 4/m m m | (123) | tetragonal primitive | 75720 |
| $Sr_2CuO_2(CO_3)$ | P 4 21 2 | (90) | tetragonal primitive | 83096 |
| $KCeSe_4$ | P 4/n b m Z | (125) | tetragonal primitive | 67656 |
| $NdLi_2Sb_2$ | P 4/n m m Z | (129) | tetragonal primitive | 36020 |
| $HgBa_2Ca_2Cu_3O_8$ | P 4/m m m | (123) | tetragonal primitive | 75730 |
| $HoBa_2Cu_3O_6$ | P 4/m m m | (123) | tetragonal primitive | 68047 |
| $LuBa_2Cu_3O_6$ | P 4/m m m | (123) | tetragonal primitive | 98113 |
| $NdBa_2Cu_3O_6$ | P 4/m m m | (123) | tetragonal primitive | 83074 |
| $Cs(Cu_4Se_3)$ | P 4/m m m | (123) | tetragonal primitive | 75196 |
| $KCu_4S_3$ | P 4/m m m | (123) | tetragonal primitive | 23336 |
| $KCu_4Se_3$ | P 4/m m m | (123) | tetragonal primitive | 280072 |

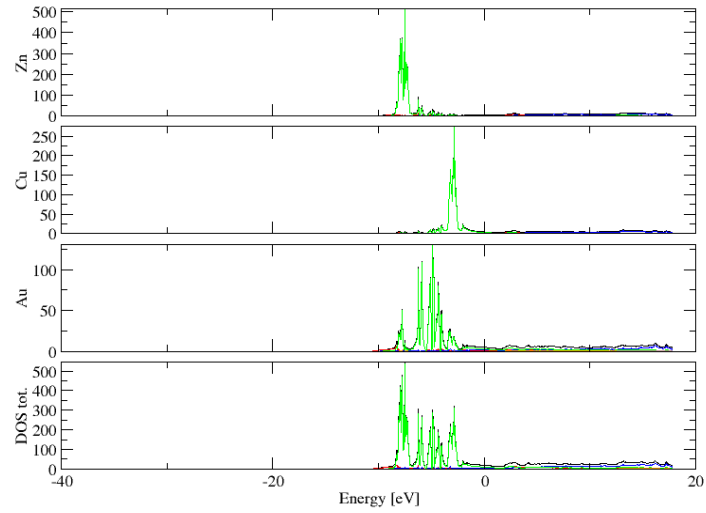

FIG. 1: (Color online) PDOS of AuCuZn$_2$ (ICSD #150571). The $s$-, $p$- and $d$-projected states are in red, blue and green, respectively. AuCuZn$_2$ crystallizes in space group F m -3 m (#225), in a cubic face-centred structure.

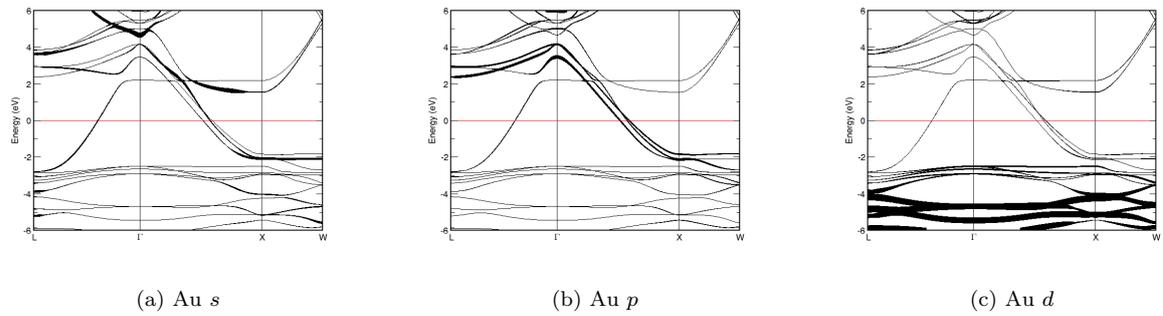

(a) Au $s$        (b) Au $p$        (c) Au $d$

FIG. 2: Fat band representation of Au in AuCuZn$_2$

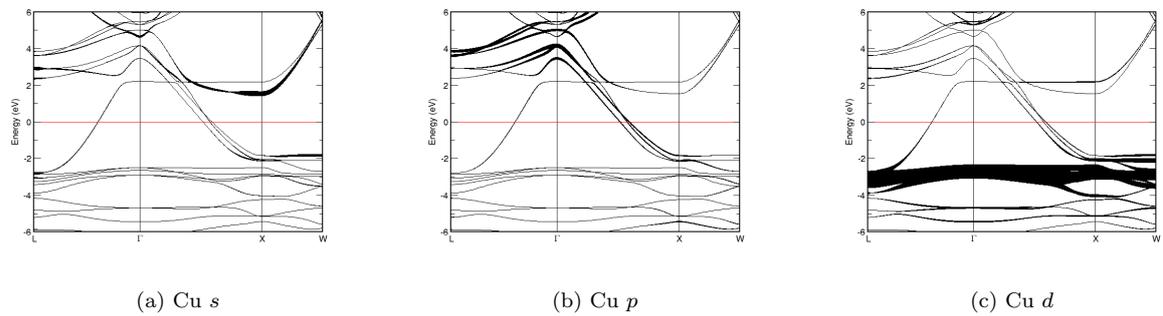

(a) Cu $s$        (b) Cu $p$        (c) Cu $d$

FIG. 3: Fat band representation of Cu in AuCuZn$_2$

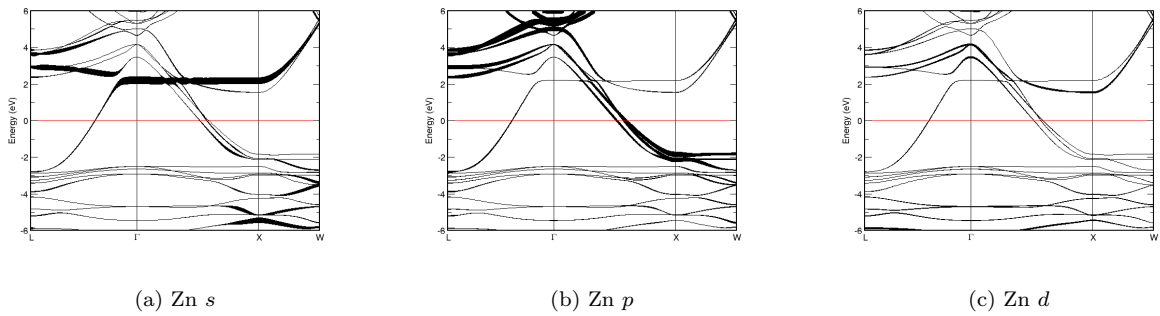

(a) Zn $s$　　　(b) Zn $p$　　　(c) Zn $d$

FIG. 4: Fat band representation of Zn in AuCuZn$_2$

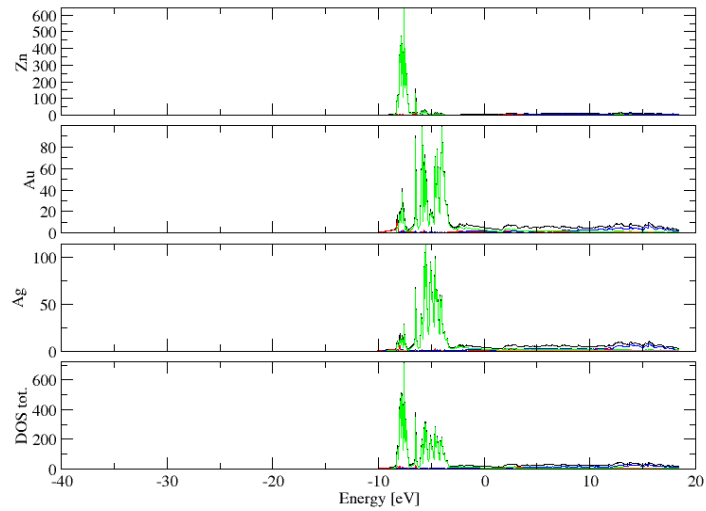

FIG. 5: (Color online) PDOS of AgAuZn$_2$ (ICSD #604792). The $s$-, $p$- and $d$-projected states are in red, blue and green, respectively. AgAuZn$_2$ crystallizes in space group F m -3 m (#225), in a cubic face-centred structure.

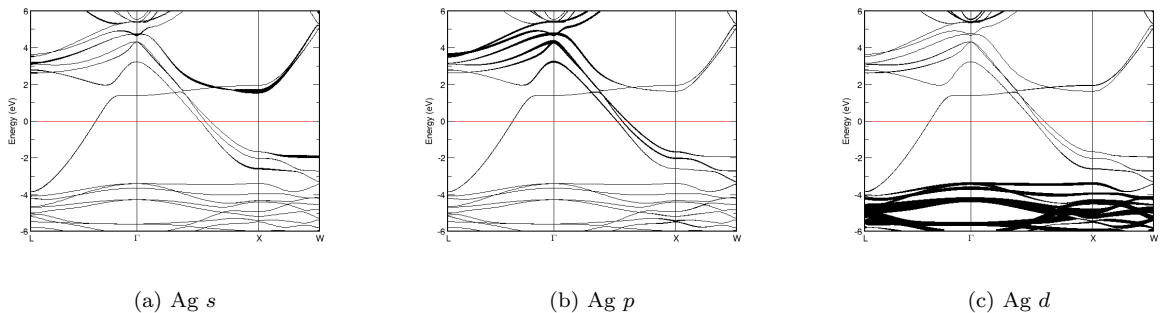

(a) Ag $s$　　　(b) Ag $p$　　　(c) Ag $d$

FIG. 6: Fat band representation of Ag in AgAuZn$_2$

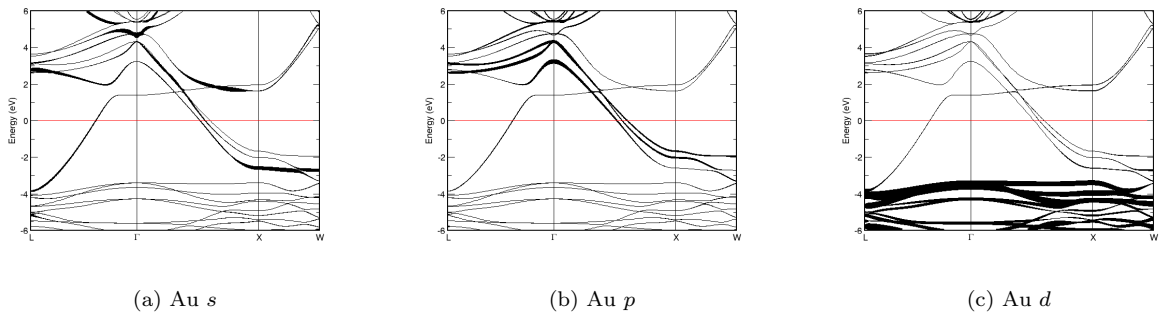

(a) Au *s*  (b) Au *p*  (c) Au *d*

FIG. 7: Fat band representation of Au in AgAuZn$_2$

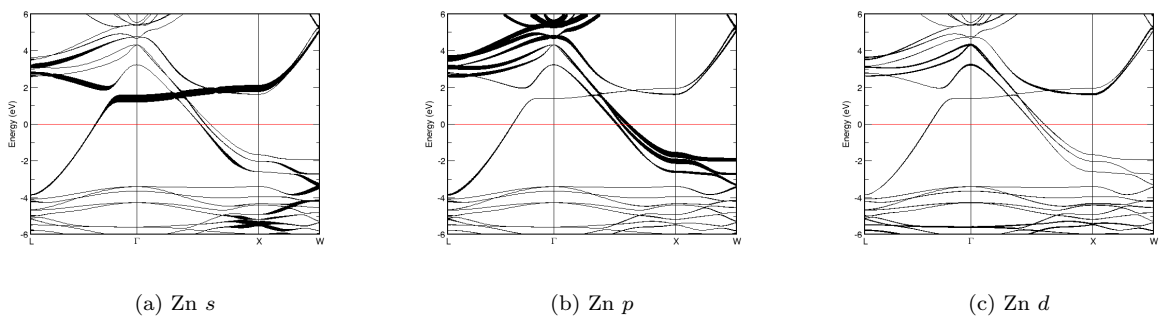

(a) Zn *s*  (b) Zn *p*  (c) Zn *d*

FIG. 8: Fat band representation of Zn in AgAuZn$_2$

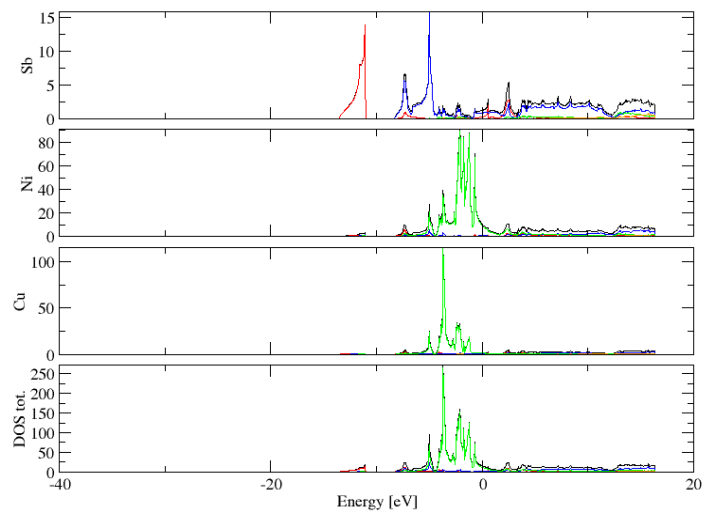

FIG. 9: (Color online) PDOS of CuNi$_2$Sb (ICSD #53320). The *s*-, *p*- and *d*-projected states are in red, blue and green, respectively. CuNi$_2$Sb crystallizes in space group F m -3 m (#225), in a cubic face-centred structure.

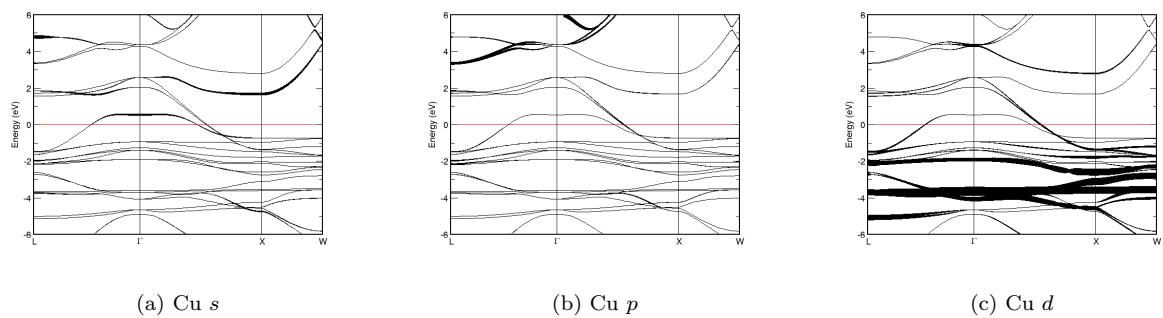

(a) Cu $s$        (b) Cu $p$        (c) Cu $d$

FIG. 10: Fat band representation of Cu in CuNi$_2$Sb

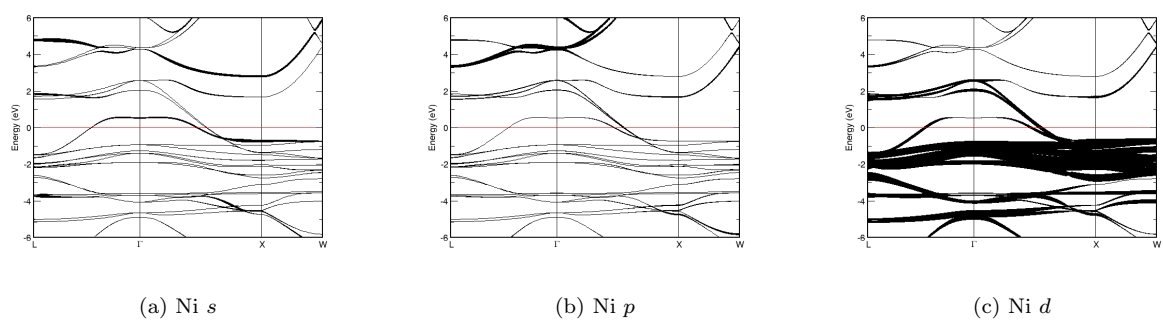

(a) Ni $s$        (b) Ni $p$        (c) Ni $d$

FIG. 11: Fat band representation of Ni in CuNi$_2$Sb

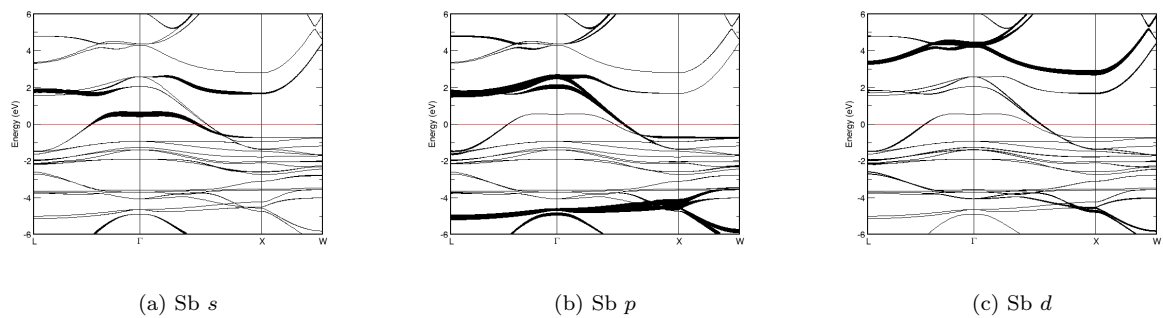

(a) Sb $s$        (b) Sb $p$        (c) Sb $d$

FIG. 12: Fat band representation of Sb in CuNi$_2$Sb

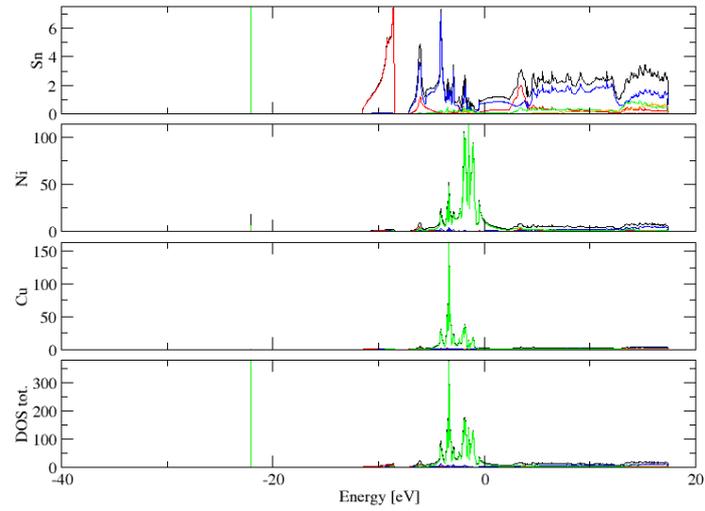

FIG. 13: (Color online) PDOS of CuNi$_2$Sn (ICSD #103068). The $s$-, $p$- and $d$-projected states are in red, blue and green, respectively. CuNi$_2$Sn crystallizes in space group F m -3 m (#225), in a cubic face-centred structure.

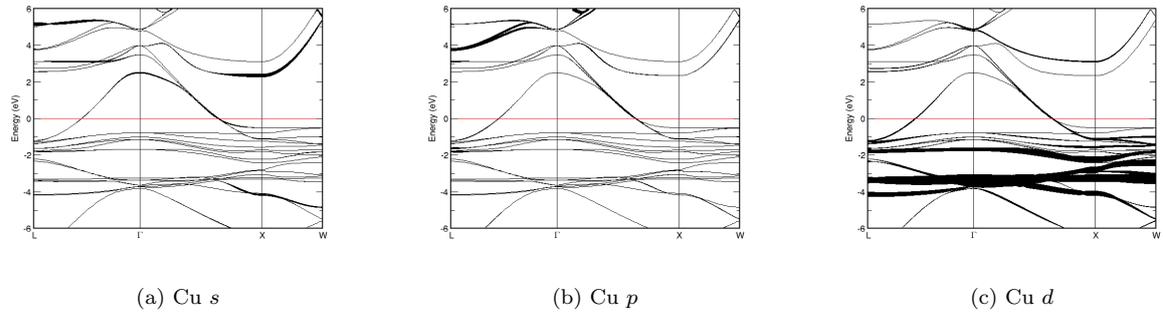

(a) Cu $s$        (b) Cu $p$        (c) Cu $d$

FIG. 14: Fat band representation of Cu in CuNi$_2$Sn

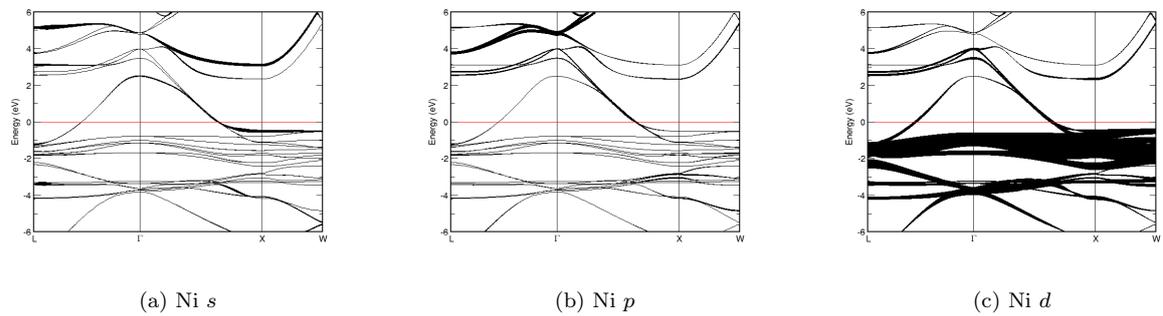

(a) Ni $s$        (b) Ni $p$        (c) Ni $d$

FIG. 15: Fat band representation of Ni in CuNi$_2$Sn

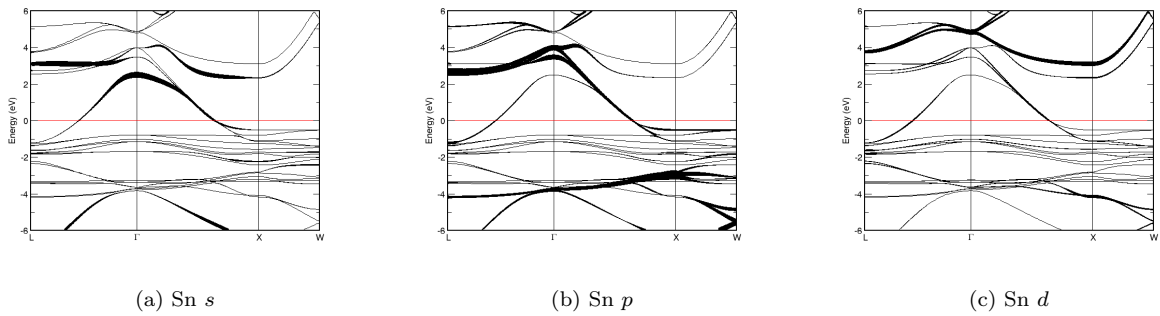

(a) Sn $s$        (b) Sn $p$        (c) Sn $d$

FIG. 16: Fat band representation of Sn in CuNi$_2$Sn

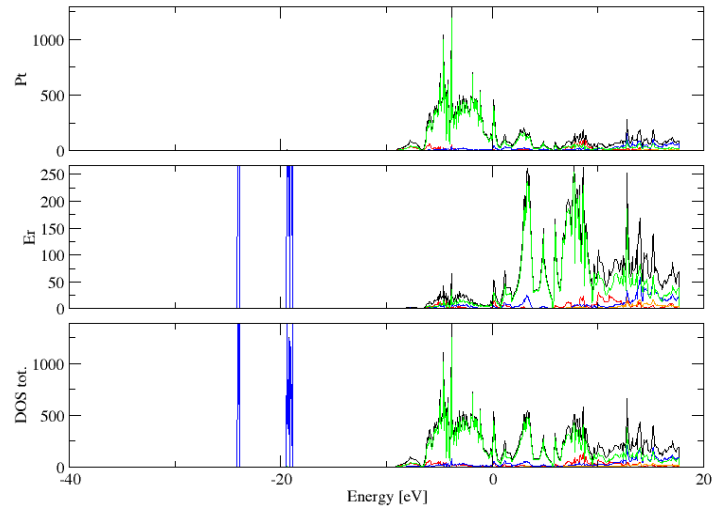

FIG. 17: (Color online) PDOS of ErPt$_2$ (ICSD #103287). The $s$-, $p$- and $d$-projected states are in red, blue and green, respectively. ErPt$_2$ crystallizes in space group F d -3 m S (#227), in a cubic face-centred structure.

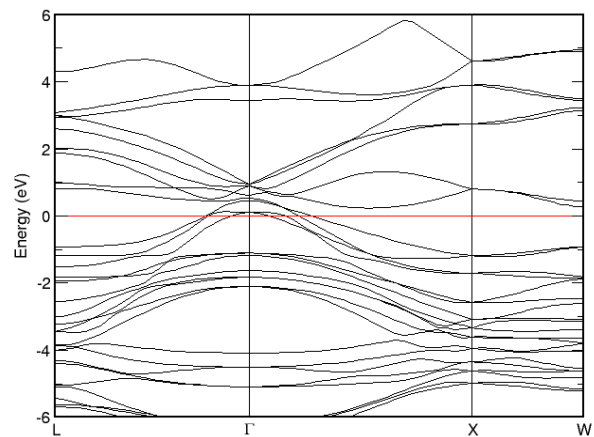

(a) E $vs.$ k

FIG. 18: Band structure of ErPt$_2$

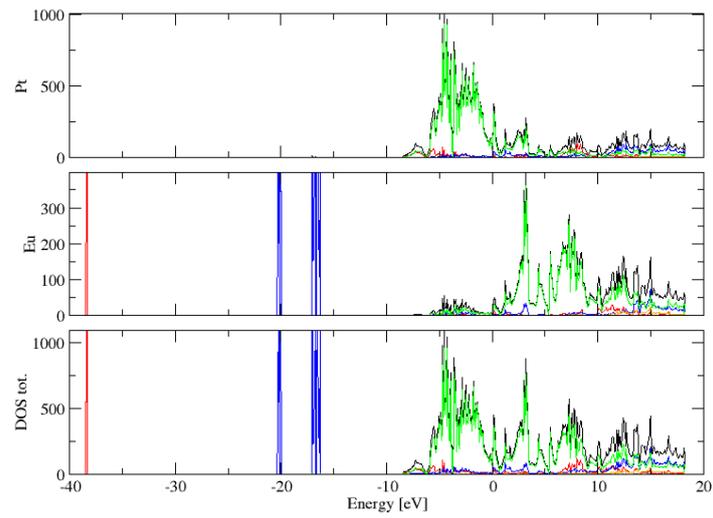

FIG. 19: (Color online) PDOS of EuPt$_2$ (ICSD #103430). The $s$-, $p$- and $d$-projected states are in red, blue and green, respectively. EuPt$_2$ crystallizes in space group F d -3 m S (#227), in a cubic face-centred structure.

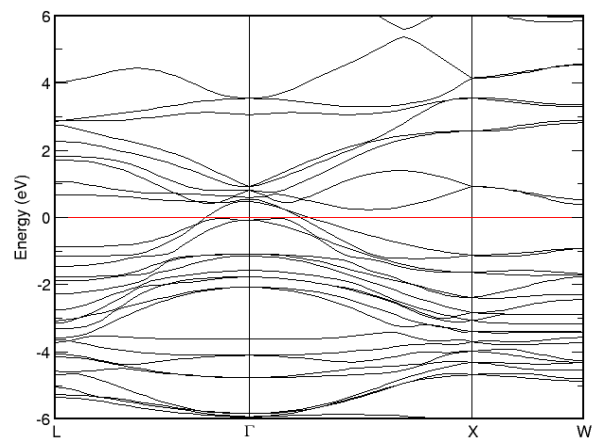

(a) E $vs.$ k

FIG. 20: Band structure of EuPt$_2$

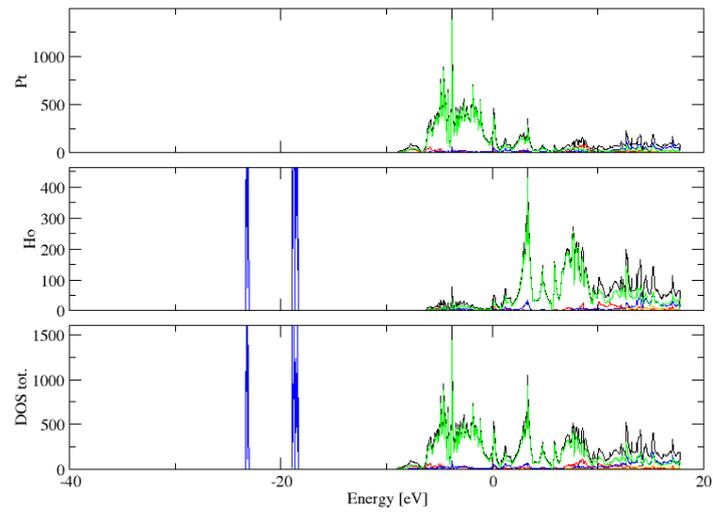

FIG. 21: (Color online) PDOS of HoPt$_2$ (ICSD #104441). The $s$-, $p$- and $d$-projected states are in red, blue and green, respectively. HoPt$_2$ crystallizes in space group F d -3 m S (#227), in a cubic face-centred structure.

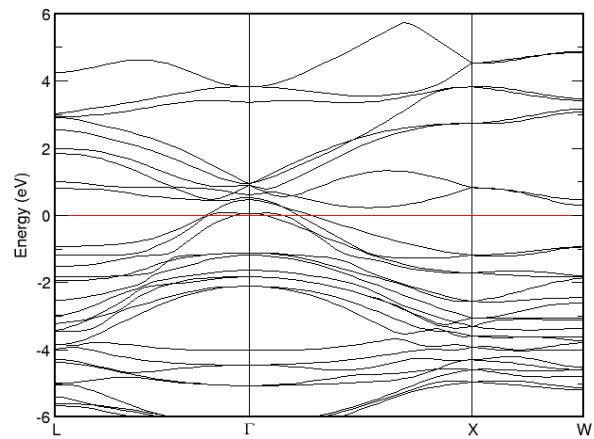

(a) E *vs.* k

FIG. 22: Band structure of HoPt$_2$

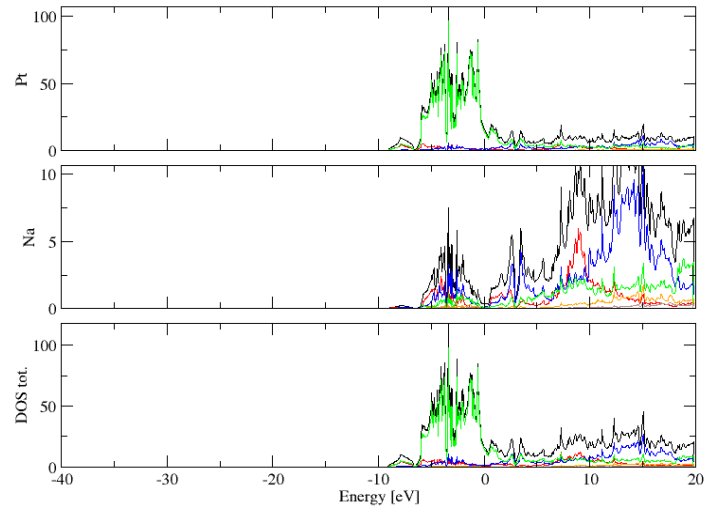

FIG. 23: (Color online) PDOS of NaPt$_2$ (ICSD #644945). The $s$-, $p$- and $d$-projected states are in red, blue and green, respectively. NaPt$_2$ crystallizes in space group F d -3 m S (#227), in a cubic face-centred structure.

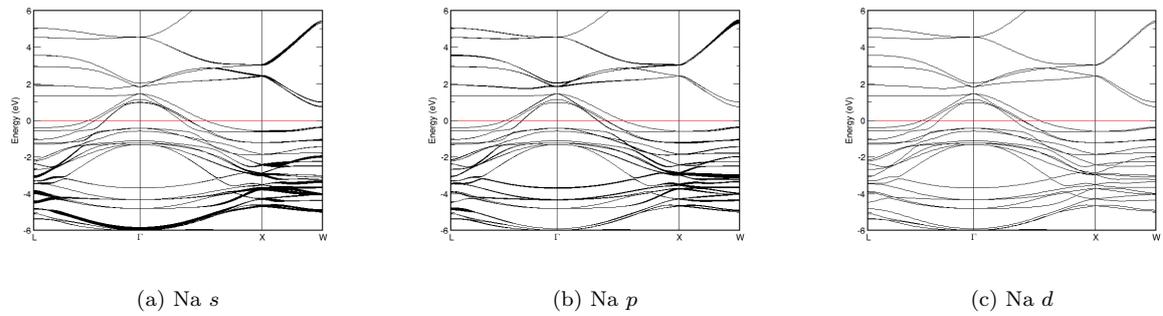

(a) Na $s$          (b) Na $p$          (c) Na $d$

FIG. 24: Fat band representation of Na in NaPt$_2$

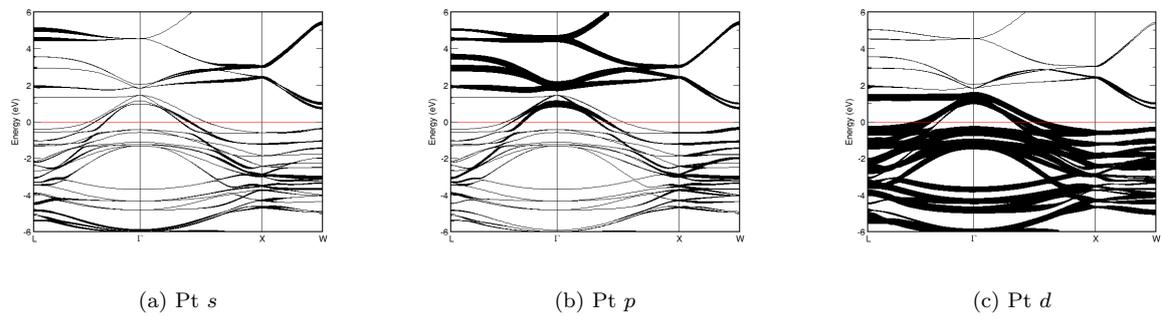

(a) Pt $s$          (b) Pt $p$          (c) Pt $d$

FIG. 25: Fat band representation of Pt in NaPt$_2$

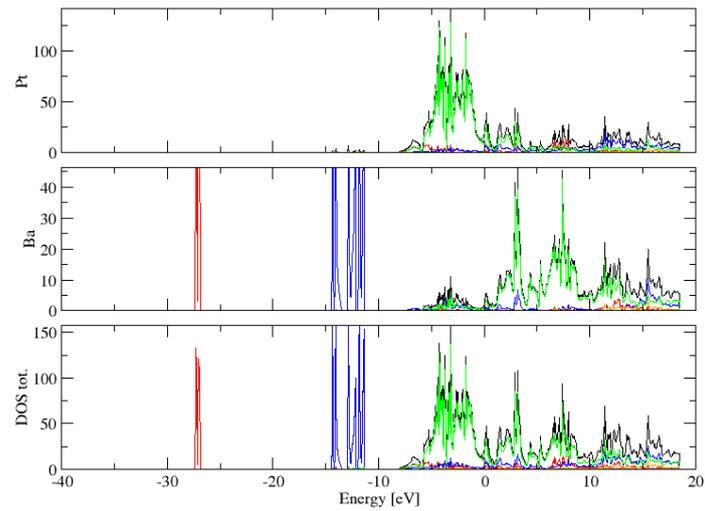

FIG. 26: (Color online) PDOS of BaPt$_2$ (ICSD #616039). The $s$-, $p$- and $d$-projected states are in red, blue and green, respectively. BaPt$_2$ crystallizes in space group F d -3 m S (#227), in a cubic face-centred structure.

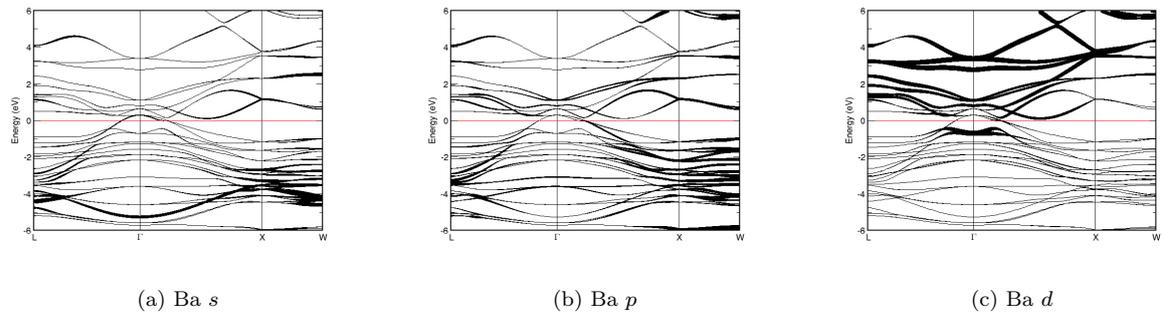

(a) Ba $s$          (b) Ba $p$          (c) Ba $d$

FIG. 27: Fat band representation of Ba in BaPt$_2$

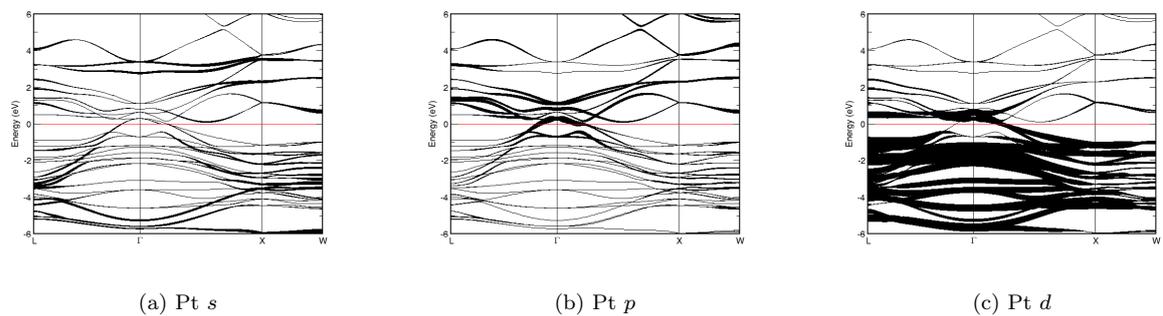

(a) Pt $s$          (b) Pt $p$          (c) Pt $d$

FIG. 28: Fat band representation of Pt in BaPt$_2$

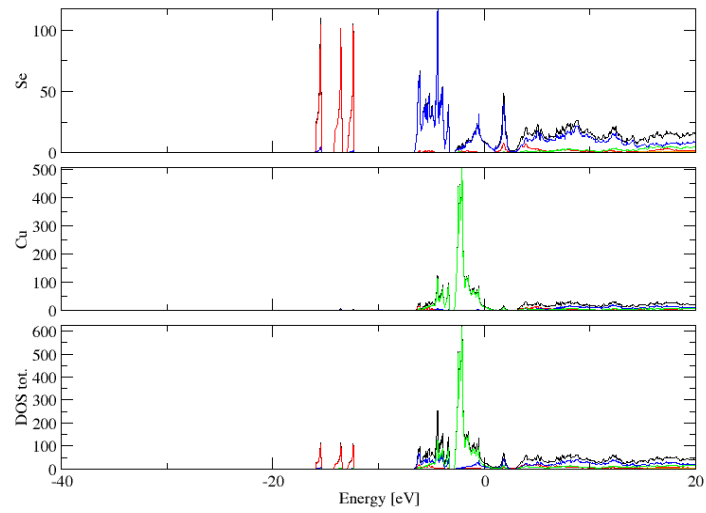

FIG. 29: (Color online) PDOS of CuSe (ICSD #240). The *s*-, *p*- and *d*-projected states are in red, blue and green, respectively. CuSe crystallizes in space group P 63/m m c (#194), in a hexagonal primitive structure.

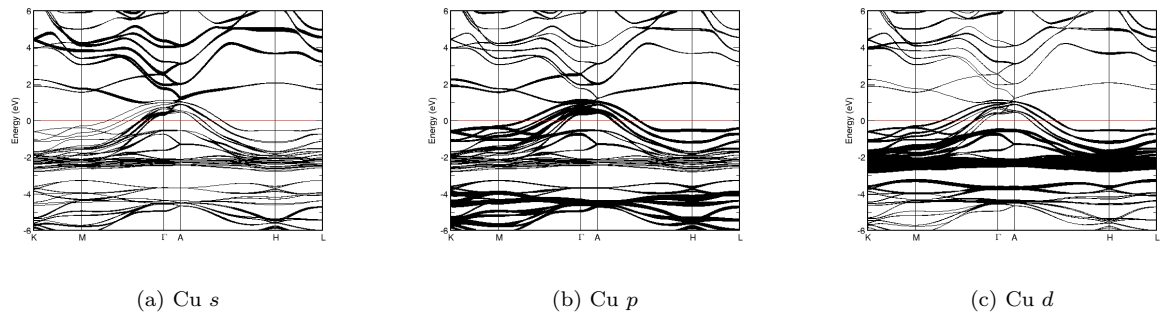

(a) Cu *s*           (b) Cu *p*          (c) Cu *d*

FIG. 30: Fat band representation of Cu in CuSe

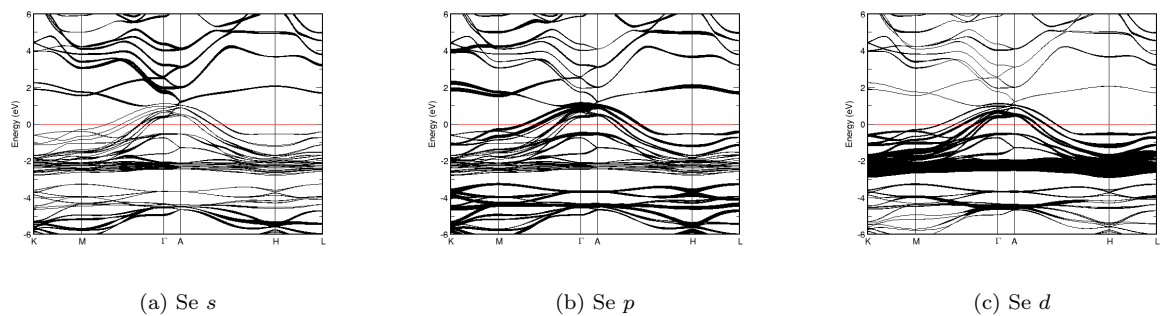

(a) Se *s*           (b) Se *p*          (c) Se *d*

FIG. 31: Fat band representation of Se in CuSe

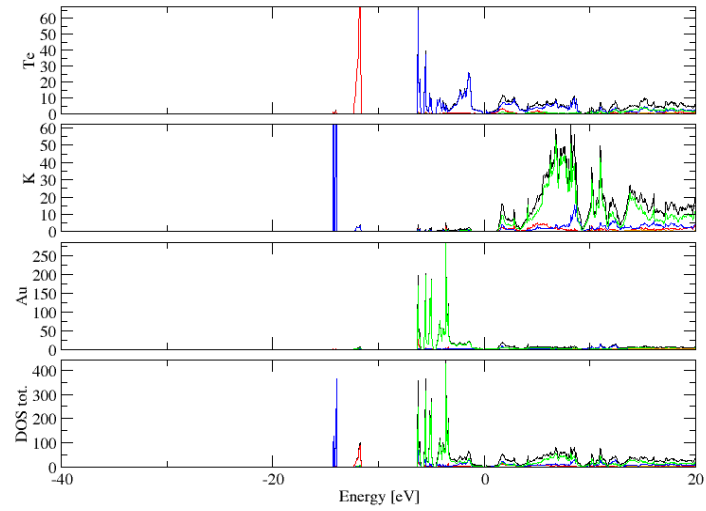

FIG. 32: (Color online) PDOS of KAuTe (ICSD #40165). The *s*-, *p*- and *d*-projected states are in red, blue and green, respectively. KAuTe crystallizes in space group P 63/m m c (#194), in a hexagonal primitive structure.

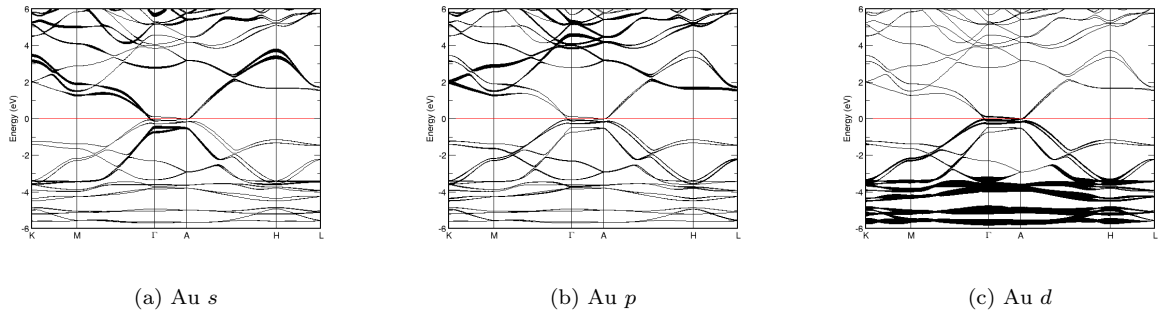

(a) Au *s*                     (b) Au *p*                     (c) Au *d*

FIG. 33: Fat band representation of Au in KAuTe

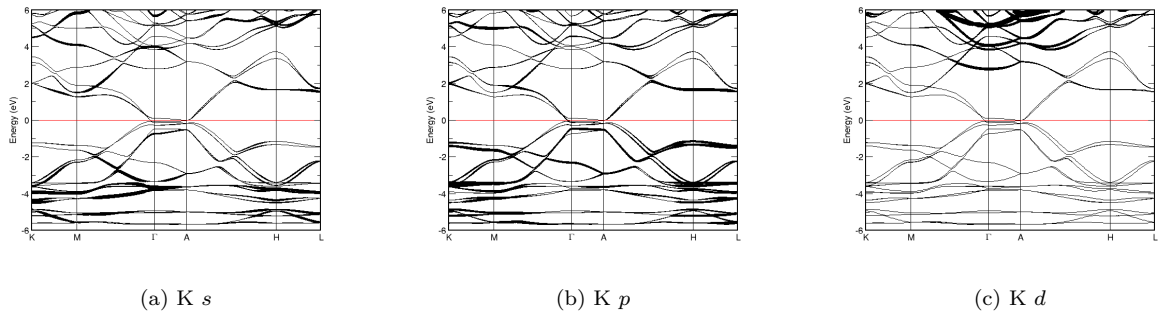

(a) K *s*                      (b) K *p*                      (c) K *d*

FIG. 34: Fat band representation of K in KAuTe

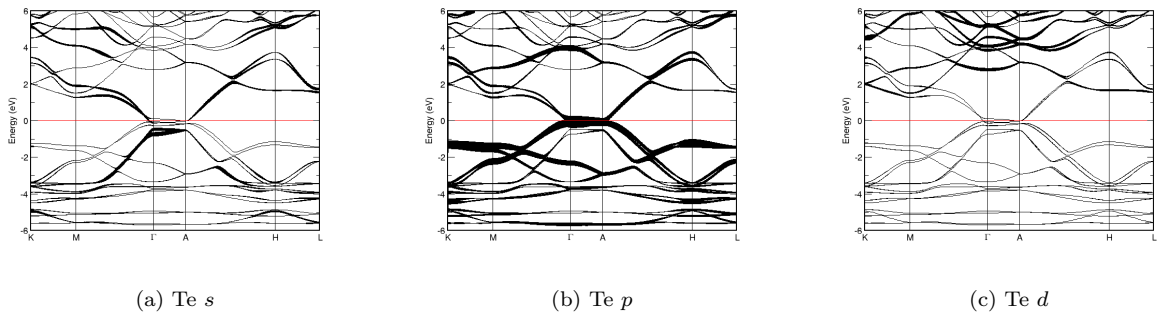

(a) Te *s*       (b) Te *p*       (c) Te *d*

FIG. 35: Fat band representation of Te in KAuTe

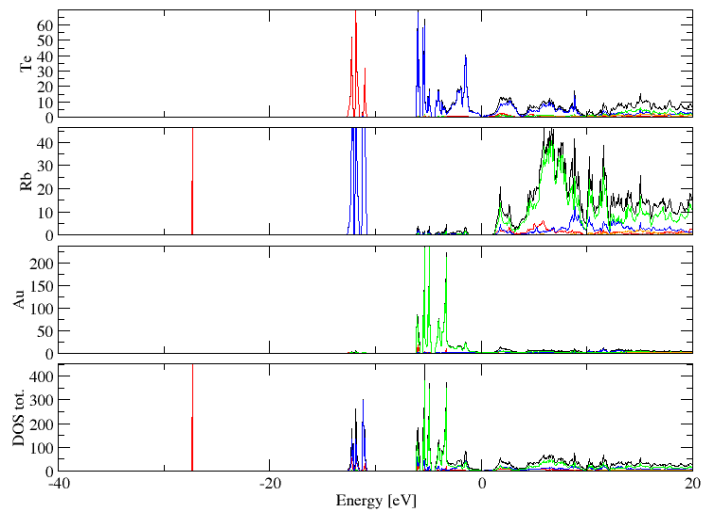

FIG. 36: (Color online) PDOS of RbAuTe (ICSD #75026). The *s*-, *p*- and *d*-projected states are in red, blue and green, respectively. RbAuTe crystallizes in space group P 63/m m c (#194), in a hexagonal primitive structure.

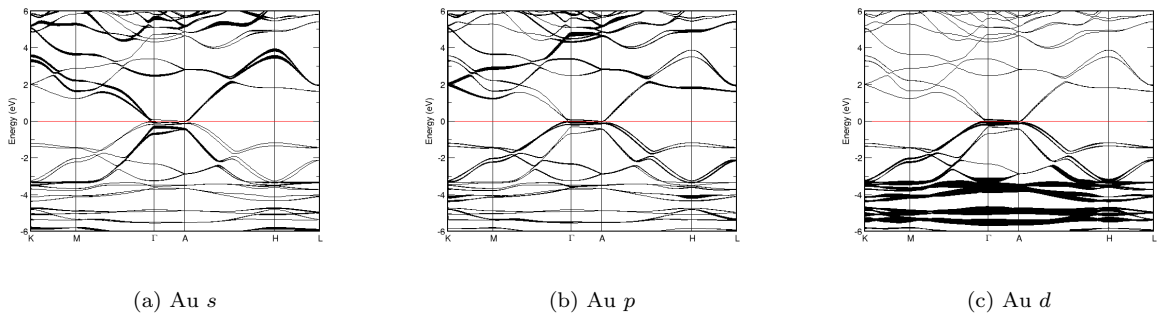

(a) Au *s*       (b) Au *p*       (c) Au *d*

FIG. 37: Fat band representation of Au in RbAuTe

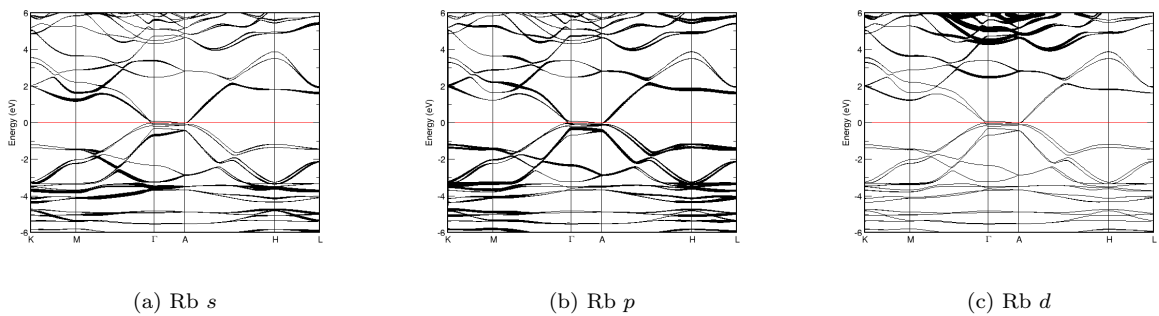

(a) Rb $s$          (b) Rb $p$          (c) Rb $d$

FIG. 38: Fat band representation of Rb in RbAuTe

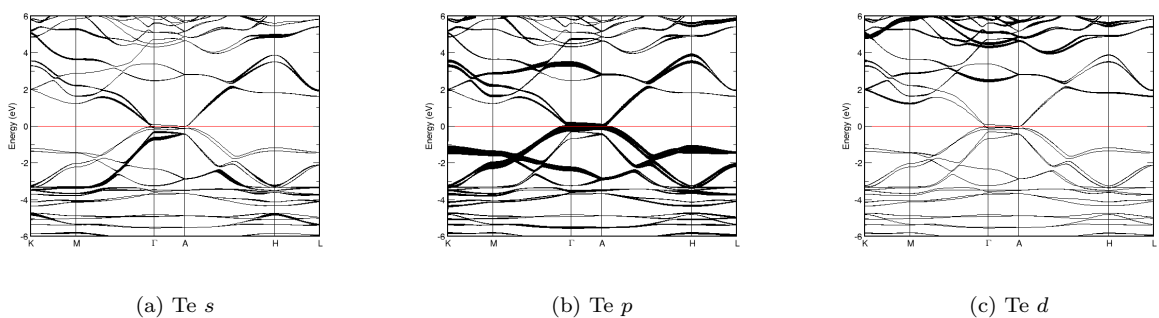

(a) Te $s$          (b) Te $p$          (c) Te $d$

FIG. 39: Fat band representation of Te in RbAuTe

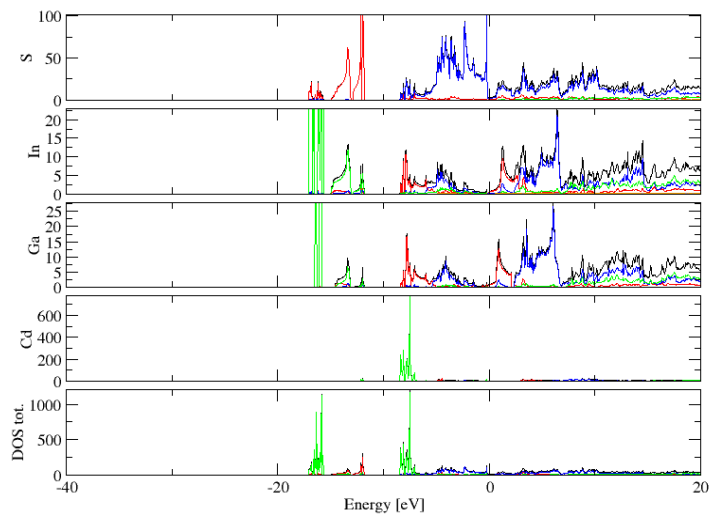

FIG. 40: (Color online) PDOS of CdInGaS$_4$ (ICSD #20785). The $s$-, $p$- and $d$-projected states are in red, blue and green, respectively. CdInGaS$_4$ crystallizes in space group P -3 m 1 (#164), in a hexagonal primitive structure.

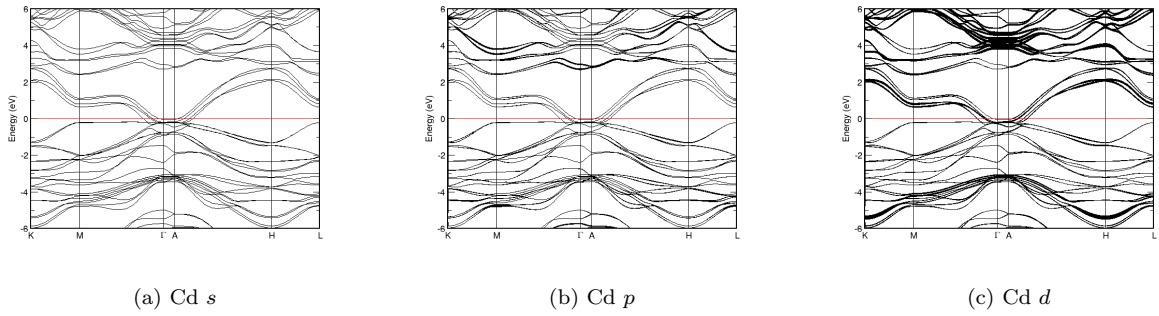

(a) Cd $s$

(b) Cd $p$

(c) Cd $d$

FIG. 41: Fat band representation of Cd in CdInGaS$_4$

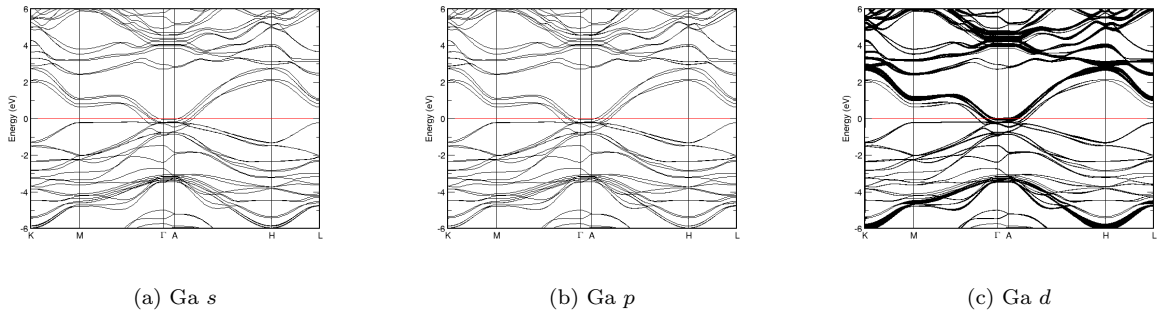

(a) Ga $s$

(b) Ga $p$

(c) Ga $d$

FIG. 42: Fat band representation of Ga in CdInGaS$_4$

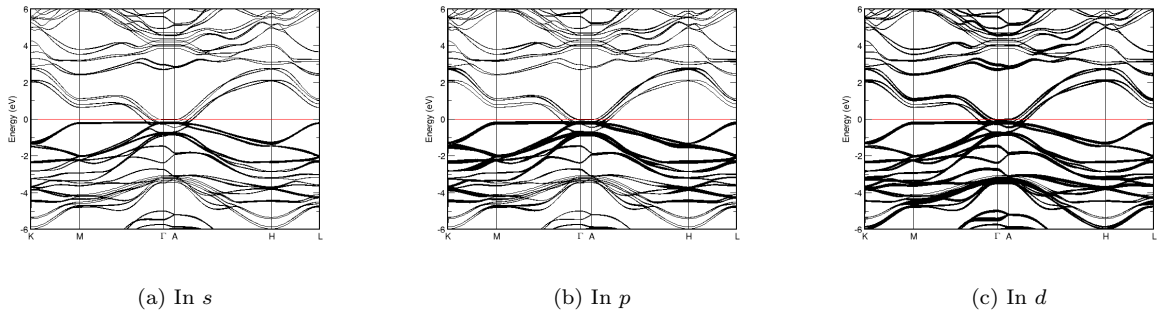

(a) In $s$

(b) In $p$

(c) In $d$

FIG. 43: Fat band representation of In in CdInGaS$_4$

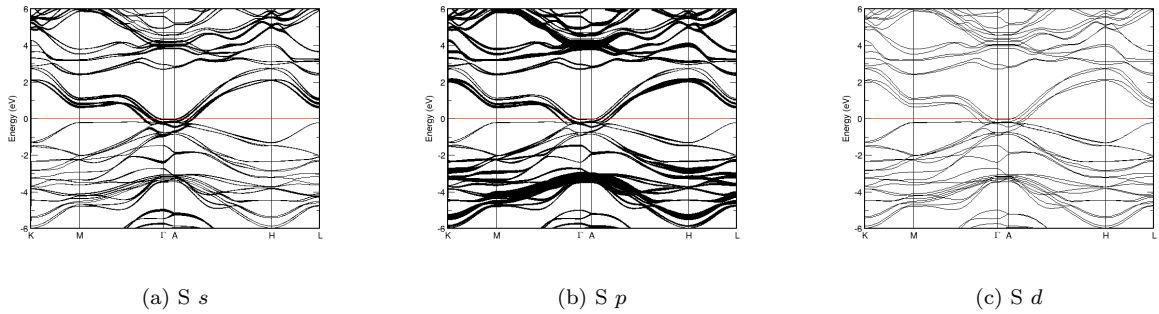

(a) S $s$

(b) S $p$

(c) S $d$

FIG. 44: Fat band representation of S in CdInGaS$_4$

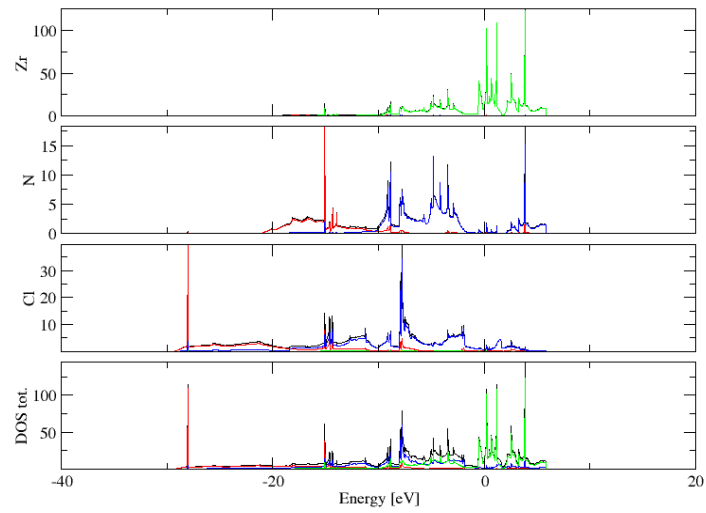

FIG. 45: (Color online) PDOS of ZrNCl (ICSD #25506). The *s*-, *p*- and *d*-projected states are in red, blue and green, respectively. ZrNCl crystallizes in space group P -3 m 1 (#164), in a hexagonal primitive structure.

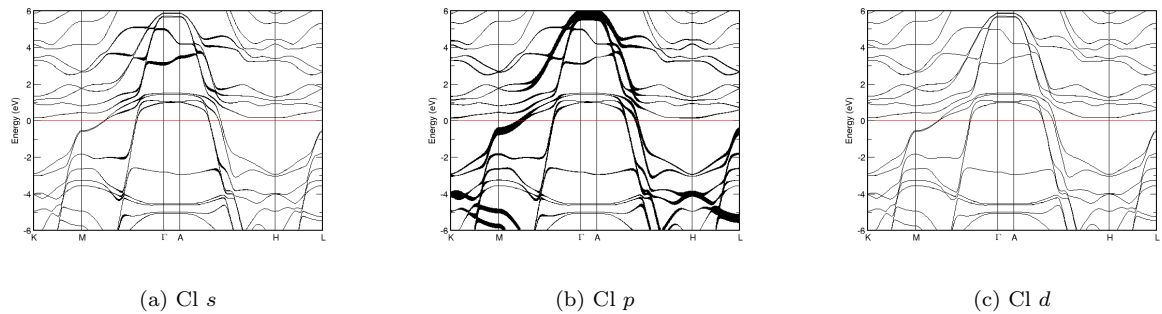

(a) Cl *s*         (b) Cl *p*        (c) Cl *d*

FIG. 46: Fat band representation of Cl in ZrNCl

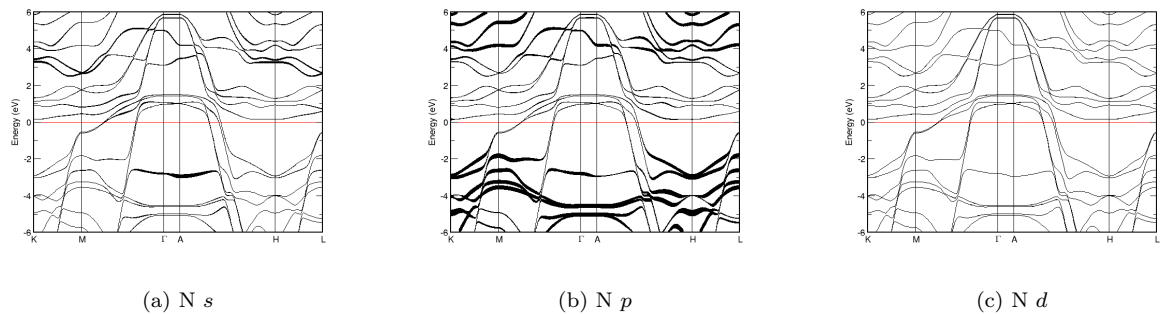

(a) N *s*        (b) N *p*        (c) N *d*

FIG. 47: Fat band representation of N in ZrNCl

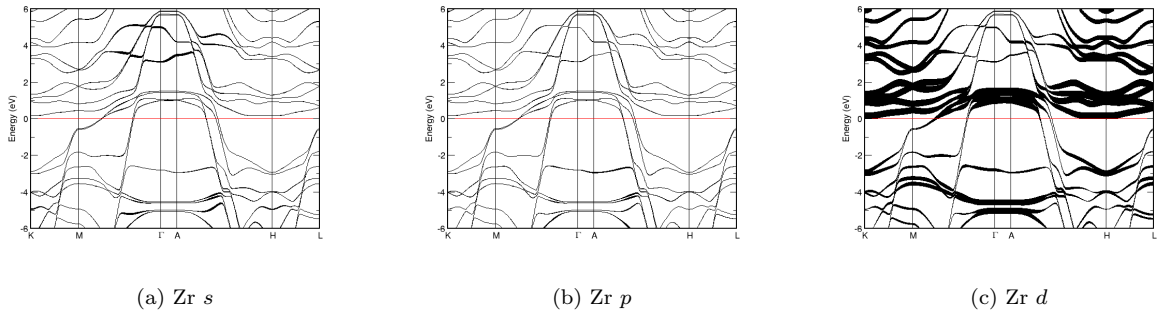

(a) Zr $s$                (b) Zr $p$                (c) Zr $d$

FIG. 48: Fat band representation of Zr in ZrNCl

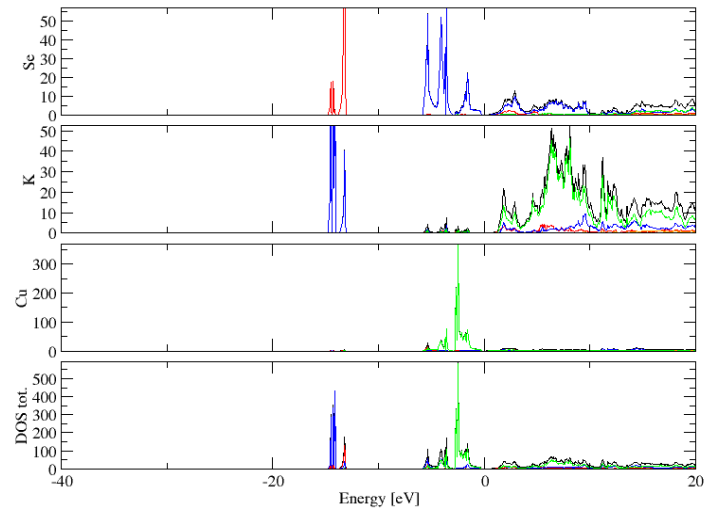

FIG. 49: (Color online) PDOS of KCuSe (ICSD #12157). The $s$-, $p$- and $d$-projected states are in red, blue and green, respectively. KCuSe crystallizes in space group P 63/m m c (#194), in a hexagonal primitive structure.

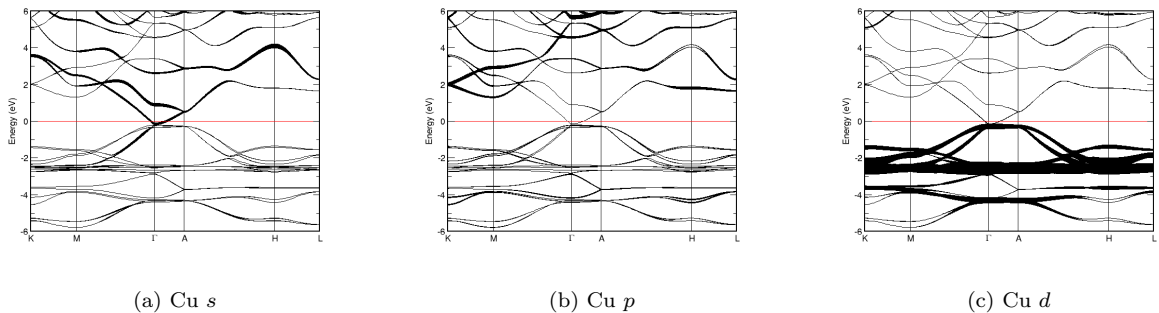

(a) Cu $s$                (b) Cu $p$                (c) Cu $d$

FIG. 50: Fat band representation of Cu in KCuSe

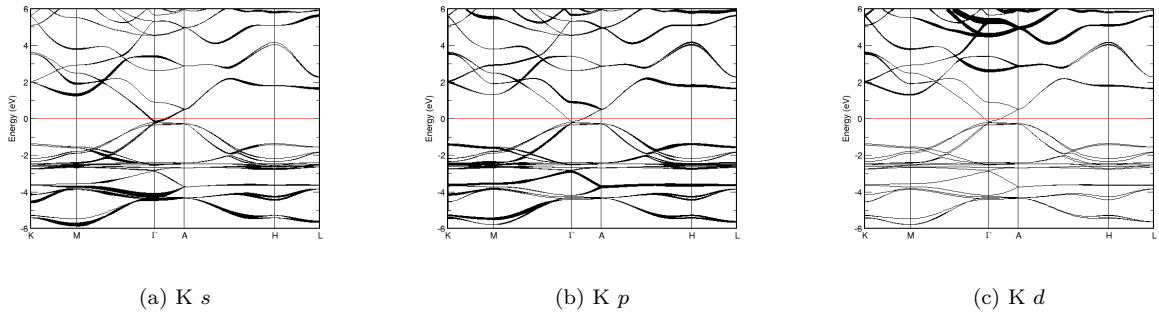

(a) K *s*  (b) K *p*  (c) K *d*

FIG. 51: Fat band representation of K in KCuSe

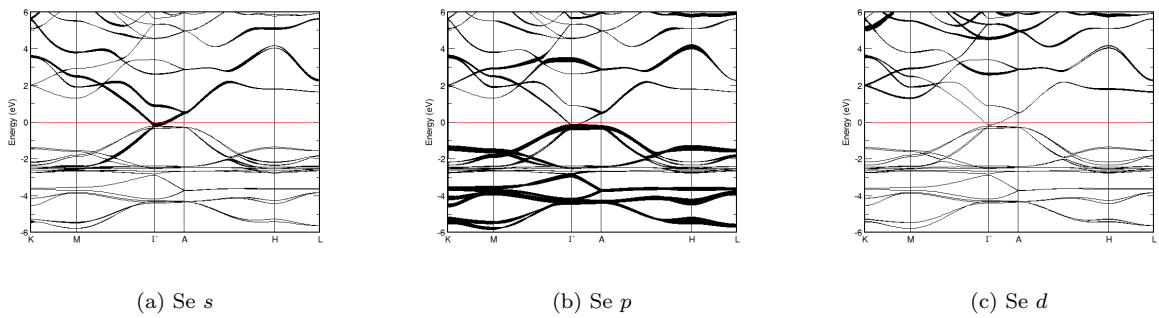

(a) Se *s*  (b) Se *p*  (c) Se *d*

FIG. 52: Fat band representation of Se in KCuSe

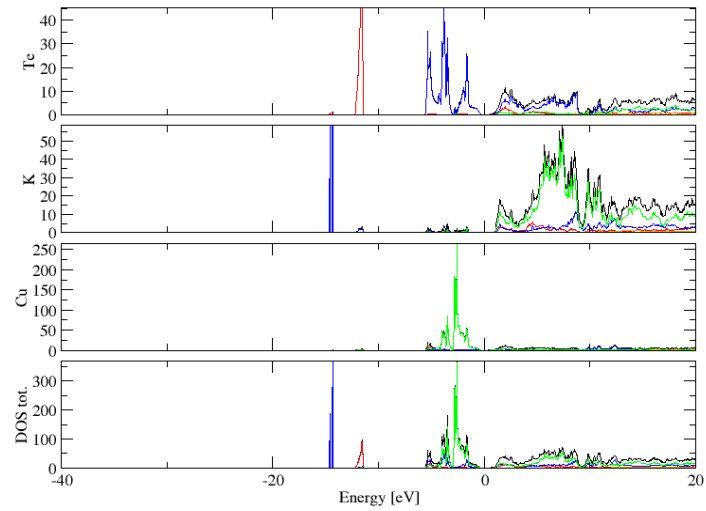

FIG. 53: (Color online) PDOS of KCuTe (ICSD #12158). The *s*-, *p*- and *d*-projected states are in red, blue and green, respectively. KCuTe crystallizes in space group P 63/m m c (#194), in a hexagonal primitive structure.

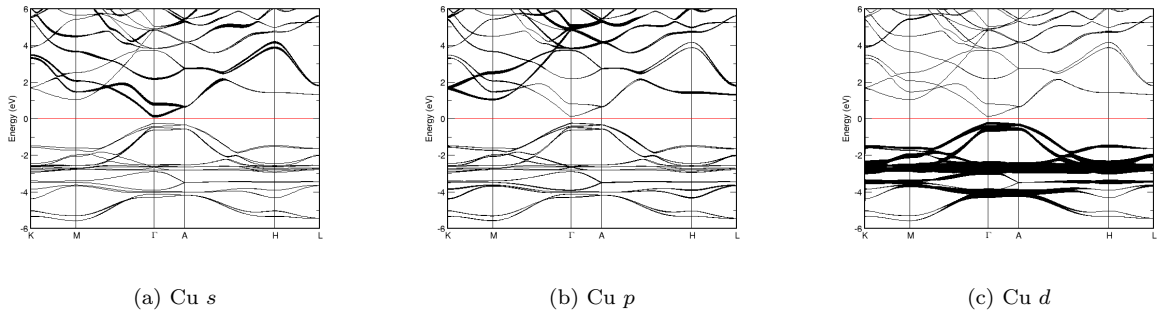

(a) Cu $s$      (b) Cu $p$      (c) Cu $d$

FIG. 54: Fat band representation of Cu in KCuTe

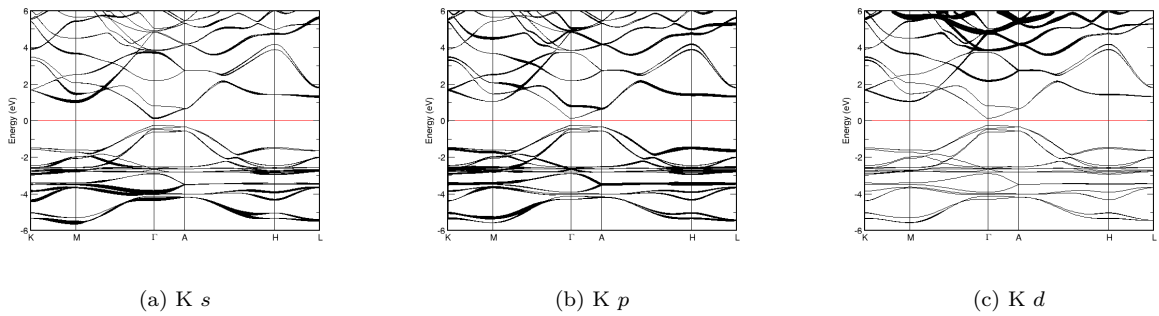

(a) K $s$      (b) K $p$      (c) K $d$

FIG. 55: Fat band representation of K in KCuTe

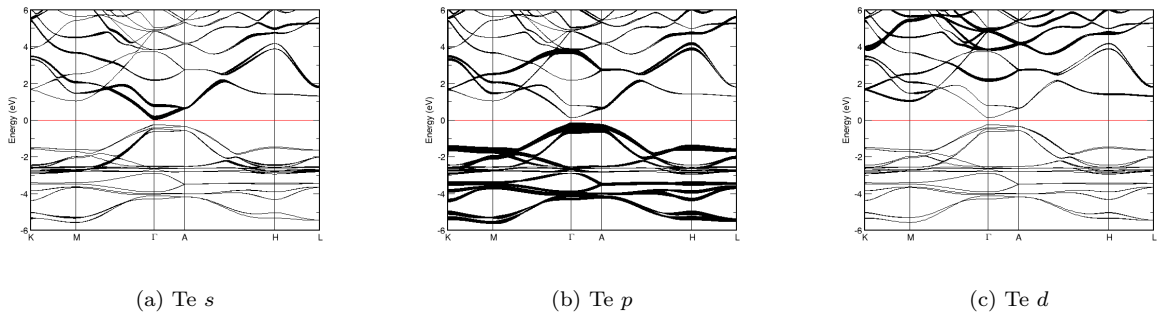

(a) Te $s$      (b) Te $p$      (c) Te $d$

FIG. 56: Fat band representation of Te in KCuTe

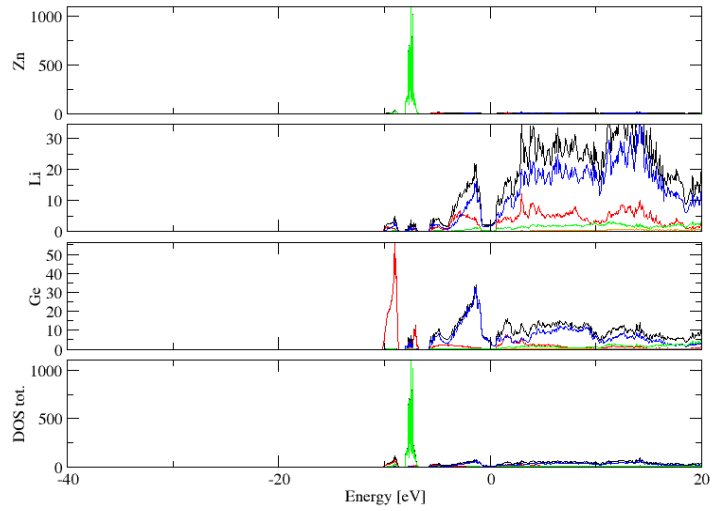

FIG. 57: (Color online) PDOS of Li$_2$ZnGe (ICSD #53678). The $s$-, $p$- and $d$-projected states are in red, blue and green, respectively. Li$_2$ZnGe crystallizes in space group P -3 m 1 (#164), in a hexagonal primitive structure.

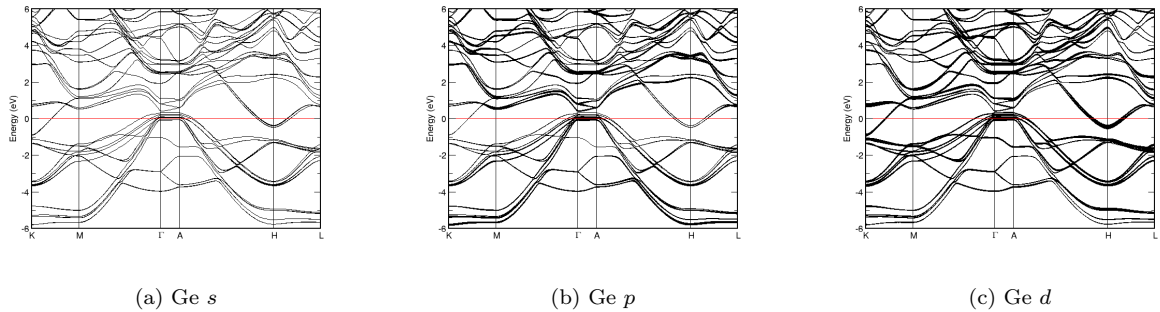

(a) Ge $s$        (b) Ge $p$        (c) Ge $d$

FIG. 58: Fat band representation of Ge in Li$_2$ZnGe

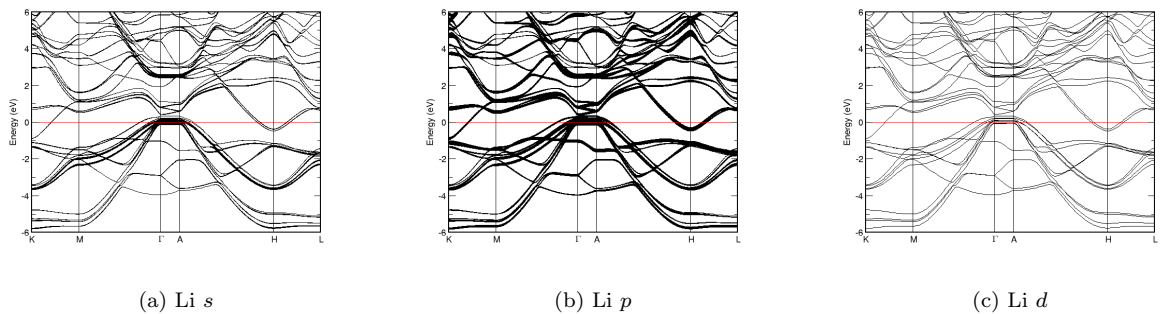

(a) Li $s$        (b) Li $p$        (c) Li $d$

FIG. 59: Fat band representation of Li in Li$_2$ZnGe

(a) Zn $s$

(b) Zn $p$

(c) Zn $d$

FIG. 60: Fat band representation of Zn in $Li_2ZnGe$

FIG. 61: (Color online) PDOS of $Li_2ZnSi$ (ICSD #16221). The $s$-, $p$- and $d$-projected states are in red, blue and green, respectively. $Li_2ZnSi$ crystallizes in space group P -3 m 1 (#164), in a hexagonal primitive structure.

(a) Li $s$

(b) Li $p$

(c) Li $d$

FIG. 62: Fat band representation of Li in $Li_2ZnSi$

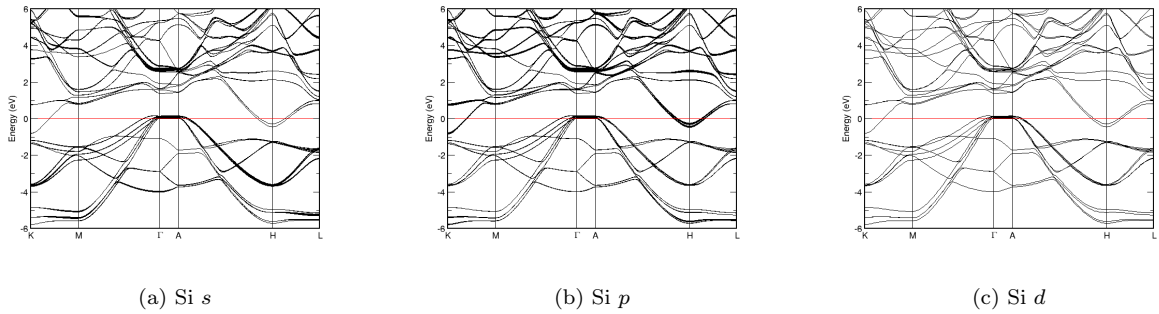

(a) Si $s$        (b) Si $p$        (c) Si $d$

FIG. 63: Fat band representation of Si in Li$_2$ZnSi

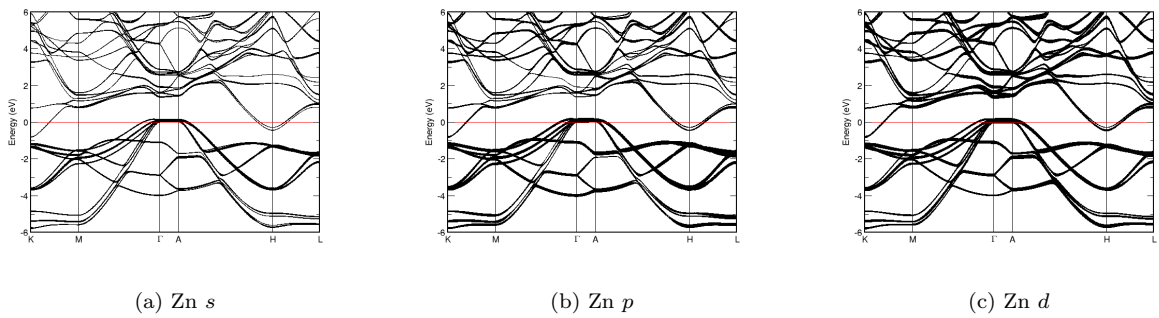

(a) Zn $s$        (b) Zn $p$        (c) Zn $d$

FIG. 64: Fat band representation of Zn in Li$_2$ZnSi

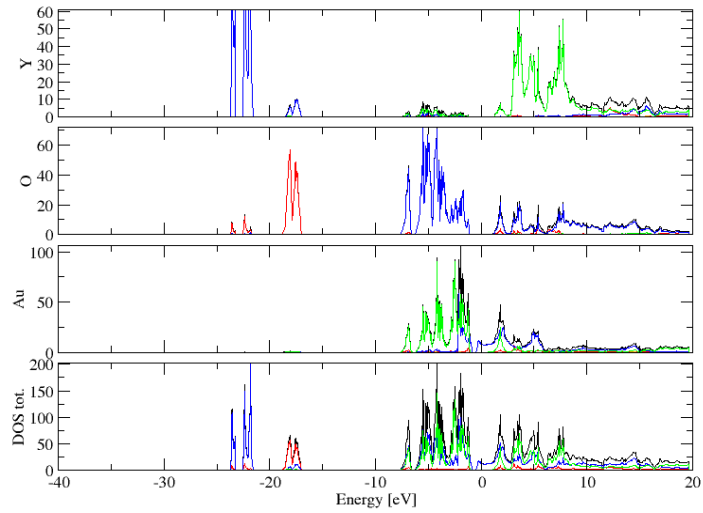

FIG. 65: (Color online) PDOS of AuYO$_2$ (ICSD #95675). The $s$-, $p$- and $d$-projected states are in red, blue and green, respectively. AuYO$_2$ crystallizes in space group P 63/m m c (#194), in a hexagonal primitive structure.

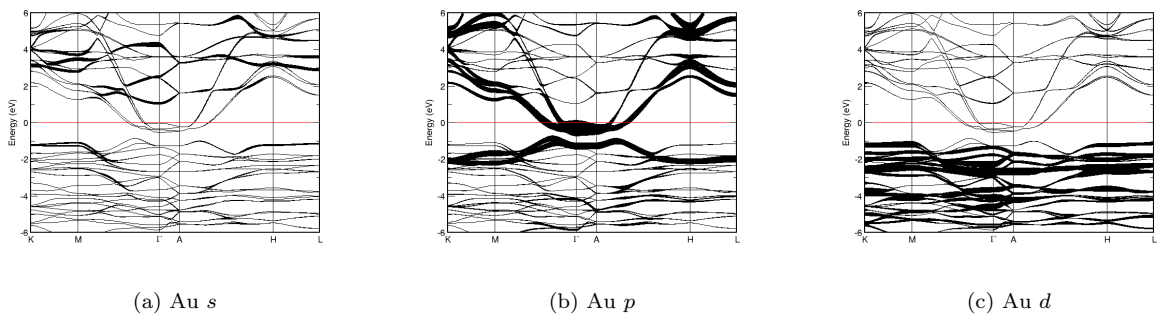

(a) Au $s$        (b) Au $p$        (c) Au $d$

FIG. 66: Fat band representation of Au in AuYO$_2$

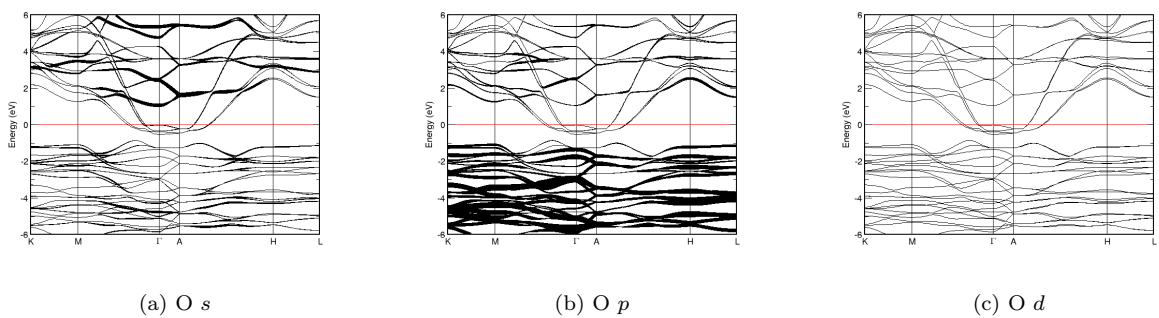

(a) O $s$        (b) O $p$        (c) O $d$

FIG. 67: Fat band representation of O in AuYO$_2$

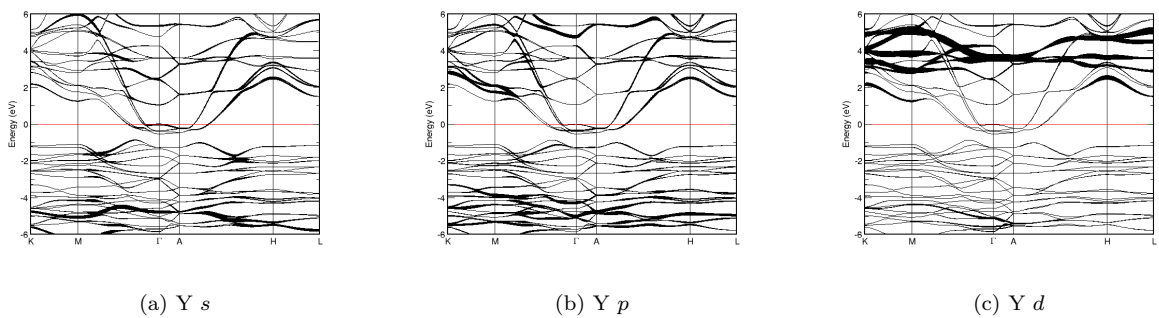

(a) Y $s$        (b) Y $p$        (c) Y $d$

FIG. 68: Fat band representation of Y in AuYO$_2$

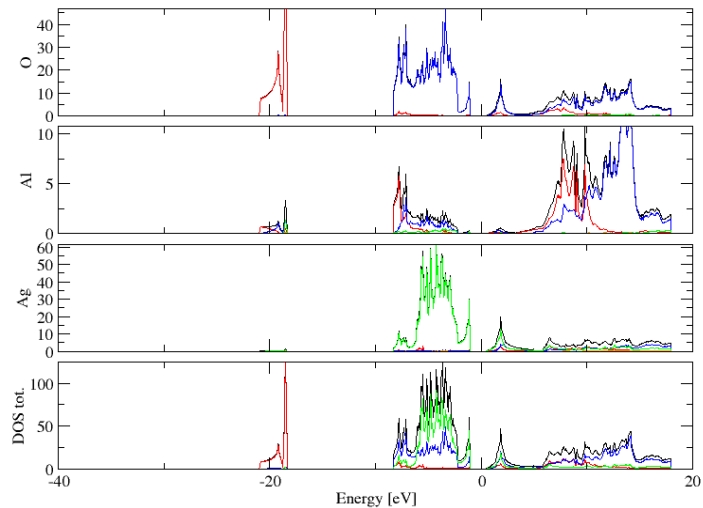

FIG. 69: (Color online) PDOS of AgAlO$_2$ (ICSD #300020). The $s$-, $p$- and $d$-projected states are in red, blue and green, respectively. AgAlO$_2$ crystallizes in space group P 63/m m c (#194), in a hexagonal primitive structure. We also note that several other members of the delafossite structure family qualify as possible high-temperature superconductors, after appropriate electron/hole doping, because of layered structure, $p - d$-hybridization and cosine-like dispersion around $E_F$.

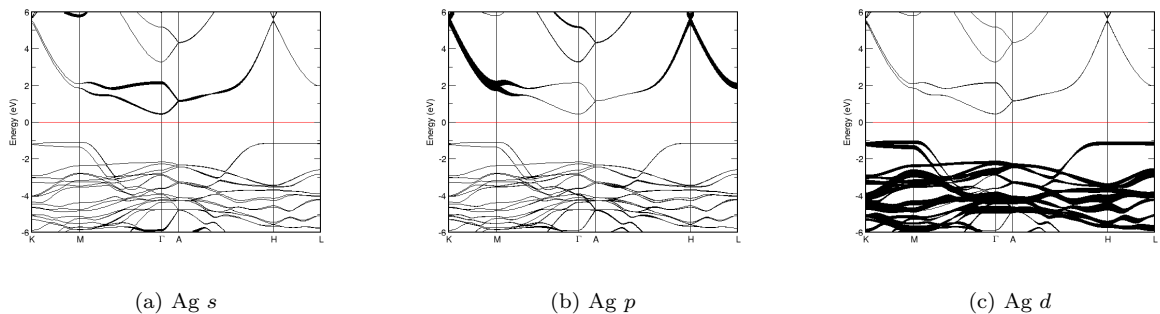

(a) Ag $s$        (b) Ag $p$        (c) Ag $d$

FIG. 70: Fat band representation of Ag in AgAlO$_2$

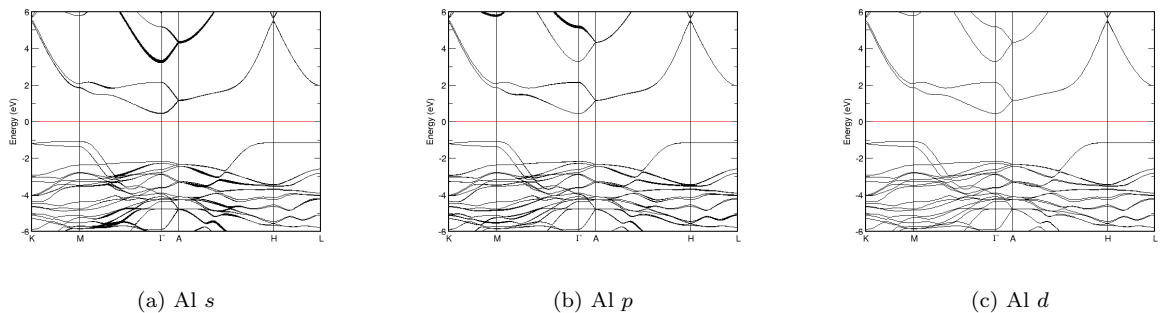

(a) Al $s$        (b) Al $p$        (c) Al $d$

FIG. 71: Fat band representation of Al in AgAlO$_2$

(a) O $s$

(b) O $p$

(c) O $d$

FIG. 72: Fat band representation of O in AgAlO$_2$

FIG. 73: (Color online) PDOS of CuBr (ICSD #30092). The $s$-, $p$- and $d$-projected states are in red, blue and green, respectively. CuBr crystallizes in space group P 63 m c (#186), in a hexagonal primitive structure.

(a) Br $s$

(b) Br $p$

(c) Br $d$

FIG. 74: Fat band representation of Br in CuBr

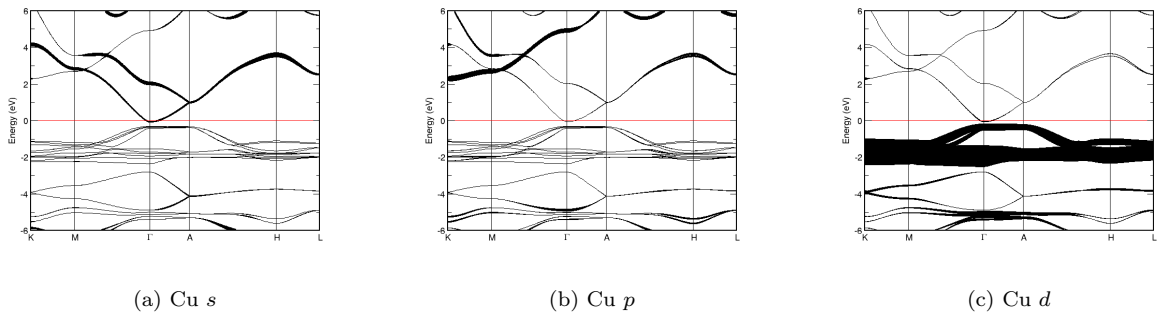

(a) Cu *s*  (b) Cu *p*  (c) Cu *d*

FIG. 75: Fat band representation of Cu in CuBr

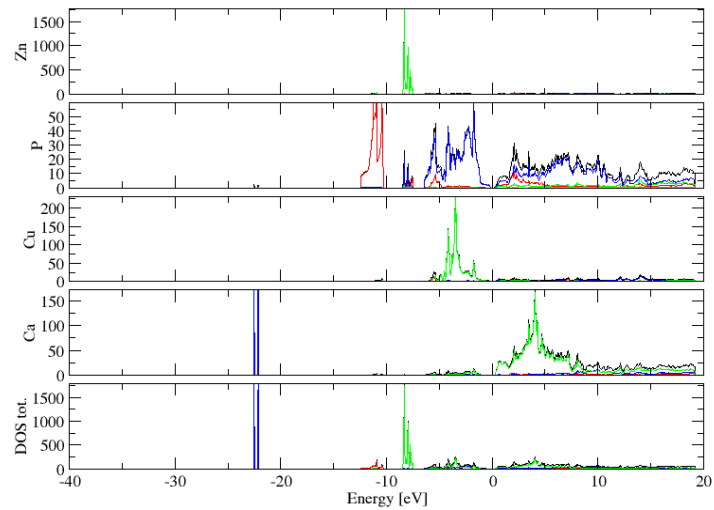

FIG. 76: (Color online) PDOS of $Ca_2CuZn_2P_3$ (ICSD #89517). The *s*-, *p*- and *d*-projected states are in red, blue and green, respectively. $Ca_2CuZn_2P_3$ crystallizes in space group P 63/m m c (#194), in a hexagonal primitive structure.

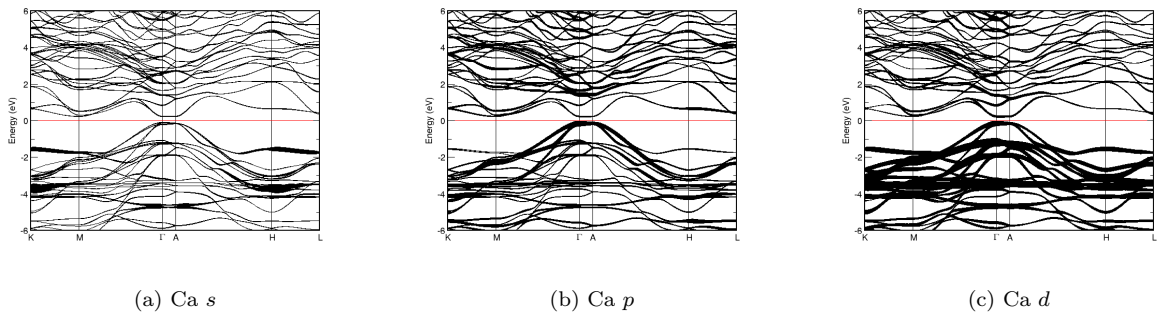

(a) Ca *s*  (b) Ca *p*  (c) Ca *d*

FIG. 77: Fat band representation of Ca in $Ca_2CuZn_2P_3$

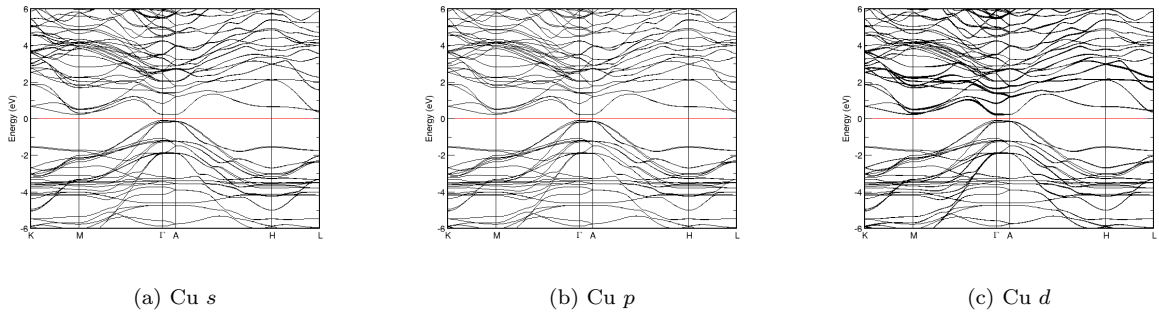

(a) Cu $s$

(b) Cu $p$

(c) Cu $d$

FIG. 78: Fat band representation of Cu in $Ca_2CuZn_2P_3$

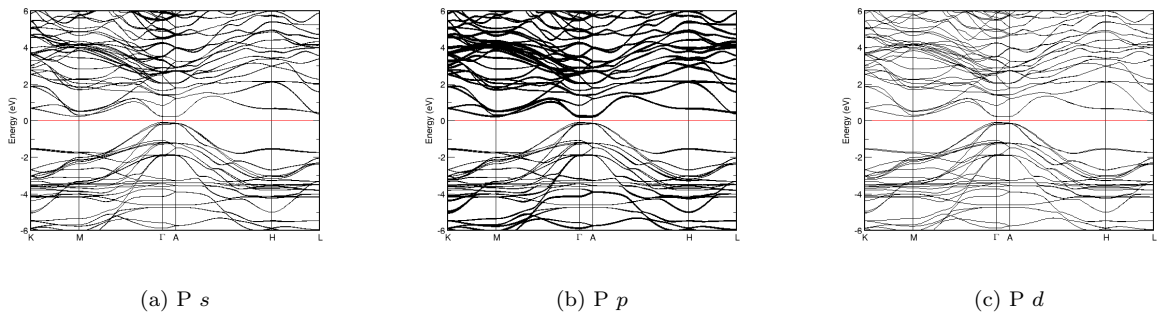

(a) P $s$

(b) P $p$

(c) P $d$

FIG. 79: Fat band representation of P in $Ca_2CuZn_2P_3$

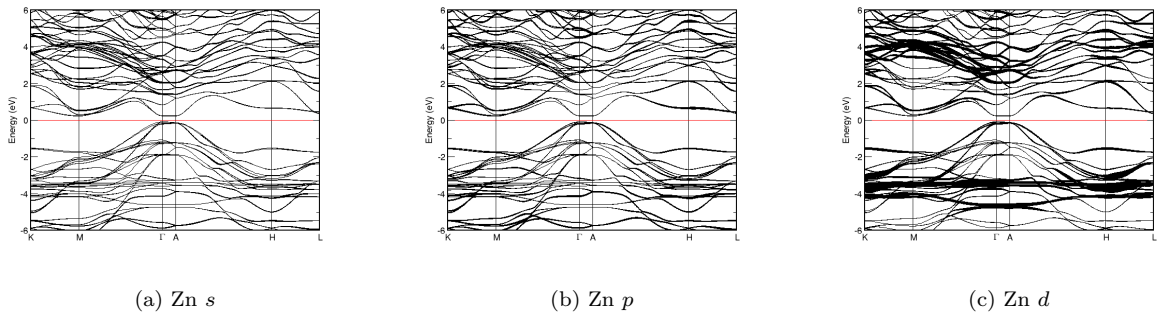

(a) Zn $s$

(b) Zn $p$

(c) Zn $d$

FIG. 80: Fat band representation of Zn in $Ca_2CuZn_2P_3$

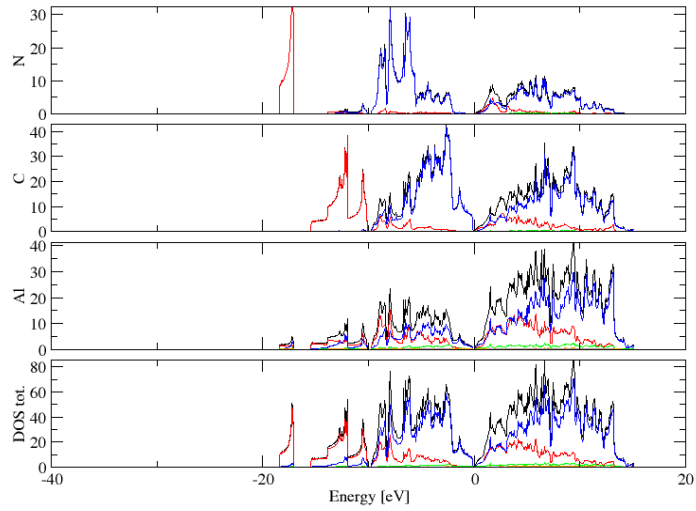

FIG. 81: (Color online) PDOS of Al$_5$C$_3$N (ICSD #26859). The $s$-, $p$- and $d$-projected states are in red, blue and green, respectively. Al$_5$C$_3$N crystallizes in space group P 63 m c (#186), in a hexagonal primitive structure.

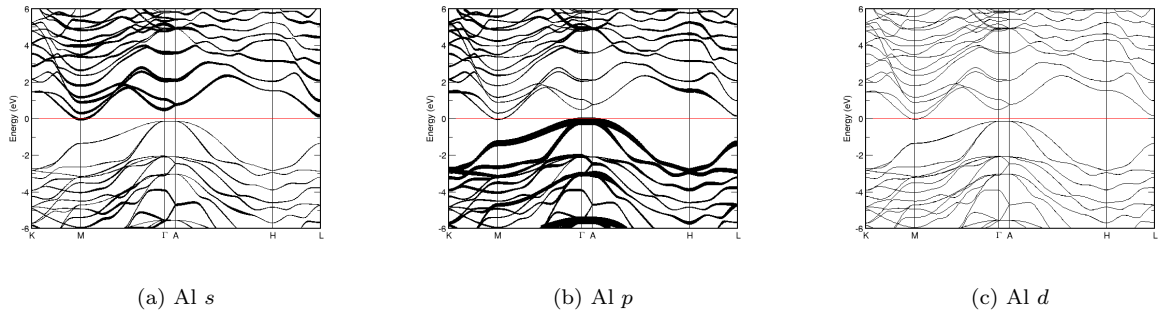

(a) Al $s$                    (b) Al $p$                    (c) Al $d$

FIG. 82: Fat band representation of Al in Al$_5$C$_3$N

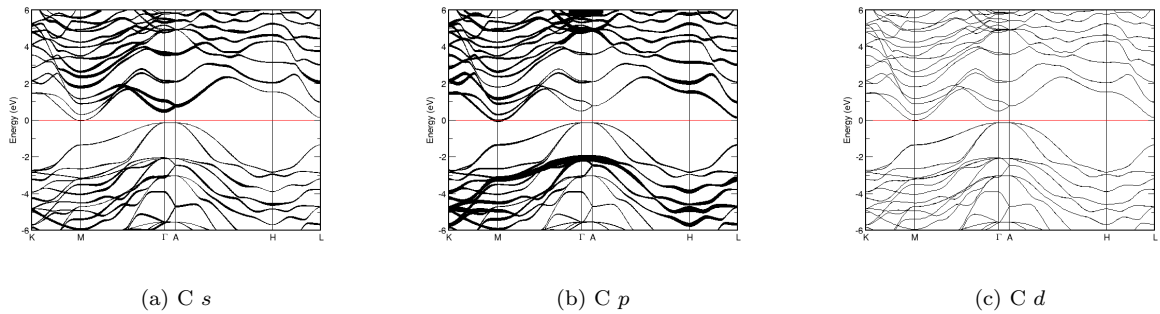

(a) C $s$                    (b) C $p$                    (c) C $d$

FIG. 83: Fat band representation of C in Al$_5$C$_3$N

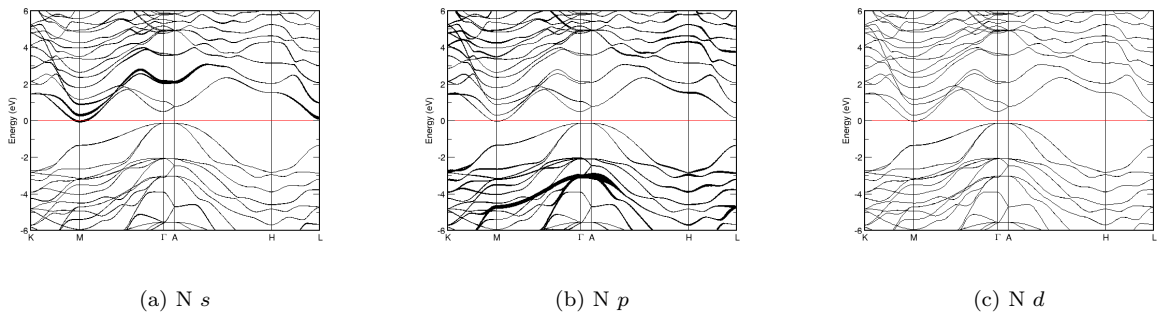

(a) N $s$  (b) N $p$  (c) N $d$

FIG. 84: Fat band representation of N in $Al_5C_3N$

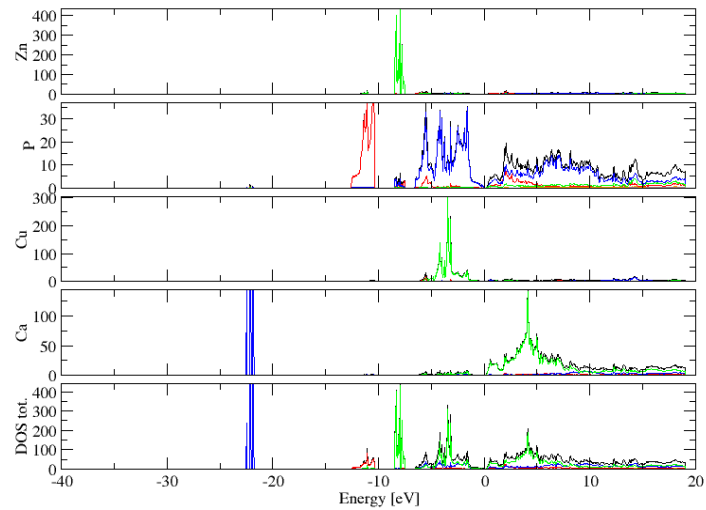

FIG. 85: (Color online) PDOS of $Ca_3Cu_2Zn_2P_4$ (ICSD #89515). The $s$-, $p$- and $d$-projected states are in red, blue and green, respectively. $Ca_3Cu_2Zn_2P_4$ crystallizes in space group P -3 m 1 (#164), in a hexagonal primitive structure.

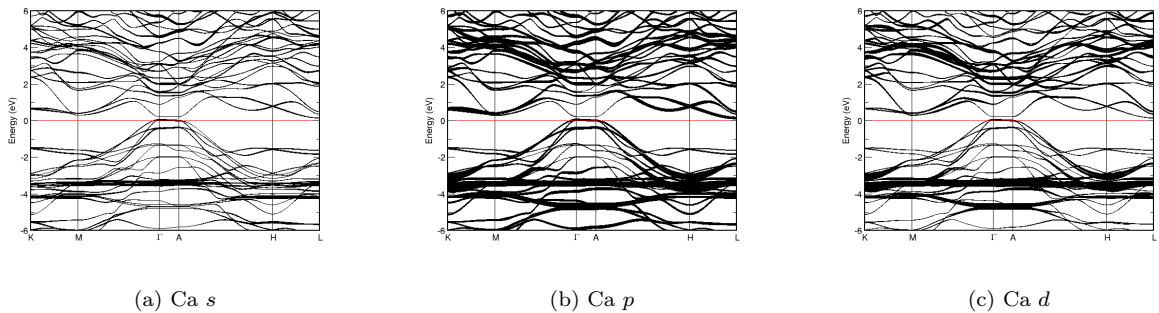

(a) Ca $s$  (b) Ca $p$  (c) Ca $d$

FIG. 86: Fat band representation of Ca in $Ca_3Cu_2Zn_2P_4$

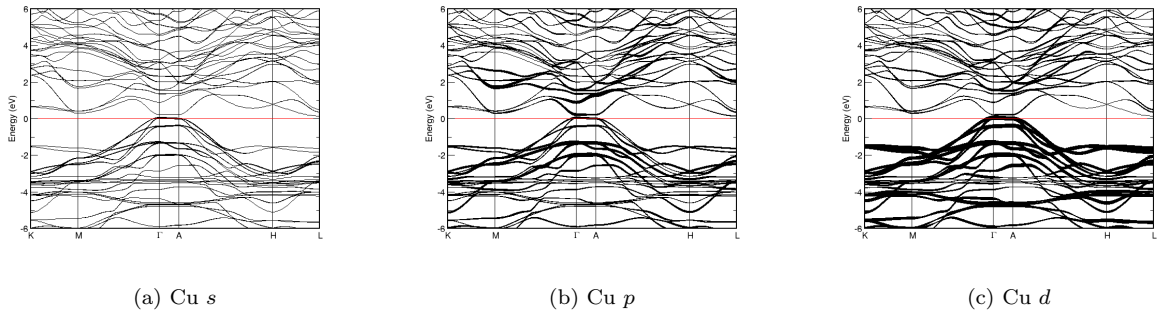

(a) Cu $s$　　　　(b) Cu $p$　　　　(c) Cu $d$

FIG. 87: Fat band representation of Cu in Ca$_3$Cu$_2$Zn$_2$P$_4$

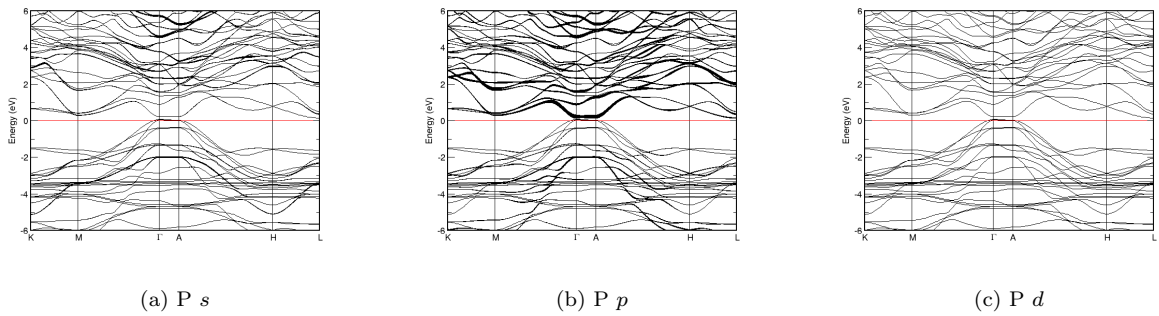

(a) P $s$　　　　(b) P $p$　　　　(c) P $d$

FIG. 88: Fat band representation of P in Ca$_3$Cu$_2$Zn$_2$P$_4$

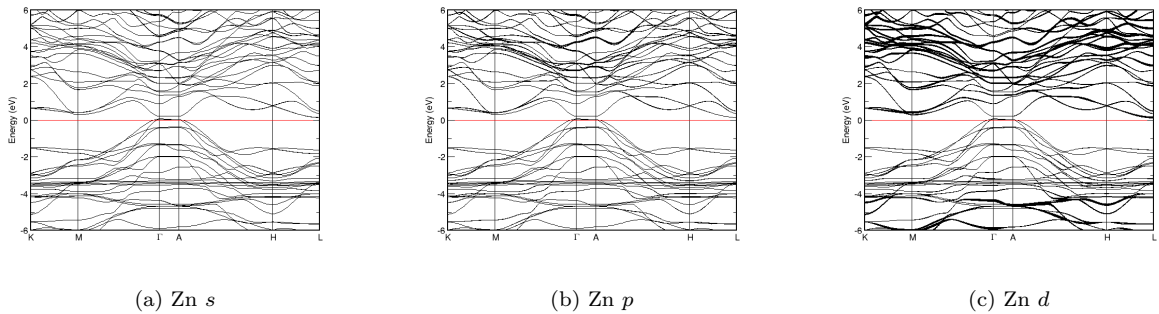

(a) Zn $s$　　　　(b) Zn $p$　　　　(c) Zn $d$

FIG. 89: Fat band representation of Zn in Ca$_3$Cu$_2$Zn$_2$P$_4$

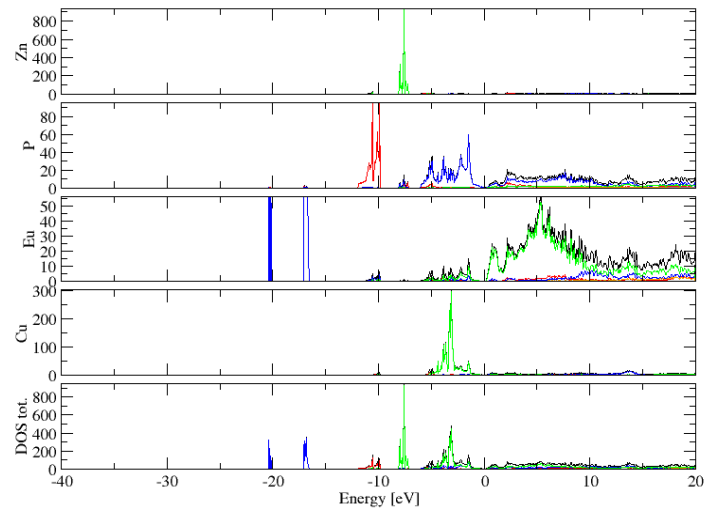

FIG. 90: (Color online) PDOS of Eu$_3$Cu$_2$Zn$_2$P$_4$ (ICSD #89516). The $s$-, $p$- and $d$-projected states are in red, blue and green, respectively. Eu$_3$Cu$_2$Zn$_2$P$_4$ crystallizes in space group P -3 m 1 (#164), in a hexagonal primitive structure.

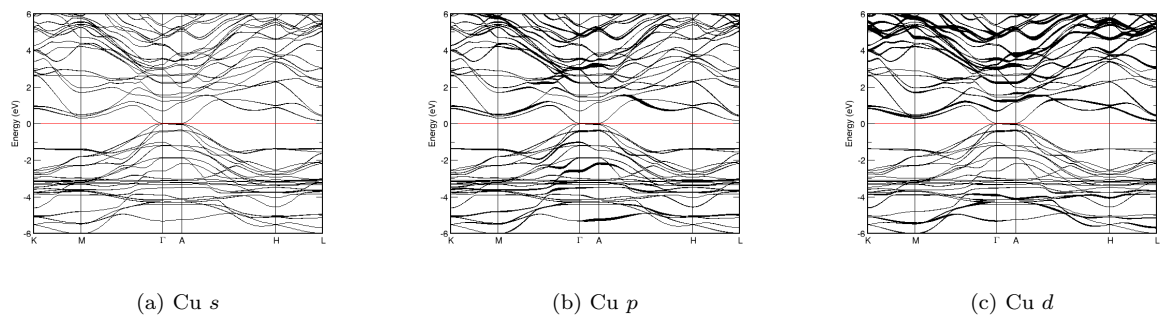

(a) Cu $s$       (b) Cu $p$       (c) Cu $d$

FIG. 91: Fat band representation of Cu in Eu$_3$Cu$_2$Zn$_2$P$_4$

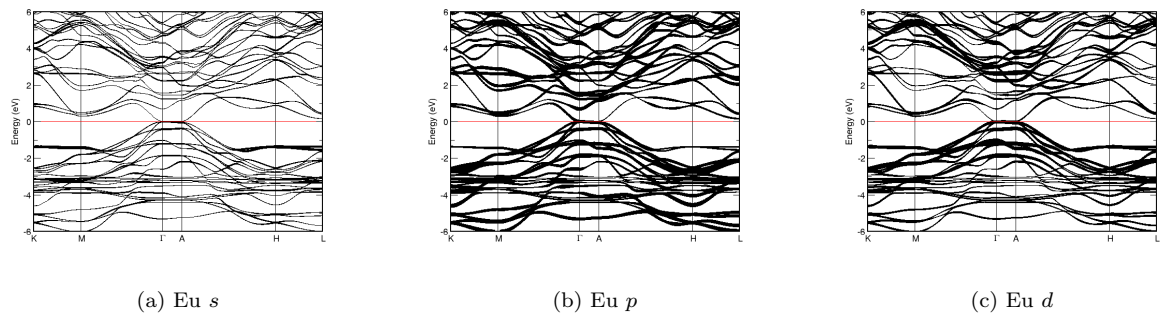

(a) Eu $s$       (b) Eu $p$       (c) Eu $d$

FIG. 92: Fat band representation of Eu in Eu$_3$Cu$_2$Zn$_2$P$_4$

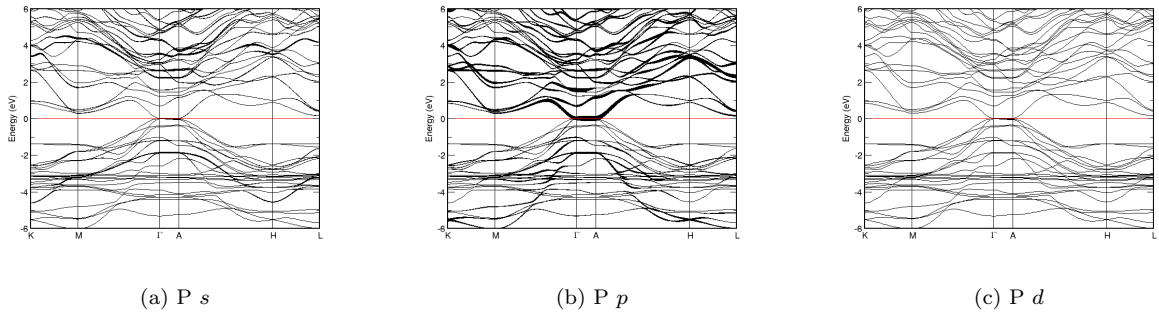

(a) P $s$

(b) P $p$

(c) P $d$

FIG. 93: Fat band representation of P in $Eu_3Cu_2Zn_2P_4$

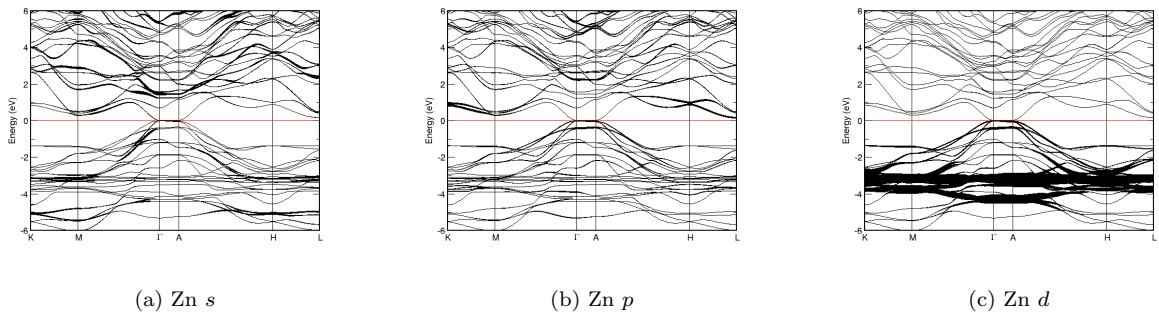

(a) Zn $s$

(b) Zn $p$

(c) Zn $d$

FIG. 94: Fat band representation of Zn in $Eu_3Cu_2Zn_2P_4$

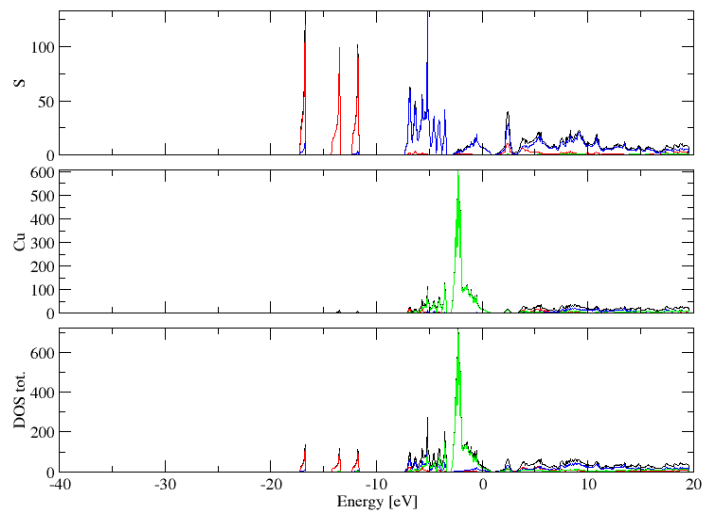

FIG. 95: (Color online) PDOS of $Cu_4(S_2)_2(CuS)_2$ (ICSD #26968). The $s$-, $p$- and $d$-projected states are in red, blue and green, respectively. $Cu_4(S_2)_2(CuS)_2$ crystallizes in space group P 63/m m c (#194), in a hexagonal primitive structure.

(a) Cu $s$  (b) Cu $p$  (c) Cu $d$

FIG. 96: Fat band representation of Cu in $Cu_4(S_2)_2(CuS)_2$

(a) S $s$  (b) S $p$  (c) S $d$

FIG. 97: Fat band representation of S in $Cu_4(S_2)_2(CuS)_2$

FIG. 98: (Color online) PDOS of $LaKPdO_3$ (ICSD #417108). The $s$-, $p$- and $d$-projected states are in red, blue and green, respectively. $LaKPdO_3$ crystallizes in space group C 1 2/m 1 (#12), in a monoclinic base-centred structure.

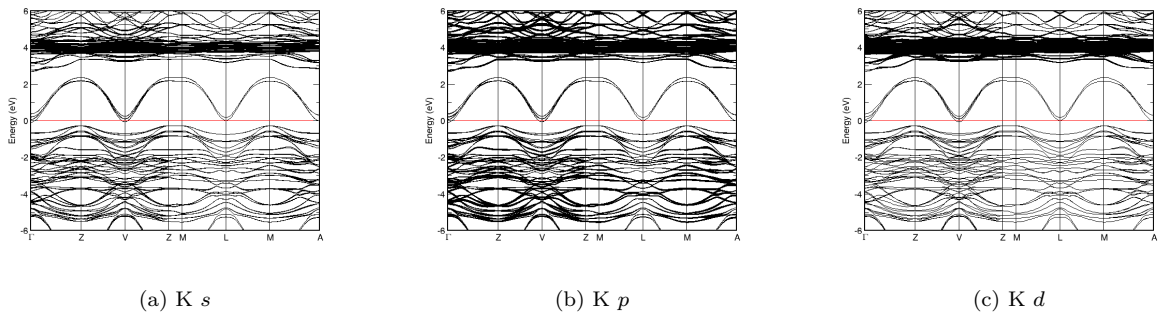

(a) K $s$  (b) K $p$  (c) K $d$

FIG. 99: Fat band representation of K in LaKPdO$_3$

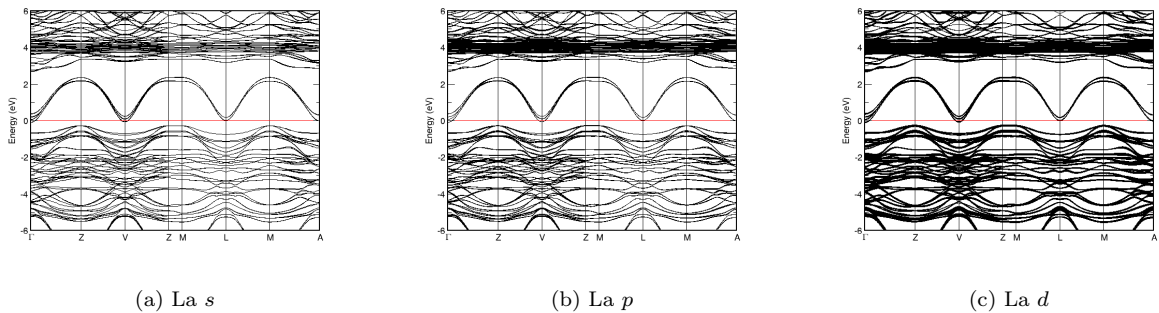

(a) La $s$  (b) La $p$  (c) La $d$

FIG. 100: Fat band representation of La in LaKPdO$_3$

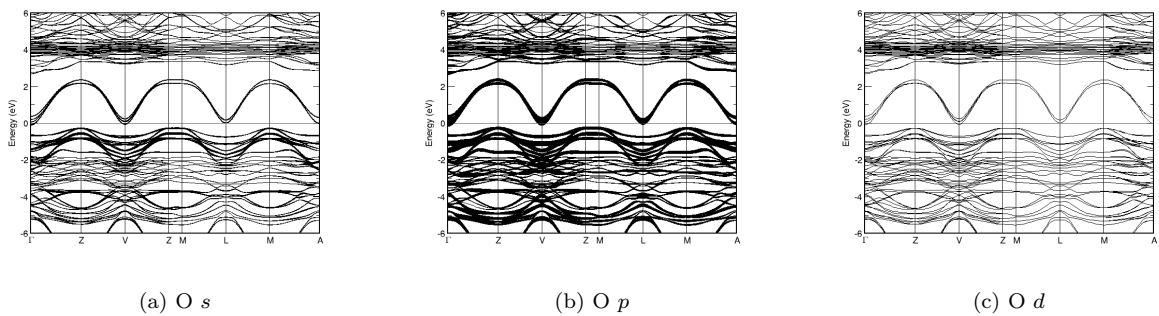

(a) O $s$  (b) O $p$  (c) O $d$

FIG. 101: Fat band representation of O in LaKPdO$_3$

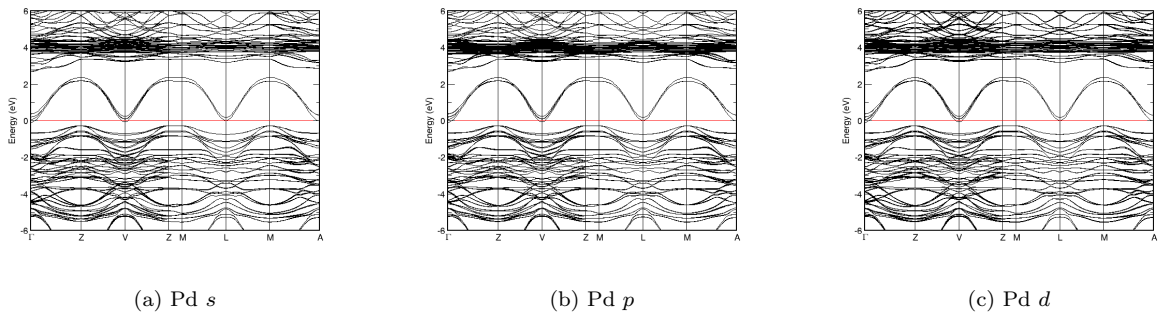

(a) Pd $s$  (b) Pd $p$  (c) Pd $d$

FIG. 102: Fat band representation of Pd in LaKPdO$_3$

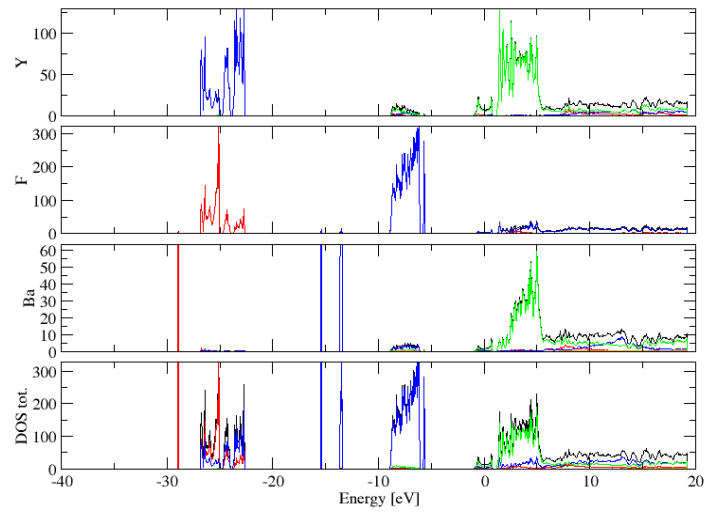

FIG. 103: (Color online) PDOS of BaY$_2$F$_8$ (ICSD #74359). The $s$-, $p$- and $d$-projected states are in red, blue and green, respectively. BaY$_2$F$_8$ crystallizes in space group C 1 2/m 1 (#12), in a monoclinic base-centred structure.

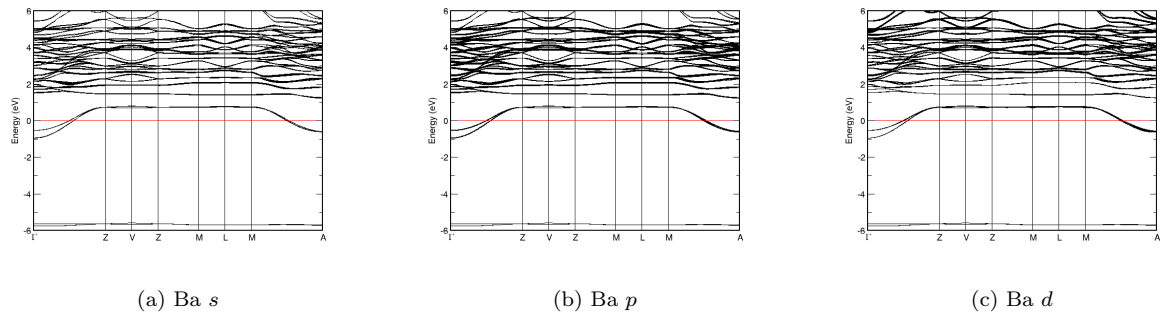

(a) Ba $s$                (b) Ba $p$                (c) Ba $d$

FIG. 104: Fat band representation of Ba in BaY$_2$F$_8$

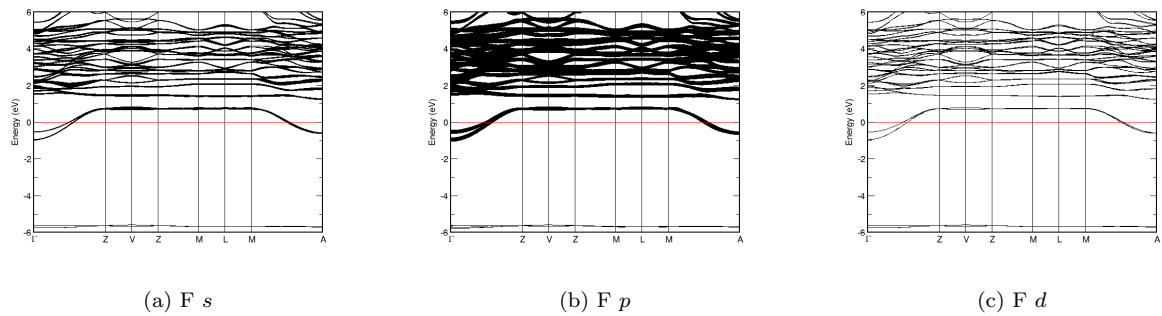

(a) F $s$                (b) F $p$                (c) F $d$

FIG. 105: Fat band representation of F in BaY$_2$F$_8$

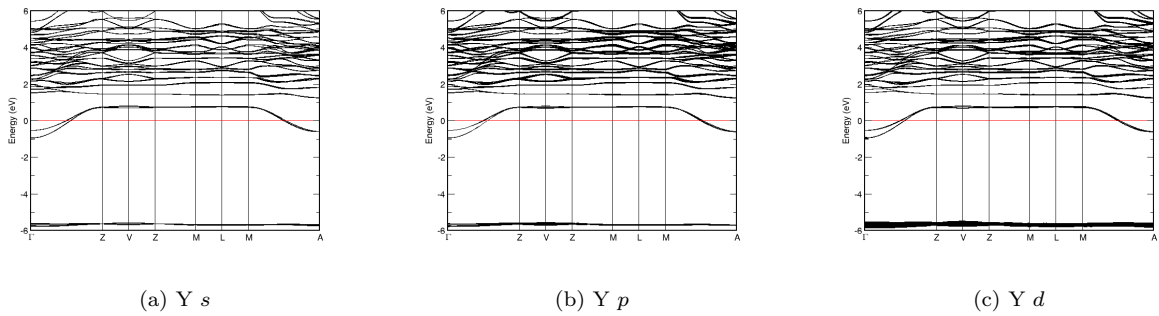

(a) Y $s$          (b) Y $p$          (c) Y $d$

FIG. 106: Fat band representation of Y in $BaY_2F_8$

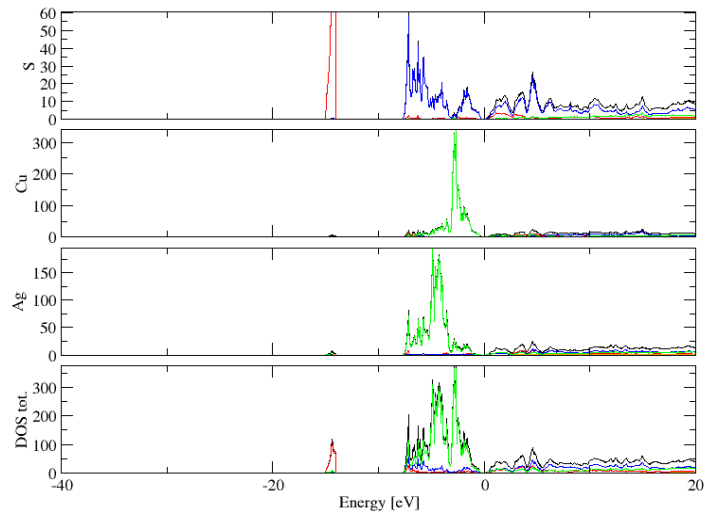

FIG. 107: (Color online) PDOS of AgCuS (ICSD #30233). The $s$-, $p$- and $d$-projected states are in red, blue and green, respectively. AgCuS crystallizes in space group C m c m (#63), in a orthorhombic base-centred structure.

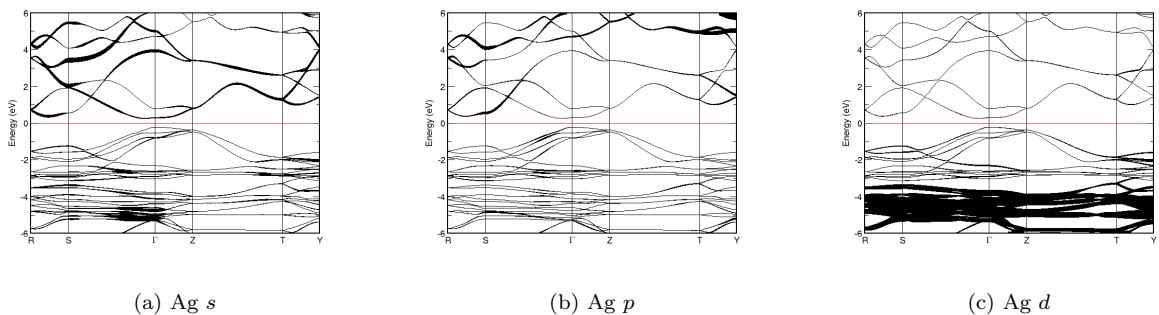

(a) Ag $s$          (b) Ag $p$          (c) Ag $d$

FIG. 108: Fat band representation of Ag in AgCuS

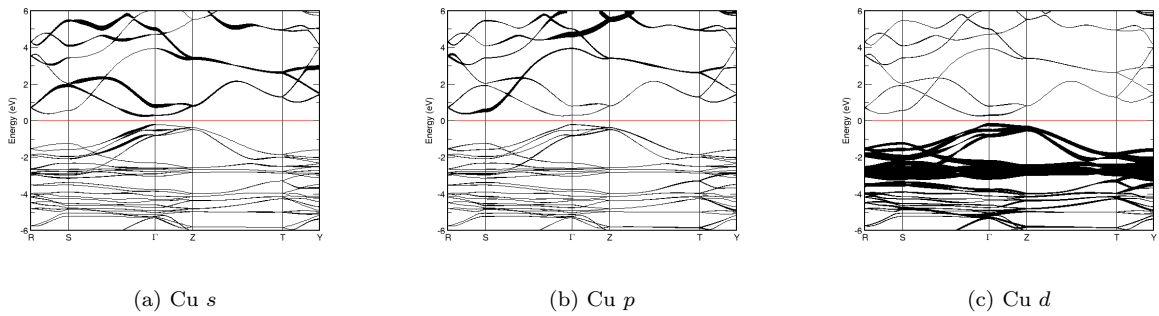

(a) Cu $s$        (b) Cu $p$        (c) Cu $d$

FIG. 109: Fat band representation of Cu in AgCuS

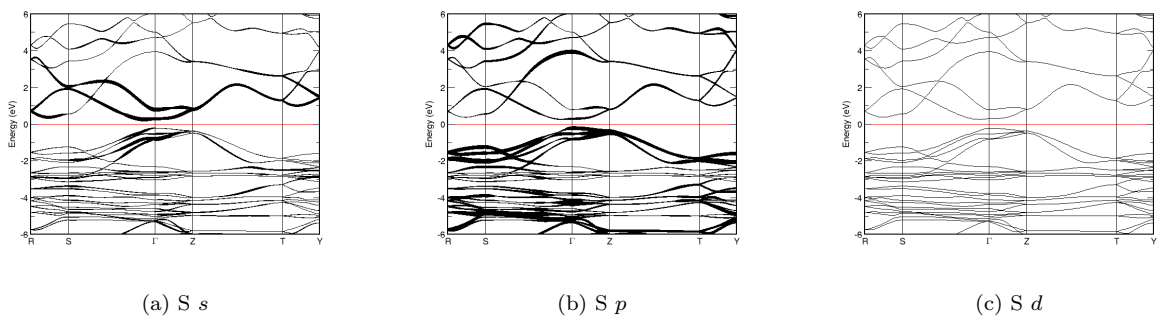

(a) S $s$        (b) S $p$        (c) S $d$

FIG. 110: Fat band representation of S in AgCuS

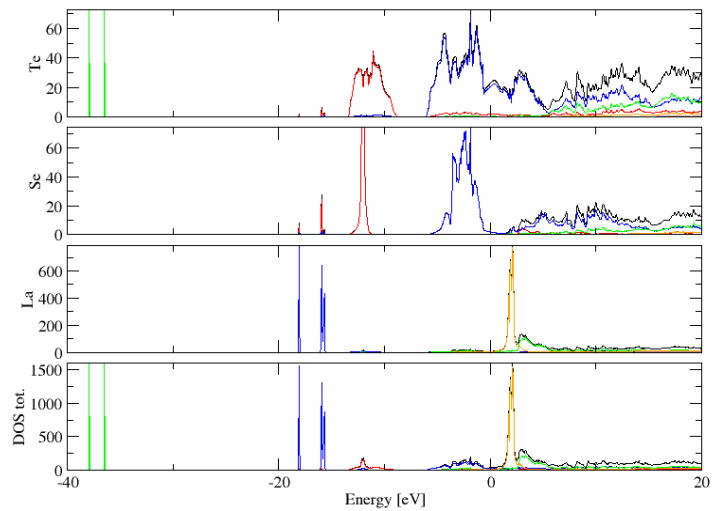

FIG. 111: (Color online) PDOS of LaSeTe$_2$ (ICSD #413171). The $s$-, $p$- and $d$-projected states are in red, blue and green, respectively. LaSeTe$_2$ crystallizes in space group C m c m (#63), in a orthorhombic base-centred structure.

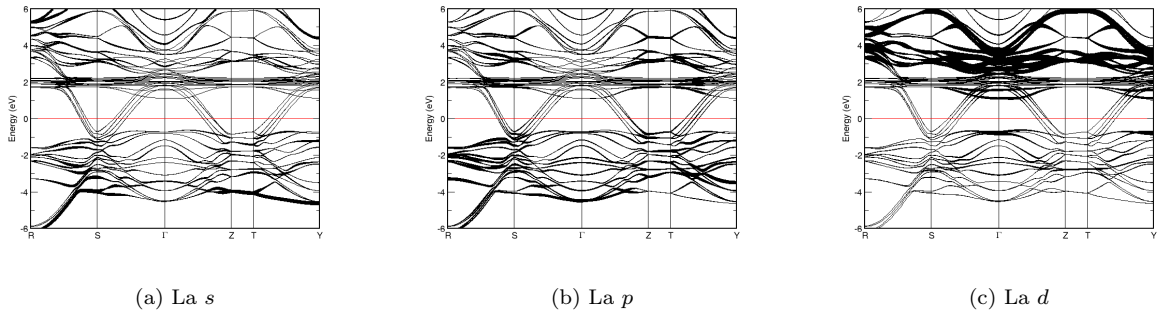

(a) La $s$                (b) La $p$                (c) La $d$

FIG. 112: Fat band representation of La in LaSeTe$_2$

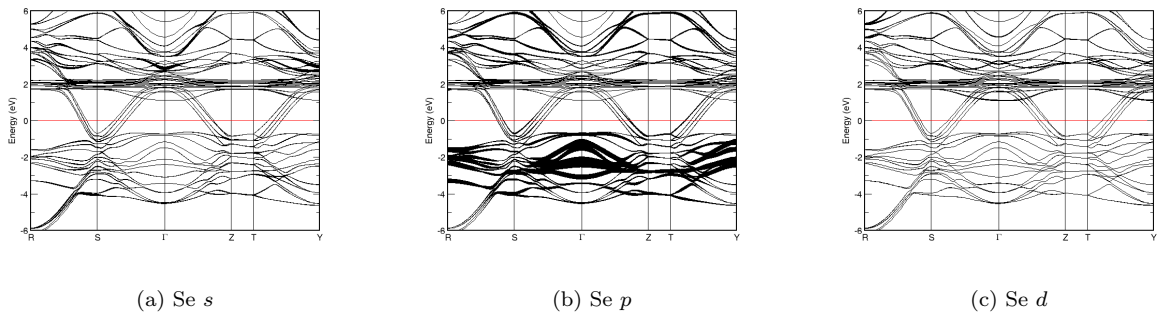

(a) Se $s$                (b) Se $p$                (c) Se $d$

FIG. 113: Fat band representation of Se in LaSeTe$_2$

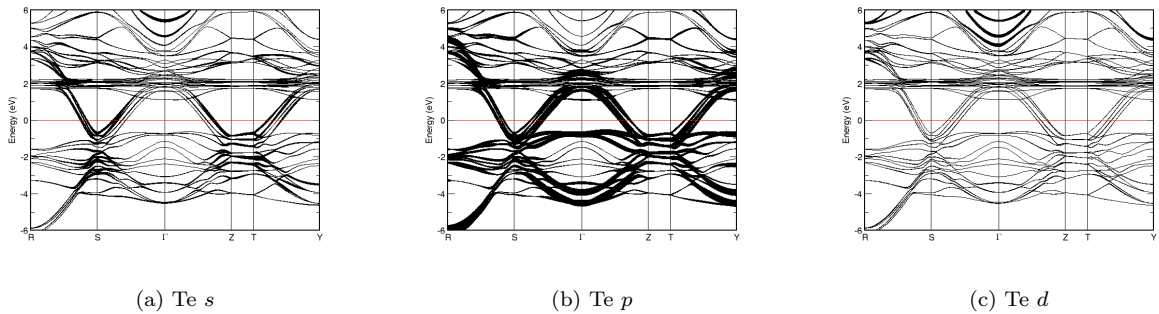

(a) Te $s$                (b) Te $p$                (c) Te $d$

FIG. 114: Fat band representation of Te in LaSeTe$_2$

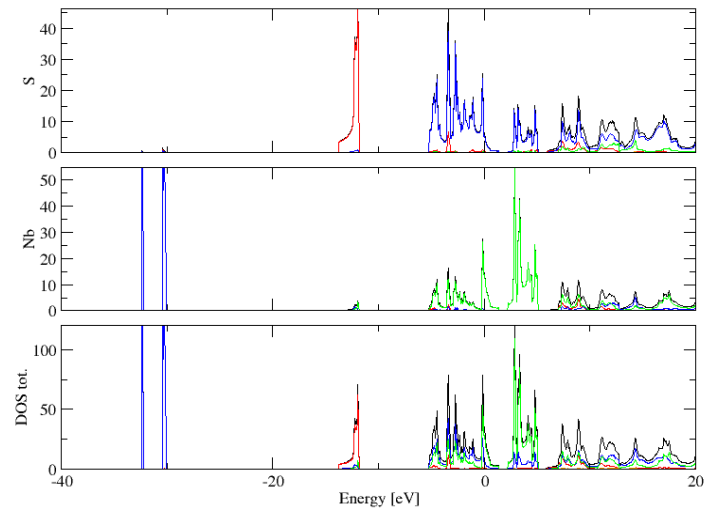

FIG. 115: (Color online) PDOS of NbS$_2$ (ICSD #67443). The $s$-, $p$- and $d$-projected states are in red, blue and green, respectively. NbS$_2$ crystallizes in space group C m 2 m (#38), in a orthorhombic base-centred structure.

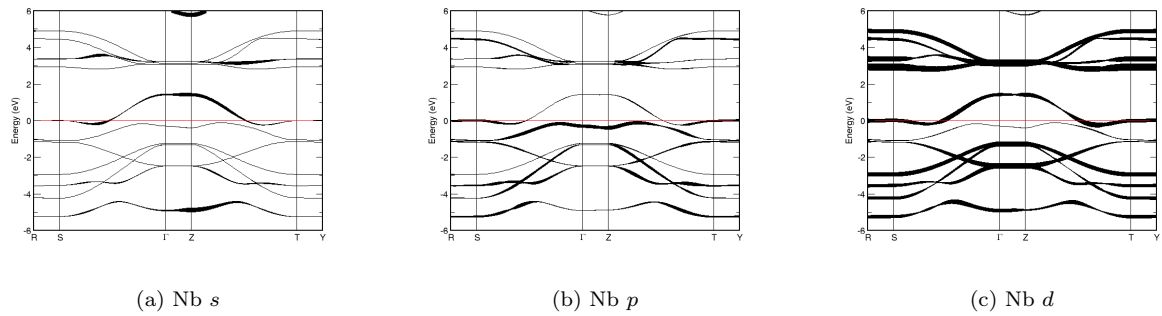

(a) Nb $s$  (b) Nb $p$  (c) Nb $d$

FIG. 116: Fat band representation of Nb in NbS$_2$

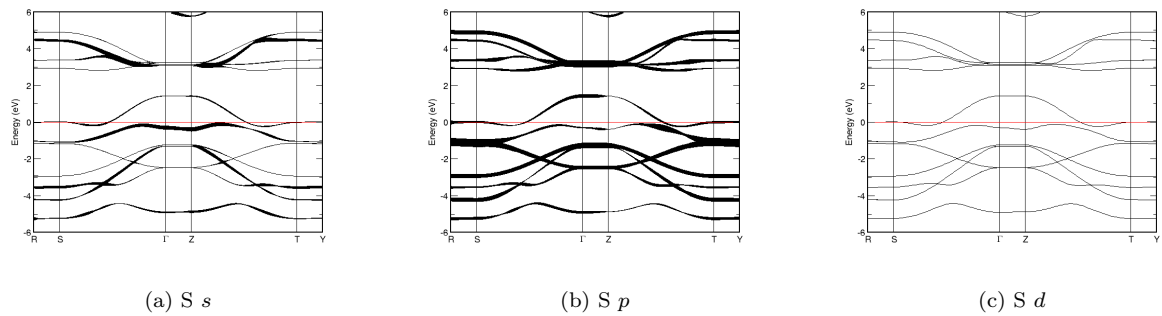

(a) S $s$  (b) S $p$  (c) S $d$

FIG. 117: Fat band representation of S in NbS$_2$

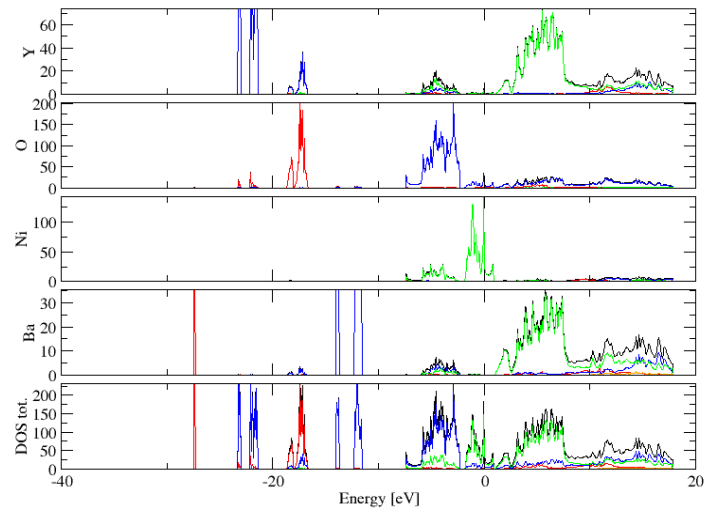

FIG. 118: (Color online) PDOS of BaNiY$_2$O$_5$ (ICSD #68795). The $s$-, $p$- and $d$-projected states are in red, blue and green, respectively. BaNiY$_2$O$_5$ crystallizes in space group I m m m (#71), in a orthorhombic body-centred structure.

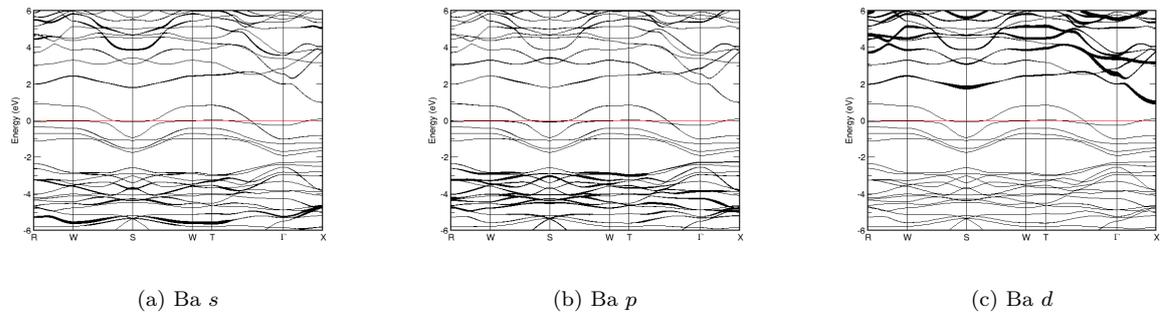

(a) Ba $s$        (b) Ba $p$        (c) Ba $d$

FIG. 119: Fat band representation of Ba in BaNiY$_2$O$_5$

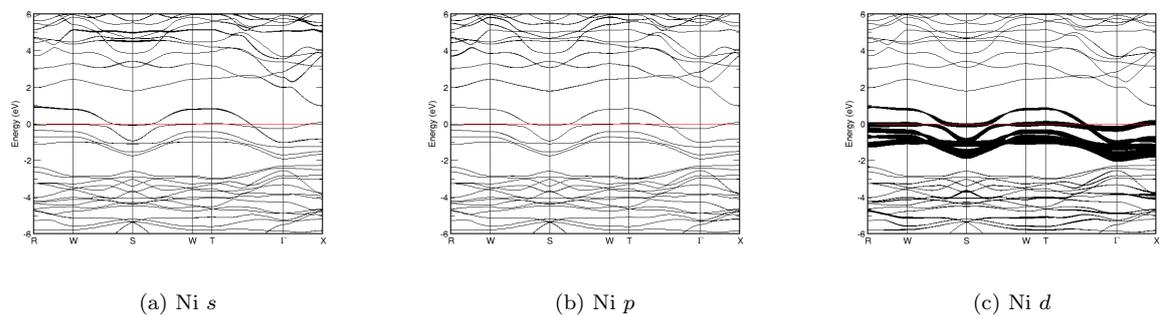

(a) Ni $s$        (b) Ni $p$        (c) Ni $d$

FIG. 120: Fat band representation of Ni in BaNiY$_2$O$_5$

(a) O $s$        (b) O $p$        (c) O $d$

FIG. 121: Fat band representation of O in BaNiY$_2$O$_5$

(a) Y $s$        (b) Y $p$        (c) Y $d$

FIG. 122: Fat band representation of Y in BaNiY$_2$O$_5$

FIG. 123: (Color online) PDOS of RuOCl$_2$ (ICSD #83883). The $s$-, $p$- and $d$-projected states are in red, blue and green, respectively. RuOCl$_2$ crystallizes in space group I m m m (#71), in a orthorhombic body-centred structure.

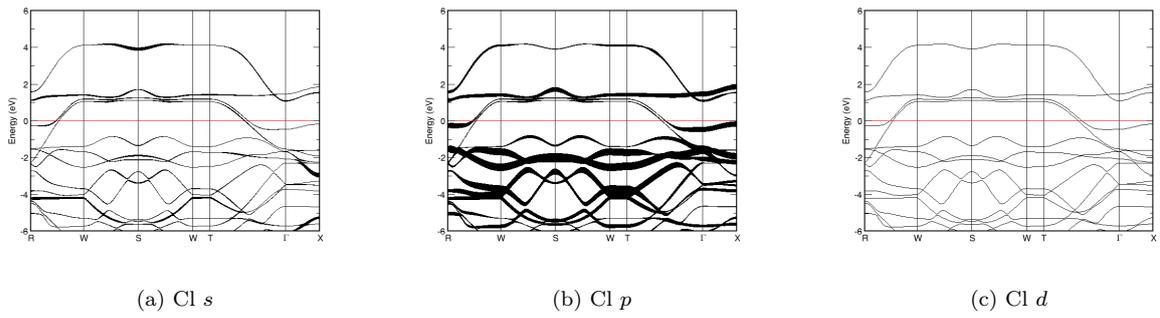

(a) Cl $s$

(b) Cl $p$

(c) Cl $d$

FIG. 124: Fat band representation of Cl in RuOCl$_2$

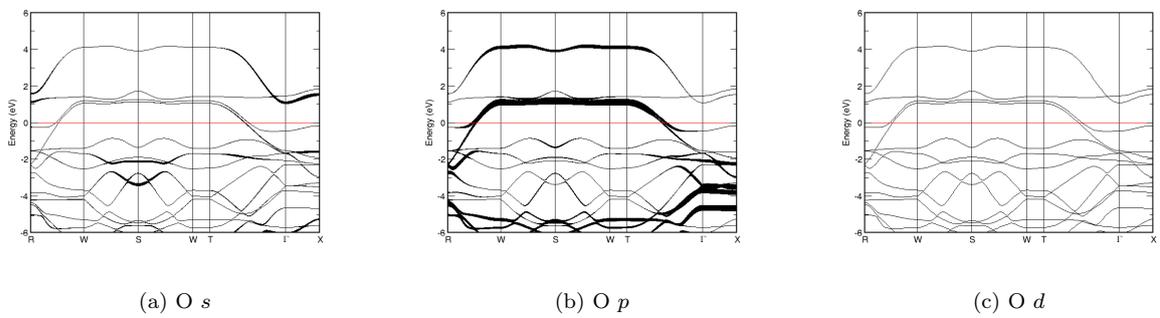

(a) O $s$

(b) O $p$

(c) O $d$

FIG. 125: Fat band representation of O in RuOCl$_2$

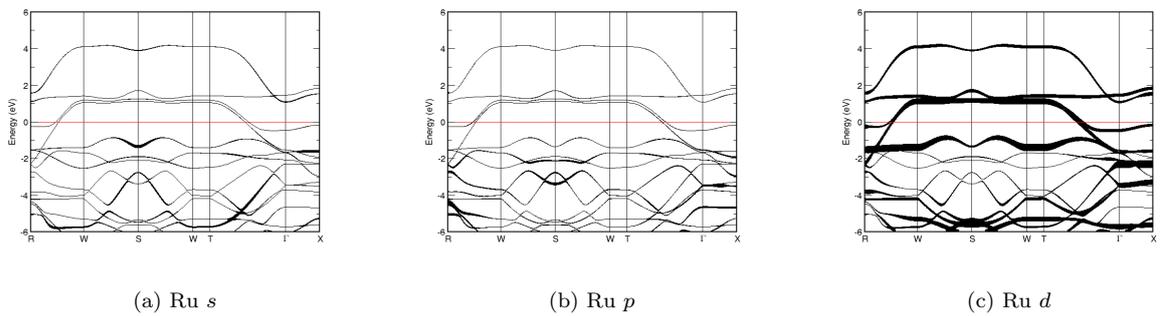

(a) Ru $s$

(b) Ru $p$

(c) Ru $d$

FIG. 126: Fat band representation of Ru in RuOCl$_2$

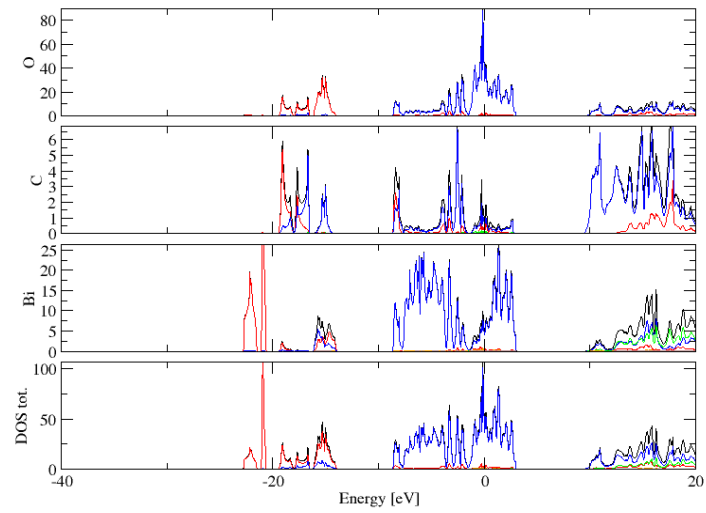

FIG. 127: (Color online) PDOS of Bi$_2$(CO$_3$)O$_2$ (ICSD #94740). The $s$-, $p$- and $d$-projected states are in red, blue and green, respectively. Bi$_2$(CO$_3$)O$_2$ crystallizes in space group I m m 2 (#44), in a orthorhombic body-centred structure.

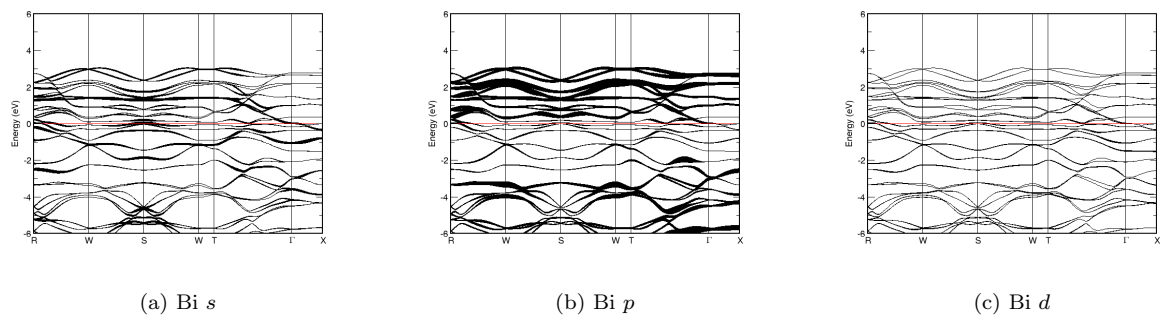

(a) Bi $s$            (b) Bi $p$            (c) Bi $d$

FIG. 128: Fat band representation of Bi in Bi$_2$(CO$_3$)O$_2$

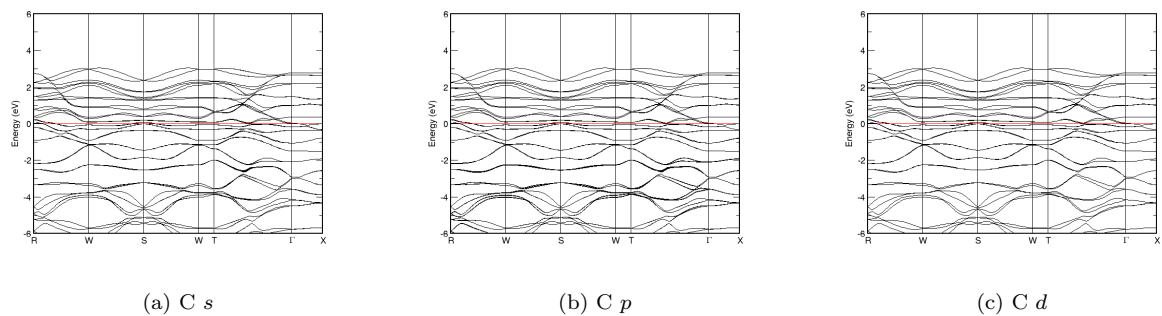

(a) C $s$            (b) C $p$            (c) C $d$

FIG. 129: Fat band representation of C in Bi$_2$(CO$_3$)O$_2$

(a) O $s$

(b) O $p$

(c) O $d$

FIG. 130: Fat band representation of O in $Bi_2(CO_3)O_2$

FIG. 131: (Color online) PDOS of $Al_2Ba_3Ge_2$ (ICSD #52612). The $s$-, $p$- and $d$-projected states are in red, blue and green, respectively. $Al_2Ba_3Ge_2$ crystallizes in space group I m m m (#71), in a orthorhombic body-centred structure.

(a) E $vs.$ k

FIG. 132: Band structure of $Ba_3Al_2Sn_2$

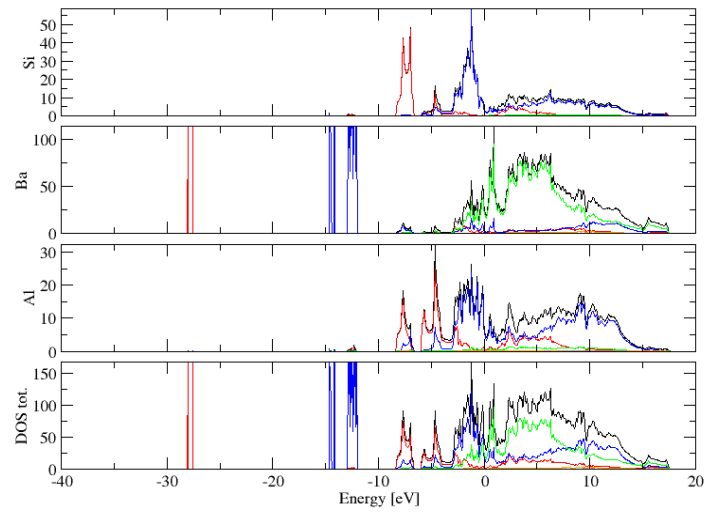

FIG. 133: (Color online) PDOS of Ba$_3$Al$_2$Si$_2$ (ICSD #100128). The $s$-, $p$- and $d$-projected states are in red, blue and green, respectively. Ba$_3$Al$_2$Si$_2$ crystallizes in space group I m m m (#71), in a orthorhombic body-centred structure.

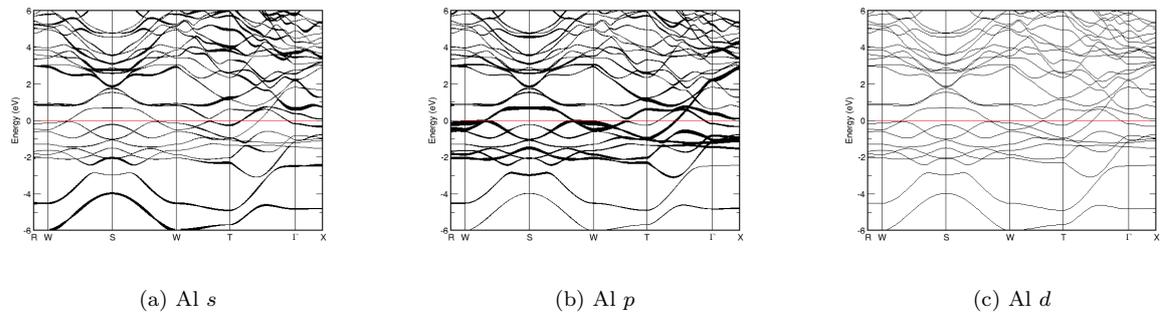

(a) Al $s$        (b) Al $p$        (c) Al $d$

FIG. 134: Fat band representation of Al in Ba$_3$Al$_2$Si$_2$

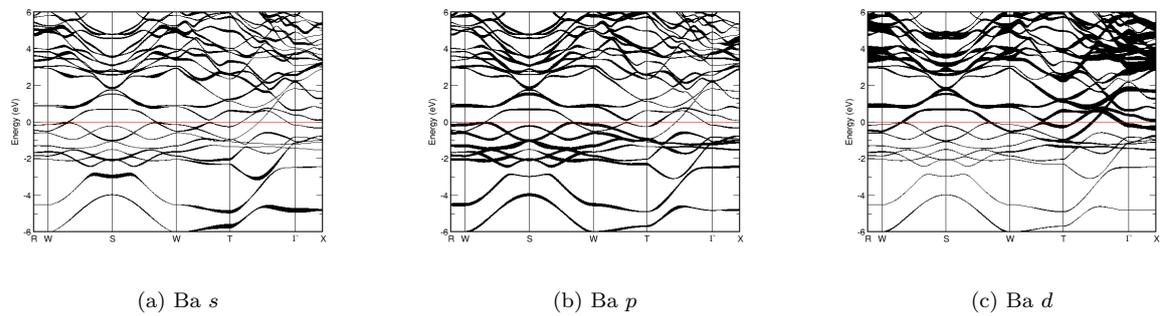

(a) Ba $s$        (b) Ba $p$        (c) Ba $d$

FIG. 135: Fat band representation of Ba in Ba$_3$Al$_2$Si$_2$

(a) Si $s$

(b) Si $p$

(c) Si $d$

FIG. 136: Fat band representation of Si in Ba$_3$Al$_2$Si$_2$

FIG. 137: (Color online) PDOS of Ba$_3$Al$_2$Sn$_2$ (ICSD #9565). The $s$-, $p$- and $d$-projected states are in red, blue and green, respectively. Ba$_3$Al$_2$Sn$_2$ crystallizes in space group I m m m (#71), in a orthorhombic body-centred structure.

(a) E $vs.$ k

FIG. 138: Band structure of Ba$_3$Al$_2$Sn$_2$

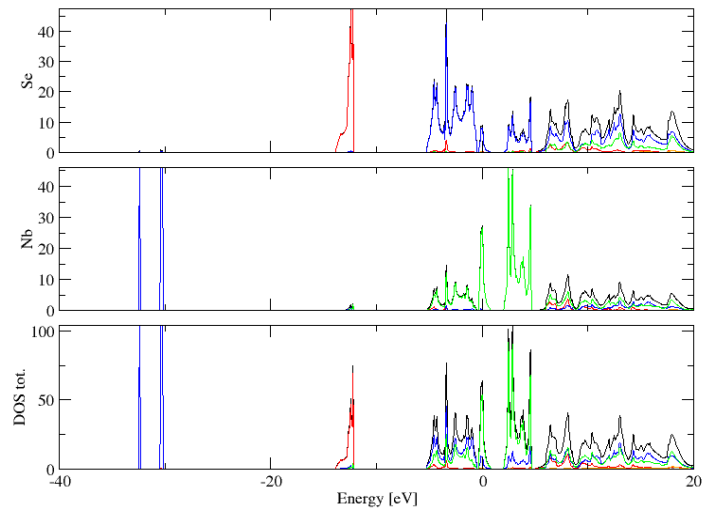

FIG. 139: (Color online) PDOS of NbSe$_2$ (ICSD #71339). The $s$-, $p$- and $d$-projected states are in red, blue and green, respectively. NbSe$_2$ crystallizes in space group F m 2 m (#42), in a orthorhombic face-centred structure.

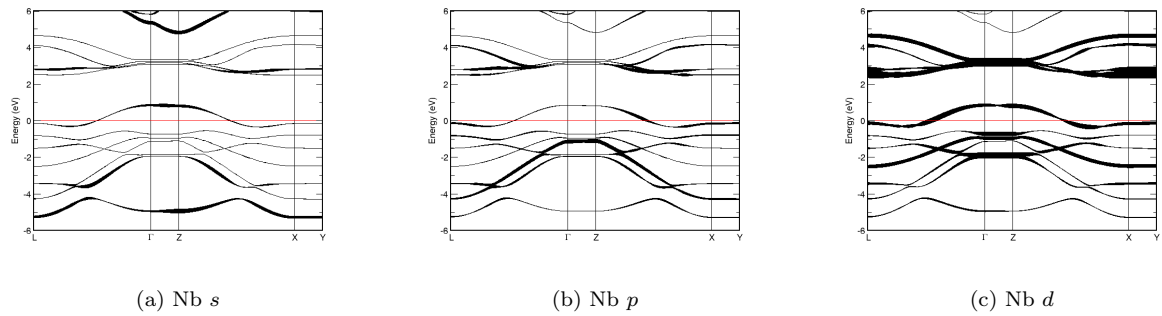

(a) Nb $s$                    (b) Nb $p$                    (c) Nb $d$

FIG. 140: Fat band representation of Nb in NbSe$_2$

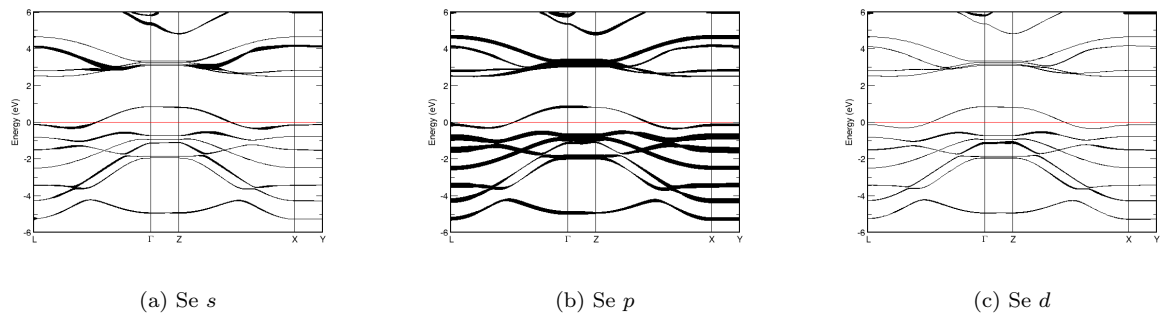

(a) Se $s$                    (b) Se $p$                    (c) Se $d$

FIG. 141: Fat band representation of Se in NbSe$_2$

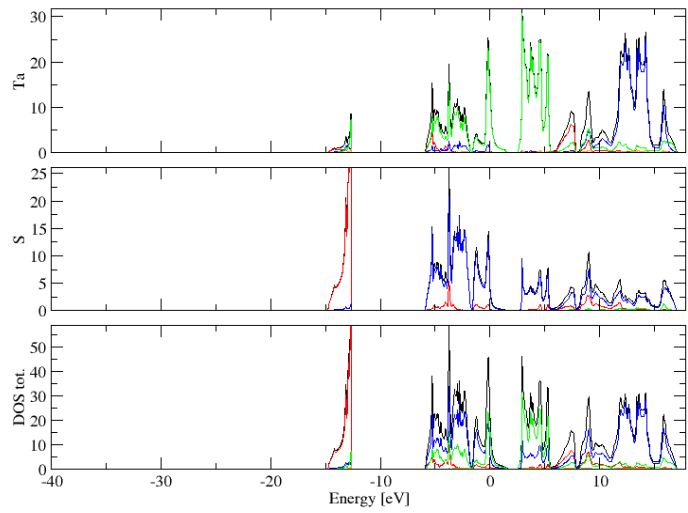

FIG. 142: (Color online) PDOS of TaS$_2$ (ICSD #280988). The $s$-, $p$- and $d$-projected states are in red, blue and green, respectively. TaS$_2$ crystallizes in space group F m 2 m (#42), in a orthorhombic face-centred structure.

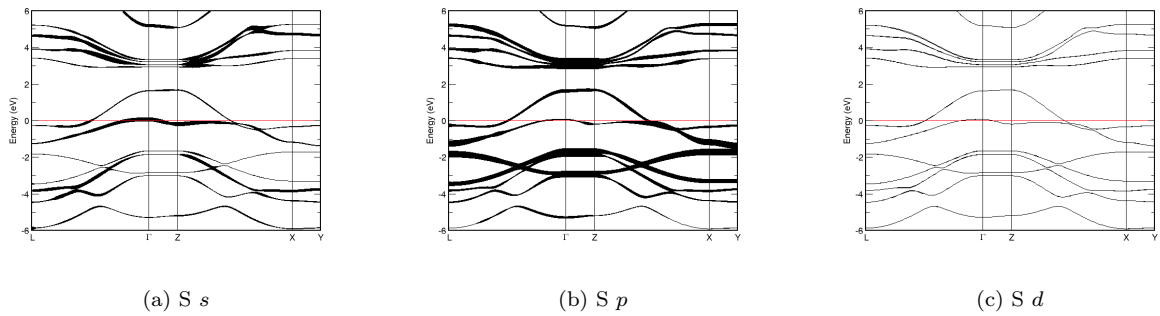

(a) S $s$            (b) S $p$            (c) S $d$

FIG. 143: Fat band representation of S in TaS$_2$

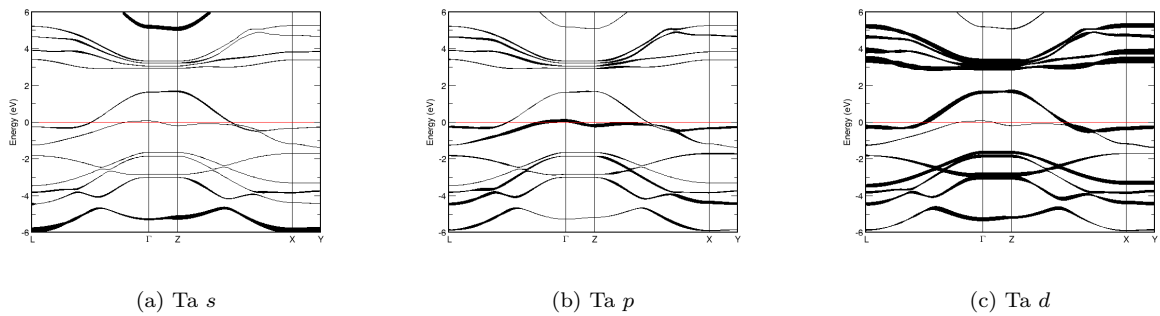

(a) Ta $s$            (b) Ta $p$            (c) Ta $d$

FIG. 144: Fat band representation of Ta in TaS$_2$

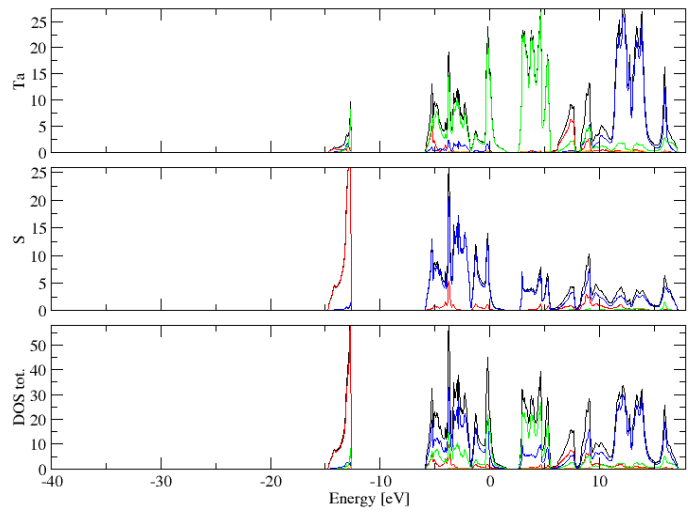

FIG. 145: (Color online) PDOS of TaS$_2$ (ICSD #67651). The $s$-, $p$- and $d$-projected states are in red, blue and green, respectively. TaS$_2$ crystallizes in space group F 2 m m (#42), in a orthorhombic face-centred structure.

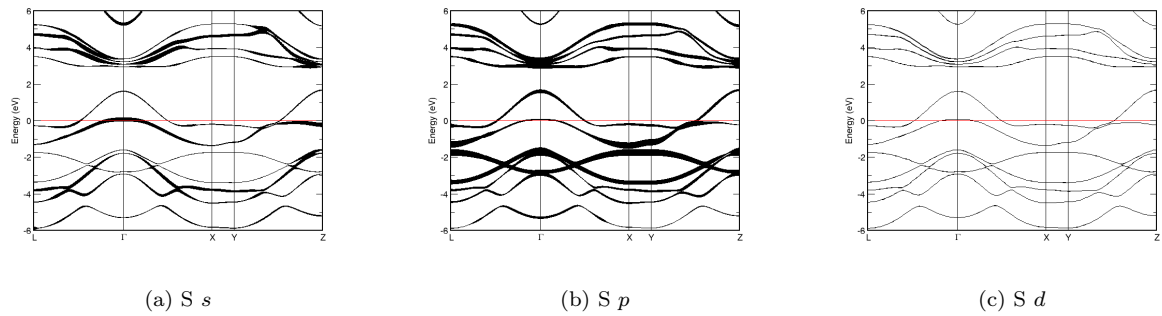

(a) S $s$      (b) S $p$      (c) S $d$

FIG. 146: Fat band representation of S in TaS$_2$

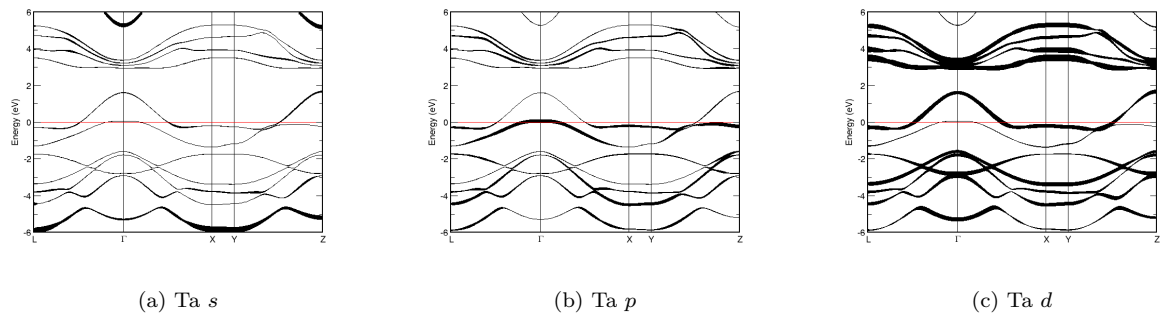

(a) Ta $s$      (b) Ta $p$      (c) Ta $d$

FIG. 147: Fat band representation of Ta in TaS$_2$

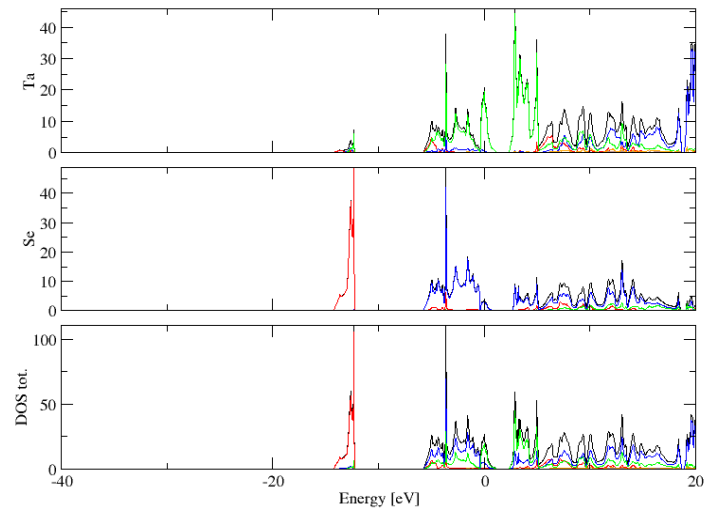

FIG. 148: (Color online) PDOS of TaSe$_2$ (ICSD #72198). The $s$-, $p$- and $d$-projected states are in red, blue and green, respectively. TaSe$_2$ crystallizes in space group F m 2 m (#42), in a orthorhombic face-centred structure.

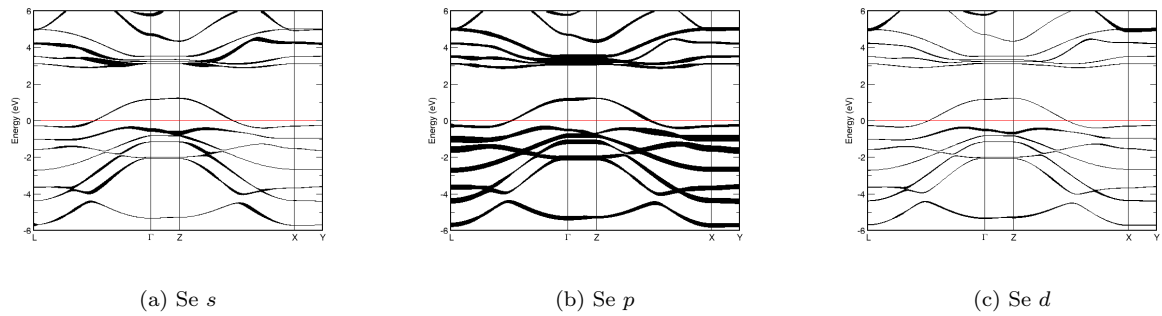

(a) Se $s$        (b) Se $p$        (c) Se $d$

FIG. 149: Fat band representation of Se in TaSe$_2$

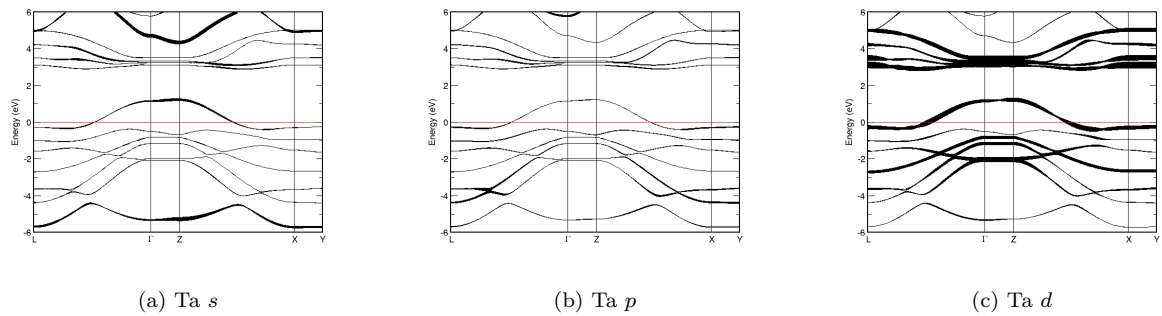

(a) Ta $s$        (b) Ta $p$        (c) Ta $d$

FIG. 150: Fat band representation of Ta in TaSe$_2$

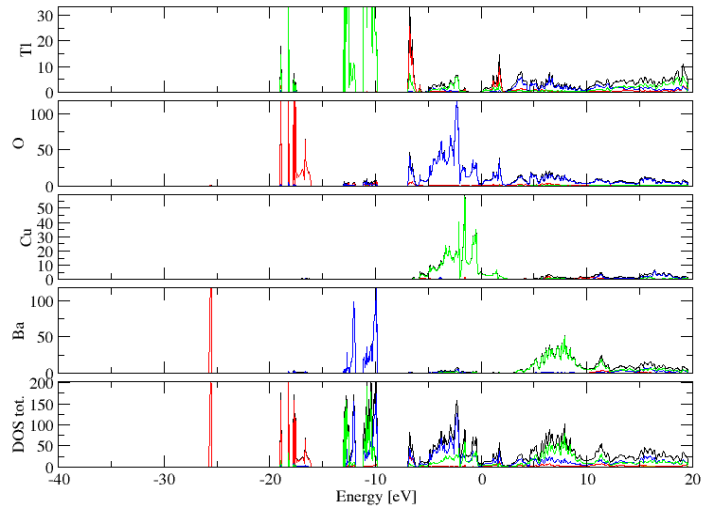

FIG. 151: (Color online) PDOS of Tl$_2$Ba$_2$CuO$_6$ (ICSD #41569). The $s$-, $p$- and $d$-projected states are in red, blue and green, respectively. Tl$_2$Ba$_2$CuO$_6$ crystallizes in space group F m m m (#69), in a orthorhombic face-centred structure.

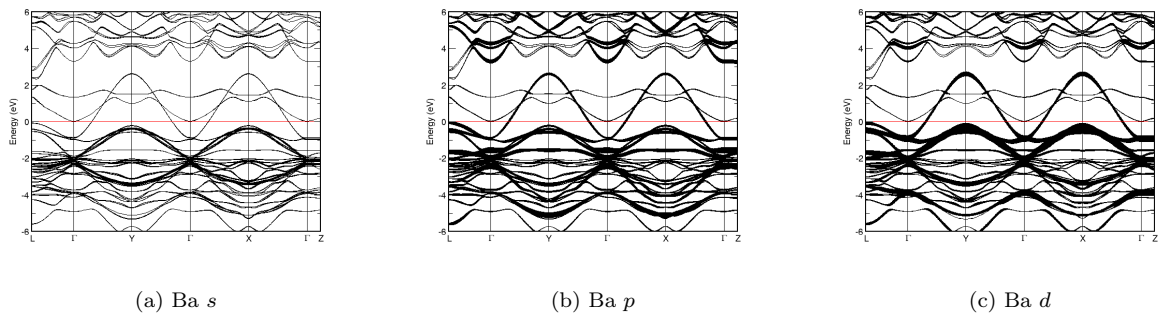

(a) Ba $s$         (b) Ba $p$         (c) Ba $d$

FIG. 152: Fat band representation of Ba in Tl$_2$Ba$_2$CuO$_6$

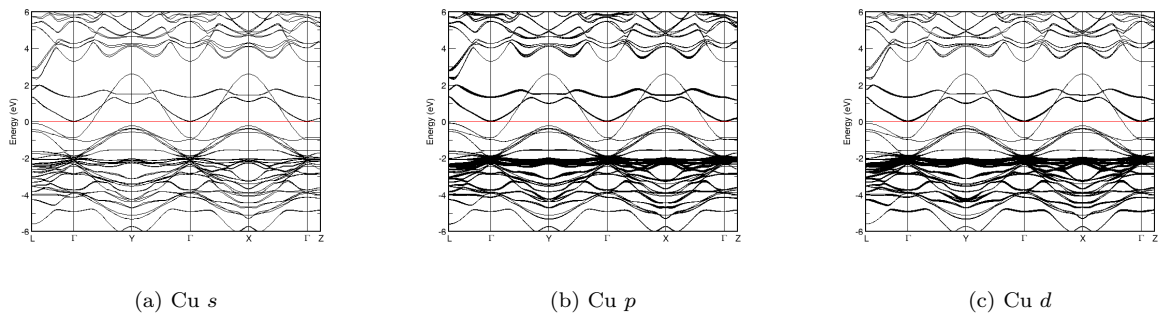

(a) Cu $s$         (b) Cu $p$         (c) Cu $d$

FIG. 153: Fat band representation of Cu in Tl$_2$Ba$_2$CuO$_6$

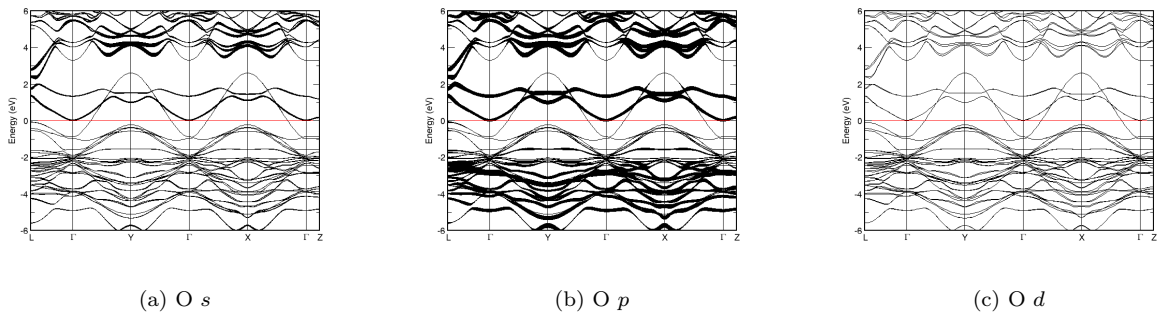

(a) O $s$    (b) O $p$    (c) O $d$

FIG. 154: Fat band representation of O in $Tl_2Ba_2CuO_6$

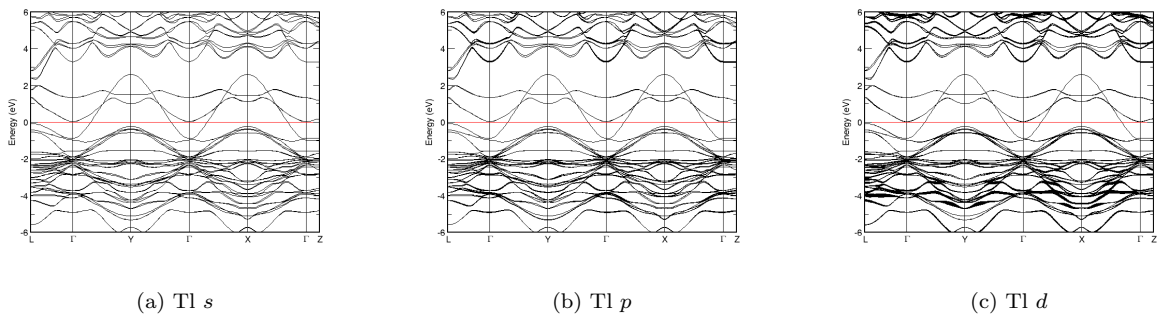

(a) Tl $s$    (b) Tl $p$    (c) Tl $d$

FIG. 155: Fat band representation of Tl in $Tl_2Ba_2CuO_6$

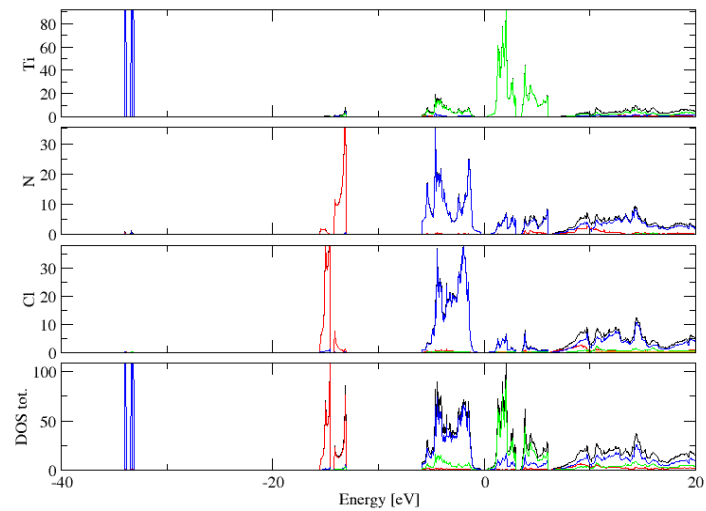

FIG. 156: (Color online) PDOS of TiNCl (ICSD #27396). The $s$-, $p$- and $d$-projected states are in red, blue and green, respectively. TiNCl crystallizes in space group P m m n S (#59), in a orthorhombic primitive structure.

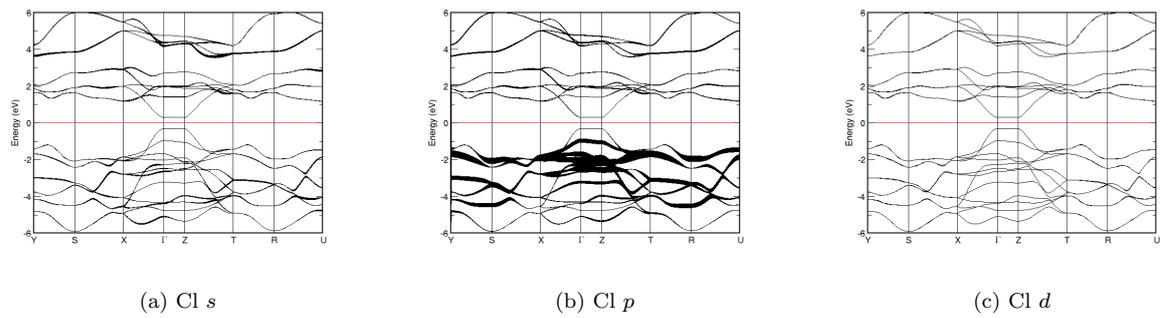

(a) Cl $s$

(b) Cl $p$

(c) Cl $d$

FIG. 157: Fat band representation of Cl in TiNCl

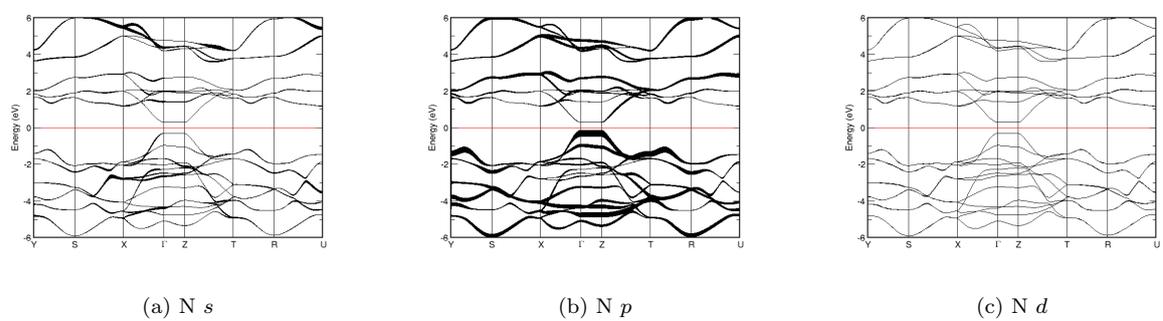

(a) N $s$

(b) N $p$

(c) N $d$

FIG. 158: Fat band representation of N in TiNCl

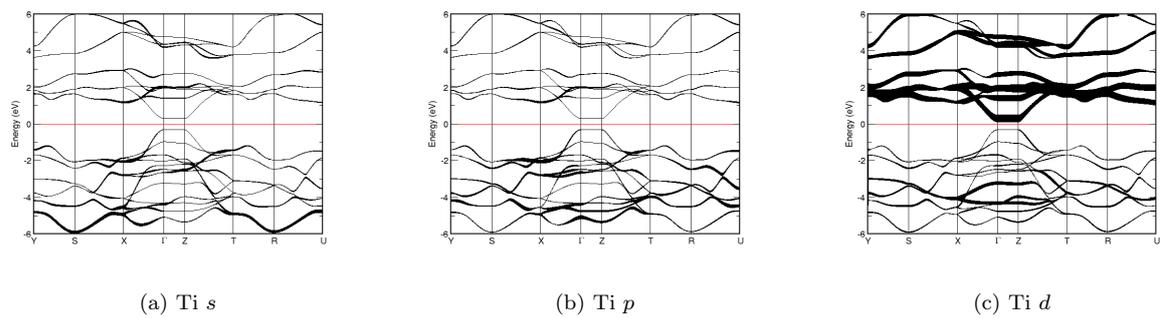

(a) Ti $s$

(b) Ti $p$

(c) Ti $d$

FIG. 159: Fat band representation of Ti in TiNCl

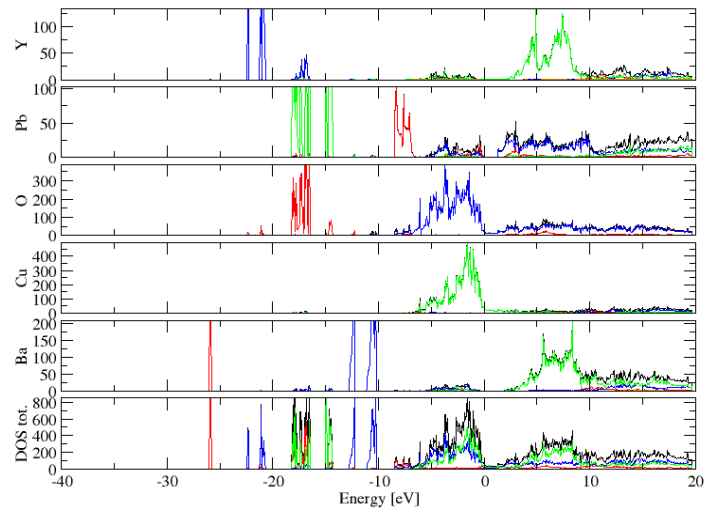

FIG. 160: (Color online) PDOS of $Pb_2Ba_2YCuCu_2O_8$ (ICSD #66088). The $s$-, $p$- and $d$-projected states are in red, blue and green, respectively. $Pb_2Ba_2YCuCu_2O_8$ crystallizes in space group P 2 21 2 (#17), in a orthorhombic primitive structure.

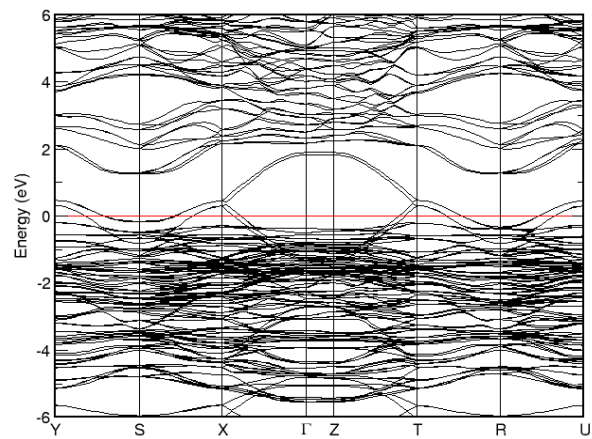

(a) E *vs.* k

FIG. 161: Band structure of $Pb_2Ba_2YCuCu_2O_8$

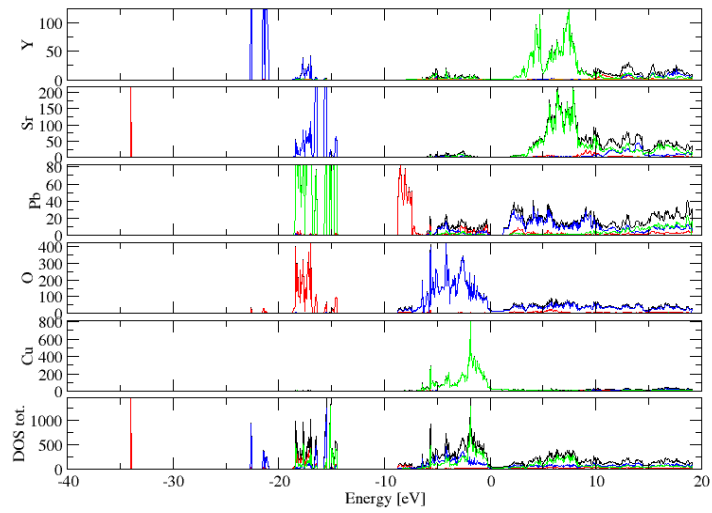

FIG. 162: (Color online) PDOS of Pb$_2$Sr$_2$YCu$_3$O$_8$ (ICSD #66587). The $s$-, $p$- and $d$-projected states are in red, blue and green, respectively. Pb$_2$Sr$_2$YCu$_3$O$_8$ crystallizes in space group P 2 21 2 (#17), in a orthorhombic primitive structure.

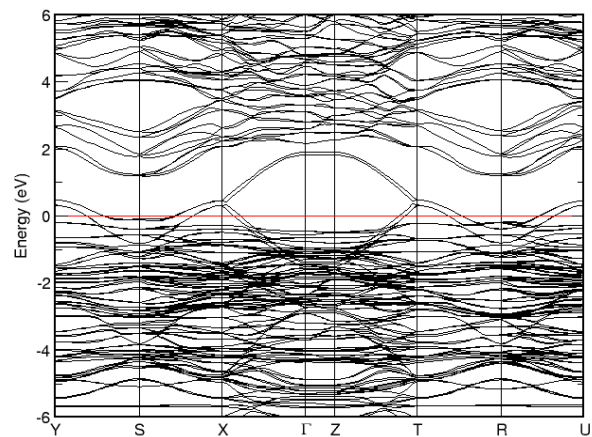

(a) E $vs.$ k

FIG. 163: Band structure of Pb$_2$Sr$_2$YCu$_3$O$_8$

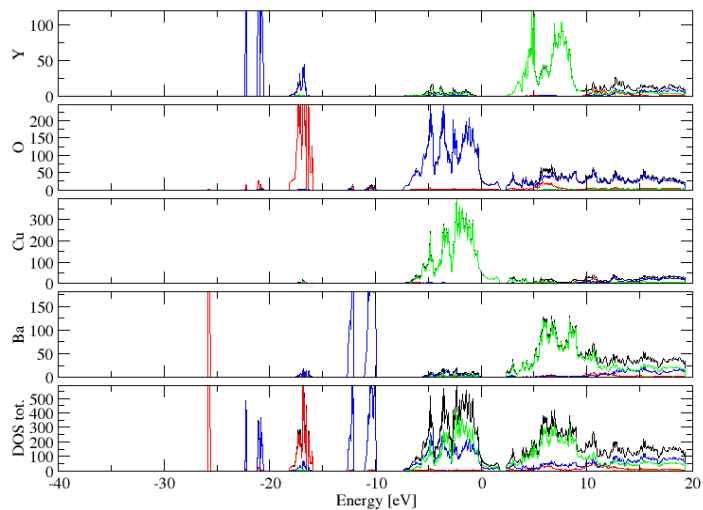

FIG. 164: (Color online) PDOS of YBa$_2$Cu$_3$O$_{6.5}$ (ICSD #75697). The $s$-, $p$- and $d$-projected states are in red, blue and green, respectively. YBa$_2$Cu$_3$O$_{6.5}$ crystallizes in space group P m m m (#47), in a orthorhombic primitive structure.

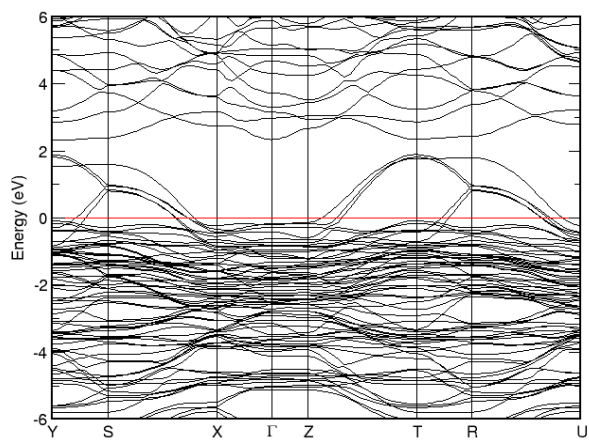

(a) E $vs.$ k

FIG. 165: Band structure of YBa$_2$Cu$_3$O$_{6.5}$

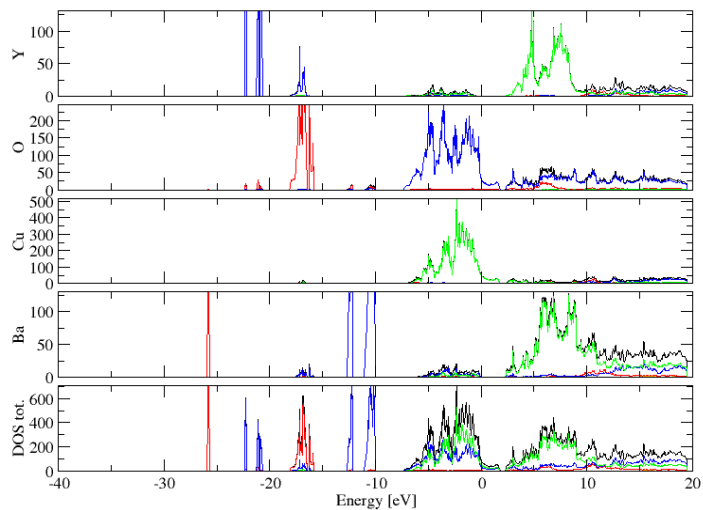

FIG. 166: (Color online) PDOS of YBa$_2$Cu$_3$O$_{6.5}$ (ICSD #96016). The $s$-, $p$- and $d$-projected states are in red, blue and green, respectively. YBa$_2$Cu$_3$O$_{6.5}$ crystallizes in space group P m m m (#47), in a orthorhombic primitive structure.

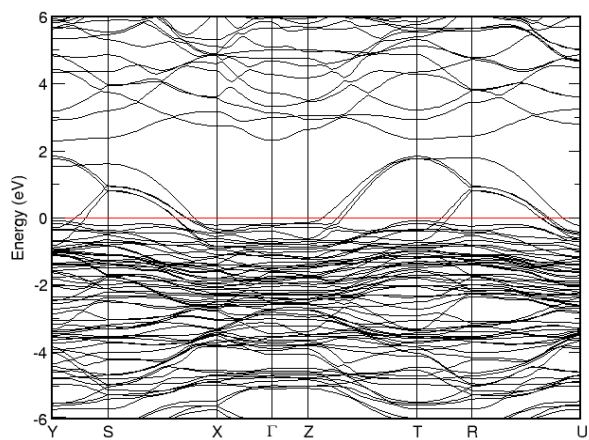

(a) E *vs.* k

FIG. 167: Band structure of YBa$_2$Cu$_3$O$_{6.5}$

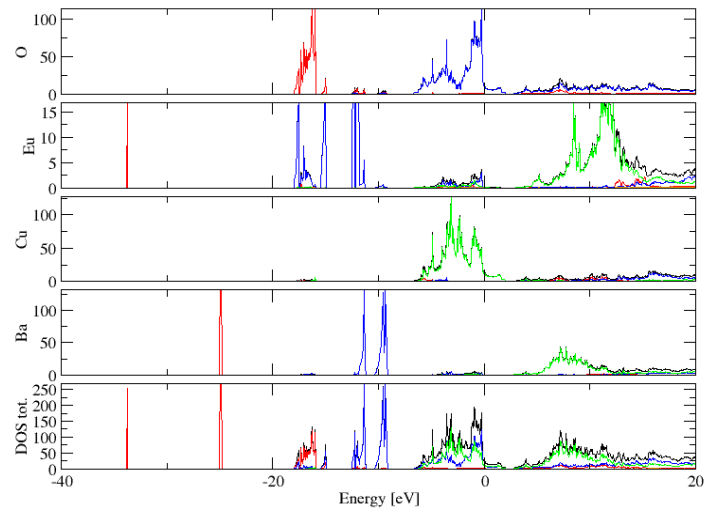

FIG. 168: (Color online) PDOS of EuBa$_2$Cu$_3$O$_7$ (ICSD #81171). The $s$-, $p$- and $d$-projected states are in red, blue and green, respectively. EuBa$_2$Cu$_3$O$_7$ crystallizes in space group P m m m (#47), in a orthorhombic primitive structure.

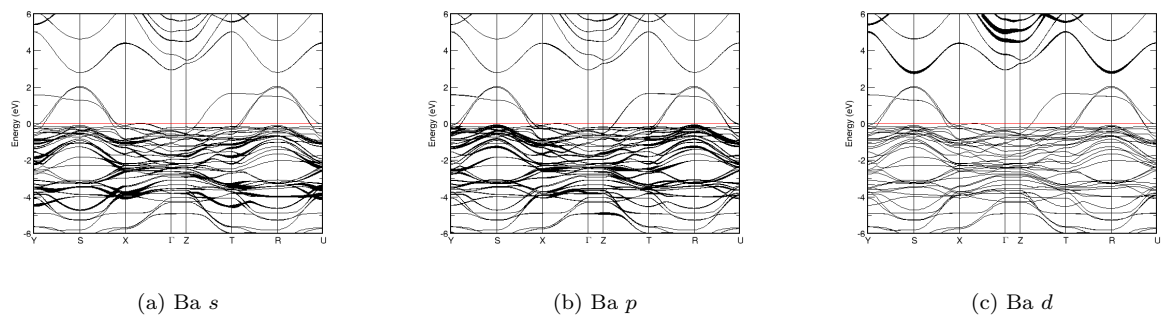

(a) Ba $s$          (b) Ba $p$          (c) Ba $d$

FIG. 169: Fat band representation of Ba in EuBa$_2$Cu$_3$O$_7$

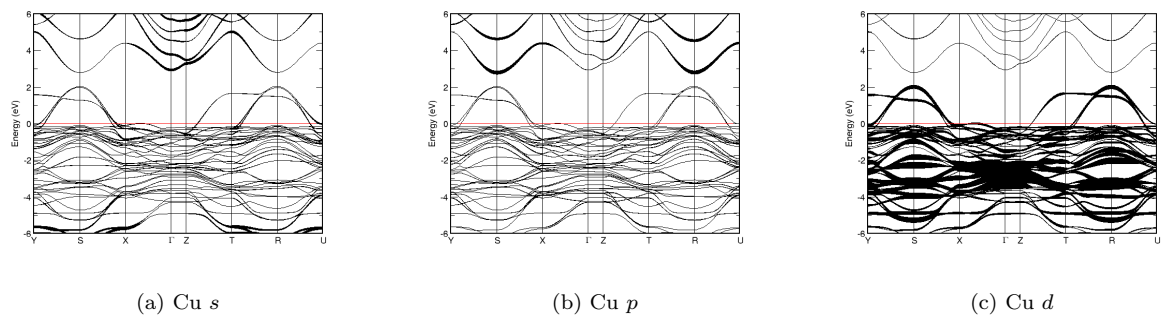

(a) Cu $s$          (b) Cu $p$          (c) Cu $d$

FIG. 170: Fat band representation of Cu in EuBa$_2$Cu$_3$O$_7$

(a) Eu $s$       (b) Eu $p$       (c) Eu $d$

FIG. 171: Fat band representation of Eu in EuBa$_2$Cu$_3$O$_7$

(a) O $s$       (b) O $p$       (c) O $d$

FIG. 172: Fat band representation of O in EuBa$_2$Cu$_3$O$_7$

FIG. 173: (Color online) PDOS of Ba$_2$GdCu$_3$O$_7$ (ICSD #56514). The $s$-, $p$- and $d$-projected states are in red, blue and green, respectively. Ba$_2$GdCu$_3$O$_7$ crystallizes in space group P m m m (#47), in a orthorhombic primitive structure.

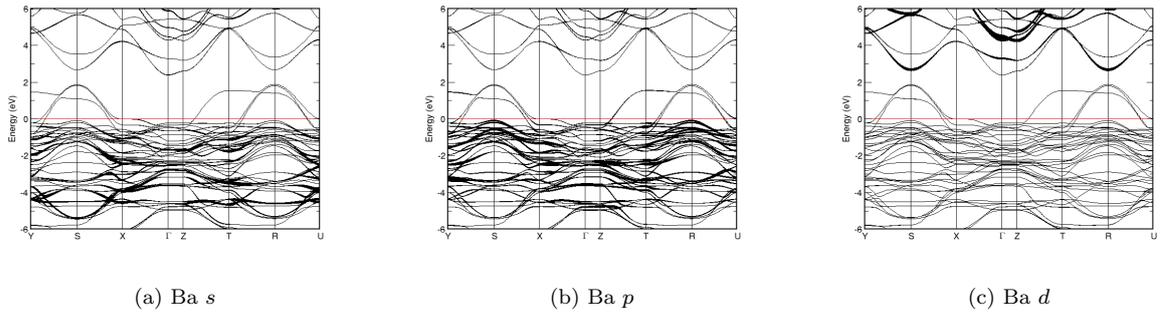

(a) Ba $s$        (b) Ba $p$        (c) Ba $d$

FIG. 174: Fat band representation of Ba in Ba$_2$GdCu$_3$O$_7$

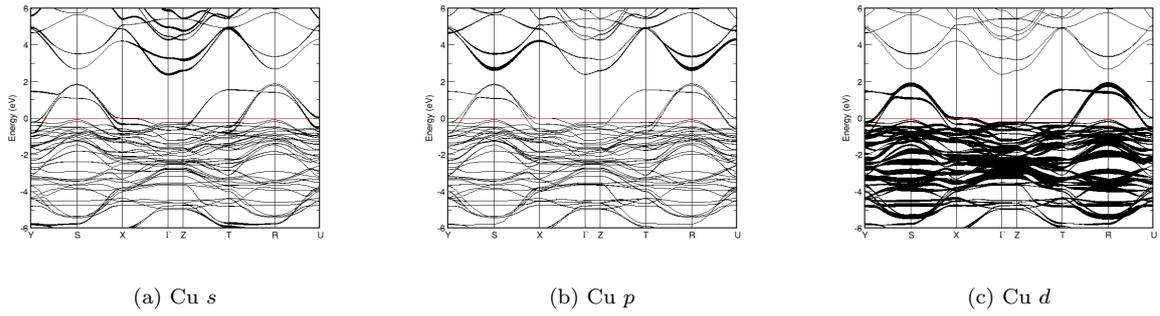

(a) Cu $s$        (b) Cu $p$        (c) Cu $d$

FIG. 175: Fat band representation of Cu in Ba$_2$GdCu$_3$O$_7$

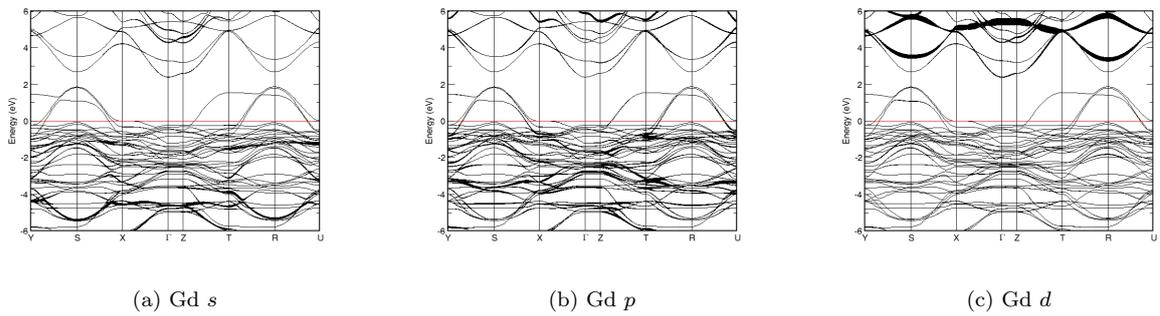

(a) Gd $s$        (b) Gd $p$        (c) Gd $d$

FIG. 176: Fat band representation of Gd in Ba$_2$GdCu$_3$O$_7$

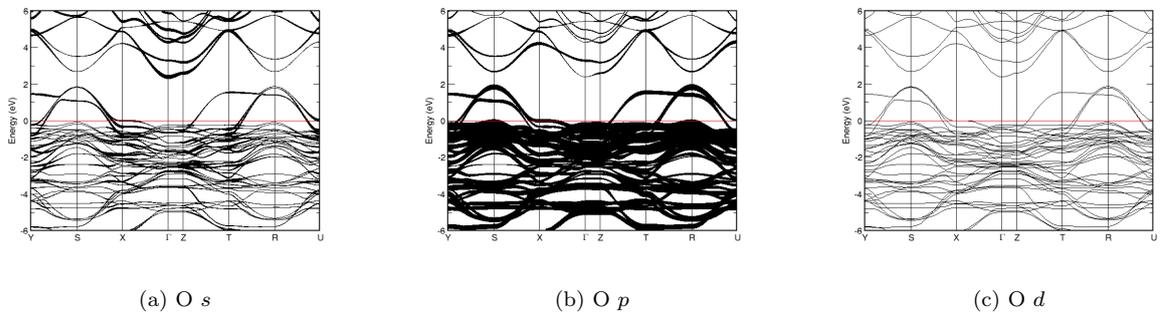

(a) O $s$        (b) O $p$        (c) O $d$

FIG. 177: Fat band representation of O in Ba$_2$GdCu$_3$O$_7$

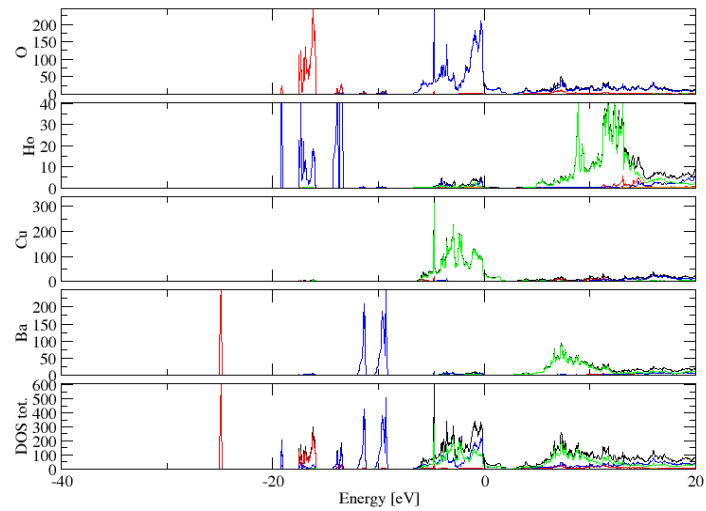

FIG. 178: (Color online) PDOS of HoBa$_2$Cu$_3$O$_7$ (ICSD #68044). The $s$-, $p$- and $d$-projected states are in red, blue and green, respectively. HoBa$_2$Cu$_3$O$_7$ crystallizes in space group P m m m (#47), in a orthorhombic primitive structure.

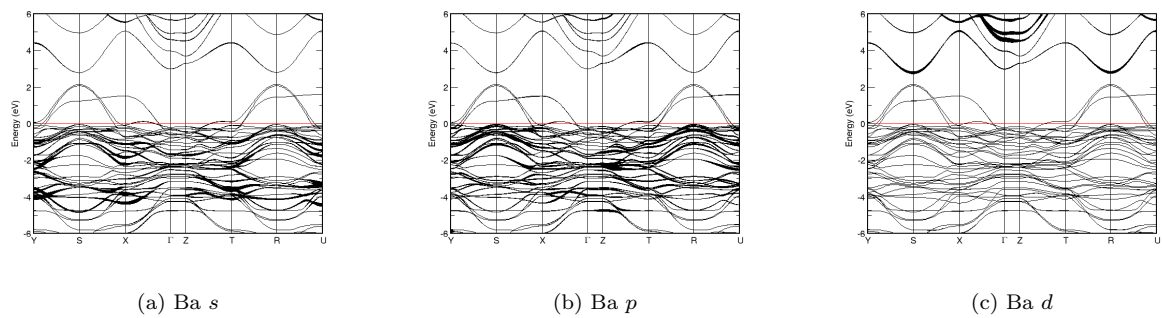

(a) Ba $s$              (b) Ba $p$              (c) Ba $d$

FIG. 179: Fat band representation of Ba in HoBa$_2$Cu$_3$O$_7$

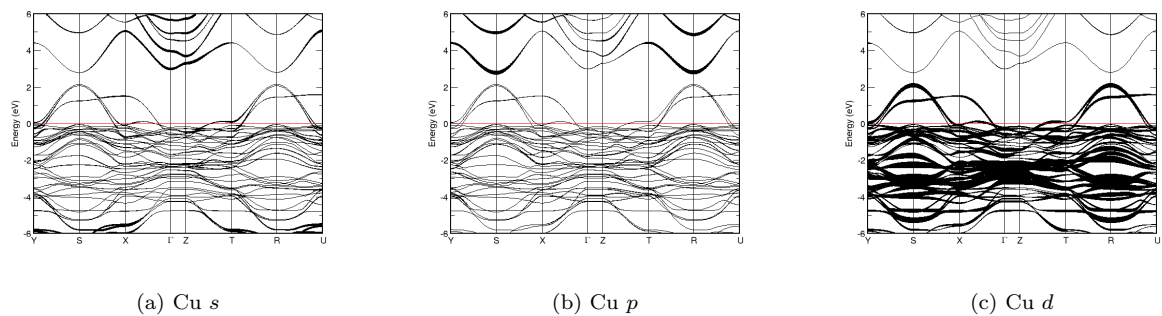

(a) Cu $s$              (b) Cu $p$              (c) Cu $d$

FIG. 180: Fat band representation of Cu in HoBa$_2$Cu$_3$O$_7$

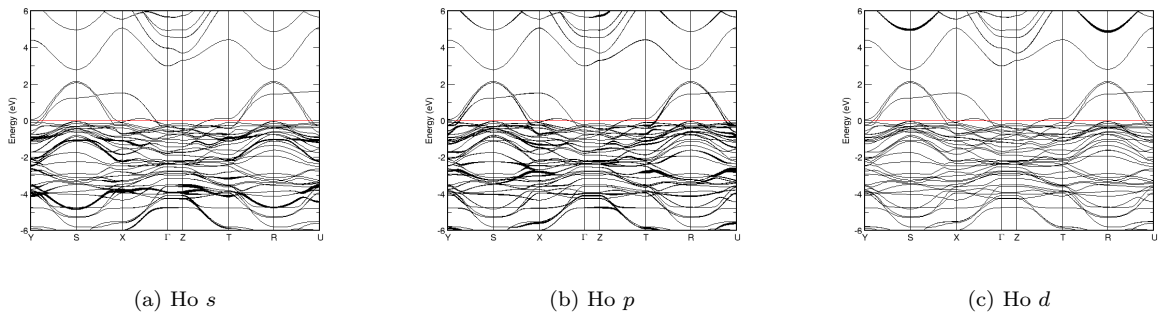

(a) Ho *s*        (b) Ho *p*       (c) Ho *d*

FIG. 181: Fat band representation of Ho in HoBa$_2$Cu$_3$O$_7$

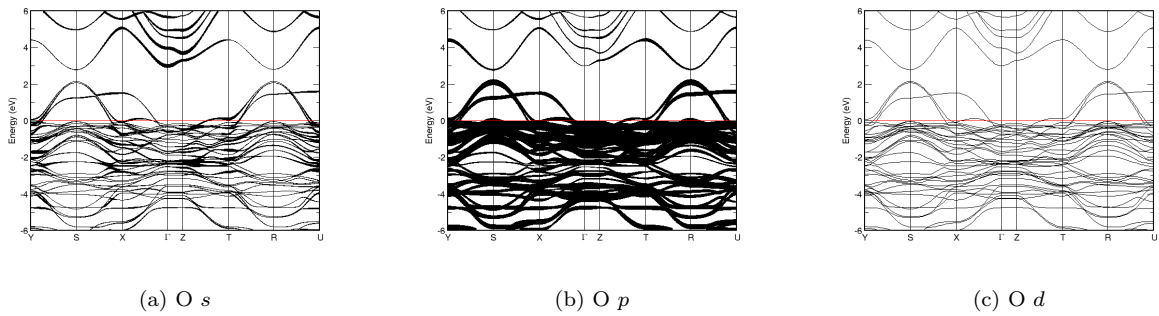

(a) O *s*       (b) O *p*       (c) O *d*

FIG. 182: Fat band representation of O in HoBa$_2$Cu$_3$O$_7$

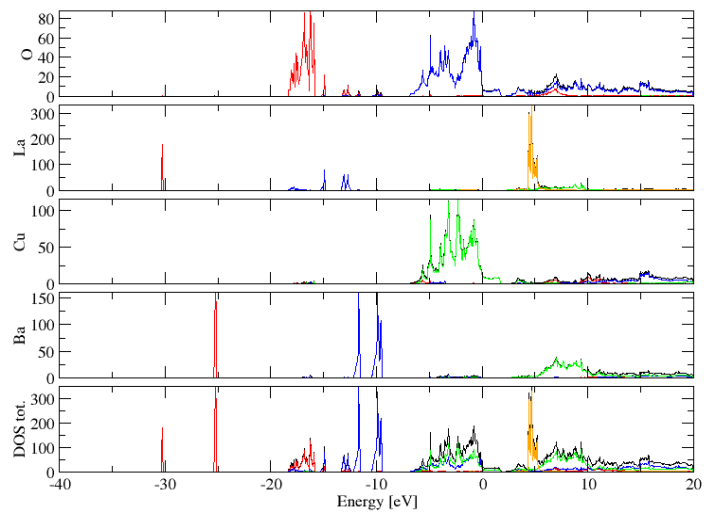

FIG. 183: (Color online) PDOS of LaBa$_2$Cu$_3$O$_7$ (ICSD #81167). The *s*-, *p*- and *d*-projected states are in red, blue and green, respectively. LaBa$_2$Cu$_3$O$_7$ crystallizes in space group P m m m (#47), in a orthorhombic primitive structure.

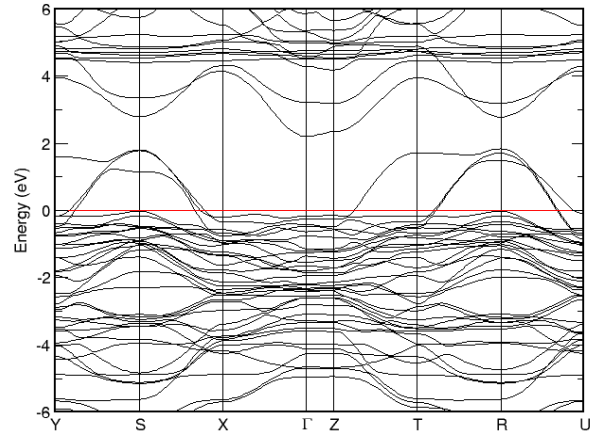

(a) E *vs.* k

FIG. 184: Band structure of LaBa$_2$Cu$_3$O$_7$

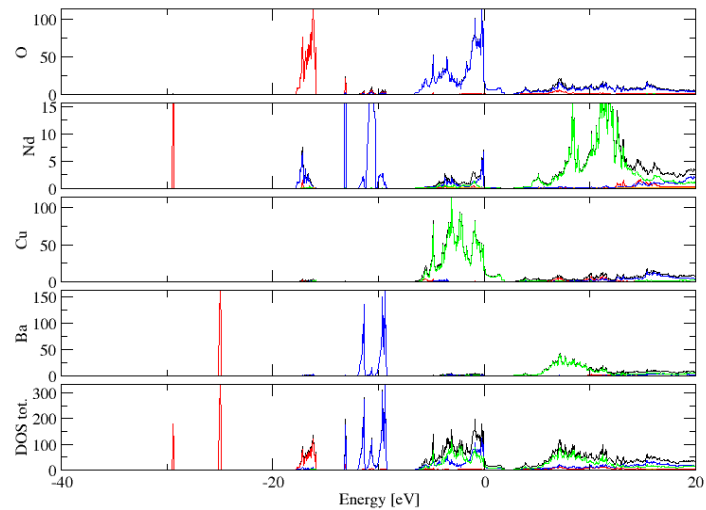

FIG. 185: (Color online) PDOS of NdBa$_2$Cu$_3$O$_7$ (ICSD #81169). The *s*-, *p*- and *d*-projected states are in red, blue and green, respectively. NdBa$_2$Cu$_3$O$_7$ crystallizes in space group P m m m (#47), in a orthorhombic primitive structure.

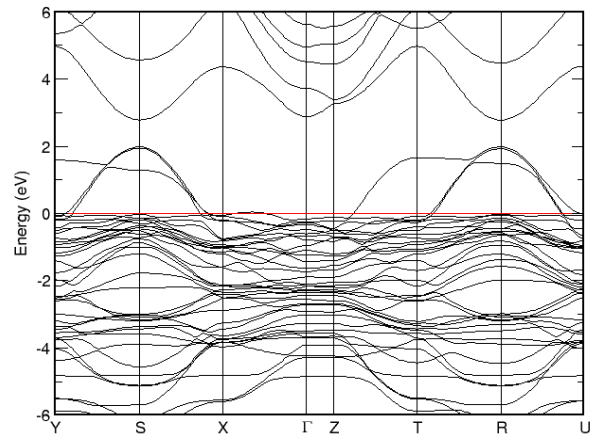

(a) E *vs.* k

FIG. 186: Band structure of NdBa$_2$Cu$_3$O$_7$

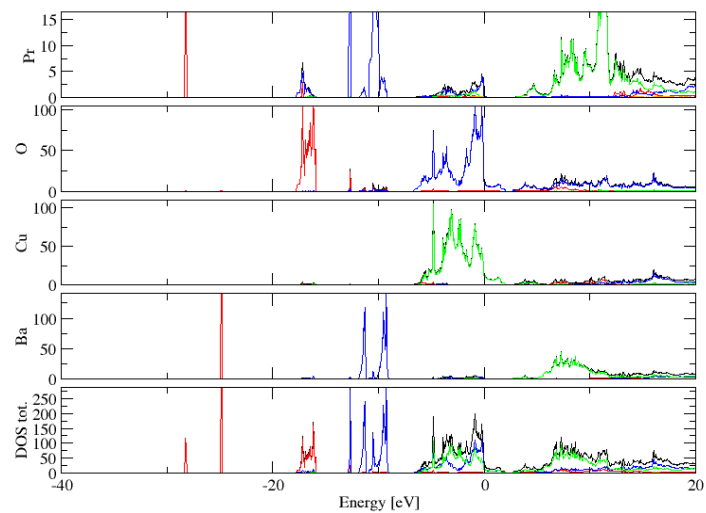

FIG. 187: (Color online) PDOS of PrBa$_2$Cu$_3$O$_7$ (ICSD #81168). The *s*-, *p*- and *d*-projected states are in red, blue and green, respectively. PrBa$_2$Cu$_3$O$_7$ crystallizes in space group P m m m (#47), in a orthorhombic primitive structure.

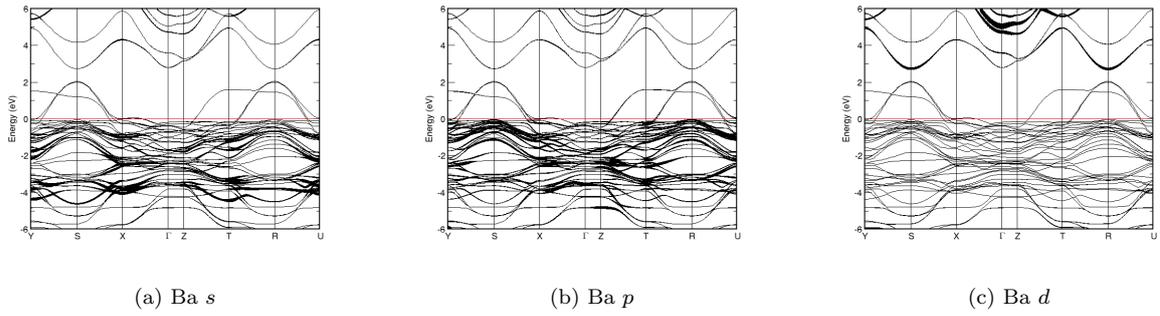

(a) Ba *s*          (b) Ba *p*          (c) Ba *d*

FIG. 188: Fat band representation of Ba in PrBa$_2$Cu$_3$O$_7$

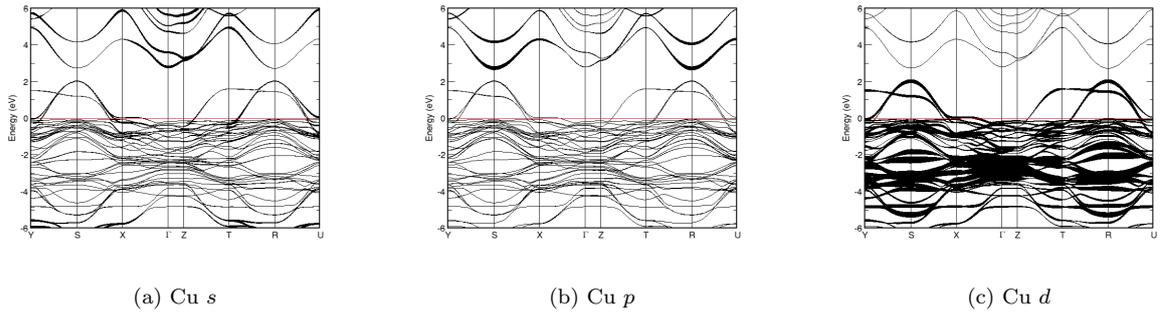

(a) Cu *s*          (b) Cu *p*          (c) Cu *d*

FIG. 189: Fat band representation of Cu in PrBa$_2$Cu$_3$O$_7$

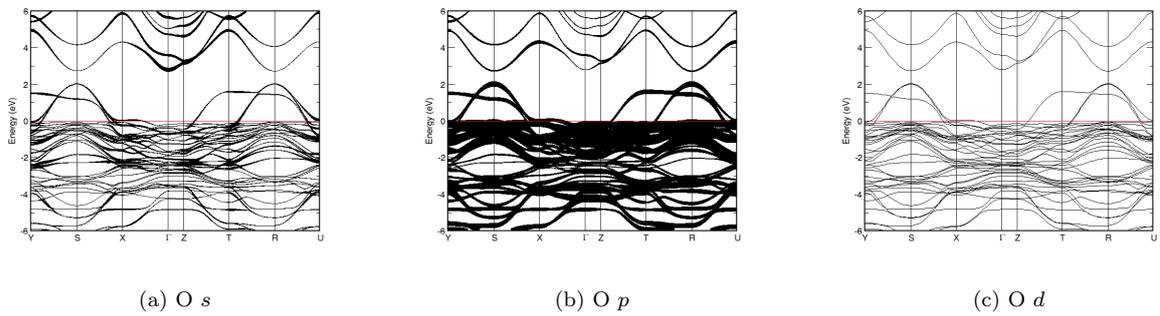

(a) O *s*          (b) O *p*          (c) O *d*

FIG. 190: Fat band representation of O in PrBa$_2$Cu$_3$O$_7$

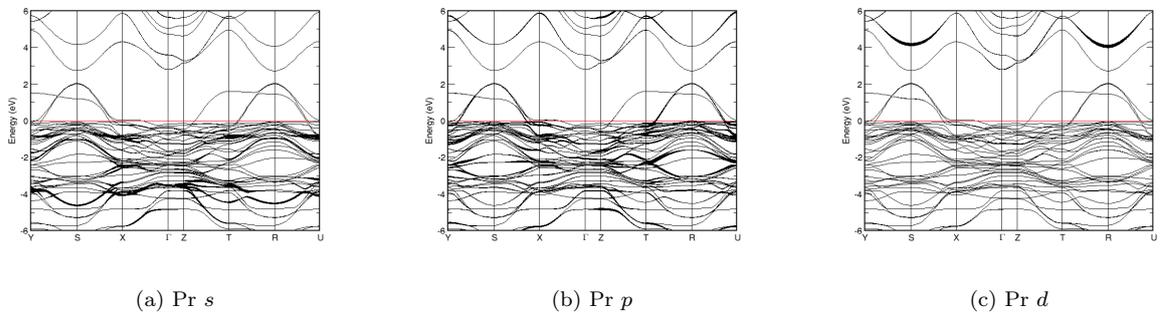

(a) Pr *s*          (b) Pr *p*          (c) Pr *d*

FIG. 191: Fat band representation of Pr in PrBa$_2$Cu$_3$O$_7$

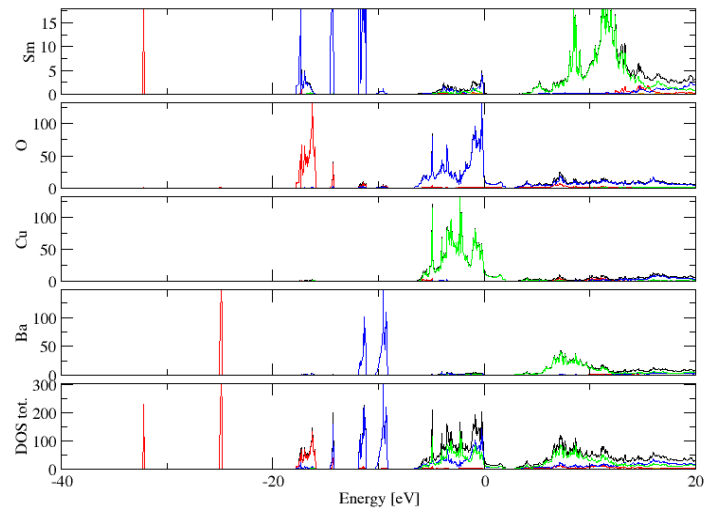

FIG. 192: (Color online) PDOS of SmBa$_2$Cu$_3$O$_7$ (ICSD #71705). The $s$-, $p$- and $d$-projected states are in red, blue and green, respectively. SmBa$_2$Cu$_3$O$_7$ crystallizes in space group P m m m (#47), in a orthorhombic primitive structure.

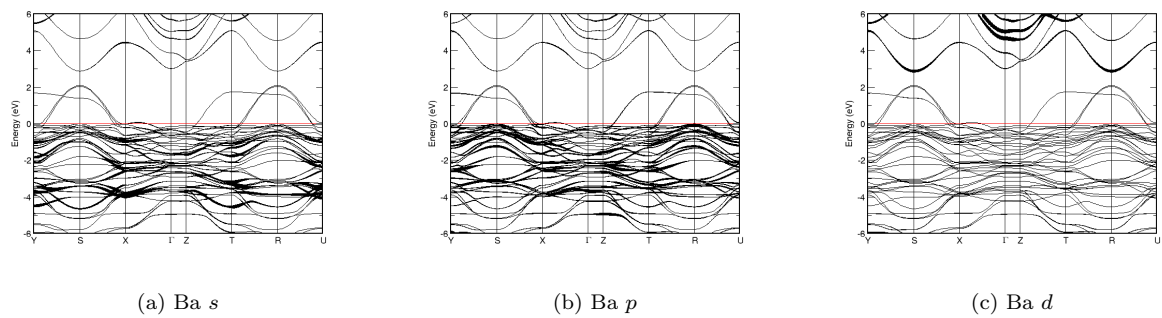

(a) Ba $s$                    (b) Ba $p$                    (c) Ba $d$

FIG. 193: Fat band representation of Ba in SmBa$_2$Cu$_3$O$_7$

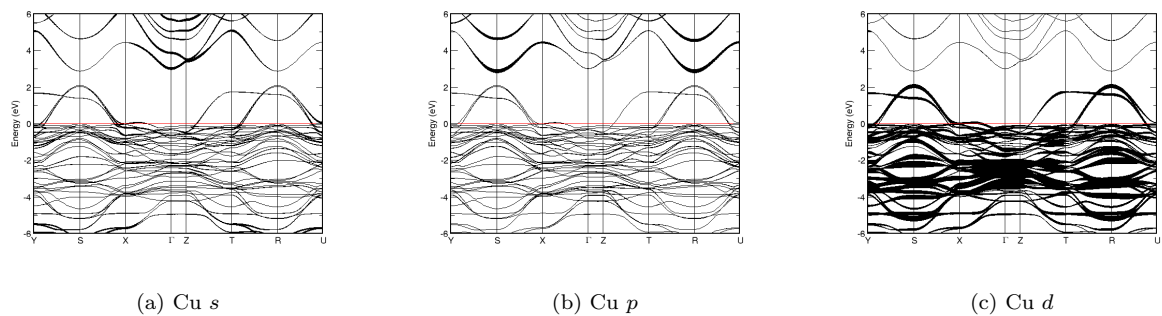

(a) Cu $s$                    (b) Cu $p$                    (c) Cu $d$

FIG. 194: Fat band representation of Cu in SmBa$_2$Cu$_3$O$_7$

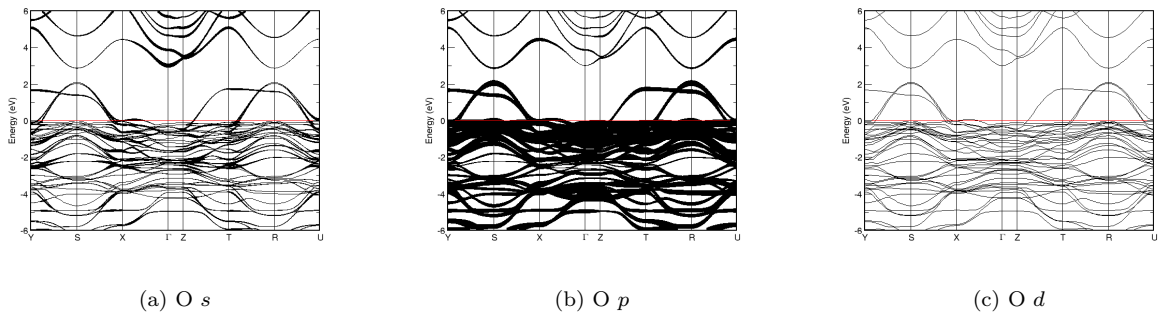

(a) O $s$      (b) O $p$      (c) O $d$

FIG. 195: Fat band representation of O in SmBa$_2$Cu$_3$O$_7$

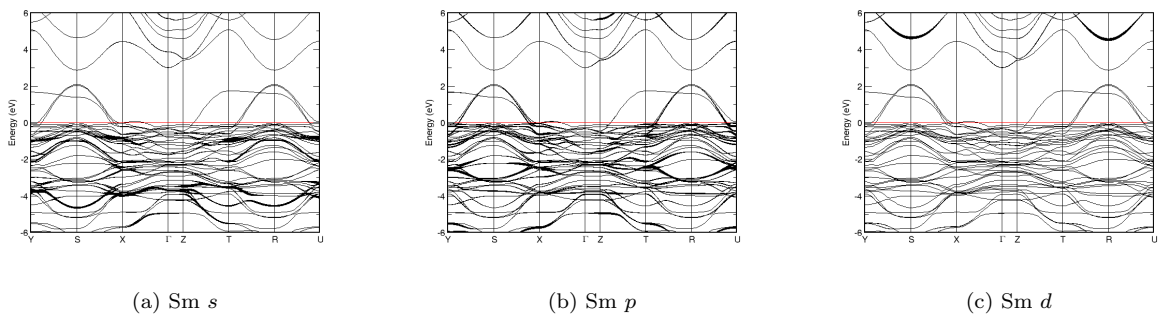

(a) Sm $s$      (b) Sm $p$      (c) Sm $d$

FIG. 196: Fat band representation of Sm in SmBa$_2$Cu$_3$O$_7$

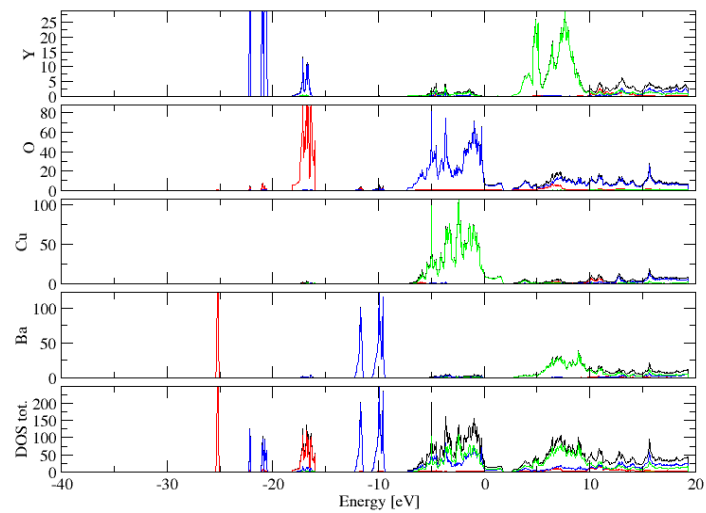

FIG. 197: (Color online) PDOS of Ba$_2$YCu$_3$O$_7$ (ICSD #202770). The $s$-, $p$- and $d$-projected states are in red, blue and green, respectively. Ba$_2$YCu$_3$O$_7$ crystallizes in space group P m m m (#47), in a orthorhombic primitive structure.

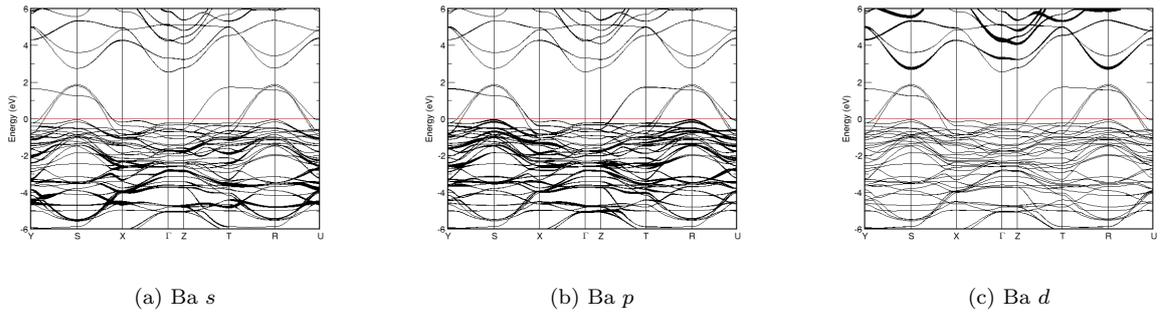

(a) Ba $s$        (b) Ba $p$        (c) Ba $d$

FIG. 198: Fat band representation of Ba in $Ba_2YCu_3O_7$

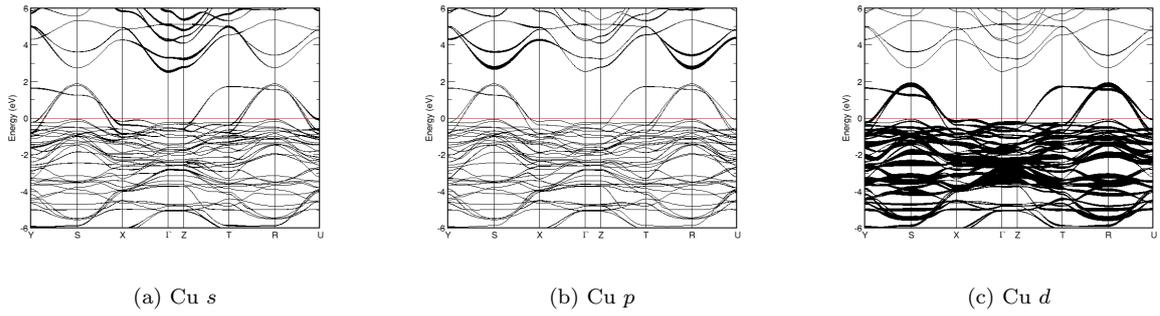

(a) Cu $s$        (b) Cu $p$        (c) Cu $d$

FIG. 199: Fat band representation of Cu in $Ba_2YCu_3O_7$

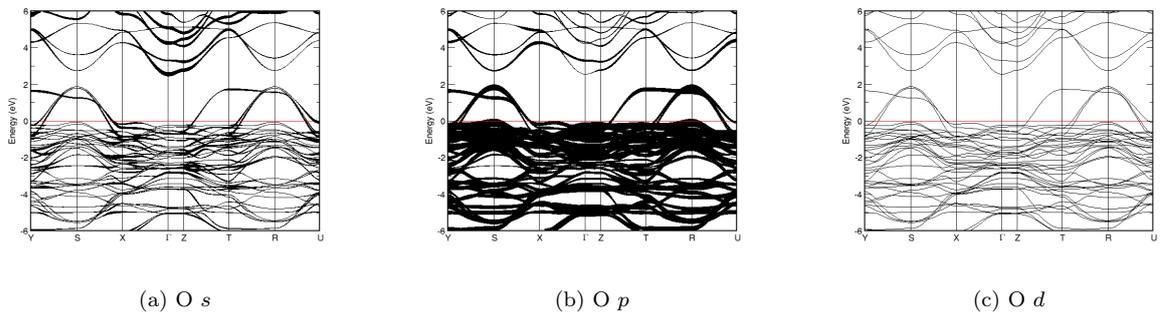

(a) O $s$        (b) O $p$        (c) O $d$

FIG. 200: Fat band representation of O in $Ba_2YCu_3O_7$

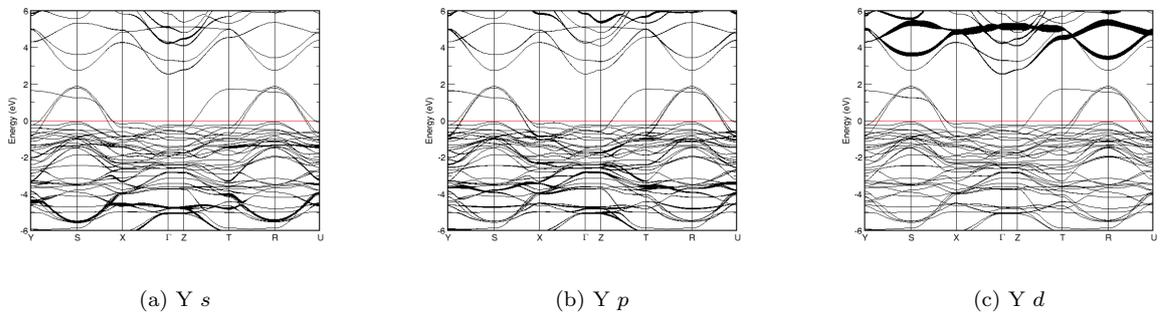

(a) Y $s$        (b) Y $p$        (c) Y $d$

FIG. 201: Fat band representation of Y in $Ba_2YCu_3O_7$

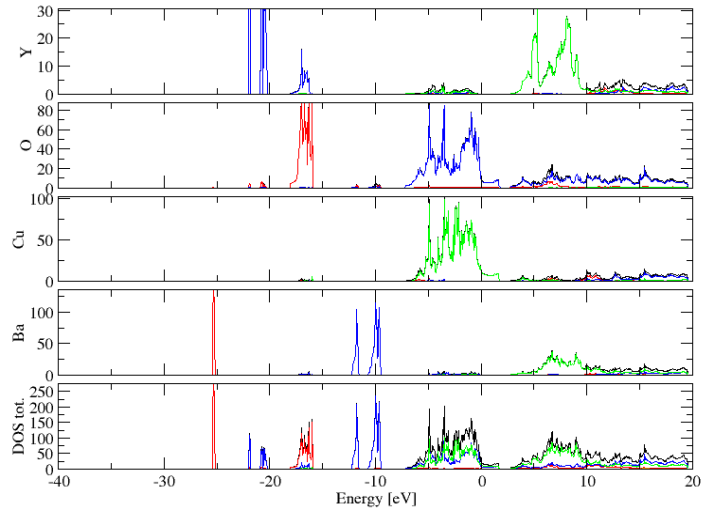

FIG. 202: (Color online) PDOS of $Ba_2YCu_3O_7$ (ICSD #77737). The $s$-, $p$- and $d$-projected states are in red, blue and green, respectively. $Ba_2YCu_3O_7$ crystallizes in space group P m m m (#47), in a orthorhombic primitive structure.

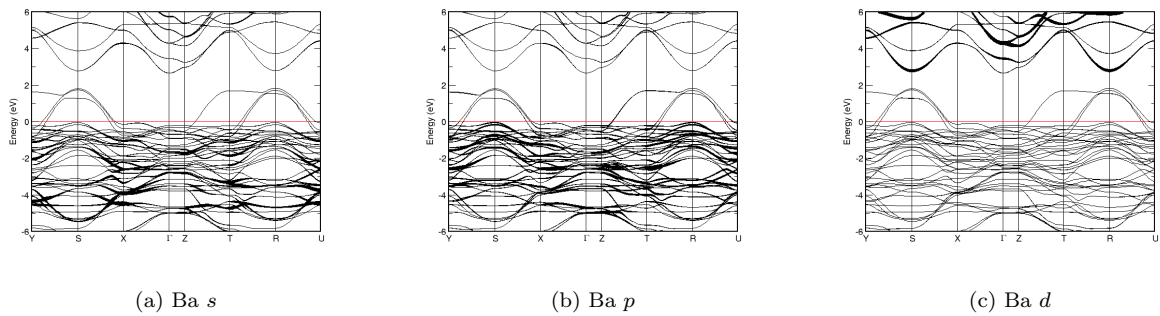

(a) Ba $s$                    (b) Ba $p$                    (c) Ba $d$

FIG. 203: Fat band representation of Ba in $Ba_2YCu_3O_7$

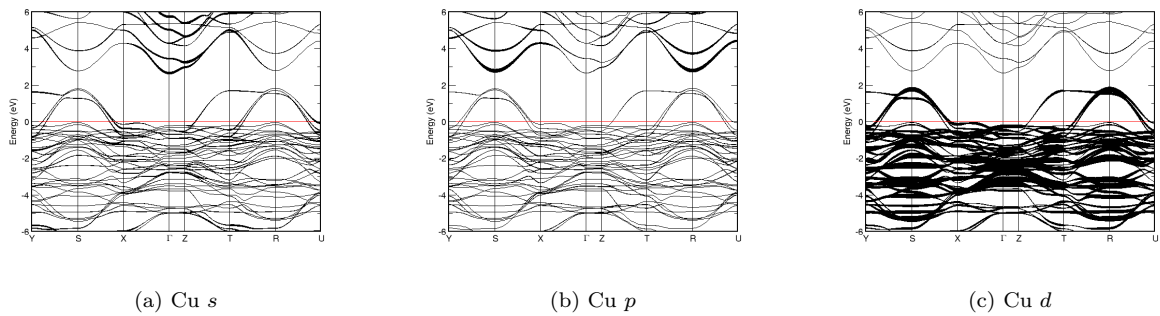

(a) Cu $s$                    (b) Cu $p$                    (c) Cu $d$

FIG. 204: Fat band representation of Cu in $Ba_2YCu_3O_7$

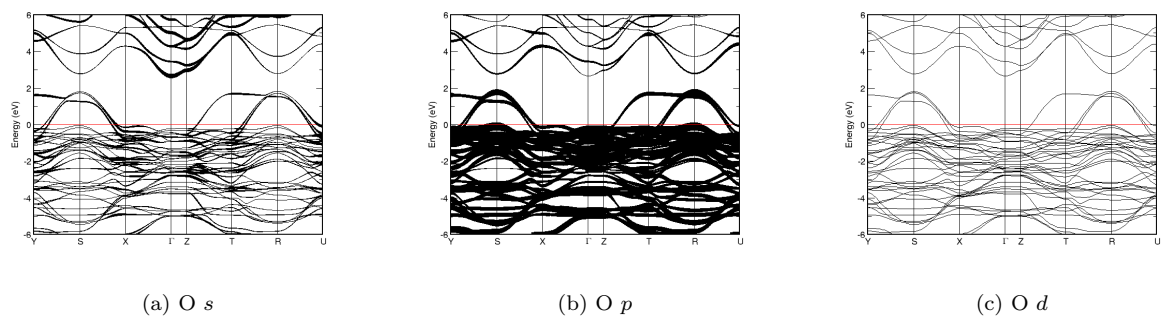

(a) O $s$        (b) O $p$        (c) O $d$

FIG. 205: Fat band representation of O in $Ba_2YCu_3O_7$

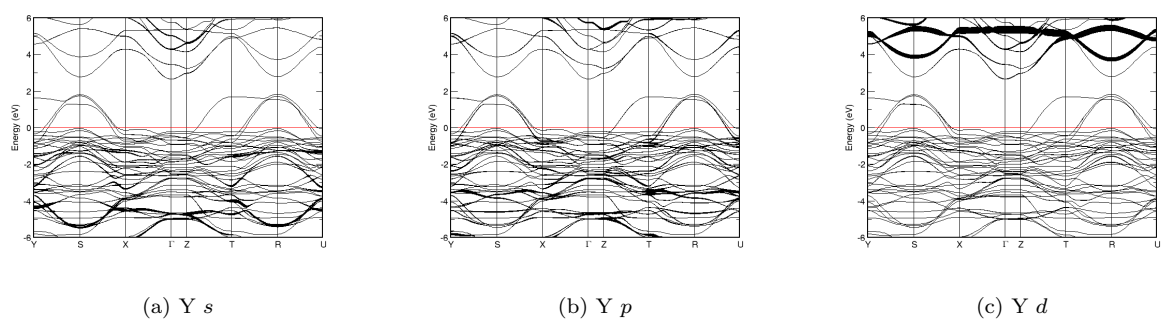

(a) Y $s$        (b) Y $p$        (c) Y $d$

FIG. 206: Fat band representation of Y in $Ba_2YCu_3O_7$

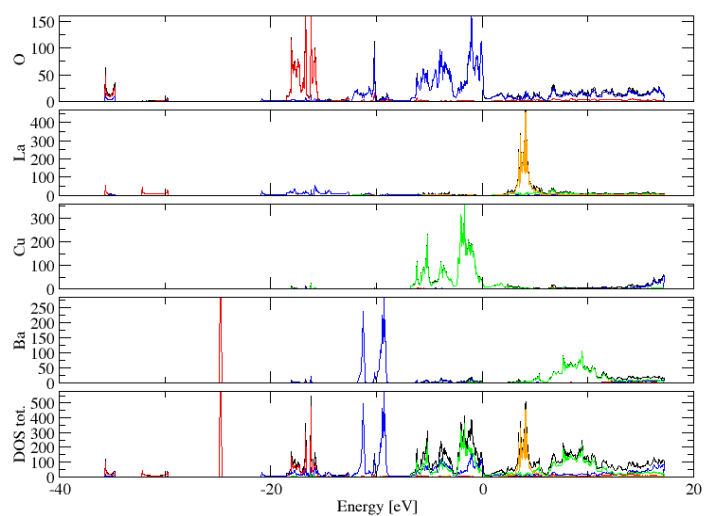

FIG. 207: (Color online) PDOS of $LaBa_2Cu_3O_8$ (ICSD #85291). The $s$-, $p$- and $d$-projected states are in red, blue and green, respectively. $LaBa_2Cu_3O_8$ crystallizes in space group P m m m (#47), in a orthorhombic primitive structure.

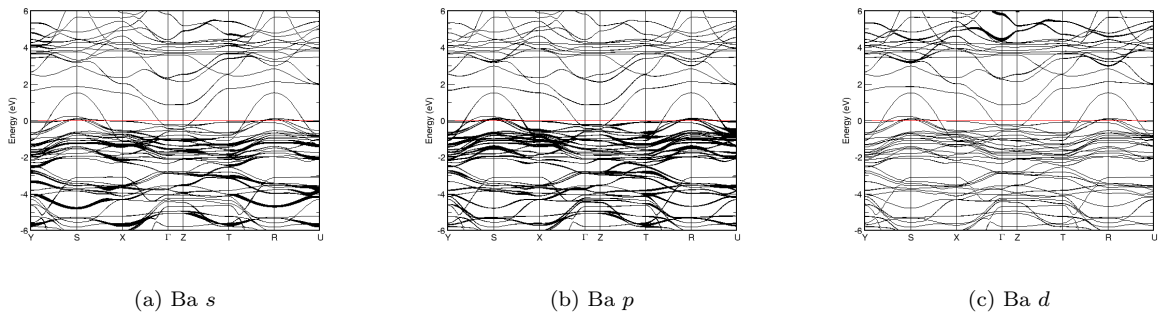

(a) Ba $s$

(b) Ba $p$

(c) Ba $d$

FIG. 208: Fat band representation of Ba in LaBa$_2$Cu$_3$O$_8$

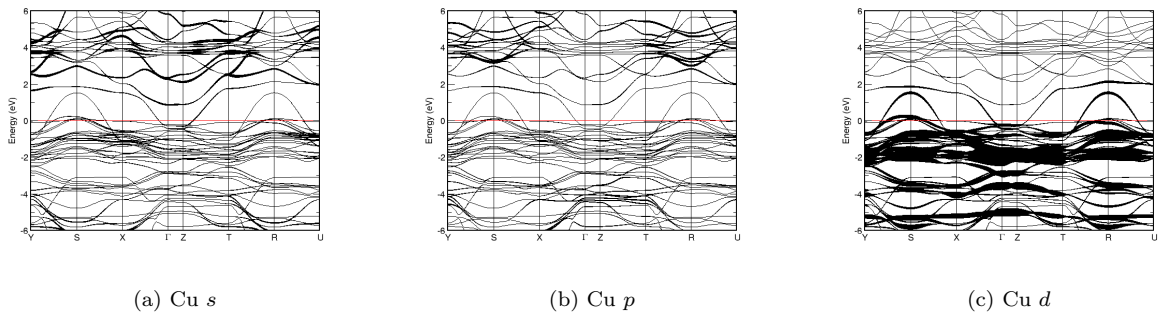

(a) Cu $s$

(b) Cu $p$

(c) Cu $d$

FIG. 209: Fat band representation of Cu in LaBa$_2$Cu$_3$O$_8$

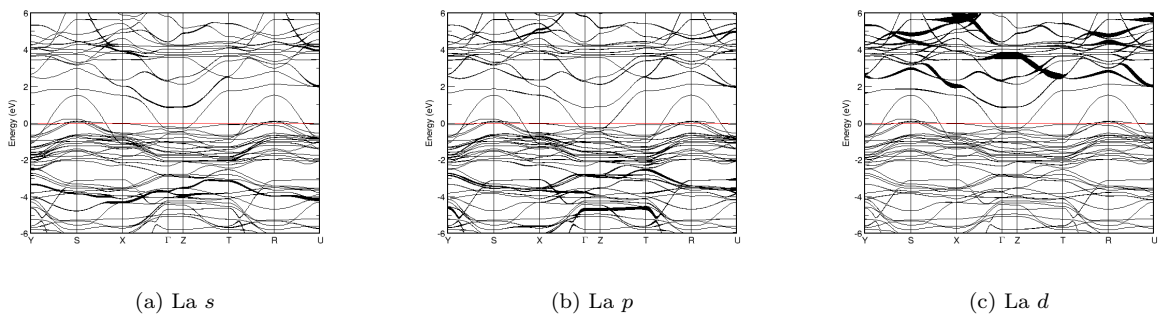

(a) La $s$

(b) La $p$

(c) La $d$

FIG. 210: Fat band representation of La in LaBa$_2$Cu$_3$O$_8$

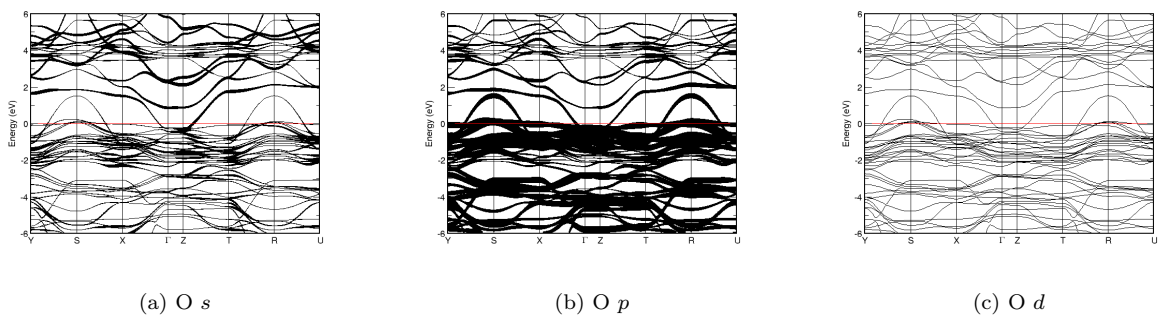

(a) O $s$

(b) O $p$

(c) O $d$

FIG. 211: Fat band representation of O in LaBa$_2$Cu$_3$O$_8$

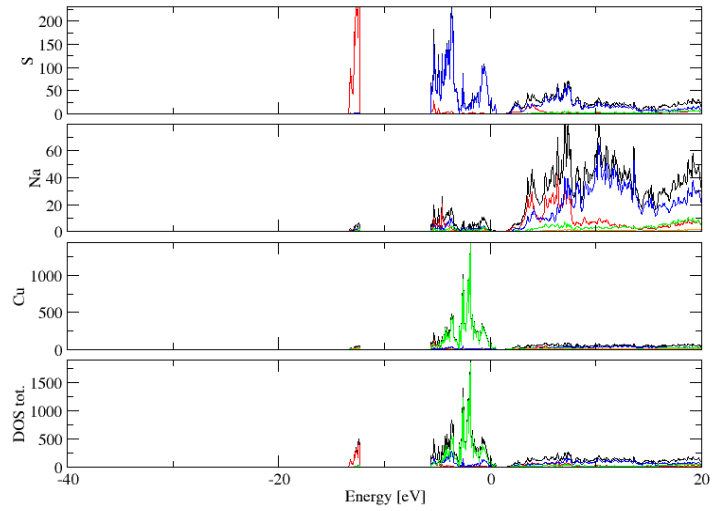

FIG. 212: (Color online) PDOS of Na$_3$Cu$_4$S$_4$ (ICSD #10004). The *s*-, *p*- and *d*-projected states are in red, blue and green, respectively. Na$_3$Cu$_4$S$_4$ crystallizes in space group P b a m (#55), in a orthorhombic primitive structure.

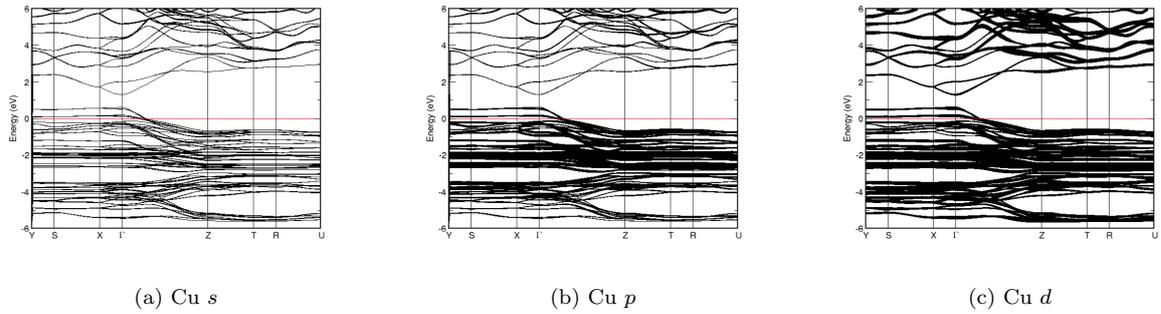

(a) Cu *s*          (b) Cu *p*          (c) Cu *d*

FIG. 213: Fat band representation of Cu in Na$_3$Cu$_4$S$_4$

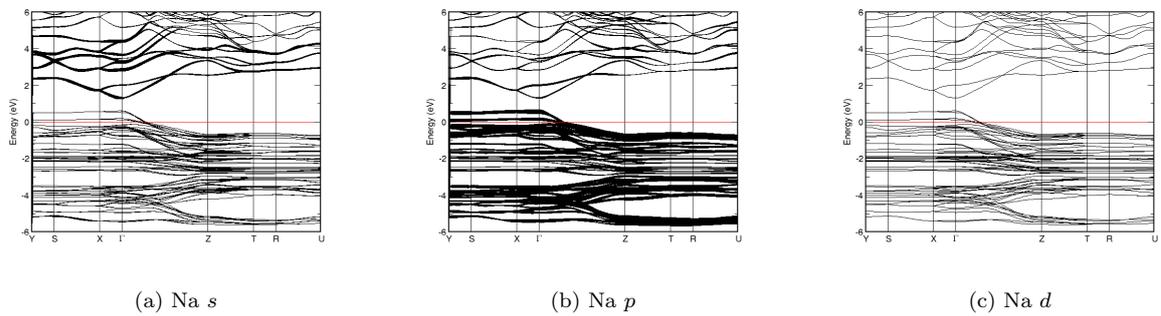

(a) Na *s*          (b) Na *p*          (c) Na *d*

FIG. 214: Fat band representation of Na in Na$_3$Cu$_4$S$_4$

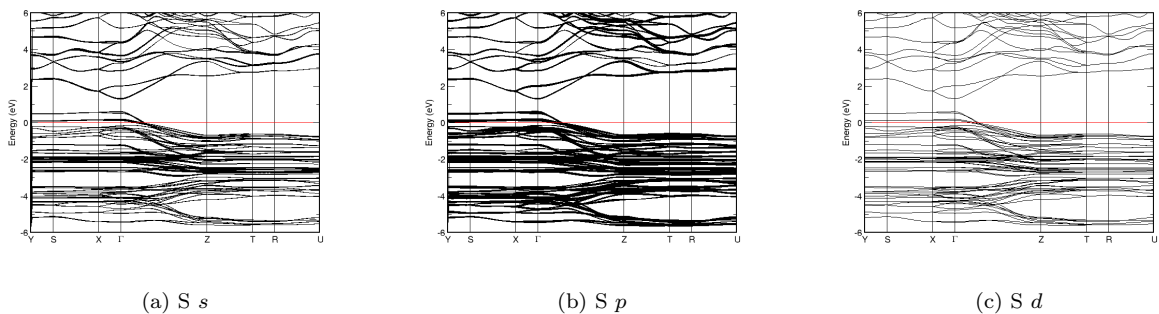

(a) S *s*

(b) S *p*

(c) S *d*

FIG. 215: Fat band representation of S in Na$_3$Cu$_4$S$_4$

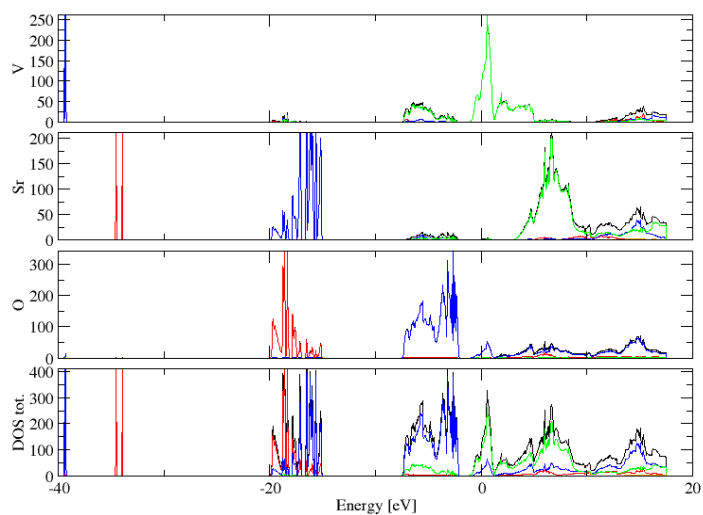

FIG. 216: (Color online) PDOS of Sr$_4$V$_3$O$_{10}$ (ICSD #73698). The *s*-, *p*- and *d*-projected states are in red, blue and green, respectively. Sr$_4$V$_3$O$_{10}$ crystallizes in space group I 4/m m m (#139), in a tetragonal body-centred structure.

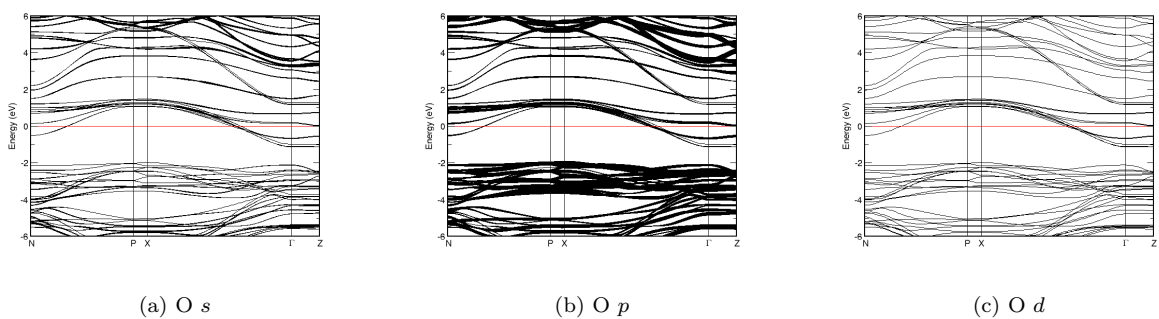

(a) O *s*

(b) O *p*

(c) O *d*

FIG. 217: Fat band representation of O in Sr$_4$V$_3$O$_{10}$

(a) Sr $s$       (b) Sr $p$       (c) Sr $d$

FIG. 218: Fat band representation of Sr in $Sr_4V_3O_{10}$

(a) V $s$       (b) V $p$       (c) V $d$

FIG. 219: Fat band representation of V in $Sr_4V_3O_{10}$

FIG. 220: (Color online) PDOS of MoB (ICSD #24280). The $s$-, $p$- and $d$-projected states are in red, blue and green, respectively. MoB crystallizes in space group I 41/a m d S (#141), in a tetragonal body-centred structure.

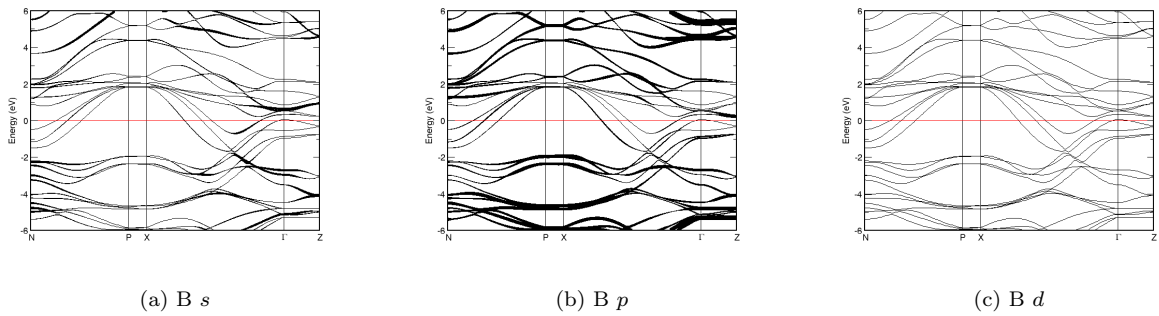

(a) B $s$             (b) B $p$             (c) B $d$

FIG. 221: Fat band representation of B in MoB

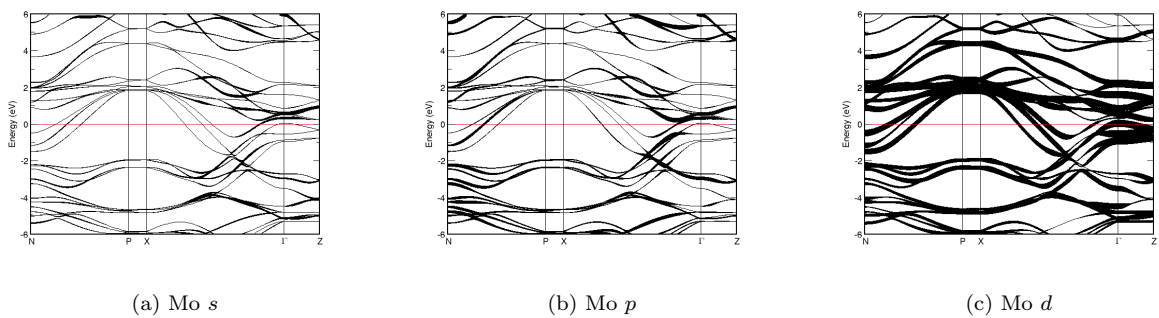

(a) Mo $s$             (b) Mo $p$             (c) Mo $d$

FIG. 222: Fat band representation of Mo in MoB

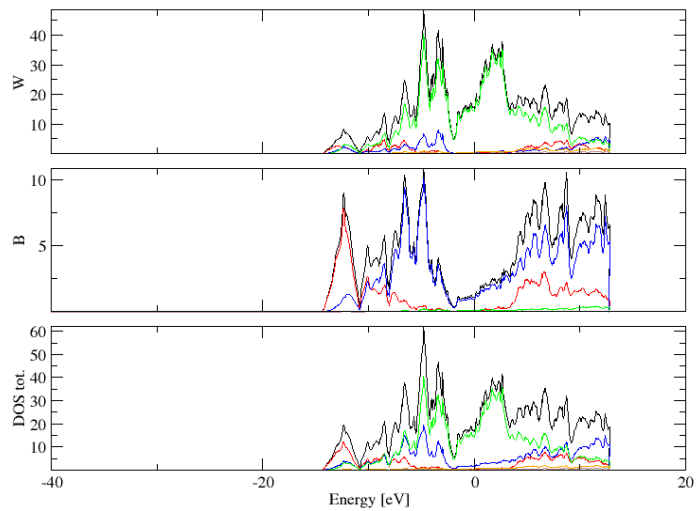

FIG. 223: (Color online) PDOS of WB (ICSD #24281). The $s$-, $p$- and $d$-projected states are in red, blue and green, respectively. WB crystallizes in space group I 41/a m d S (#141), in a tetragonal body-centred structure.

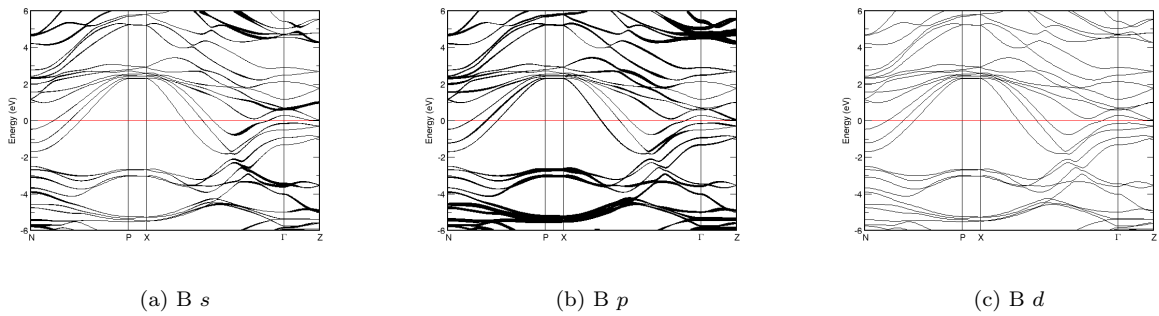

(a) B $s$        (b) B $p$        (c) B $d$

FIG. 224: Fat band representation of B in WB

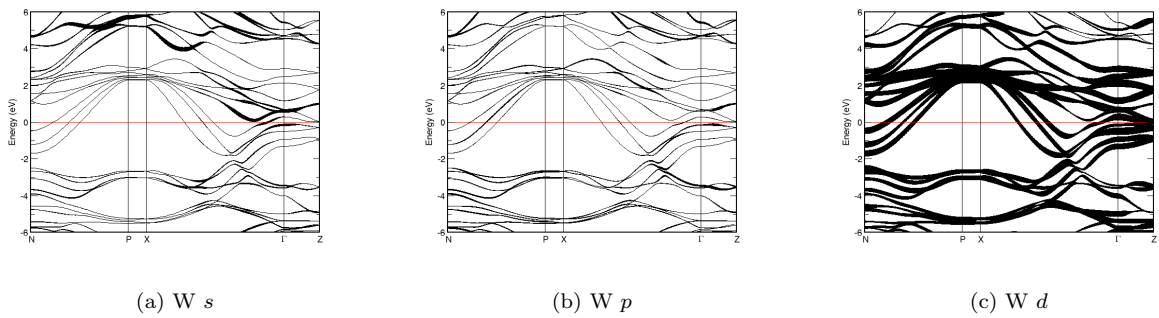

(a) W $s$        (b) W $p$        (c) W $d$

FIG. 225: Fat band representation of W in WB

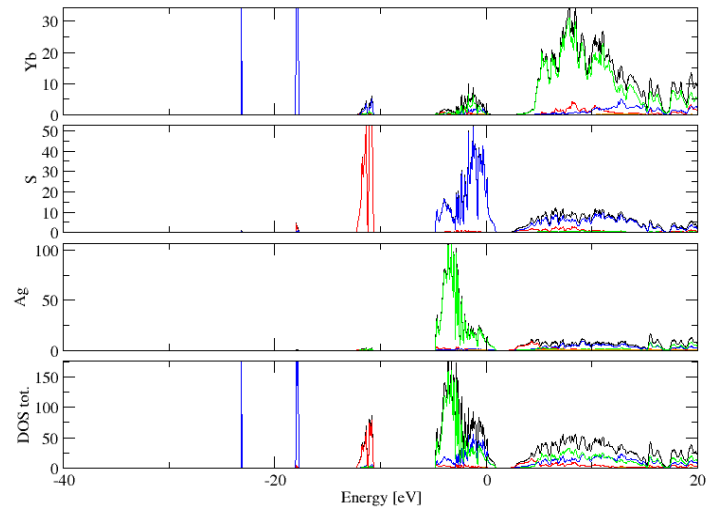

FIG. 226: (Color online) PDOS of Yb(AgS$_2$) (ICSD #27091). The $s$-, $p$- and $d$-projected states are in red, blue and green, respectively. Yb(AgS$_2$) crystallizes in space group I 41 m d (#109), in a tetragonal body-centred structure.

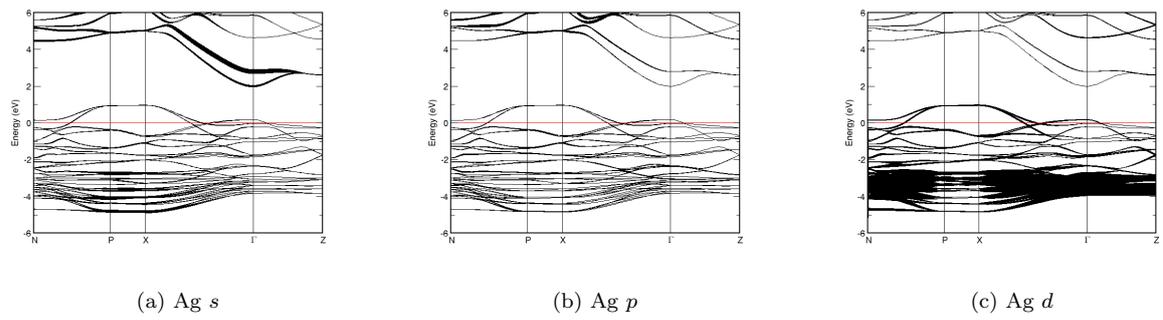

(a) Ag $s$

(b) Ag $p$

(c) Ag $d$

FIG. 227: Fat band representation of Ag in Yb(AgS$_2$)

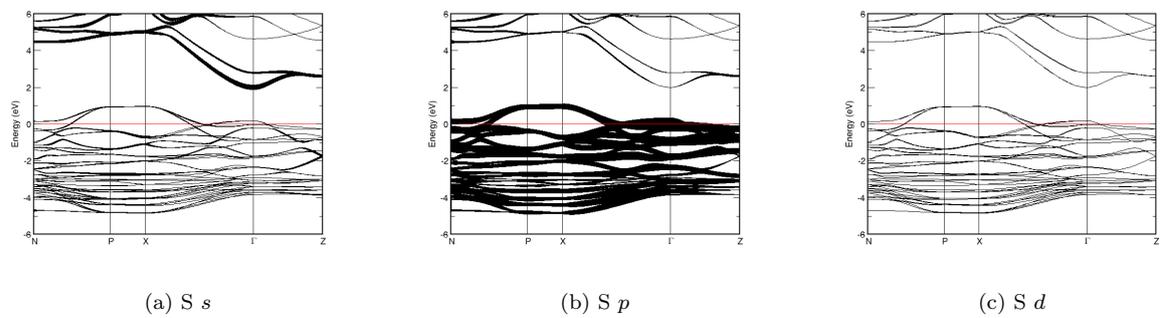

(a) S $s$

(b) S $p$

(c) S $d$

FIG. 228: Fat band representation of S in Yb(AgS$_2$)

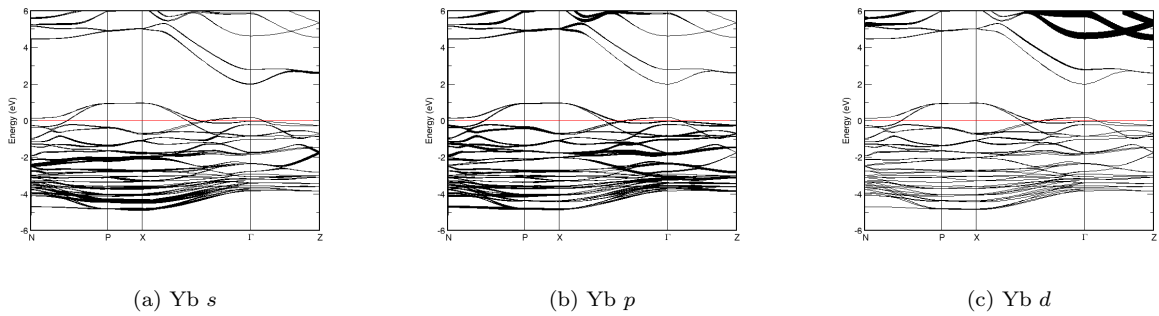

(a) Yb $s$

(b) Yb $p$

(c) Yb $d$

FIG. 229: Fat band representation of Yb in Yb(AgS$_2$)

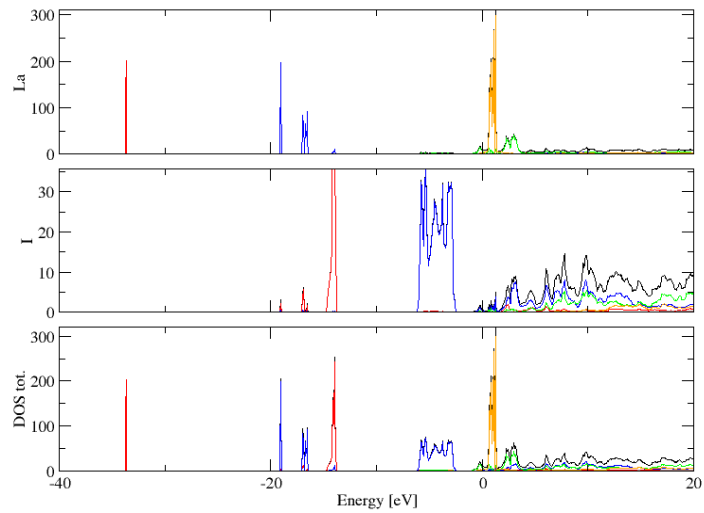

FIG. 230: (Color online) PDOS of LaI$_2$ (ICSD #202452). The $s$-, $p$- and $d$-projected states are in red, blue and green, respectively. LaI$_2$ crystallizes in space group I 4/m m m (#139), in a tetragonal body-centred structure.

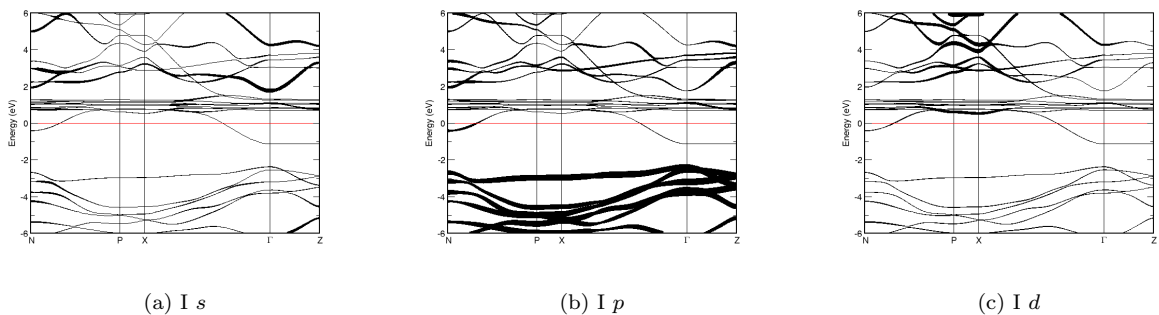

(a) I $s$

(b) I $p$

(c) I $d$

FIG. 231: Fat band representation of I in LaI$_2$

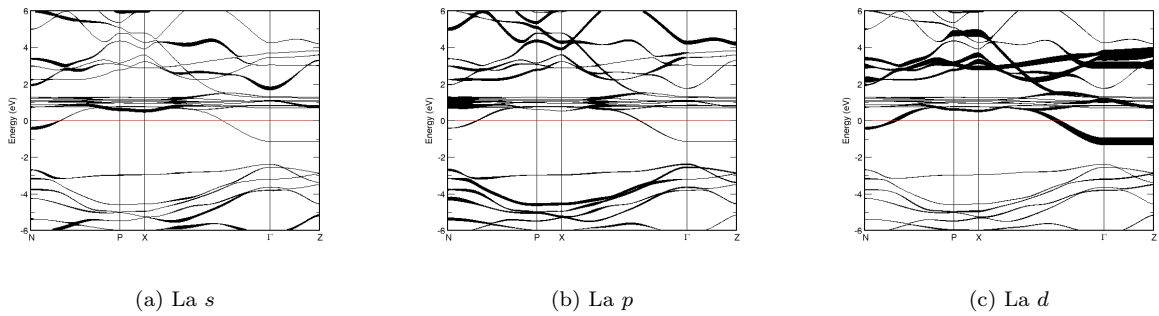

(a) La $s$        (b) La $p$        (c) La $d$

FIG. 232: Fat band representation of La in LaI$_2$

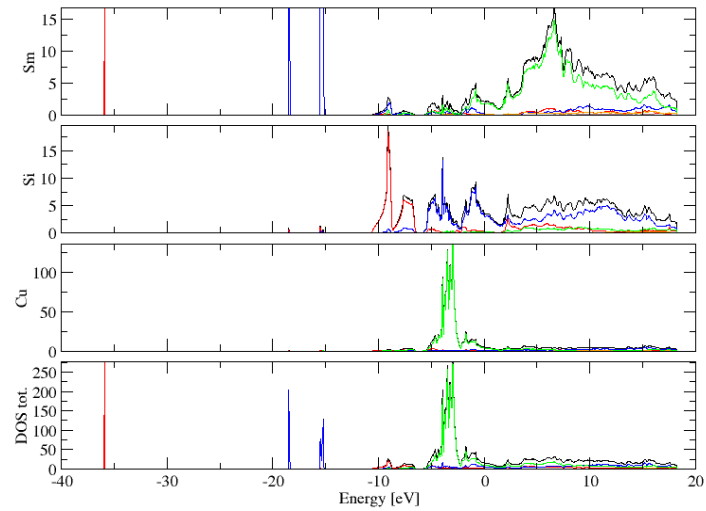

FIG. 233: (Color online) PDOS of SmCu$_2$Si$_2$ (ICSD #106843). The $s$-, $p$- and $d$-projected states are in red, blue and green, respectively. SmCu$_2$Si$_2$ crystallizes in space group I 4/m m m (#139), in a tetragonal body-centred structure.

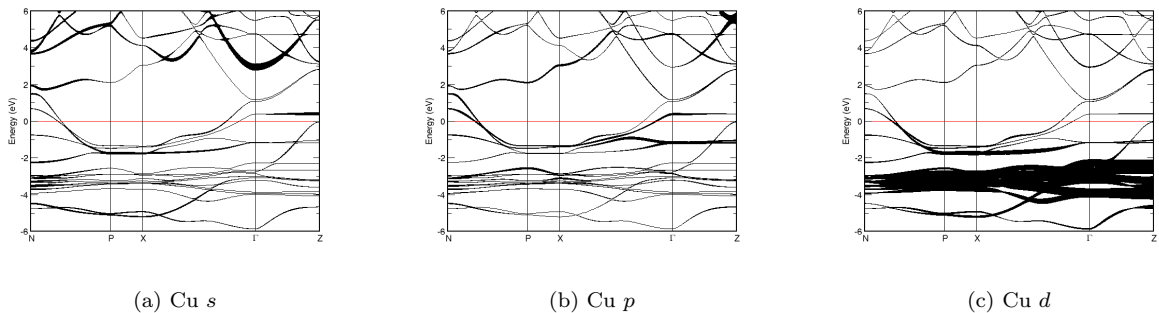

(a) Cu $s$        (b) Cu $p$        (c) Cu $d$

FIG. 234: Fat band representation of Cu in SmCu$_2$Si$_2$

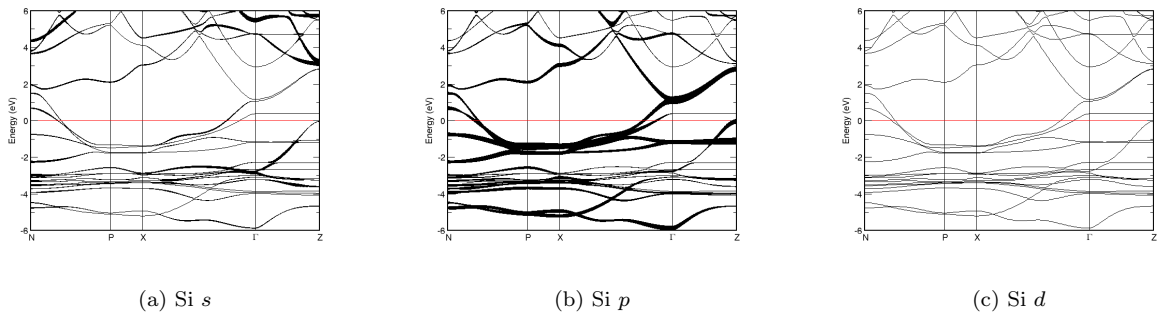

(a) Si $s$       (b) Si $p$       (c) Si $d$

FIG. 235: Fat band representation of Si in SmCu$_2$Si$_2$

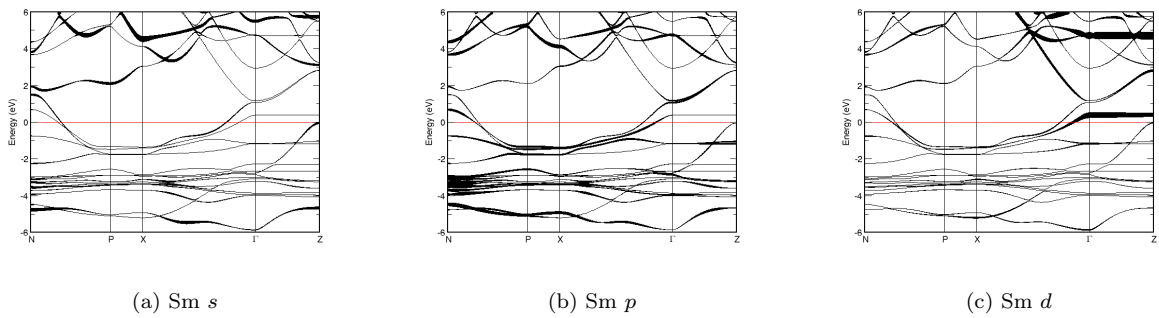

(a) Sm $s$       (b) Sm $p$       (c) Sm $d$

FIG. 236: Fat band representation of Sm in SmCu$_2$Si$_2$

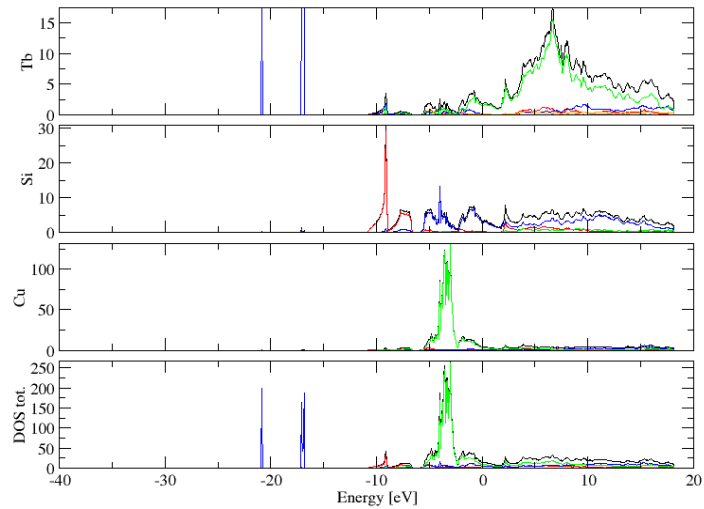

FIG. 237: (Color online) PDOS of TbCu$_2$Si$_2$ (ICSD #106844). The $s$-, $p$- and $d$-projected states are in red, blue and green, respectively. TbCu$_2$Si$_2$ crystallizes in space group I 4/m m m (#139), in a tetragonal body-centred structure.

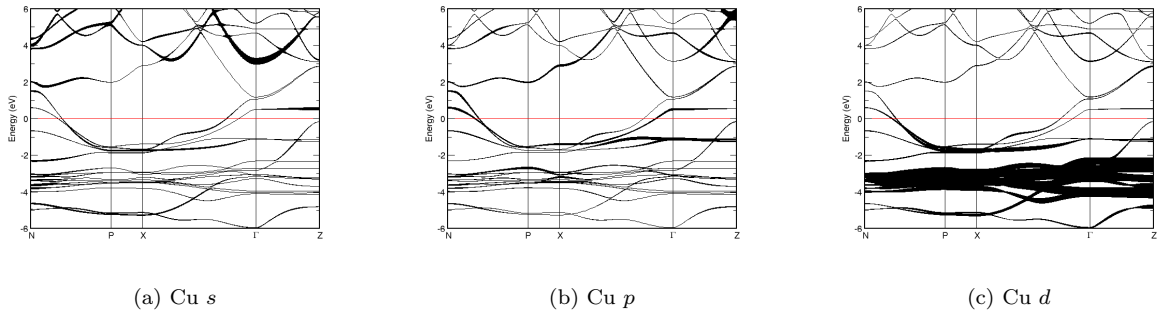

(a) Cu $s$        (b) Cu $p$        (c) Cu $d$

FIG. 238: Fat band representation of Cu in TbCu$_2$Si$_2$

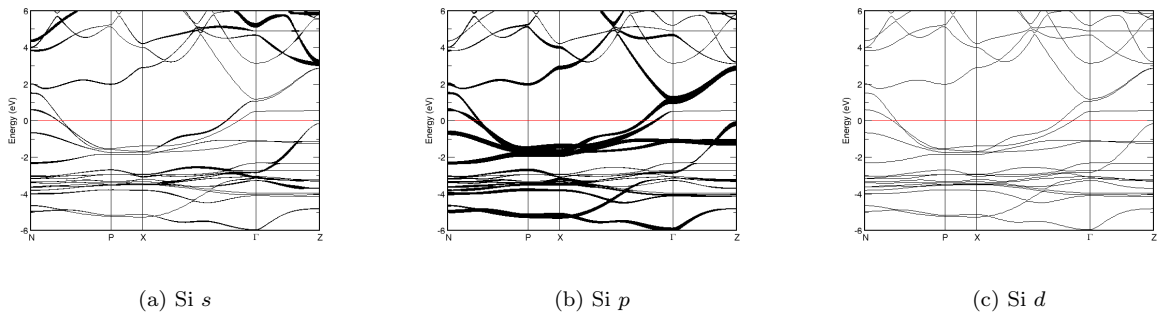

(a) Si $s$        (b) Si $p$        (c) Si $d$

FIG. 239: Fat band representation of Si in TbCu$_2$Si$_2$

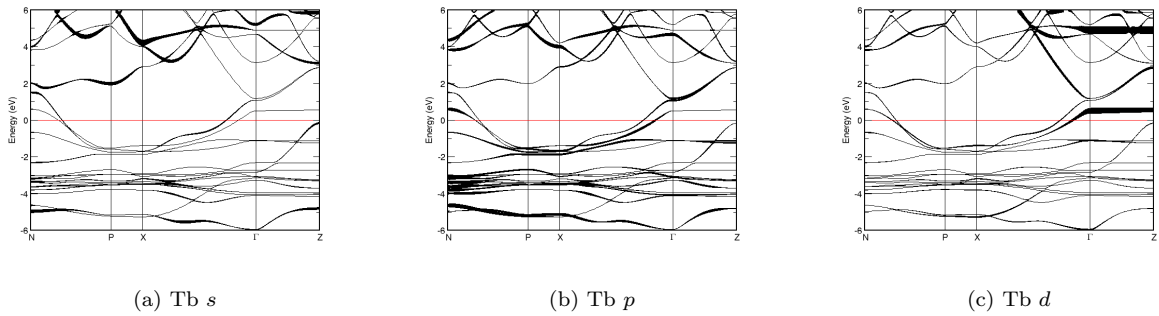

(a) Tb $s$        (b) Tb $p$        (c) Tb $d$

FIG. 240: Fat band representation of Tb in TbCu$_2$Si$_2$

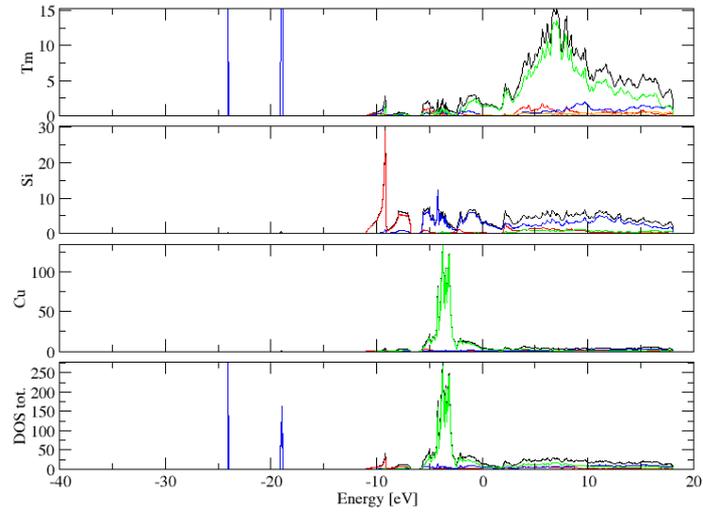

FIG. 241: (Color online) PDOS of Cu$_2$TmSi$_2$ (ICSD #53349). The $s$-, $p$- and $d$-projected states are in red, blue and green, respectively. Cu$_2$TmSi$_2$ crystallizes in space group I 4/m m m (#139), in a tetragonal body-centred structure.

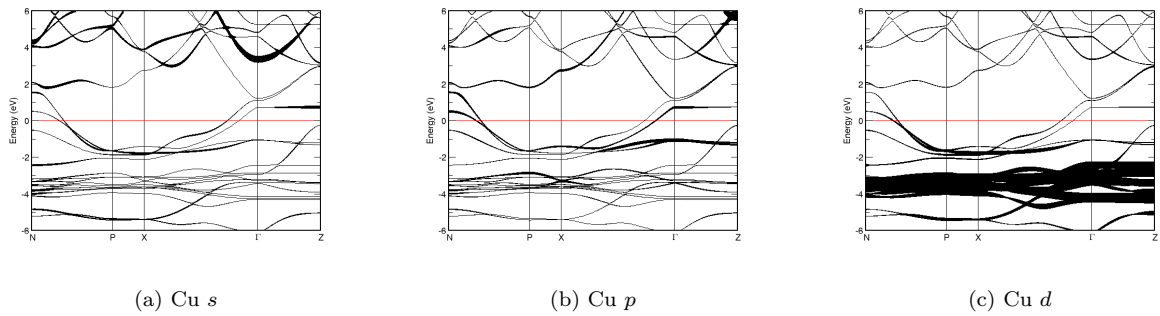

(a) Cu $s$        (b) Cu $p$        (c) Cu $d$

FIG. 242: Fat band representation of Cu in Cu$_2$TmSi$_2$

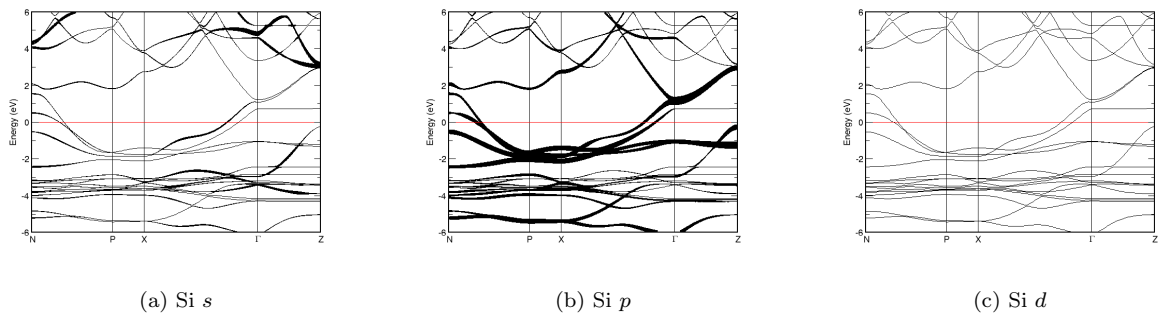

(a) Si $s$        (b) Si $p$        (c) Si $d$

FIG. 243: Fat band representation of Si in Cu$_2$TmSi$_2$

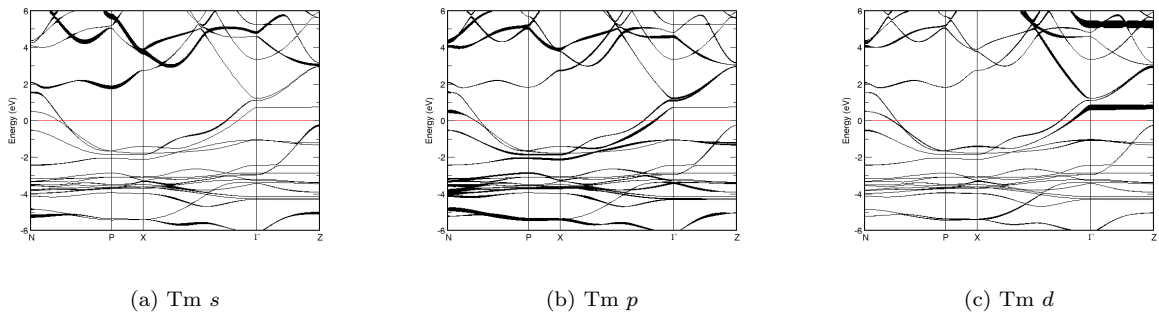

(a) Tm $s$       (b) Tm $p$       (c) Tm $d$

FIG. 244: Fat band representation of Tm in $Cu_2TmSi_2$

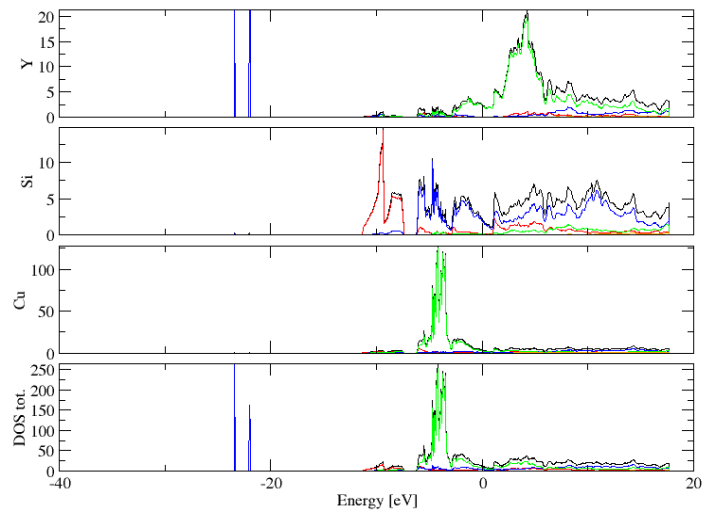

FIG. 245: (Color online) PDOS of $Cu_2YSi_2$ (ICSD #23551). The $s$-, $p$- and $d$-projected states are in red, blue and green, respectively. $Cu_2YSi_2$ crystallizes in space group I 4/m m m (#139), in a tetragonal body-centred structure.

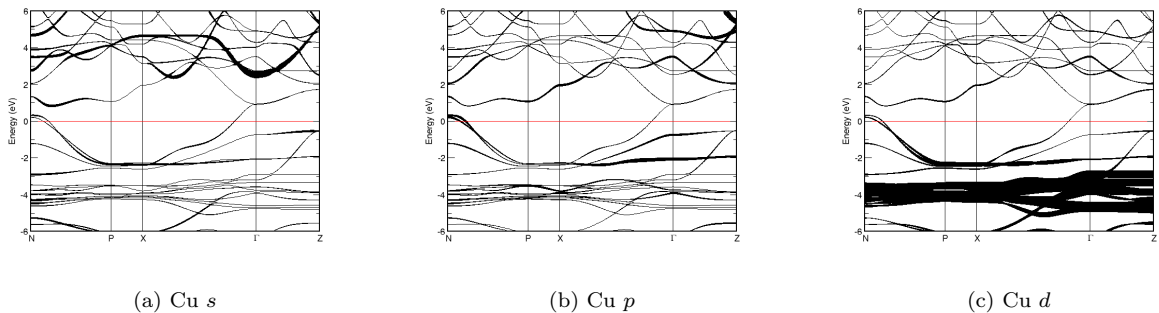

(a) Cu $s$       (b) Cu $p$       (c) Cu $d$

FIG. 246: Fat band representation of Cu in $Cu_2YSi_2$

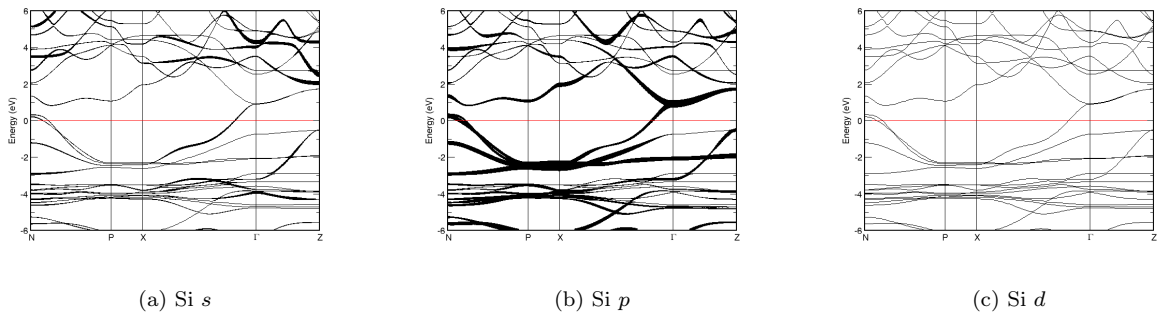

(a) Si $s$                (b) Si $p$                (c) Si $d$

FIG. 247: Fat band representation of Si in $Cu_2YSi_2$

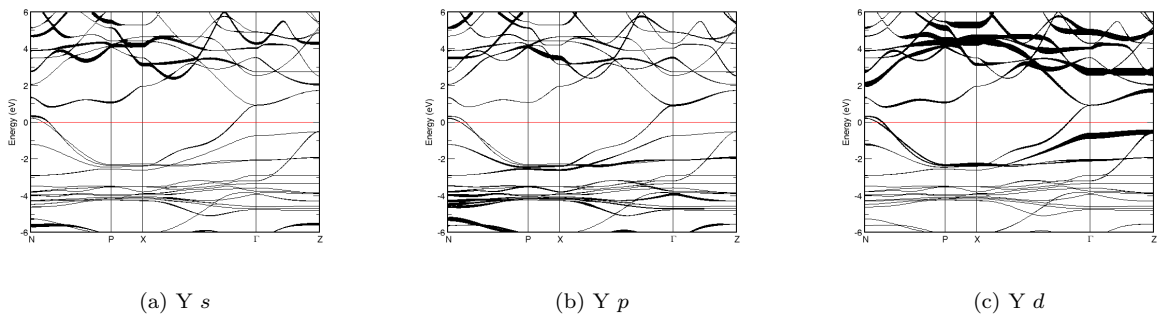

(a) Y $s$                (b) Y $p$                (c) Y $d$

FIG. 248: Fat band representation of Y in $Cu_2YSi_2$

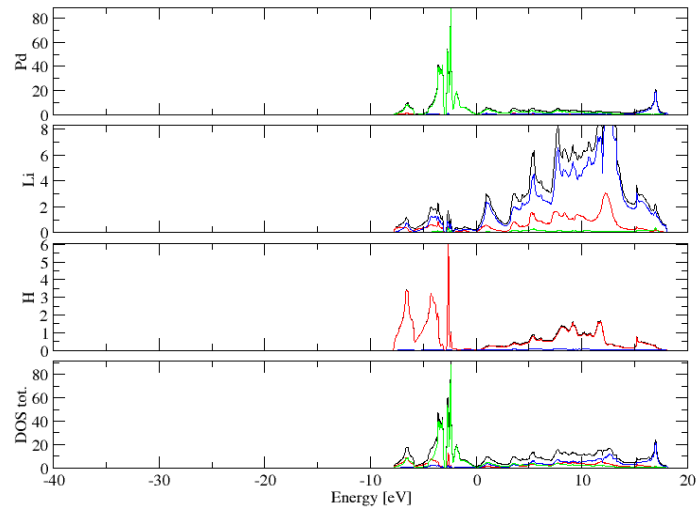

FIG. 249: (Color online) PDOS of $Li_2PdH_2$ (ICSD #108534). The $s$-, $p$- and $d$-projected states are in red, blue and green, respectively. $Li_2PdH_2$ crystallizes in space group I 4/m m m (#139), in a tetragonal body-centred structure.

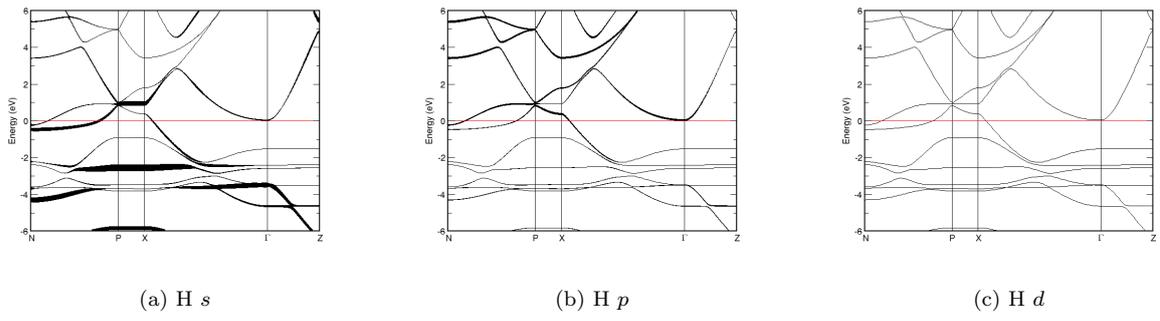

(a) H $s$        (b) H $p$        (c) H $d$

FIG. 250: Fat band representation of H in Li$_2$PdH$_2$

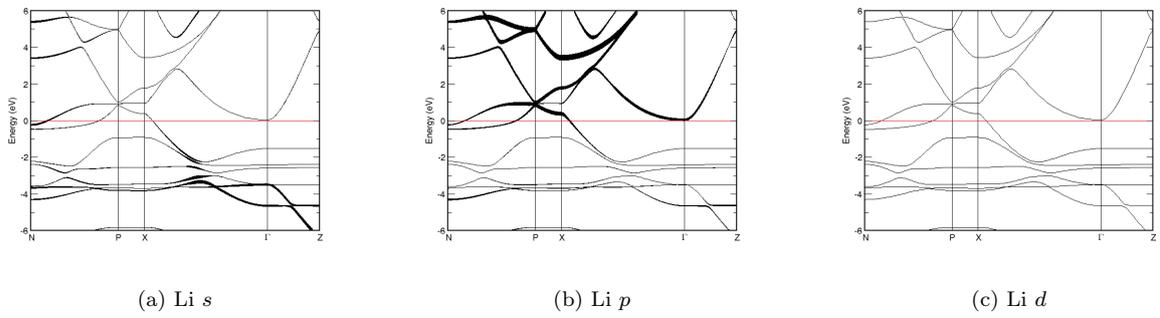

(a) Li $s$        (b) Li $p$        (c) Li $d$

FIG. 251: Fat band representation of Li in Li$_2$PdH$_2$

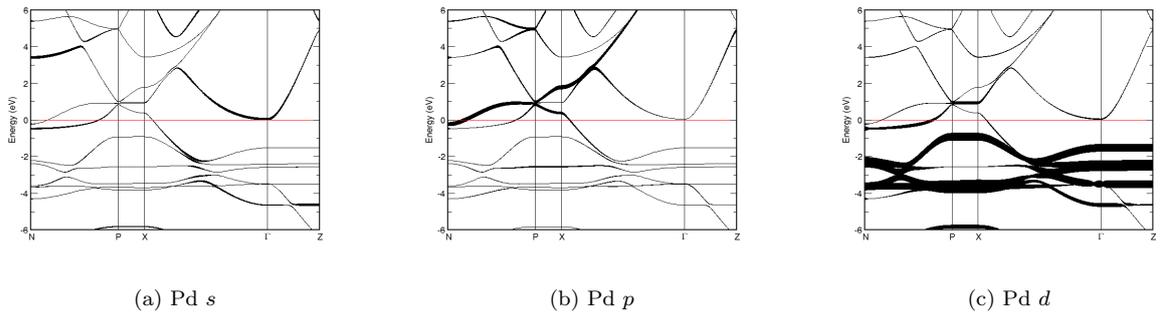

(a) Pd $s$        (b) Pd $p$        (c) Pd $d$

FIG. 252: Fat band representation of Pd in Li$_2$PdH$_2$

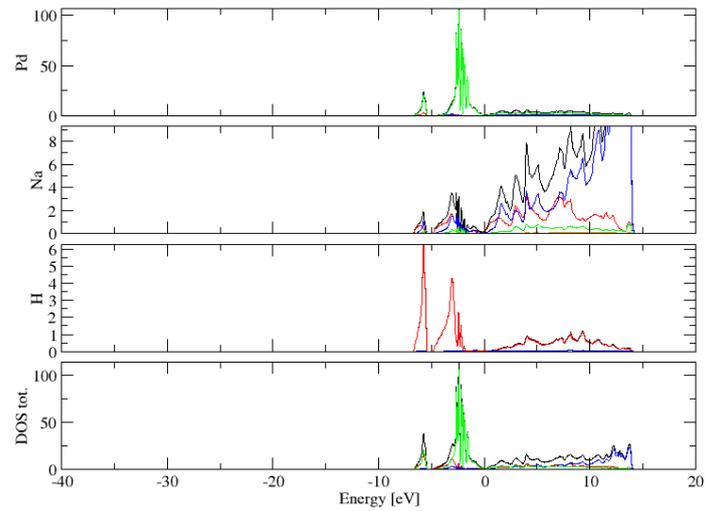

FIG. 253: (Color online) PDOS of Na$_2$PdH$_2$ (ICSD #68071). The $s$-, $p$- and $d$-projected states are in red, blue and green, respectively. Na$_2$PdH$_2$ crystallizes in space group I 4/m m m (#139), in a tetragonal body-centred structure.

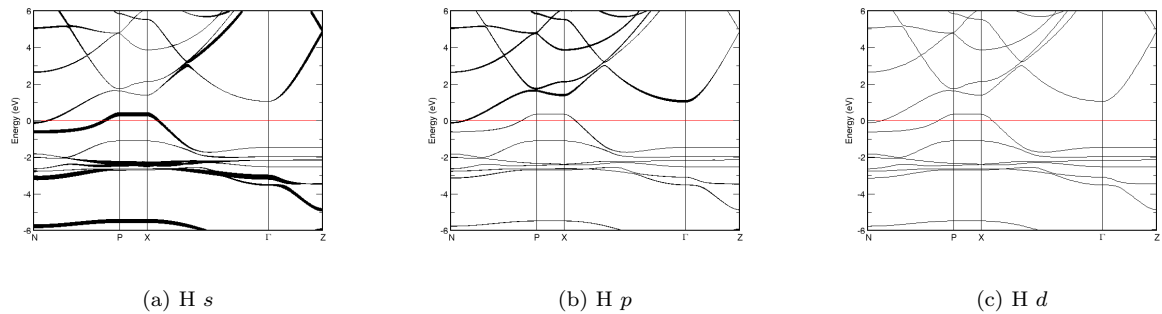

(a) H $s$        (b) H $p$        (c) H $d$

FIG. 254: Fat band representation of H in Na$_2$PdH$_2$

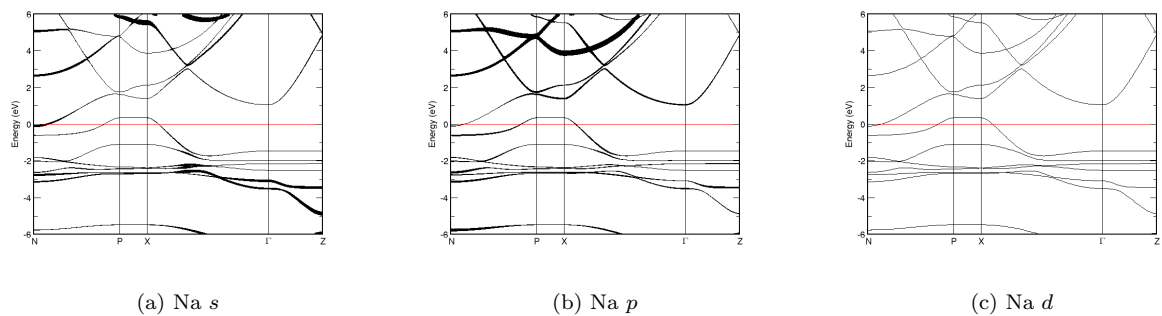

(a) Na $s$        (b) Na $p$        (c) Na $d$

FIG. 255: Fat band representation of Na in Na$_2$PdH$_2$

(a) Pd *s*

(b) Pd *p*

(c) Pd *d*

FIG. 256: Fat band representation of Pd in Na$_2$PdH$_2$

FIG. 257: (Color online) PDOS of (Cu$_2$S$_2$)(Sr$_2$NiO$_2$) (ICSD #88424). The *s*-, *p*- and *d*-projected states are in red, blue and green, respectively. (Cu$_2$S$_2$)(Sr$_2$NiO$_2$) crystallizes in space group I 4/m m m (#139), in a tetragonal body-centred structure.

(a) Cu *s*

(b) Cu *p*

(c) Cu *d*

FIG. 258: Fat band representation of Cu in (Cu$_2$S$_2$)(Sr$_2$NiO$_2$)

(a) Ni $s$

(b) Ni $p$

(c) Ni $d$

FIG. 259: Fat band representation of Ni in $(Cu_2S_2)(Sr_2NiO_2)$

(a) O $s$

(b) O $p$

(c) O $d$

FIG. 260: Fat band representation of O in $(Cu_2S_2)(Sr_2NiO_2)$

(a) S $s$

(b) S $p$

(c) S $d$

FIG. 261: Fat band representation of S in $(Cu_2S_2)(Sr_2NiO_2)$

(a) Sr $s$

(b) Sr $p$

(c) Sr $d$

FIG. 262: Fat band representation of Sr in $(Cu_2S_2)(Sr_2NiO_2)$

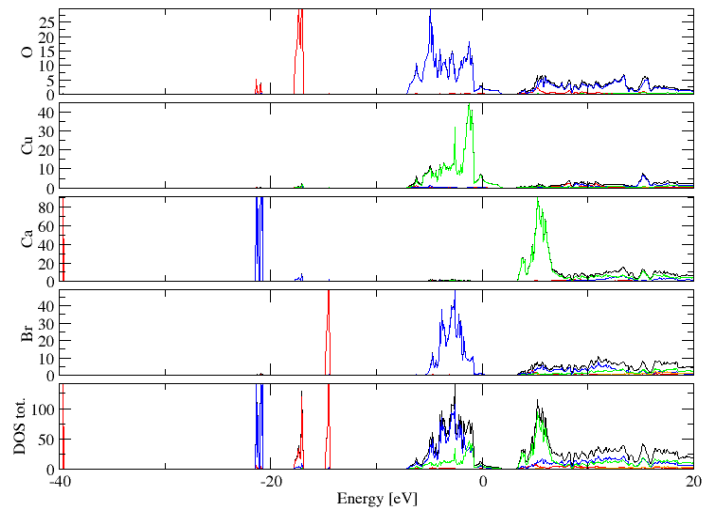

FIG. 263: (Color online) PDOS of $Ca_2(CuBr_2O_2)$ (ICSD #1028). The $s$-, $p$- and $d$-projected states are in red, blue and green, respectively. $Ca_2(CuBr_2O_2)$ crystallizes in space group I 4/m m m (#139), in a tetragonal body-centred structure.

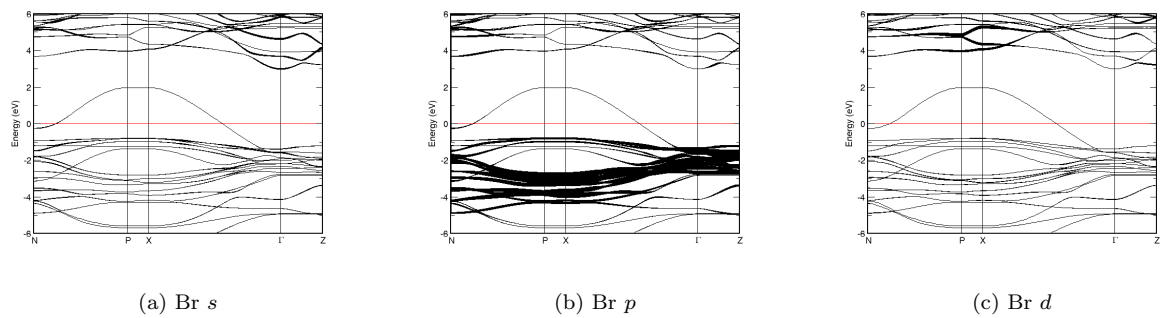

(a) Br $s$           (b) Br $p$           (c) Br $d$

FIG. 264: Fat band representation of Br in $Ca_2(CuBr_2O_2)$

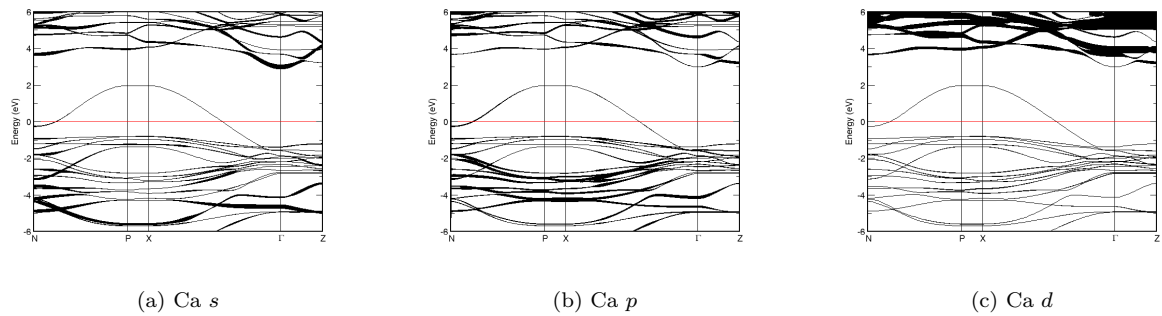

(a) Ca $s$           (b) Ca $p$           (c) Ca $d$

FIG. 265: Fat band representation of Ca in $Ca_2(CuBr_2O_2)$

(a) Cu $s$          (b) Cu $p$          (c) Cu $d$

FIG. 266: Fat band representation of Cu in $Ca_2(CuBr_2O_2)$

(a) O $s$          (b) O $p$          (c) O $d$

FIG. 267: Fat band representation of O in $Ca_2(CuBr_2O_2)$

FIG. 268: (Color online) PDOS of $Sr_2CoO_2Br_2$ (ICSD #151789). The $s$-, $p$- and $d$-projected states are in red, blue and green, respectively. $Sr_2CoO_2Br_2$ crystallizes in space group I 4/m m m (#139), in a tetragonal body-centred structure.

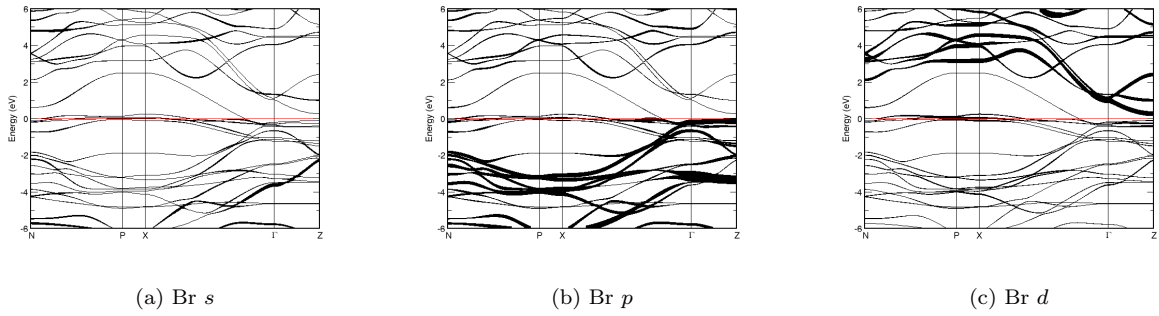

(a) Br $s$

(b) Br $p$

(c) Br $d$

FIG. 269: Fat band representation of Br in $Sr_2CoO_2Br_2$

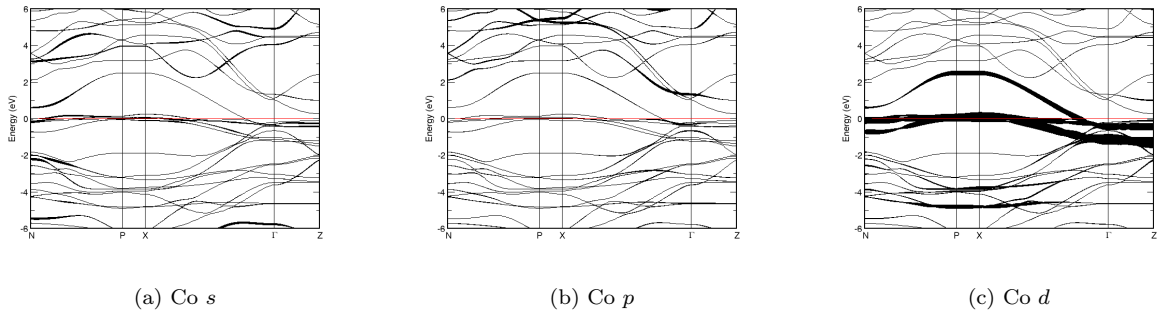

(a) Co $s$

(b) Co $p$

(c) Co $d$

FIG. 270: Fat band representation of Co in $Sr_2CoO_2Br_2$

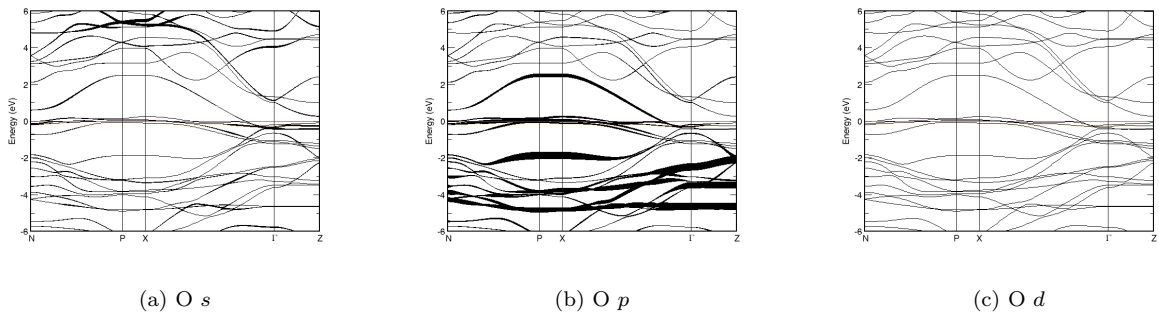

(a) O $s$

(b) O $p$

(c) O $d$

FIG. 271: Fat band representation of O in $Sr_2CoO_2Br_2$

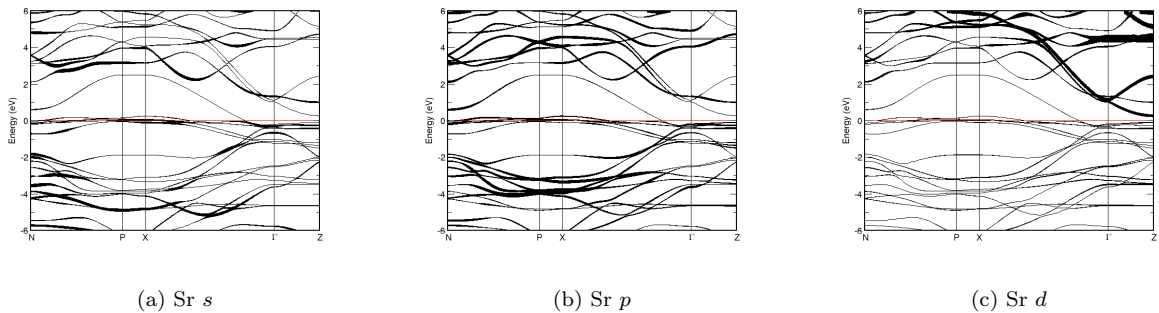

(a) Sr $s$

(b) Sr $p$

(c) Sr $d$

FIG. 272: Fat band representation of Sr in $Sr_2CoO_2Br_2$

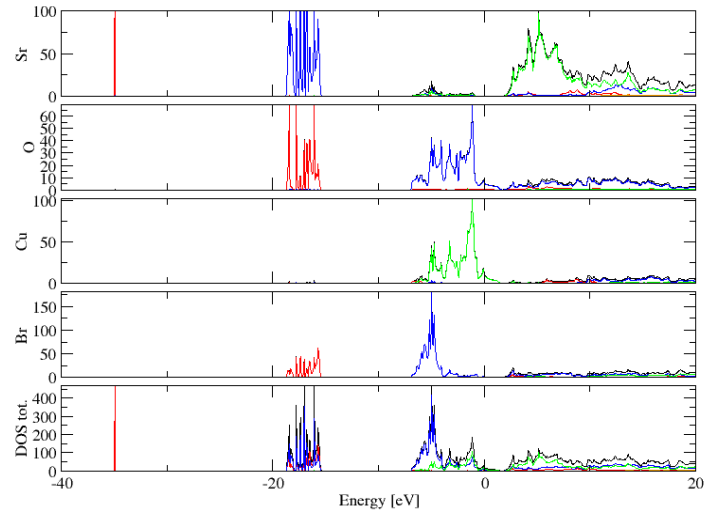

FIG. 273: (Color online) PDOS of CuSr$_2$Br$_2$O$_2$ (ICSD #1178). The $s$-, $p$- and $d$-projected states are in red, blue and green, respectively. CuSr$_2$Br$_2$O$_2$ crystallizes in space group I 4/m m m (#139), in a tetragonal body-centred structure.

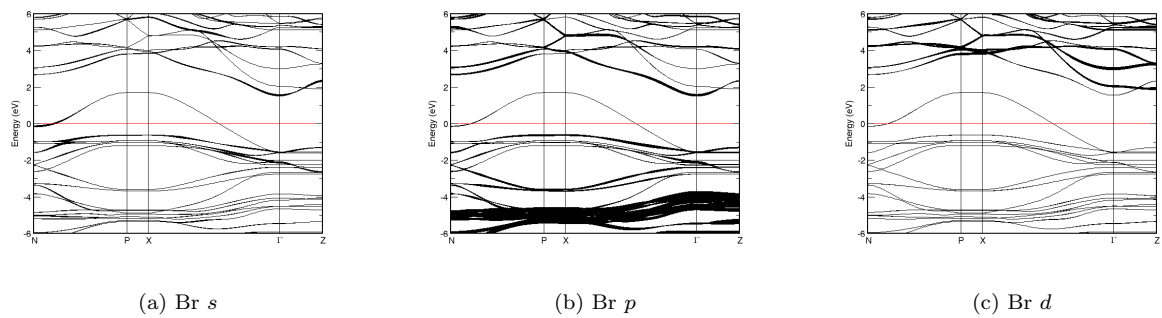

(a) Br $s$        (b) Br $p$        (c) Br $d$

FIG. 274: Fat band representation of Br in CuSr$_2$Br$_2$O$_2$

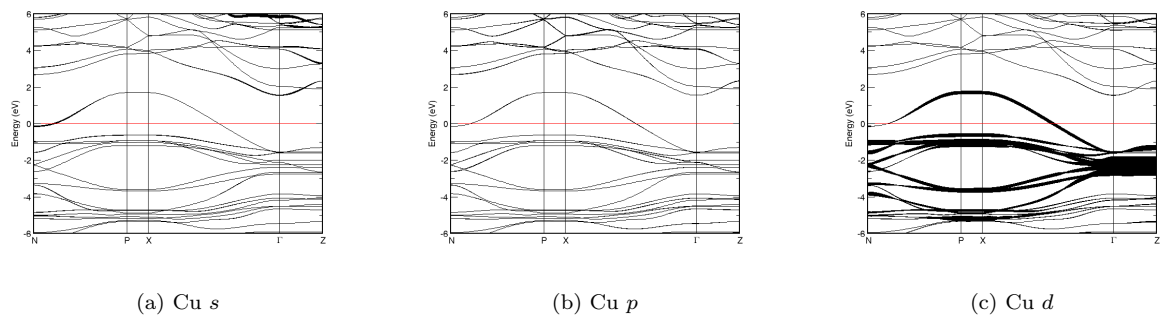

(a) Cu $s$        (b) Cu $p$        (c) Cu $d$

FIG. 275: Fat band representation of Cu in CuSr$_2$Br$_2$O$_2$

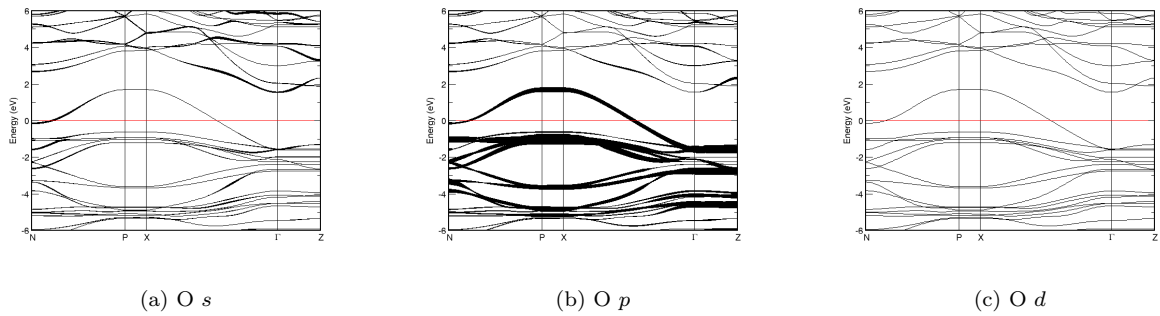

FIG. 276: Fat band representation of O in CuSr$_2$Br$_2$O$_2$

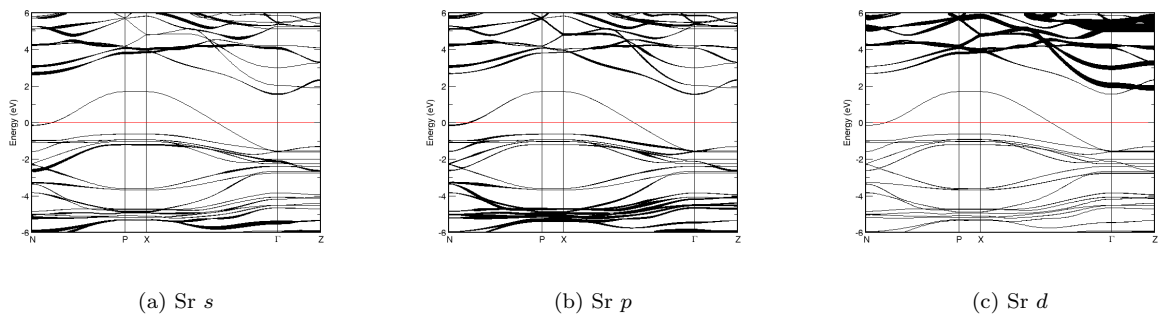

FIG. 277: Fat band representation of Sr in CuSr$_2$Br$_2$O$_2$

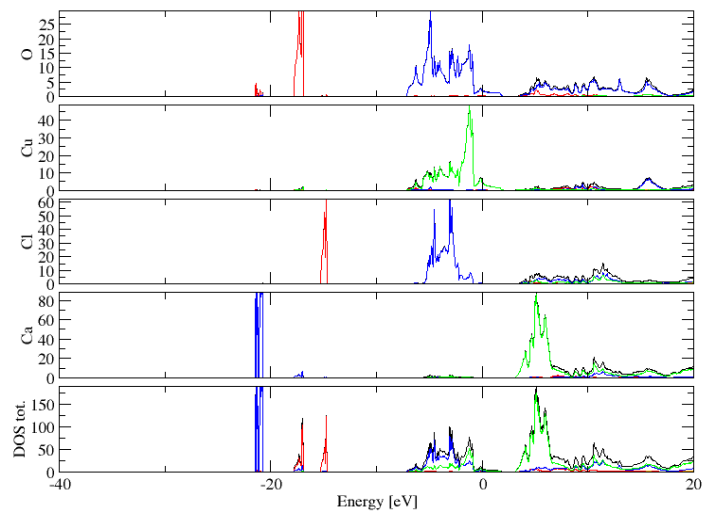

FIG. 278: (Color online) PDOS of Ca$_2$(CuCl$_2$O$_2$) (ICSD #1027). The $s$-, $p$- and $d$-projected states are in red, blue and green, respectively. Ca$_2$(CuCl$_2$O$_2$) crystallizes in space group I 4/m m m (#139), in a tetragonal body-centred structure.

(a) Ca $s$

(b) Ca $p$

(c) Ca $d$

FIG. 279: Fat band representation of Ca in $Ca_2(CuCl_2O_2)$

(a) Cl $s$

(b) Cl $p$

(c) Cl $d$

FIG. 280: Fat band representation of Cl in $Ca_2(CuCl_2O_2)$

(a) Cu $s$

(b) Cu $p$

(c) Cu $d$

FIG. 281: Fat band representation of Cu in $Ca_2(CuCl_2O_2)$

(a) O $s$

(b) O $p$

(c) O $d$

FIG. 282: Fat band representation of O in $Ca_2(CuCl_2O_2)$

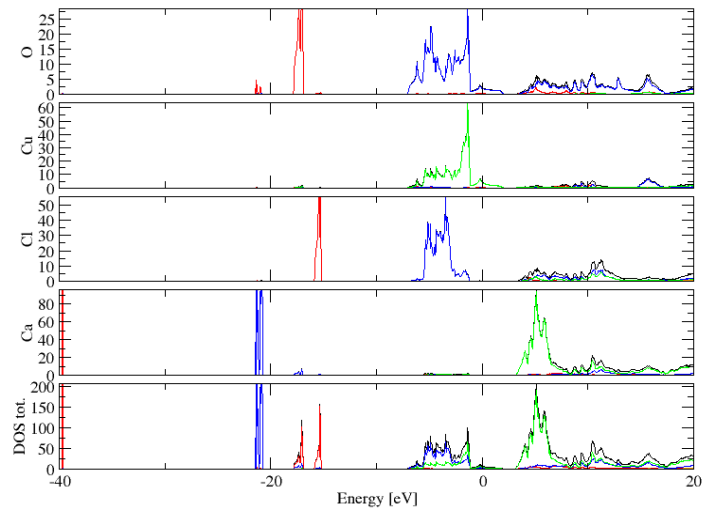

FIG. 283: (Color online) PDOS of $Ca_2CuO_2Cl_2$ (ICSD #83117). The $s$-, $p$- and $d$-projected states are in red, blue and green, respectively. $Ca_2CuO_2Cl_2$ crystallizes in space group I 4/m m m (#139), in a tetragonal body-centred structure.

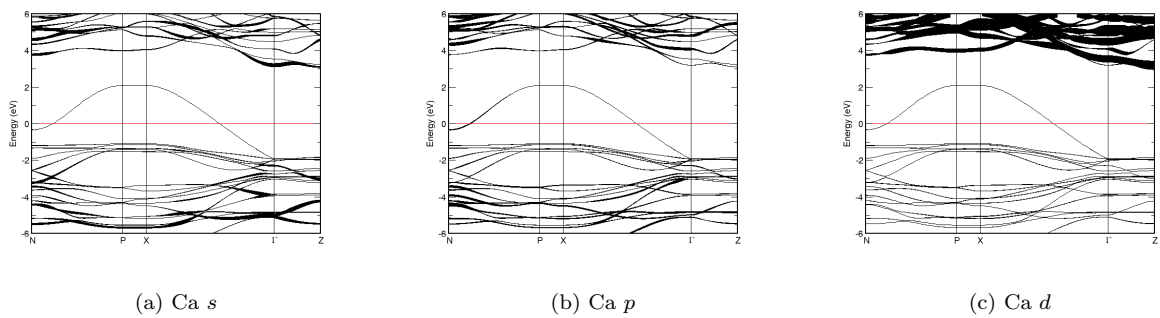

(a) Ca $s$        (b) Ca $p$        (c) Ca $d$

FIG. 284: Fat band representation of Ca in $Ca_2CuO_2Cl_2$

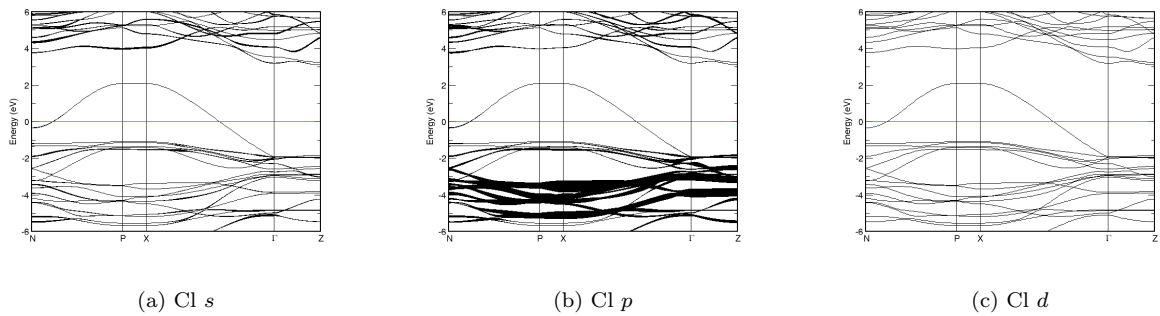

(a) Cl $s$        (b) Cl $p$        (c) Cl $d$

FIG. 285: Fat band representation of Cl in $Ca_2CuO_2Cl_2$

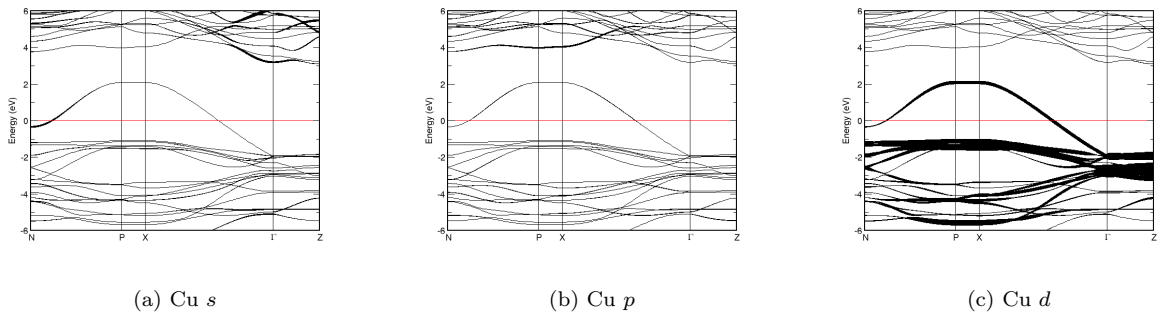

(a) Cu $s$                    (b) Cu $p$                    (c) Cu $d$

FIG. 286: Fat band representation of Cu in Ca$_2$CuO$_2$Cl$_2$

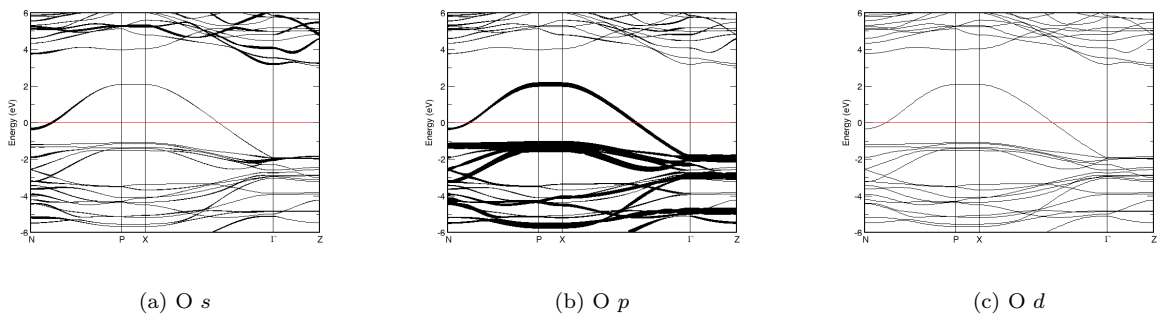

(a) O $s$                    (b) O $p$                    (c) O $d$

FIG. 287: Fat band representation of O in Ca$_2$CuO$_2$Cl$_2$

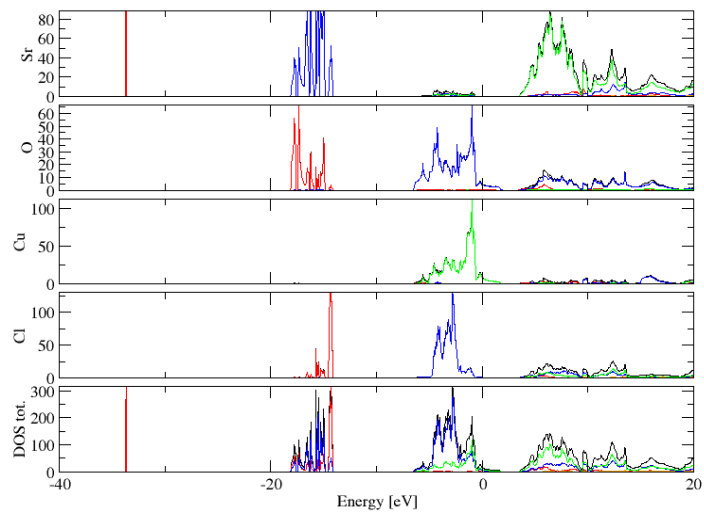

FIG. 288: (Color online) PDOS of Sr$_2$CuO$_2$Cl$_2$ (ICSD #4087). The $s$-, $p$- and $d$-projected states are in red, blue and green, respectively. Sr$_2$CuO$_2$Cl$_2$ crystallizes in space group I 4/m m m (#139), in a tetragonal body-centred structure.

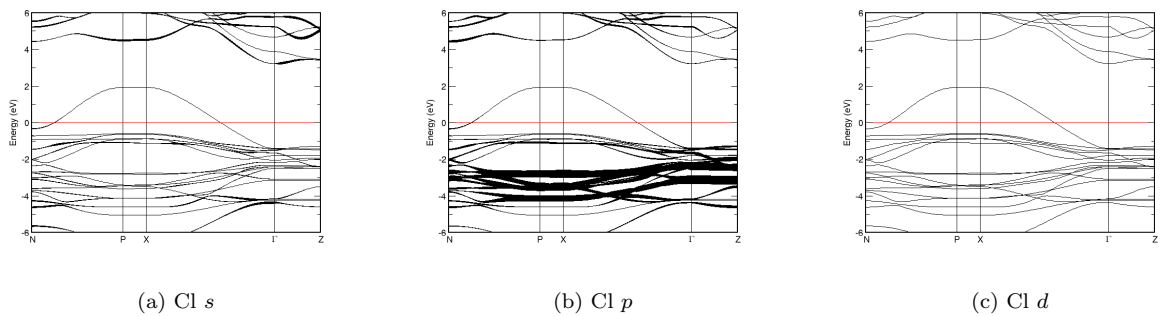

(a) Cl *s*    (b) Cl *p*    (c) Cl *d*

FIG. 289: Fat band representation of Cl in Sr$_2$CuO$_2$Cl$_2$

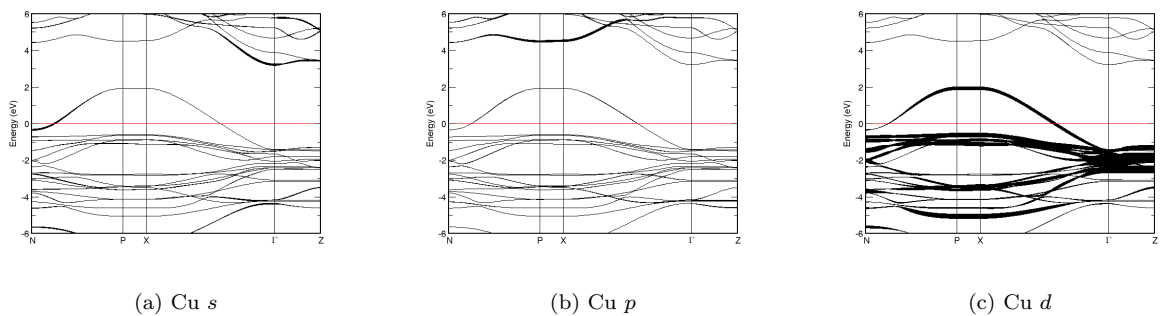

(a) Cu *s*    (b) Cu *p*    (c) Cu *d*

FIG. 290: Fat band representation of Cu in Sr$_2$CuO$_2$Cl$_2$

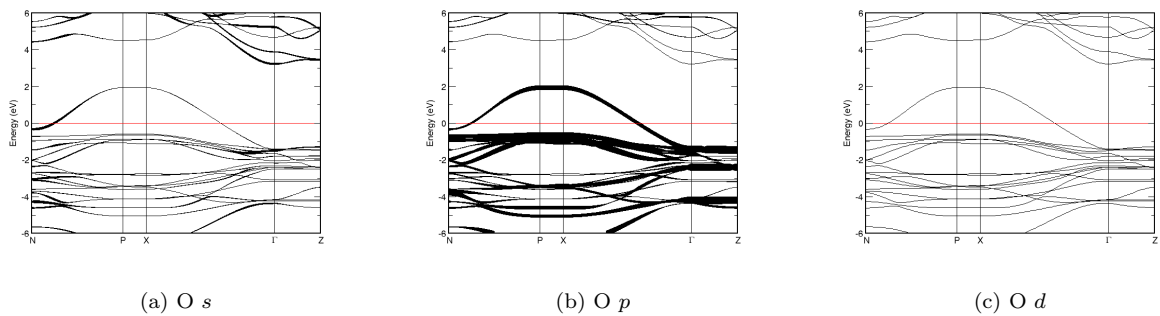

(a) O *s*    (b) O *p*    (c) O *d*

FIG. 291: Fat band representation of O in Sr$_2$CuO$_2$Cl$_2$

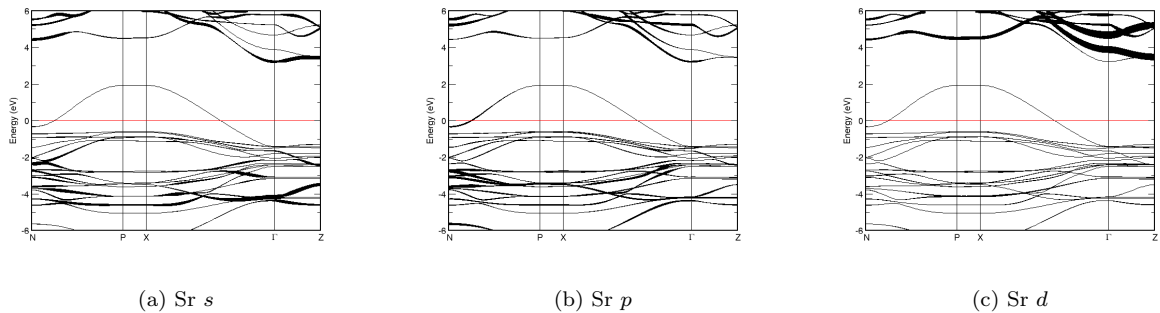

(a) Sr *s*    (b) Sr *p*    (c) Sr *d*

FIG. 292: Fat band representation of Sr in Sr$_2$CuO$_2$Cl$_2$

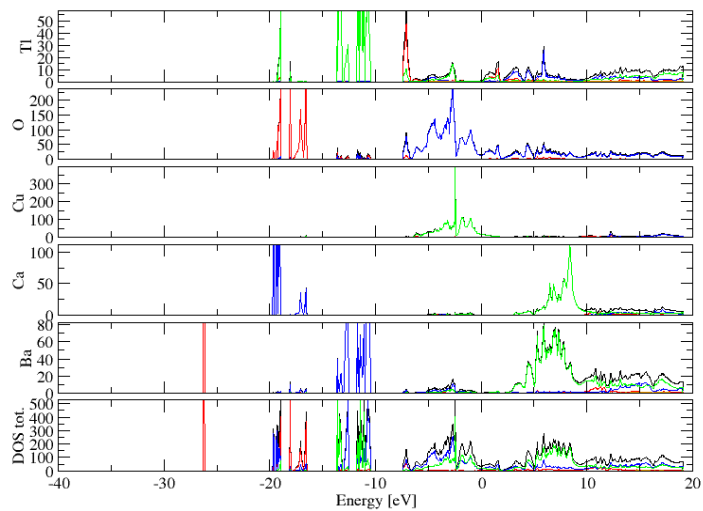

FIG. 293: (Color online) PDOS of Tl$_2$Ba$_2$CaCu$_2$O$_8$ (ICSD #78592). The $s$-, $p$- and $d$-projected states are in red, blue and green, respectively. Tl$_2$Ba$_2$CaCu$_2$O$_8$ crystallizes in space group I 4/m m m (#139), in a tetragonal body-centred structure.

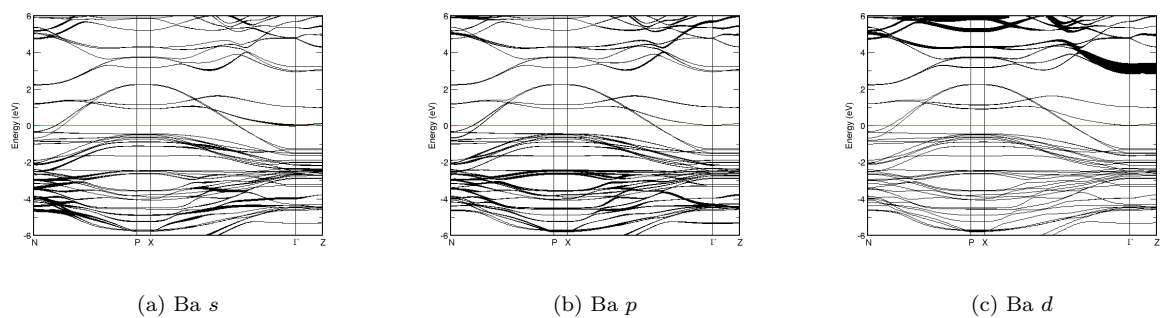

(a) Ba $s$          (b) Ba $p$          (c) Ba $d$

FIG. 294: Fat band representation of Ba in Tl$_2$Ba$_2$CaCu$_2$O$_8$

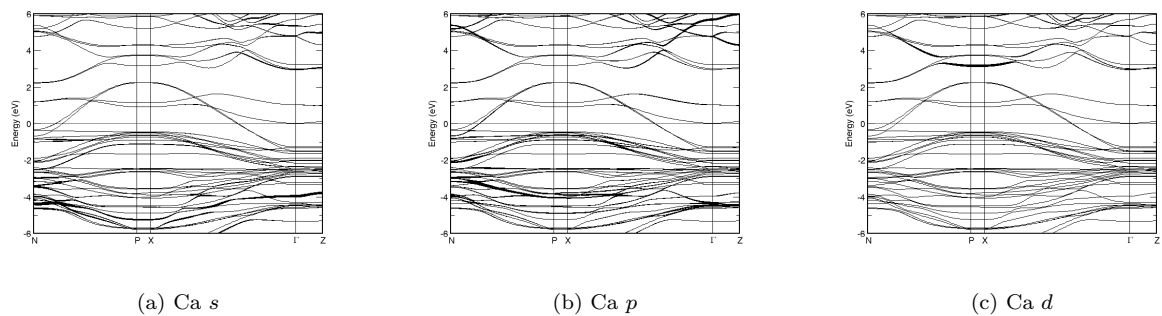

(a) Ca $s$          (b) Ca $p$          (c) Ca $d$

FIG. 295: Fat band representation of Ca in Tl$_2$Ba$_2$CaCu$_2$O$_8$

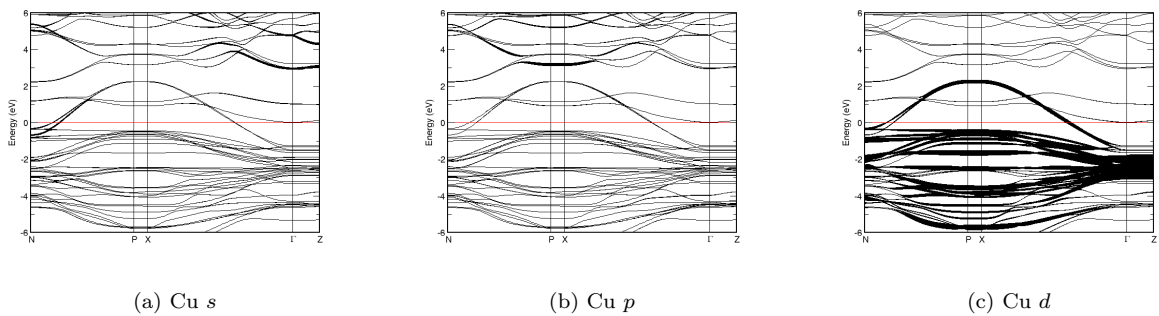

(a) Cu $s$    (b) Cu $p$    (c) Cu $d$

FIG. 296: Fat band representation of Cu in $Tl_2Ba_2CaCu_2O_8$

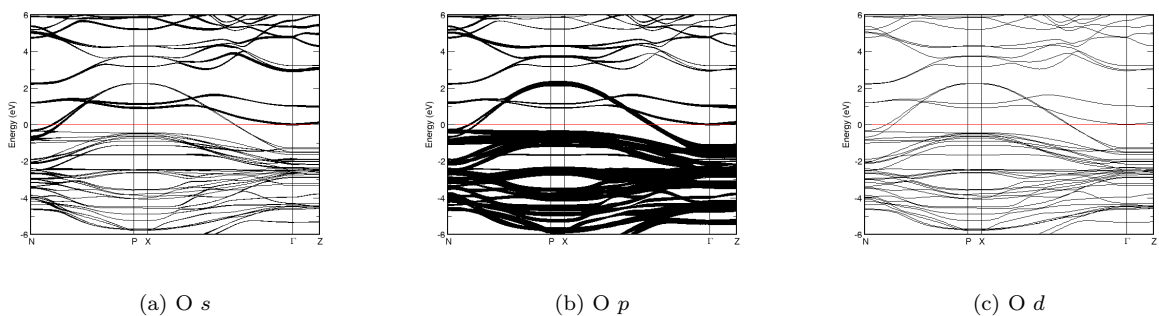

(a) O $s$    (b) O $p$    (c) O $d$

FIG. 297: Fat band representation of O in $Tl_2Ba_2CaCu_2O_8$

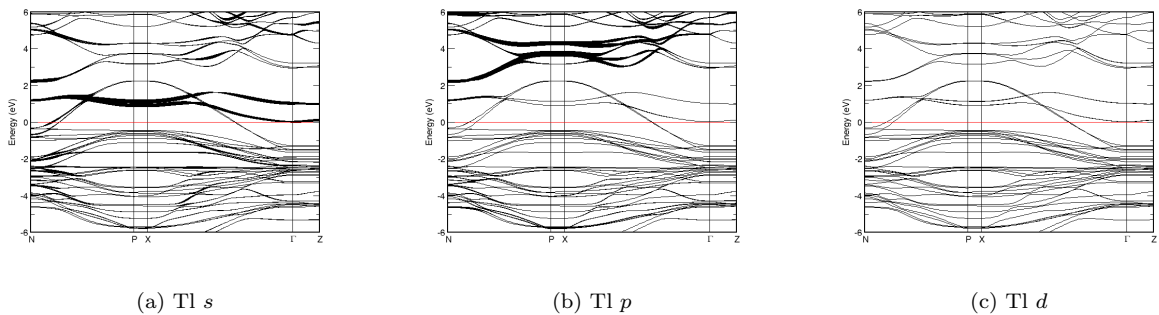

(a) Tl $s$    (b) Tl $p$    (c) Tl $d$

FIG. 298: Fat band representation of Tl in $Tl_2Ba_2CaCu_2O_8$

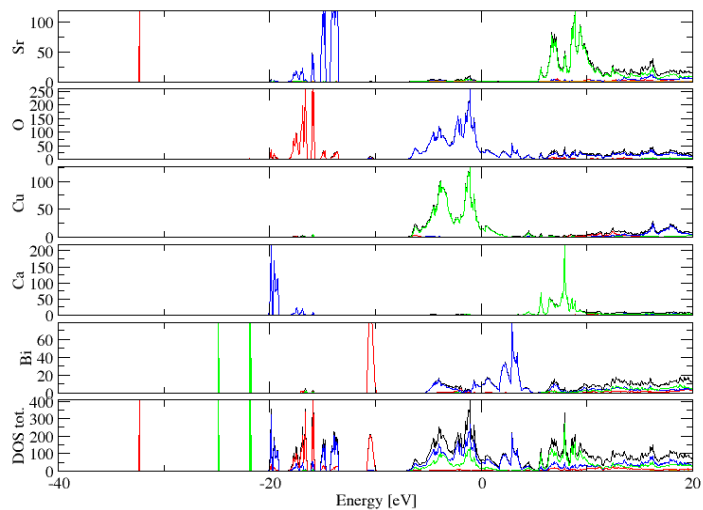

FIG. 299: (Color online) PDOS of Bi$_2$Sr$_2$CaCu$_2$O$_8$ (ICSD #68188). The $s$-, $p$- and $d$-projected states are in red, blue and green, respectively. Bi$_2$Sr$_2$CaCu$_2$O$_8$ crystallizes in space group I 4/m m m (#139), in a tetragonal body-centred structure.

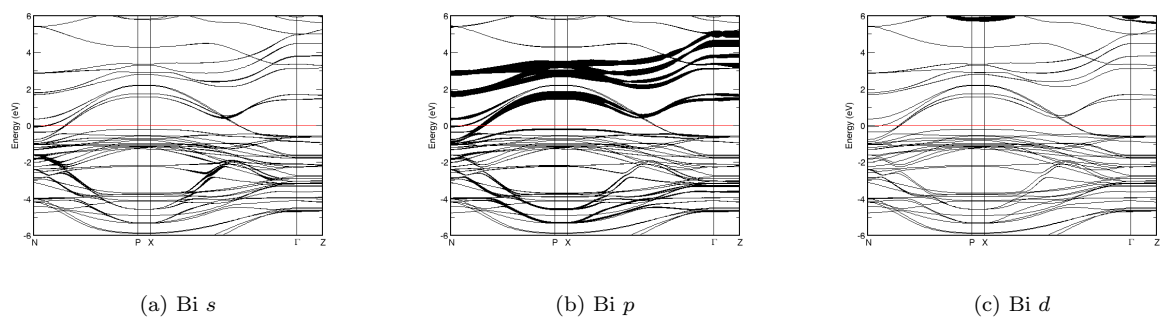

(a) Bi $s$          (b) Bi $p$          (c) Bi $d$

FIG. 300: Fat band representation of Bi in Bi$_2$Sr$_2$CaCu$_2$O$_8$

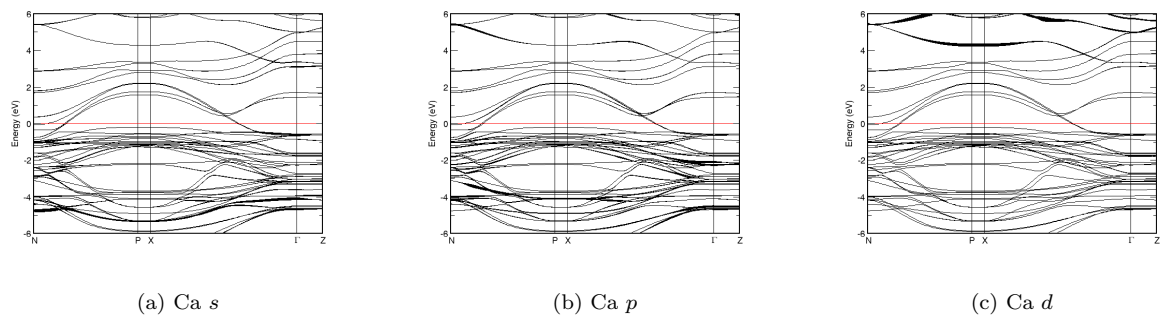

(a) Ca $s$          (b) Ca $p$          (c) Ca $d$

FIG. 301: Fat band representation of Ca in Bi$_2$Sr$_2$CaCu$_2$O$_8$

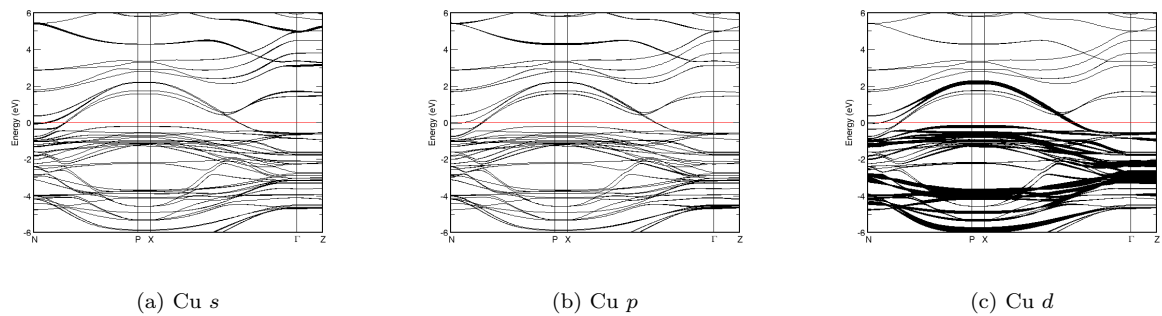

(a) Cu $s$         (b) Cu $p$         (c) Cu $d$

FIG. 302: Fat band representation of Cu in $Bi_2Sr_2CaCu_2O_8$

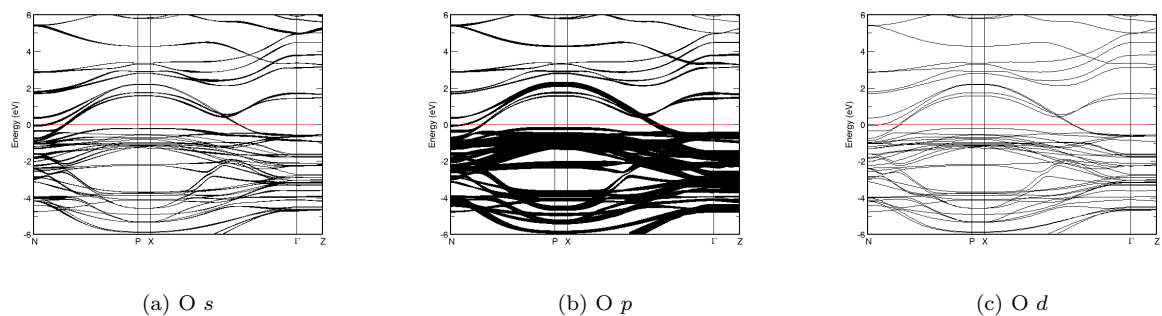

(a) O $s$         (b) O $p$         (c) O $d$

FIG. 303: Fat band representation of O in $Bi_2Sr_2CaCu_2O_8$

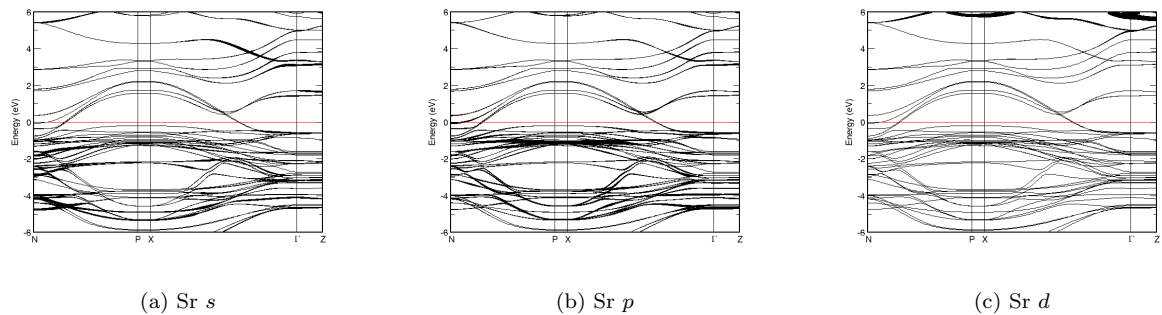

(a) Sr $s$         (b) Sr $p$         (c) Sr $d$

FIG. 304: Fat band representation of Sr in $Bi_2Sr_2CaCu_2O_8$

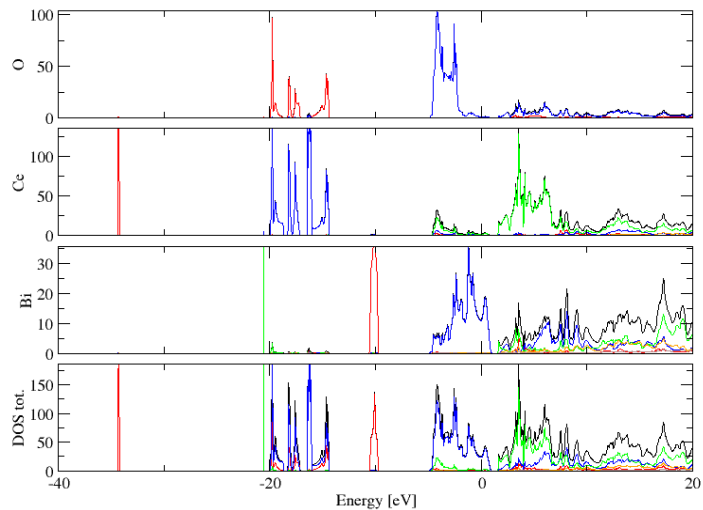

FIG. 305: (Color online) PDOS of Ce$_2$BiO$_2$ (ICSD #9099). The $s$-, $p$- and $d$-projected states are in red, blue and green, respectively. Ce$_2$BiO$_2$ crystallizes in space group I 4/m m m (#139), in a tetragonal body-centred structure.

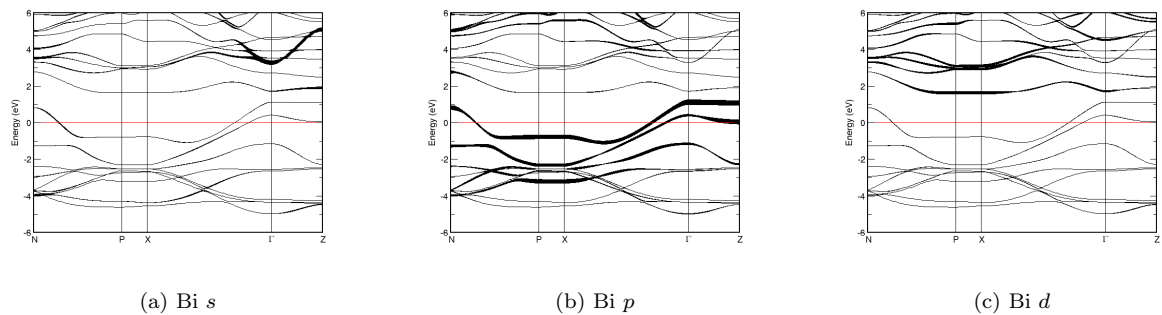

(a) Bi $s$                    (b) Bi $p$                    (c) Bi $d$

FIG. 306: Fat band representation of Bi in Ce$_2$BiO$_2$

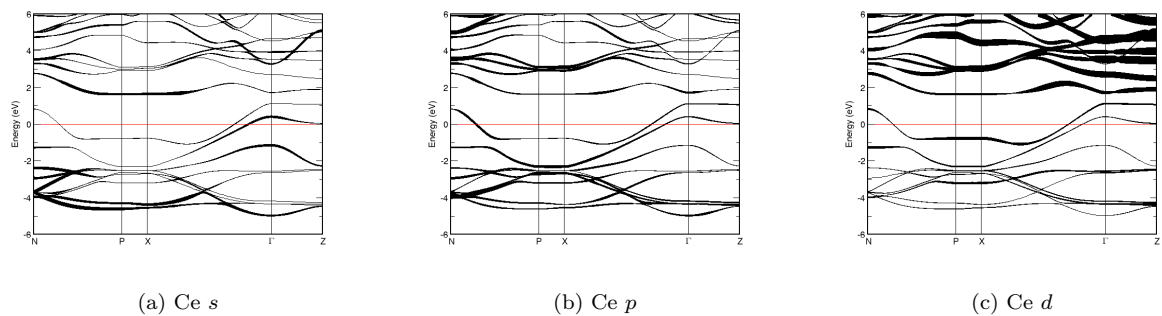

(a) Ce $s$                    (b) Ce $p$                    (c) Ce $d$

FIG. 307: Fat band representation of Ce in Ce$_2$BiO$_2$

(a) O $s$

(b) O $p$

(c) O $d$

FIG. 308: Fat band representation of O in $Ce_2BiO_2$

FIG. 309: (Color online) PDOS of $Ce_2SbO_2$ (ICSD #9100). The $s$-, $p$- and $d$-projected states are in red, blue and green, respectively. $Ce_2SbO_2$ crystallizes in space group I 4/m m m (#139), in a tetragonal body-centred structure.

(a) Ce $s$

(b) Ce $p$

(c) Ce $d$

FIG. 310: Fat band representation of Ce in $Ce_2SbO_2$

(a) O $s$

(b) O $p$

(c) O $d$

FIG. 311: Fat band representation of O in $Ce_2SbO_2$

(a) Sb $s$

(b) Sb $p$

(c) Sb $d$

FIG. 312: Fat band representation of Sb in $Ce_2SbO_2$

FIG. 313: (Color online) PDOS of $CePd_2Si_2$ (ICSD #621852). The $s$-, $p$- and $d$-projected states are in red, blue and green, respectively. $CePd_2Si_2$ crystallizes in space group I 4/m m m (#139), in a tetragonal body-centred structure.

(a) Ce $s$  (b) Ce $p$  (c) Ce $d$

FIG. 314: Fat band representation of Ce in CePd$_2$Si$_2$

(a) Pd $s$  (b) Pd $p$  (c) Pd $d$

FIG. 315: Fat band representation of Pd in CePd$_2$Si$_2$

(a) Si $s$  (b) Si $p$  (c) Si $d$

FIG. 316: Fat band representation of Si in CePd$_2$Si$_2$

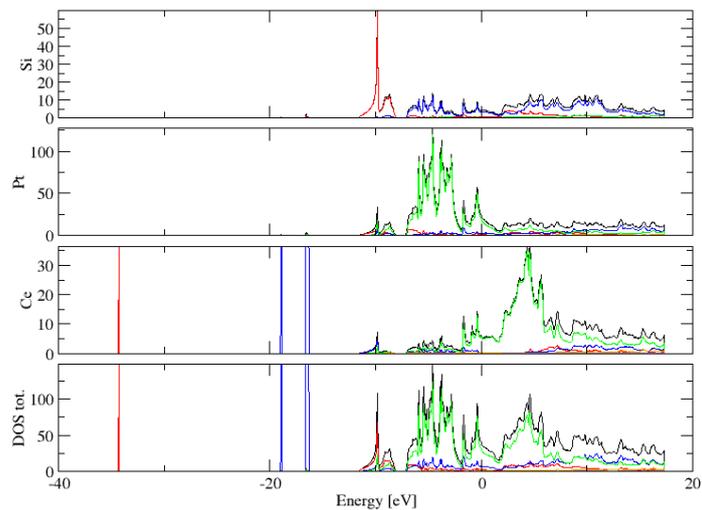

FIG. 317: (Color online) PDOS of CePt$_2$Si$_2$ (ICSD #52895). The $s$-, $p$- and $d$-projected states are in red, blue and green, respectively. CePt$_2$Si$_2$ crystallizes in space group I 4/m m m (#139), in a tetragonal body-centred structure.

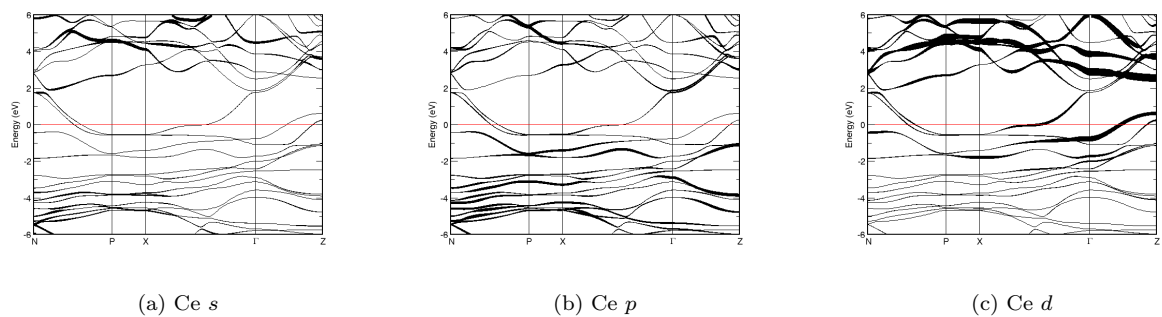

(a) Ce $s$        (b) Ce $p$        (c) Ce $d$

FIG. 318: Fat band representation of Ce in CePt$_2$Si$_2$

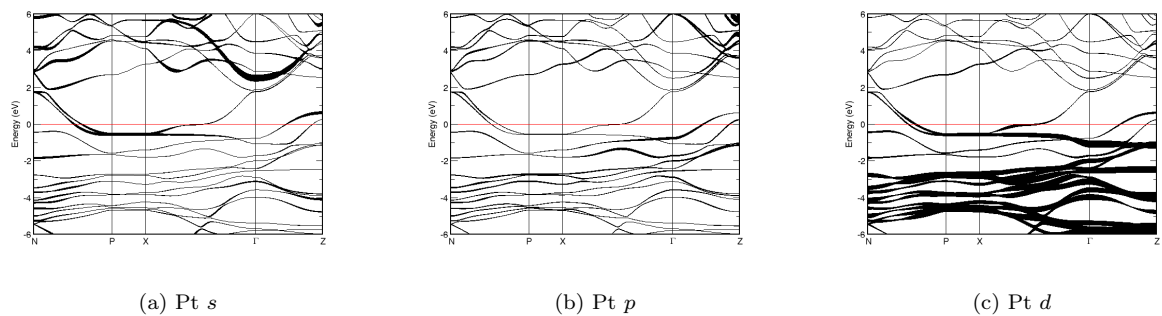

(a) Pt $s$        (b) Pt $p$        (c) Pt $d$

FIG. 319: Fat band representation of Pt in CePt$_2$Si$_2$

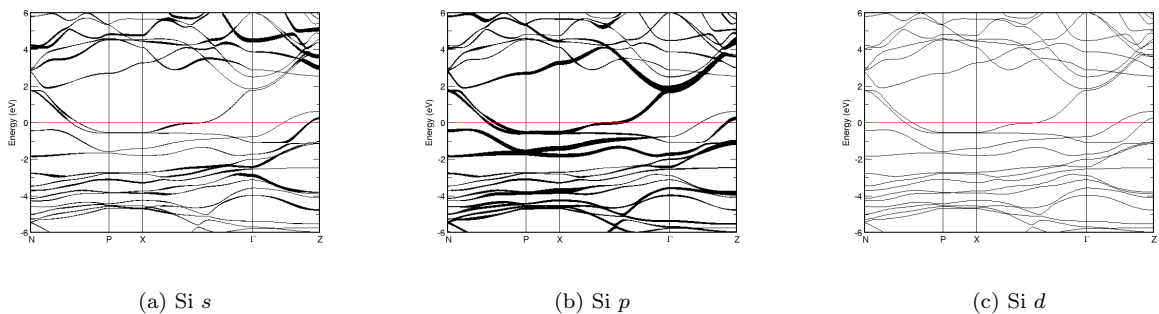

(a) Si $s$       (b) Si $p$       (c) Si $d$

FIG. 320: Fat band representation of Si in CePt$_2$Si$_2$

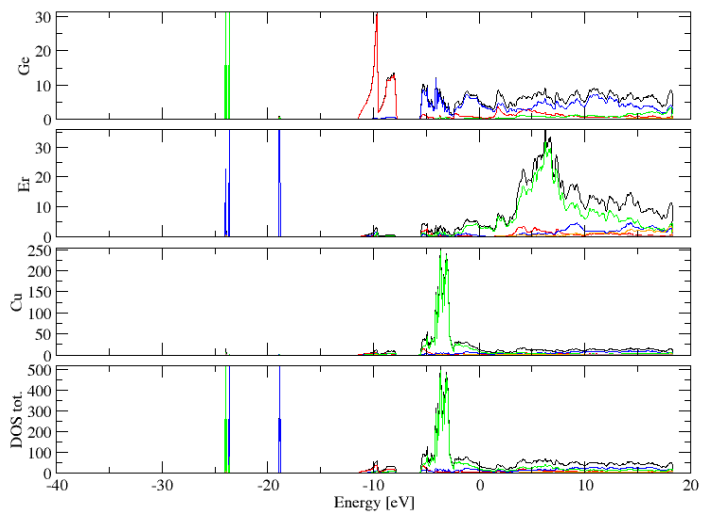

FIG. 321: (Color online) PDOS of Cu$_2$ErGe$_2$ (ICSD #53251). The $s$-, $p$- and $d$-projected states are in red, blue and green, respectively. Cu$_2$ErGe$_2$ crystallizes in space group I 4/m m m (#139), in a tetragonal body-centred structure.

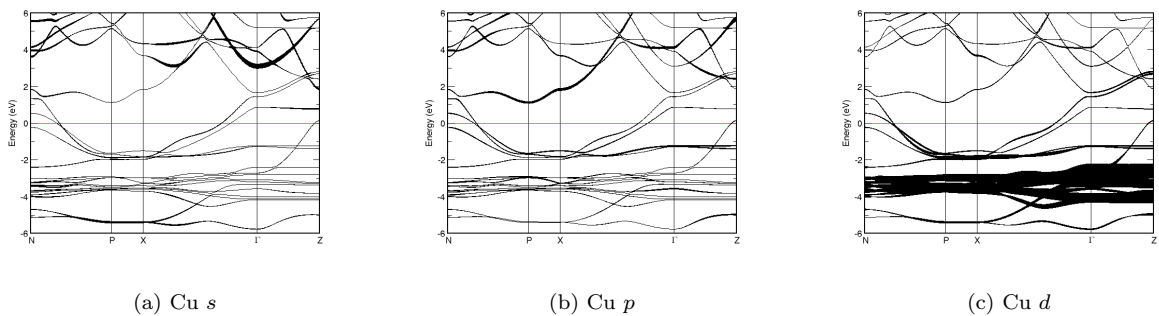

(a) Cu $s$       (b) Cu $p$       (c) Cu $d$

FIG. 322: Fat band representation of Cu in Cu$_2$ErGe$_2$

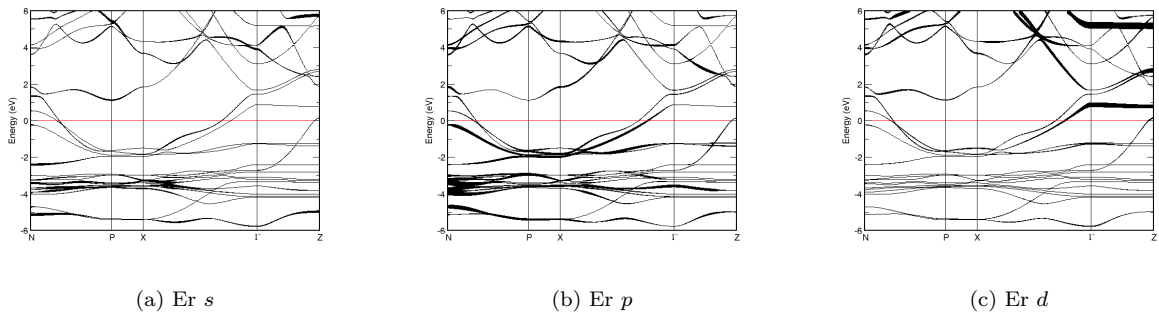

(a) Er $s$        (b) Er $p$        (c) Er $d$

FIG. 323: Fat band representation of Er in $Cu_2ErGe_2$

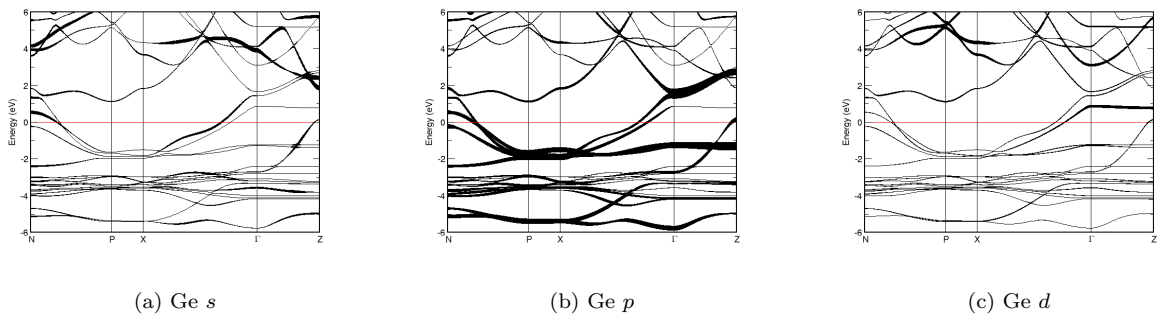

(a) Ge $s$        (b) Ge $p$        (c) Ge $d$

FIG. 324: Fat band representation of Ge in $Cu_2ErGe_2$

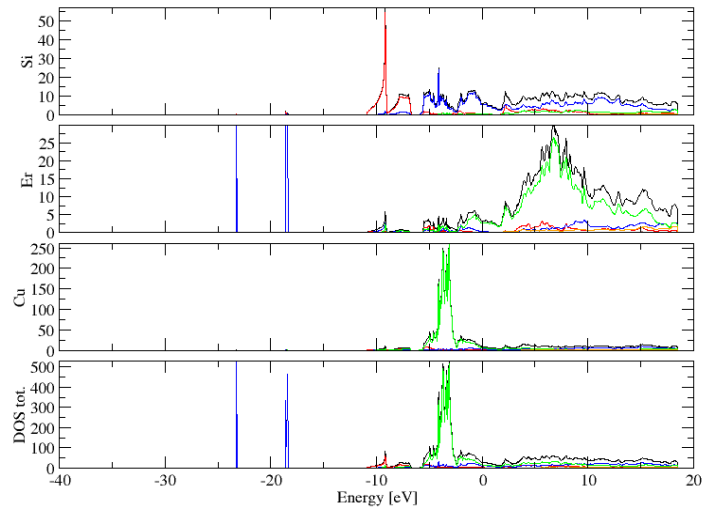

FIG. 325: (Color online) PDOS of $ErCu_2Si_2$ (ICSD #106845). The $s$-, $p$- and $d$-projected states are in red, blue and green, respectively. $ErCu_2Si_2$ crystallizes in space group I 4/m m m (#139), in a tetragonal body-centred structure.

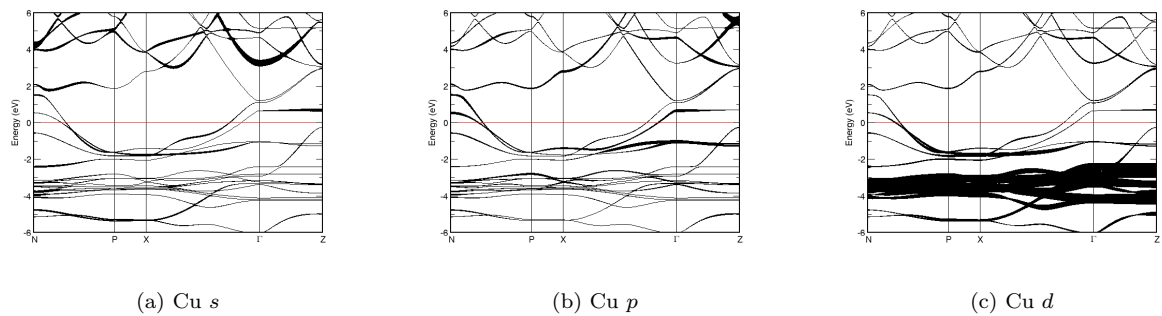

(a) Cu $s$     (b) Cu $p$     (c) Cu $d$

FIG. 326: Fat band representation of Cu in $ErCu_2Si_2$

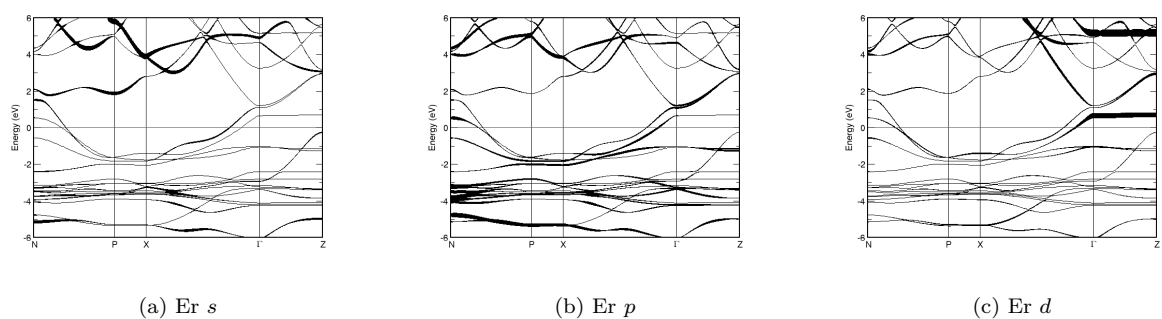

(a) Er $s$     (b) Er $p$     (c) Er $d$

FIG. 327: Fat band representation of Er in $ErCu_2Si_2$

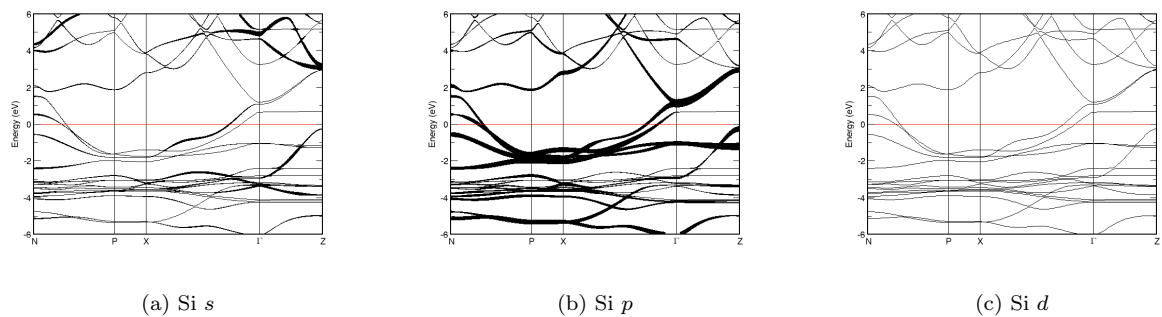

(a) Si $s$     (b) Si $p$     (c) Si $d$

FIG. 328: Fat band representation of Si in $ErCu_2Si_2$

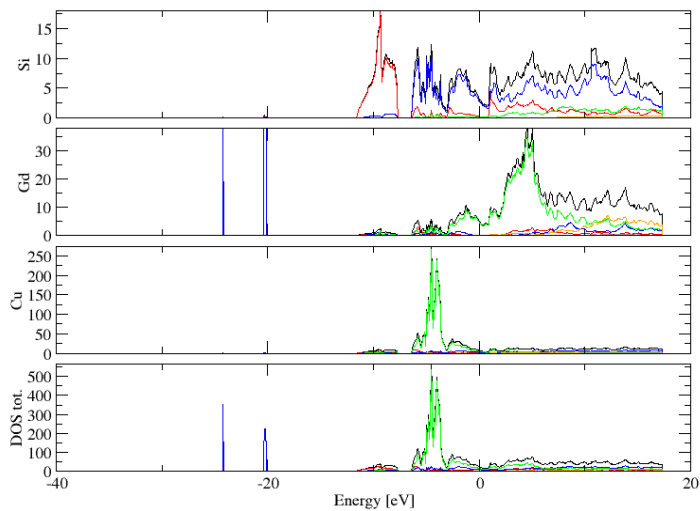

FIG. 329: (Color online) PDOS of Cu$_2$GdSi$_2$ (ICSD #64825). The $s$-, $p$- and $d$-projected states are in red, blue and green, respectively. Cu$_2$GdSi$_2$ crystallizes in space group I 4/m m m (#139), in a tetragonal body-centred structure.

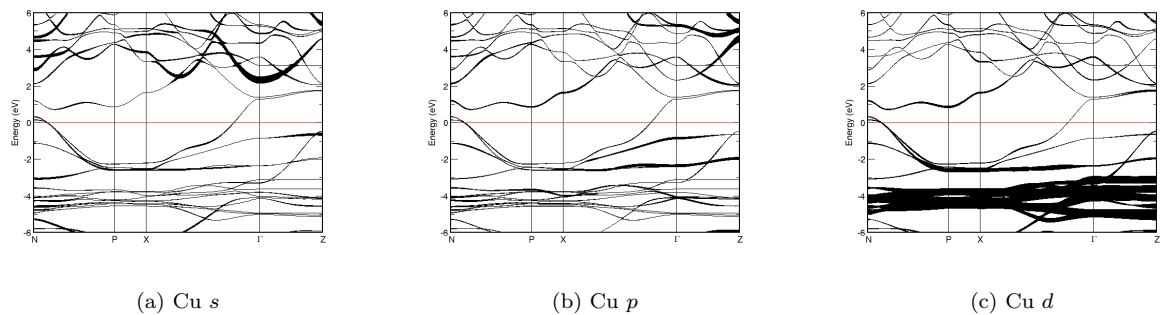

(a) Cu $s$                   (b) Cu $p$                   (c) Cu $d$

FIG. 330: Fat band representation of Cu in Cu$_2$GdSi$_2$

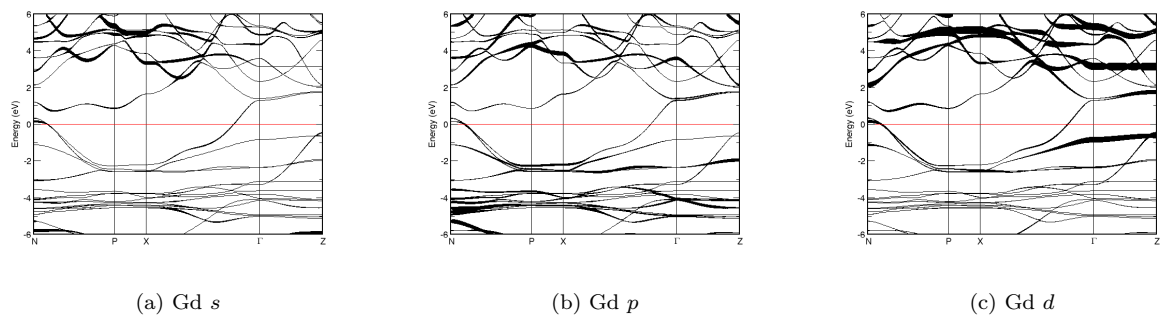

(a) Gd $s$                   (b) Gd $p$                   (c) Gd $d$

FIG. 331: Fat band representation of Gd in Cu$_2$GdSi$_2$

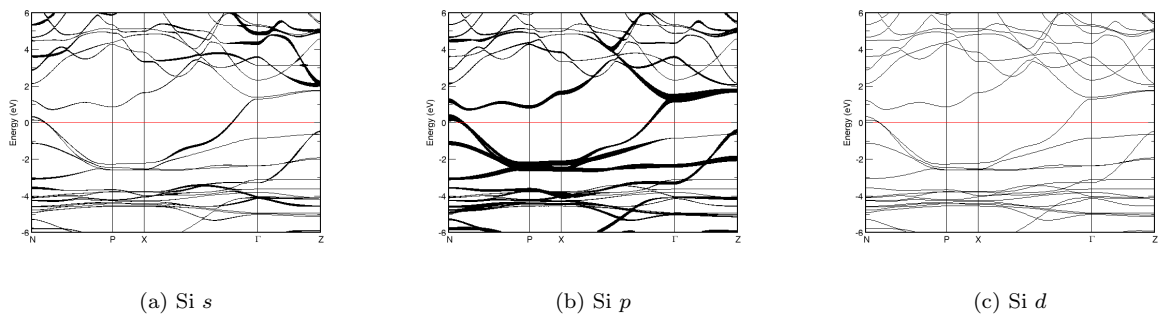

(a) Si $s$

(b) Si $p$

(c) Si $d$

FIG. 332: Fat band representation of Si in $Cu_2GdSi_2$

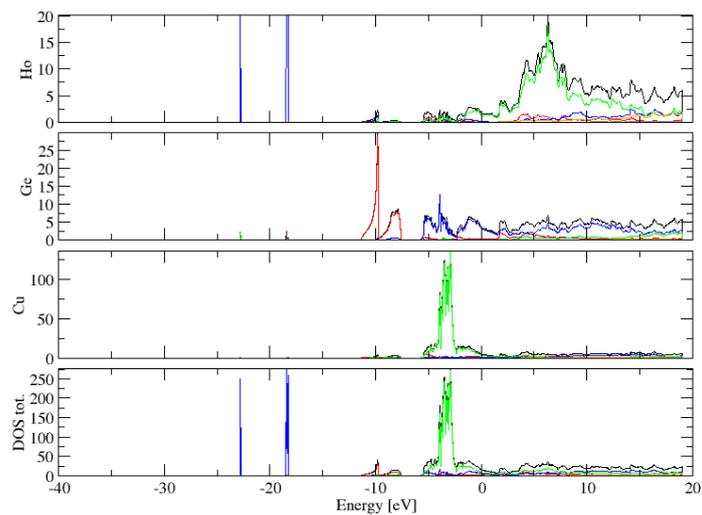

FIG. 333: (Color online) PDOS of $Cu_2HoGe_2$ (ICSD #53270). The $s$-, $p$- and $d$-projected states are in red, blue and green, respectively. $Cu_2HoGe_2$ crystallizes in space group I 4/m m m (#139), in a tetragonal body-centred structure.

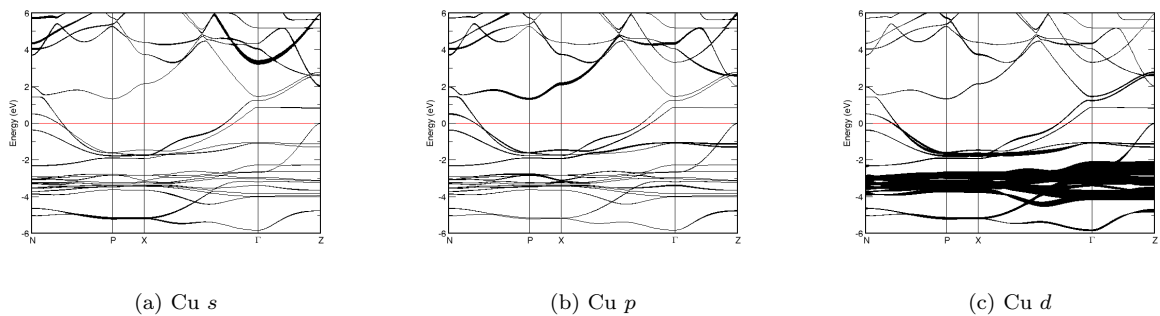

(a) Cu $s$

(b) Cu $p$

(c) Cu $d$

FIG. 334: Fat band representation of Cu in $Cu_2HoGe_2$

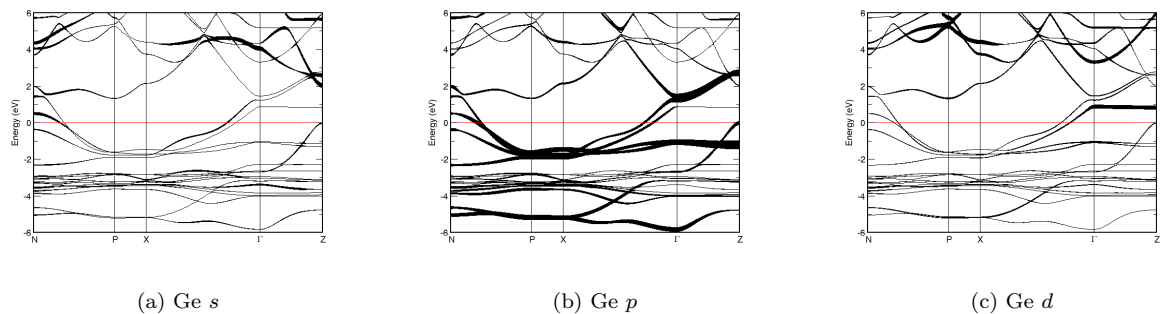

(a) Ge $s$       (b) Ge $p$       (c) Ge $d$

FIG. 335: Fat band representation of Ge in $Cu_2HoGe_2$

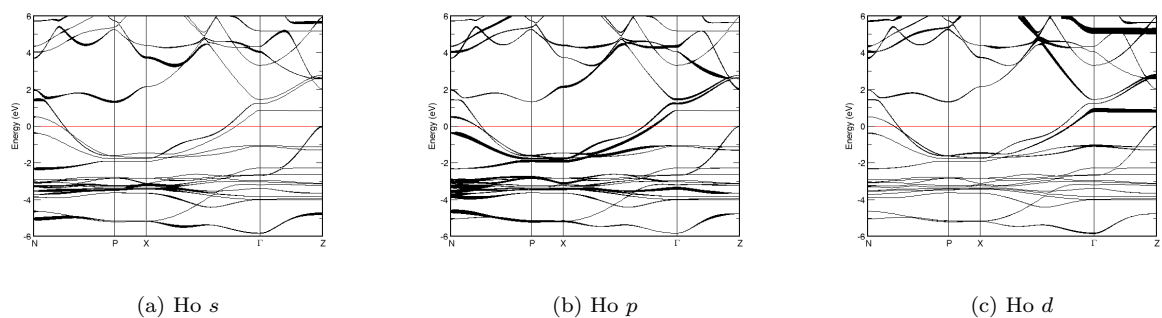

(a) Ho $s$       (b) Ho $p$       (c) Ho $d$

FIG. 336: Fat band representation of Ho in $Cu_2HoGe_2$

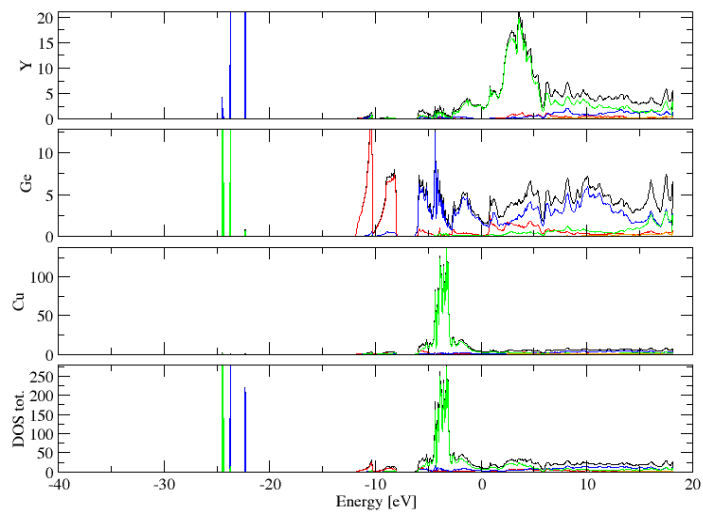

FIG. 337: (Color online) PDOS of $YCu_2Ge_2$ (ICSD #52764). The $s$-, $p$- and $d$-projected states are in red, blue and green, respectively. $YCu_2Ge_2$ crystallizes in space group I 4/m m m (#139), in a tetragonal body-centred structure.

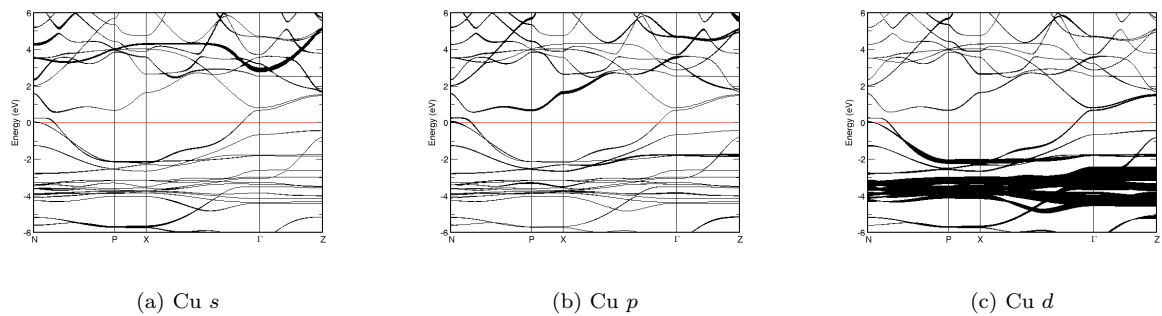

(a) Cu $s$        (b) Cu $p$        (c) Cu $d$

FIG. 338: Fat band representation of Cu in YCu$_2$Ge$_2$

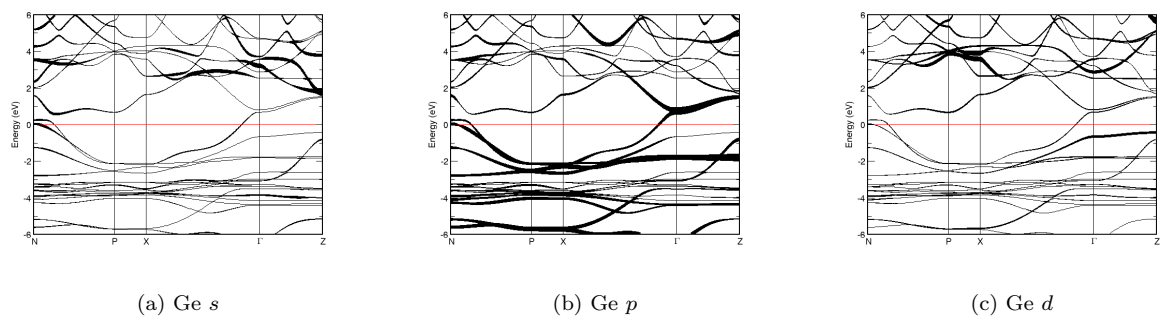

(a) Ge $s$        (b) Ge $p$        (c) Ge $d$

FIG. 339: Fat band representation of Ge in YCu$_2$Ge$_2$

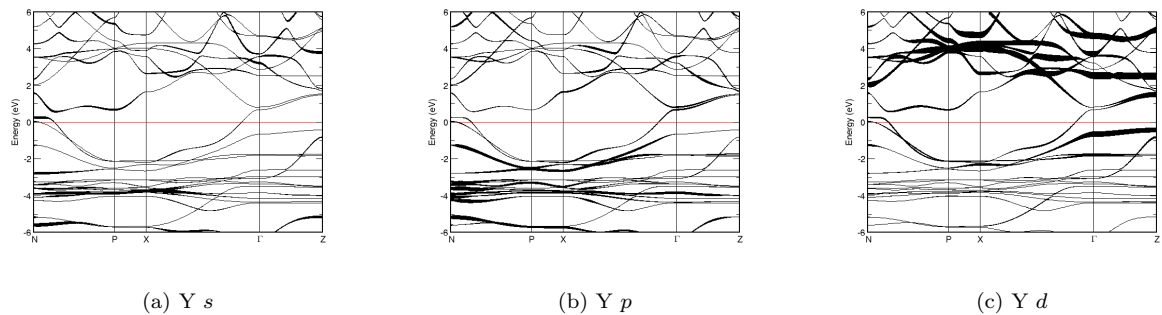

(a) Y $s$        (b) Y $p$        (c) Y $d$

FIG. 340: Fat band representation of Y in YCu$_2$Ge$_2$

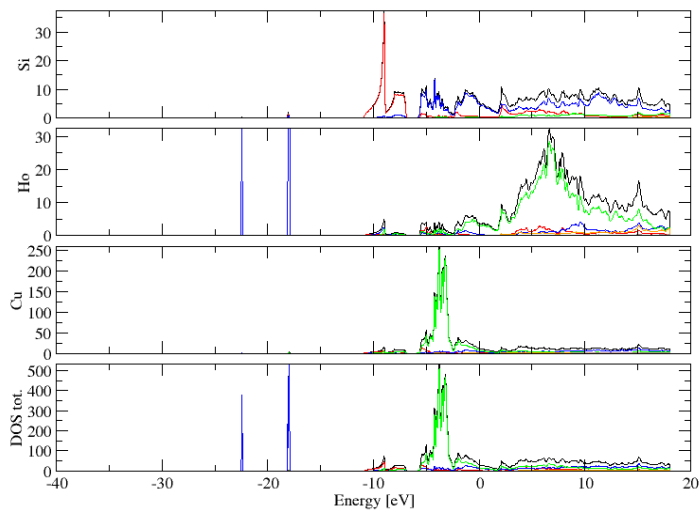

FIG. 341: (Color online) PDOS of Cu$_2$HoSi$_2$ (ICSD #53289). The $s$-, $p$- and $d$-projected states are in red, blue and green, respectively. Cu$_2$HoSi$_2$ crystallizes in space group I 4/m m m (#139), in a tetragonal body-centred structure.

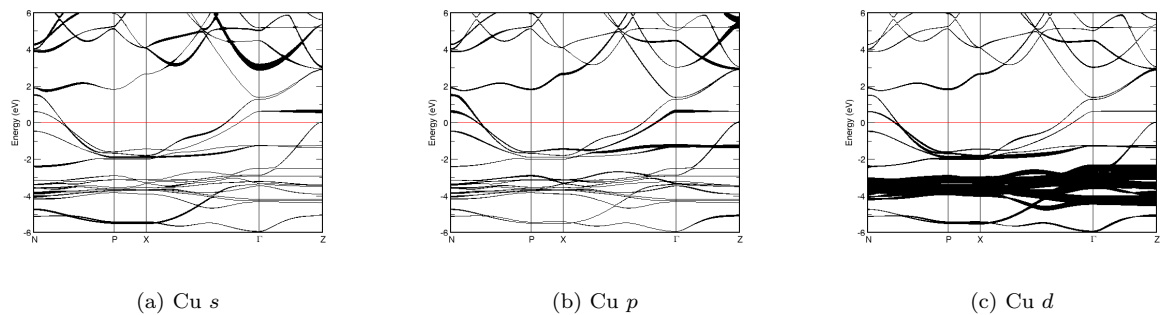

(a) Cu $s$        (b) Cu $p$        (c) Cu $d$

FIG. 342: Fat band representation of Cu in Cu$_2$HoSi$_2$

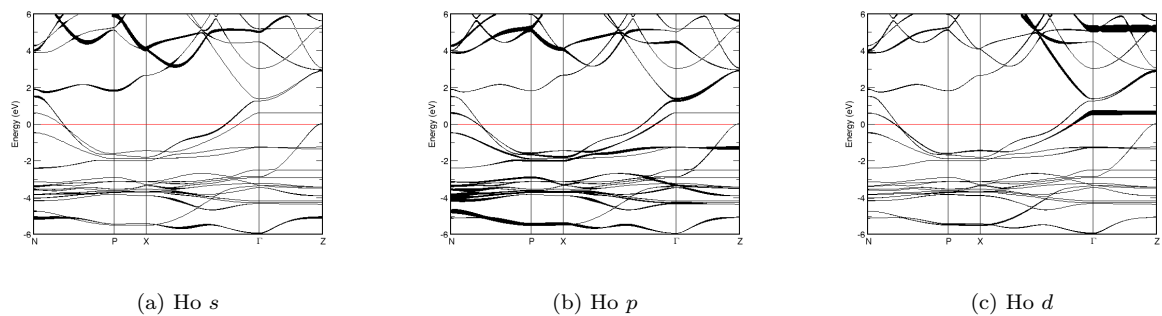

(a) Ho $s$        (b) Ho $p$        (c) Ho $d$

FIG. 343: Fat band representation of Ho in Cu$_2$HoSi$_2$

(a) Si $s$      (b) Si $p$      (c) Si $d$

FIG. 344: Fat band representation of Si in $Cu_2HoSi_2$

FIG. 345: (Color online) PDOS of $NdCu_2Si_2$ (ICSD #106842). The $s$-, $p$- and $d$-projected states are in red, blue and green, respectively. $NdCu_2Si_2$ crystallizes in space group I 4/m m m (#139), in a tetragonal body-centred structure.

(a) Cu $s$      (b) Cu $p$      (c) Cu $d$

FIG. 346: Fat band representation of Cu in $NdCu_2Si_2$

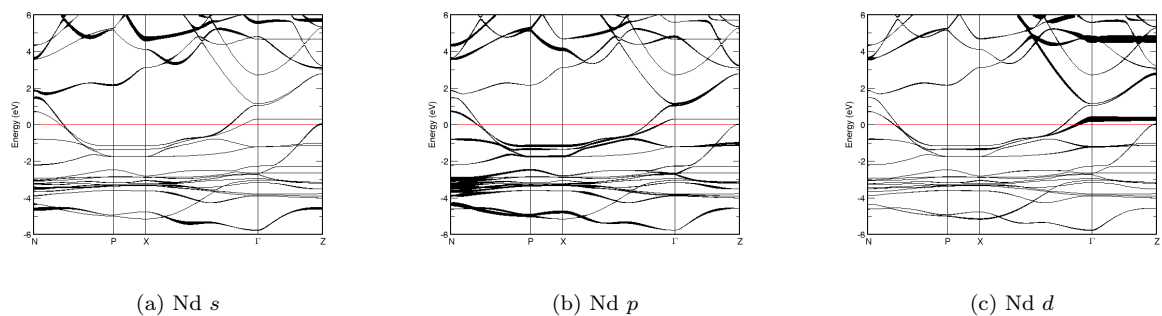

(a) Nd $s$    (b) Nd $p$    (c) Nd $d$

FIG. 347: Fat band representation of Nd in NdCu$_2$Si$_2$

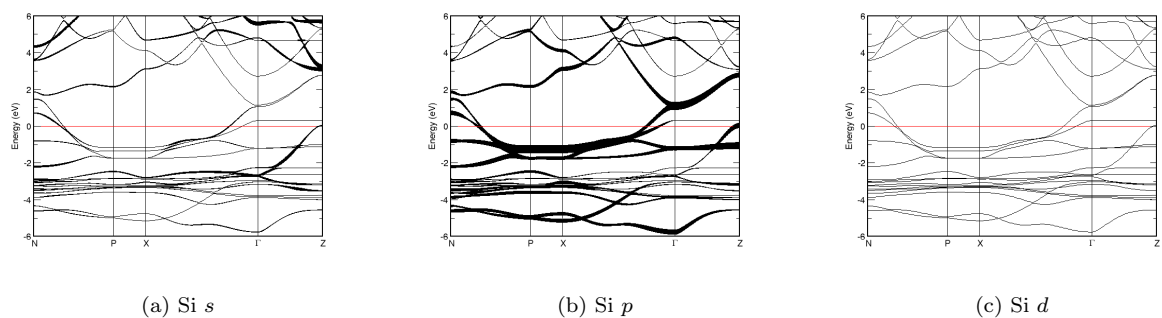

(a) Si $s$    (b) Si $p$    (c) Si $d$

FIG. 348: Fat band representation of Si in NdCu$_2$Si$_2$

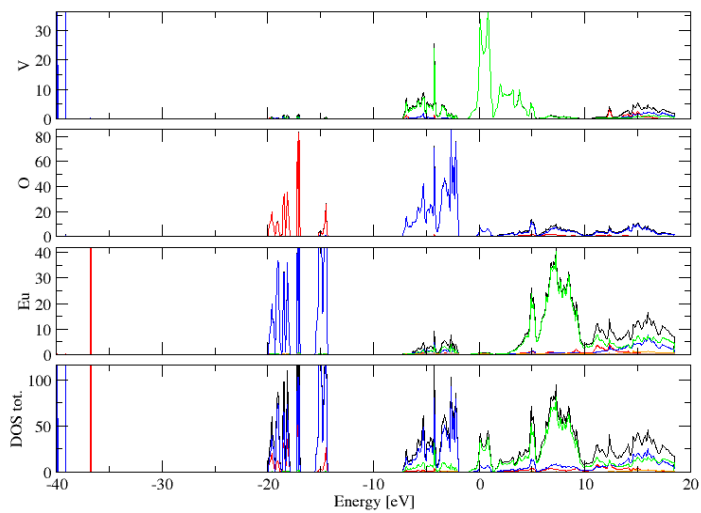

FIG. 349: (Color online) PDOS of Eu$_2$(VO$_4$) (ICSD #89000). The $s$-, $p$- and $d$-projected states are in red, blue and green, respectively. Eu$_2$(VO$_4$) crystallizes in space group I 4/m m m (#139), in a tetragonal body-centred structure.

(a) Eu $s$      (b) Eu $p$      (c) Eu $d$

FIG. 350: Fat band representation of Eu in Eu$_2$(VO$_4$)

(a) O $s$      (b) O $p$      (c) O $d$

FIG. 351: Fat band representation of O in Eu$_2$(VO$_4$)

(a) V $s$      (b) V $p$      (c) V $d$

FIG. 352: Fat band representation of V in Eu$_2$(VO$_4$)

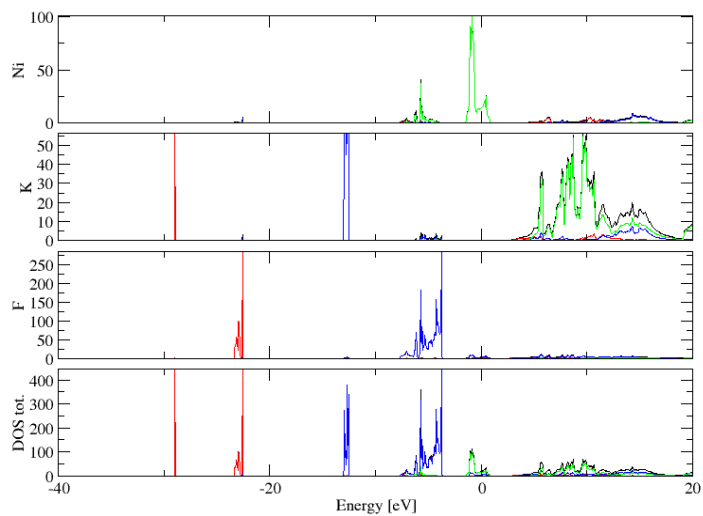

FIG. 353: (Color online) PDOS of $K_2(NiF_4)$ (ICSD #15576). The *s*-, *p*- and *d*-projected states are in red, blue and green, respectively. $K_2(NiF_4)$ crystallizes in space group I 4/m m m (#139), in a tetragonal body-centred structure.

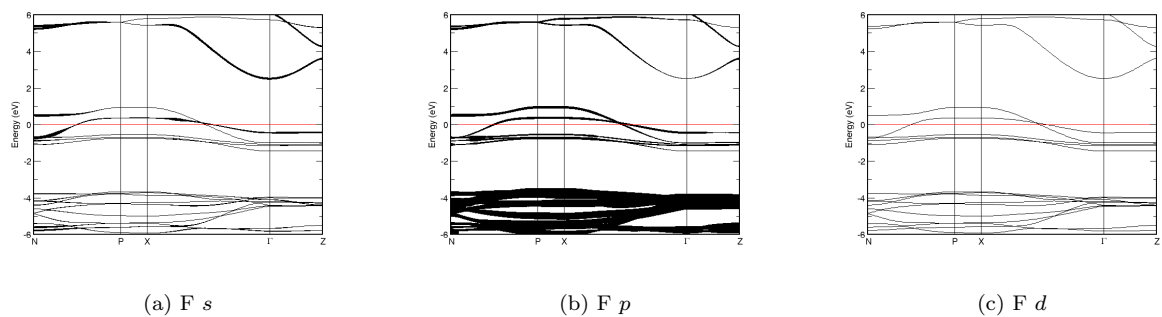

(a) F *s*                    (b) F *p*                    (c) F *d*

FIG. 354: Fat band representation of F in $K_2(NiF_4)$

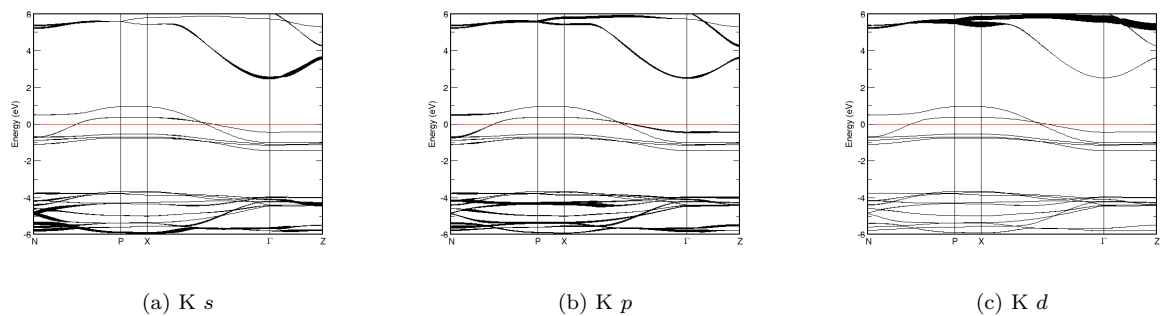

(a) K *s*                    (b) K *p*                    (c) K *d*

FIG. 355: Fat band representation of K in $K_2(NiF_4)$

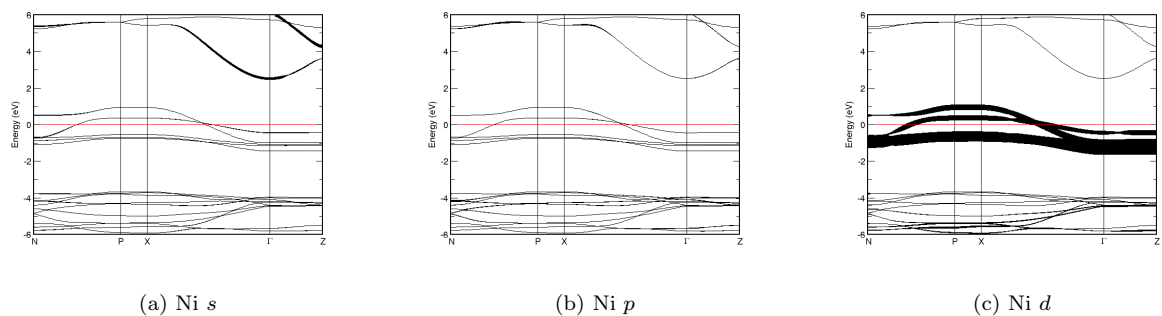

(a) Ni $s$      (b) Ni $p$      (c) Ni $d$

FIG. 356: Fat band representation of Ni in $K_2(NiF_4)$

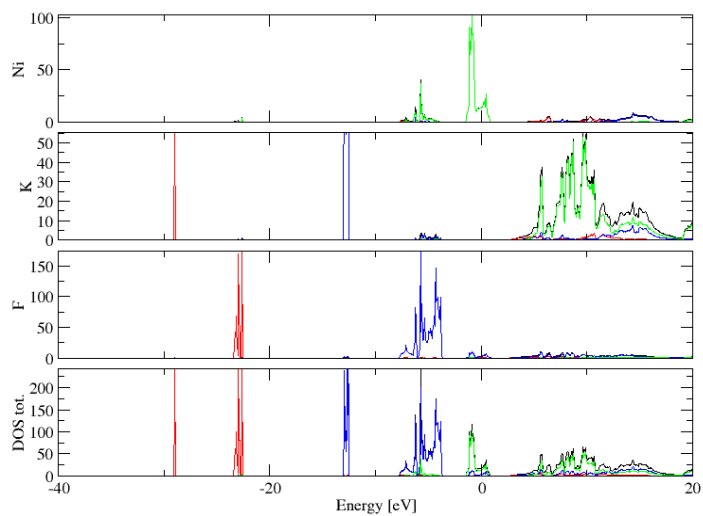

FIG. 357: (Color online) PDOS of $K_2(NiF_4)$ (ICSD #631720). The $s$-, $p$- and $d$-projected states are in red, blue and green, respectively. $K_2(NiF_4)$ crystallizes in space group I 4/m m m (#139), in a tetragonal body-centred structure.

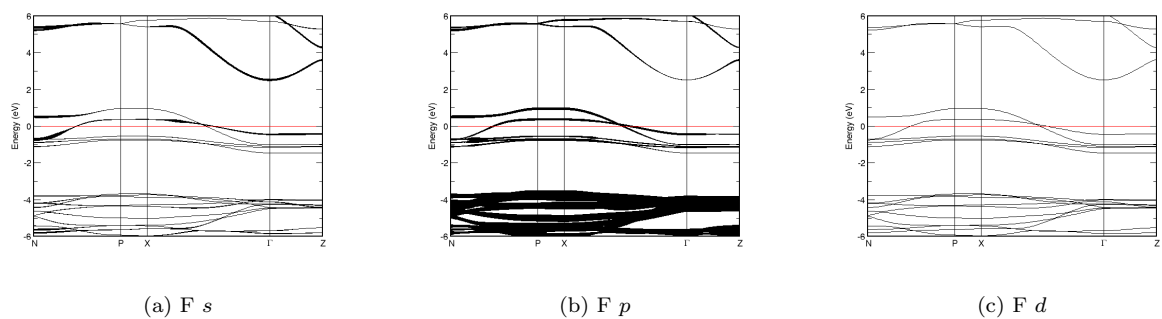

(a) F $s$      (b) F $p$      (c) F $d$

FIG. 358: Fat band representation of F in $K_2(NiF_4)$

(a) K $s$

(b) K $p$

(c) K $d$

FIG. 359: Fat band representation of K in $K_2(NiF_4)$

(a) Ni $s$

(b) Ni $p$

(c) Ni $d$

FIG. 360: Fat band representation of Ni in $K_2(NiF_4)$

FIG. 361: (Color online) PDOS of $Rb_2(NiF_4)$ (ICSD #69682). The $s$-, $p$- and $d$-projected states are in red, blue and green, respectively. $Rb_2(NiF_4)$ crystallizes in space group I 4/m m m (#139), in a tetragonal body-centred structure.

(a) F $s$

(b) F $p$

(c) F $d$

FIG. 362: Fat band representation of F in $Rb_2(NiF_4)$

(a) Ni $s$

(b) Ni $p$

(c) Ni $d$

FIG. 363: Fat band representation of Ni in $Rb_2(NiF_4)$

(a) Rb $s$

(b) Rb $p$

(c) Rb $d$

FIG. 364: Fat band representation of Rb in $Rb_2(NiF_4)$

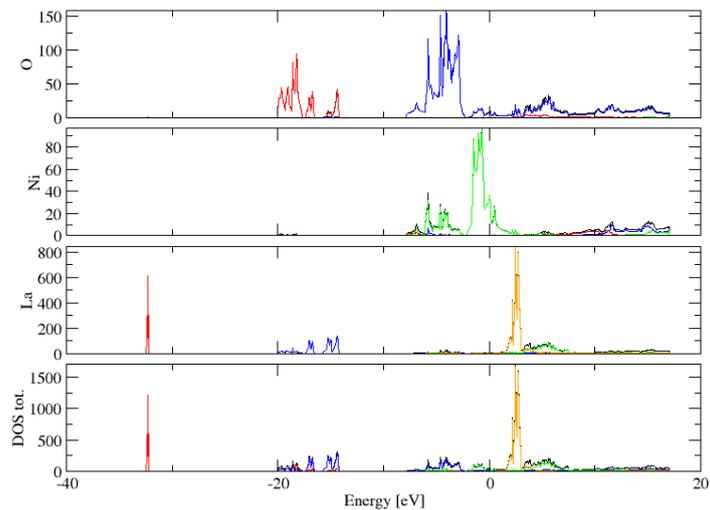

FIG. 365: (Color online) PDOS of La$_2$(NiO$_4$) (ICSD #1179). The $s$-, $p$- and $d$-projected states are in red, blue and green, respectively. La$_2$(NiO$_4$) crystallizes in space group I 4/m m m (#139), in a tetragonal body-centred structure.

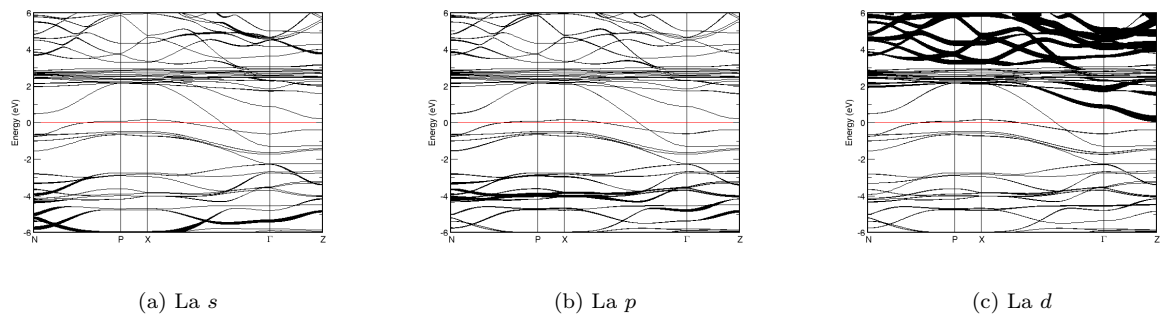

(a) La $s$            (b) La $p$            (c) La $d$

FIG. 366: Fat band representation of La in La$_2$(NiO$_4$)

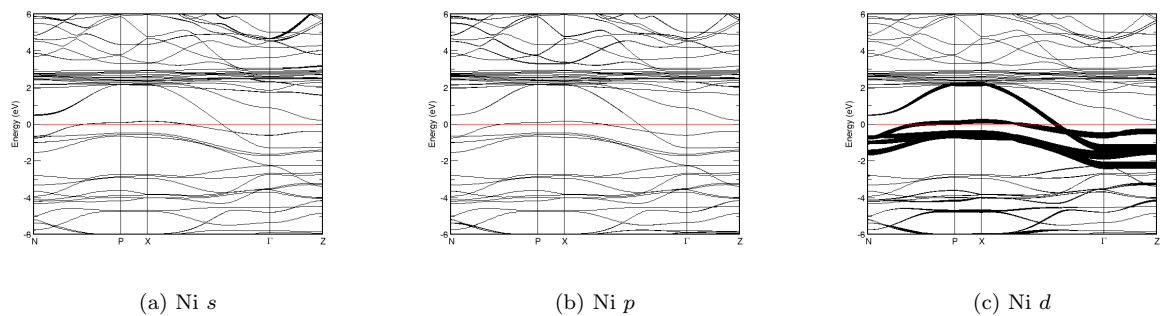

(a) Ni $s$            (b) Ni $p$            (c) Ni $d$

FIG. 367: Fat band representation of Ni in La$_2$(NiO$_4$)

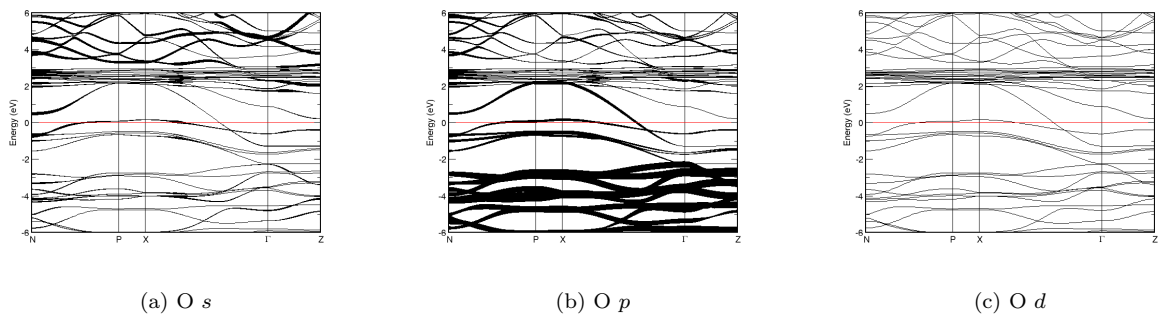

(a) O *s*    (b) O *p*    (c) O *d*

FIG. 368: Fat band representation of O in La$_2$(NiO$_4$)

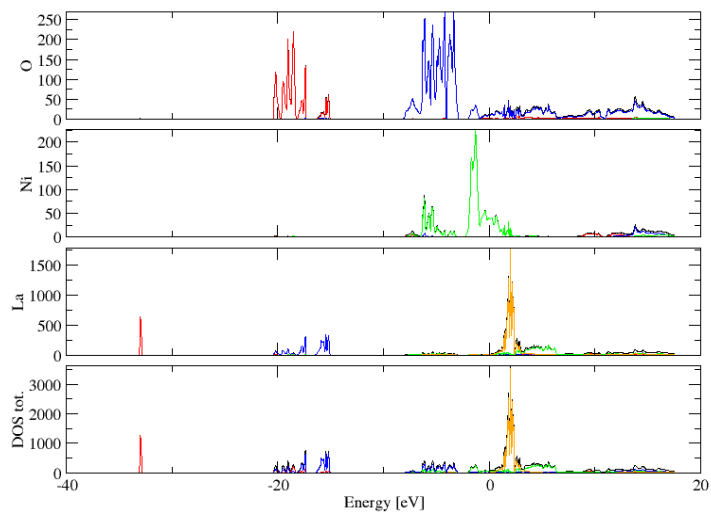

FIG. 369: (Color online) PDOS of La$_2$(NiO$_4$) (ICSD #33536). The *s*-, *p*- and *d*-projected states are in red, blue and green, respectively. La$_2$(NiO$_4$) crystallizes in space group I 4/m m m (#139), in a tetragonal body-centred structure.

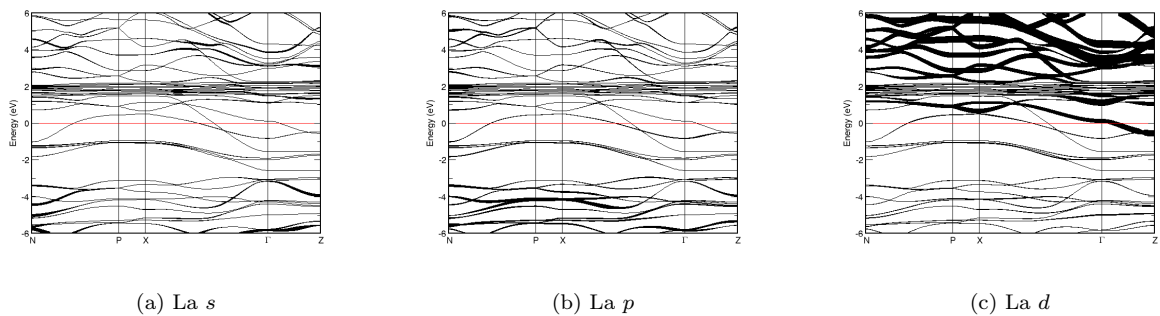

(a) La *s*    (b) La *p*    (c) La *d*

FIG. 370: Fat band representation of La in La$_2$(NiO$_4$)

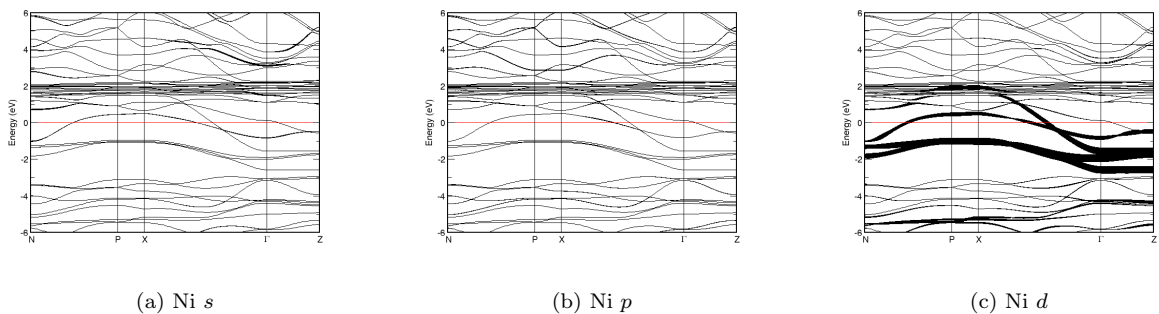

(a) Ni $s$         (b) Ni $p$         (c) Ni $d$

FIG. 371: Fat band representation of Ni in $La_2(NiO_4)$

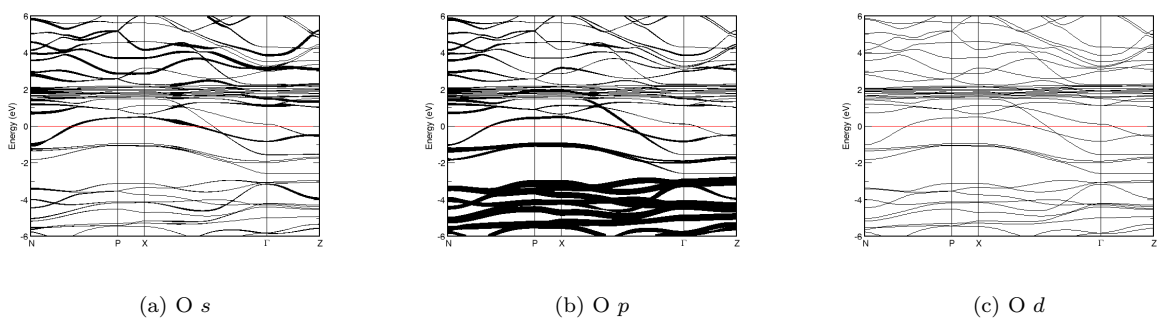

(a) O $s$         (b) O $p$         (c) O $d$

FIG. 372: Fat band representation of O in $La_2(NiO_4)$

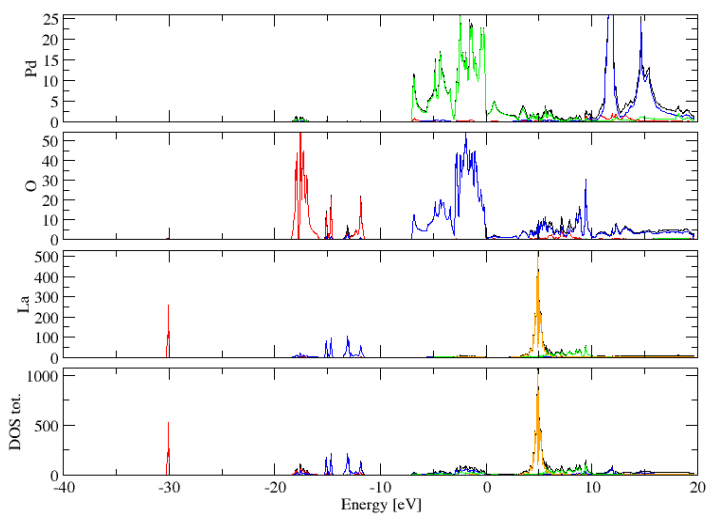

FIG. 373: (Color online) PDOS of $La_2PdO_4$ (ICSD #40262). The $s$-, $p$- and $d$-projected states are in red, blue and green, respectively. $La_2PdO_4$ crystallizes in space group I 4/m m m (#139), in a tetragonal body-centred structure.

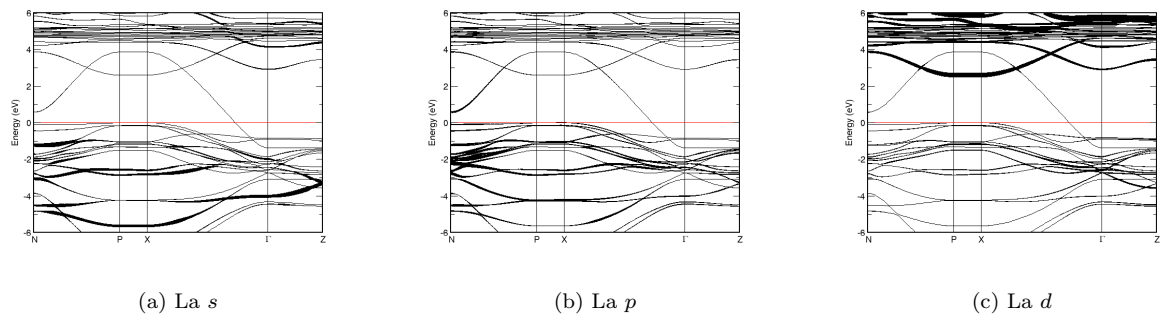

(a) La $s$      (b) La $p$      (c) La $d$

FIG. 374: Fat band representation of La in La$_2$PdO$_4$

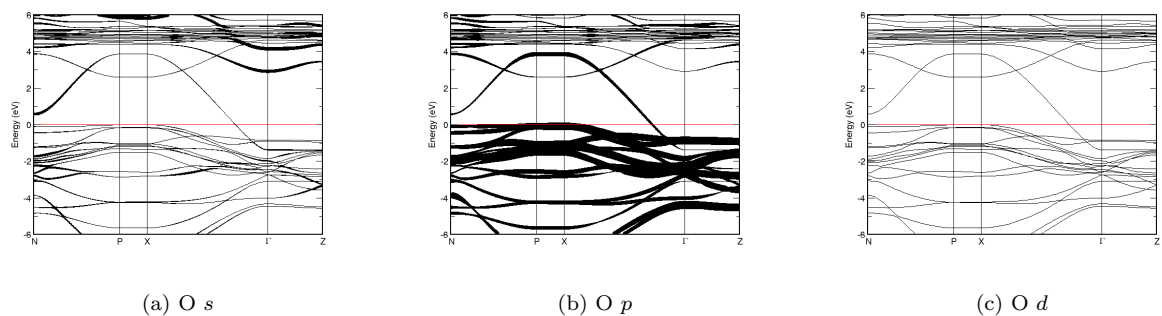

(a) O $s$      (b) O $p$      (c) O $d$

FIG. 375: Fat band representation of O in La$_2$PdO$_4$

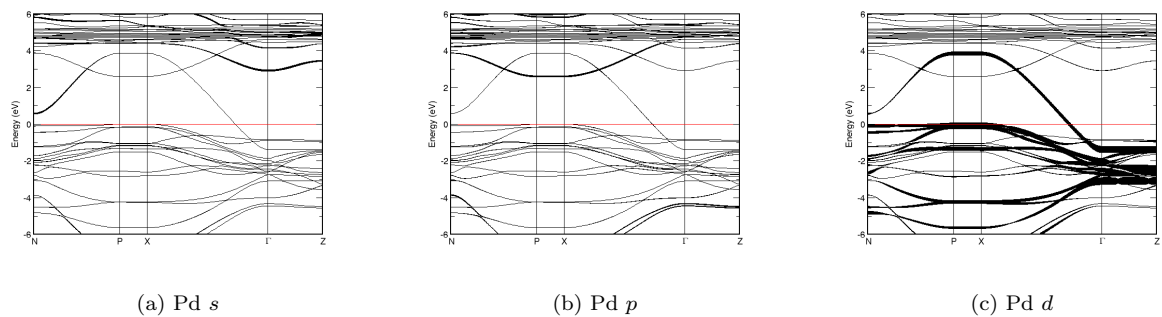

(a) Pd $s$      (b) Pd $p$      (c) Pd $d$

FIG. 376: Fat band representation of Pd in La$_2$PdO$_4$

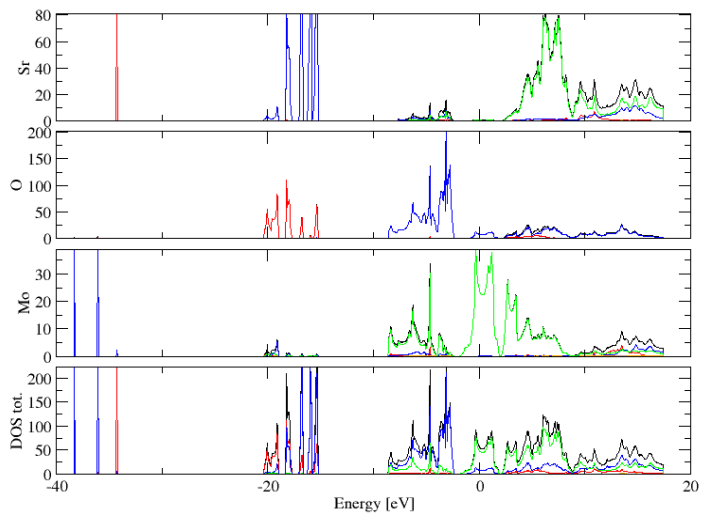

FIG. 377: (Color online) PDOS of $Sr_2(MoO_4)$ (ICSD #152123). The $s$-, $p$- and $d$-projected states are in red, blue and green, respectively. $Sr_2(MoO_4)$ crystallizes in space group I 4/m m m (#139), in a tetragonal body-centred structure.

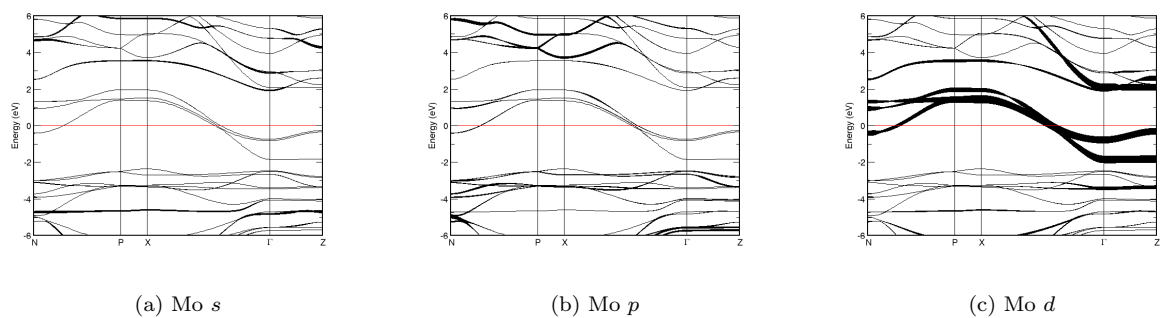

(a) Mo $s$           (b) Mo $p$           (c) Mo $d$

FIG. 378: Fat band representation of Mo in $Sr_2(MoO_4)$

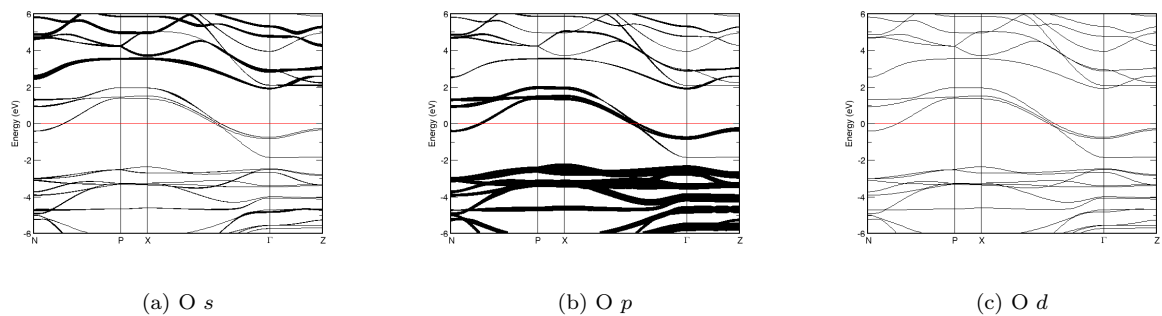

(a) O $s$           (b) O $p$           (c) O $d$

FIG. 379: Fat band representation of O in $Sr_2(MoO_4)$

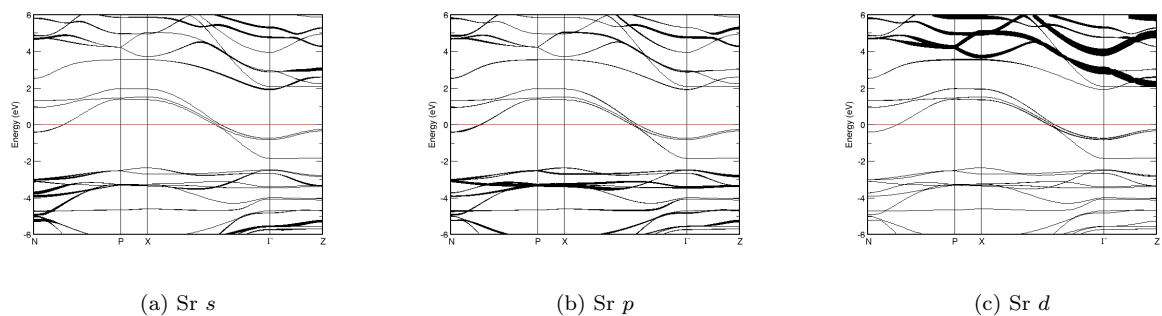

(a) Sr $s$    (b) Sr $p$    (c) Sr $d$

FIG. 380: Fat band representation of Sr in $Sr_2(MoO_4)$

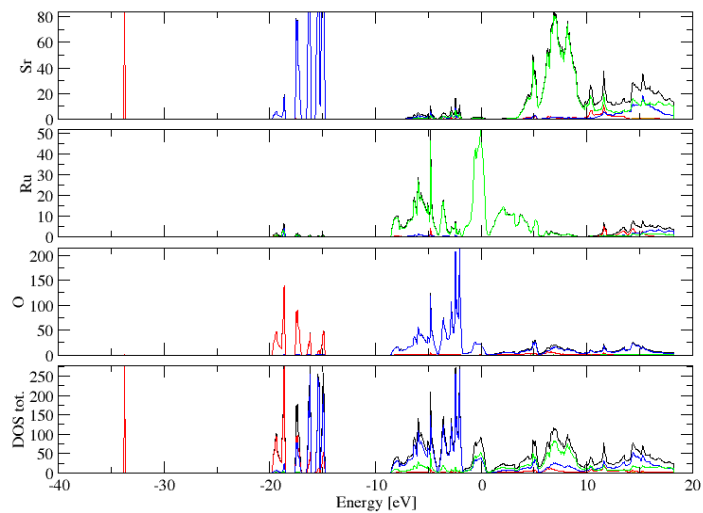

FIG. 381: (Color online) PDOS of $Sr_2(RuO_4)$ (ICSD #157401). The $s$-, $p$- and $d$-projected states are in red, blue and green, respectively. $Sr_2(RuO_4)$ crystallizes in space group I 4/m m m (#139), in a tetragonal body-centred structure.

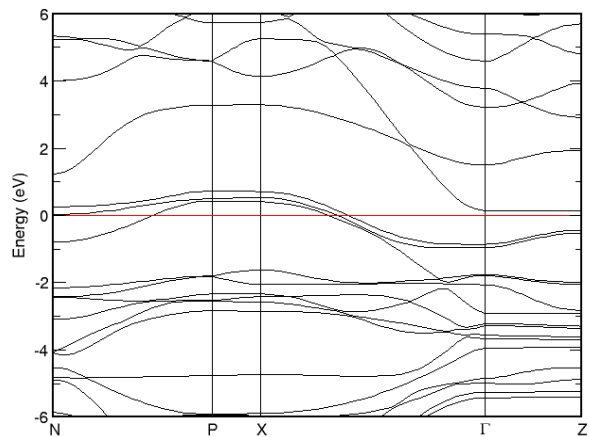

(a) E *vs.* k

FIG. 382: Band structure of $Sr_2(RuO_4)$

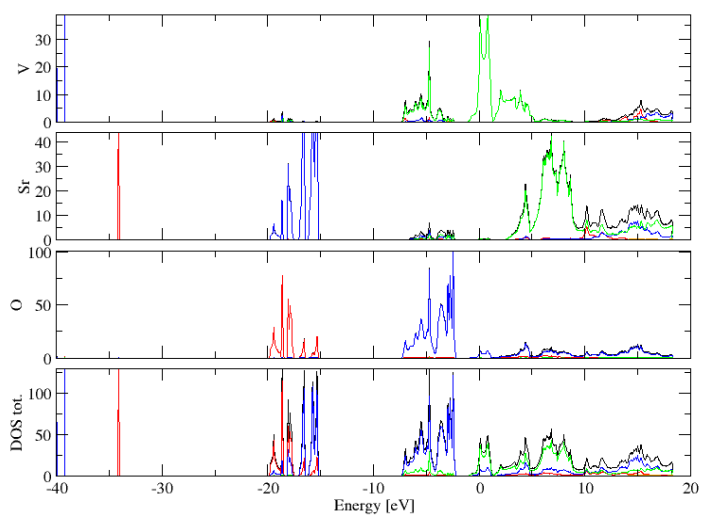

FIG. 383: (Color online) PDOS of $Sr_2VO_4$ (ICSD #72219). The *s*-, *p*- and *d*-projected states are in red, blue and green, respectively. $Sr_2VO_4$ crystallizes in space group I 4/m m m (#139), in a tetragonal body-centred structure.

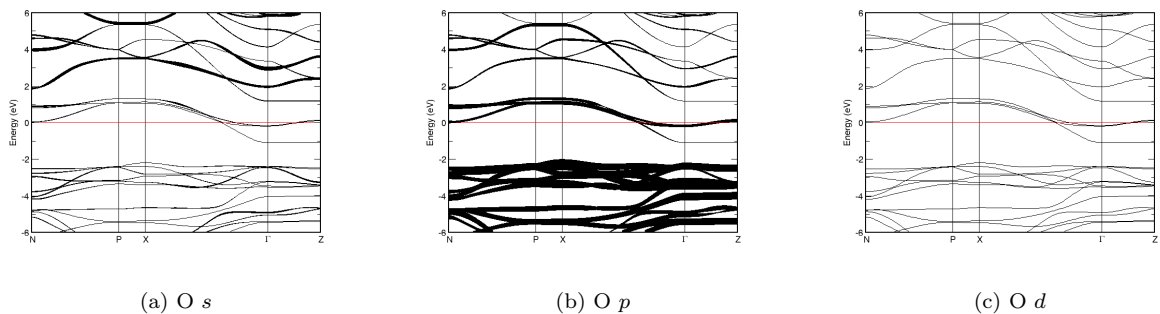

(a) O *s*          (b) O *p*          (c) O *d*

FIG. 384: Fat band representation of O in $Sr_2VO_4$

(a) Sr $s$      (b) Sr $p$      (c) Sr $d$

FIG. 385: Fat band representation of Sr in $Sr_2VO_4$

(a) V $s$      (b) V $p$      (c) V $d$

FIG. 386: Fat band representation of V in $Sr_2VO_4$

FIG. 387: (Color online) PDOS of $Cs_2AgF_4$ (ICSD #16254). The $s$-, $p$- and $d$-projected states are in red, blue and green, respectively. $Cs_2AgF_4$ crystallizes in space group I 4/m m m (#139), in a tetragonal body-centred structure.

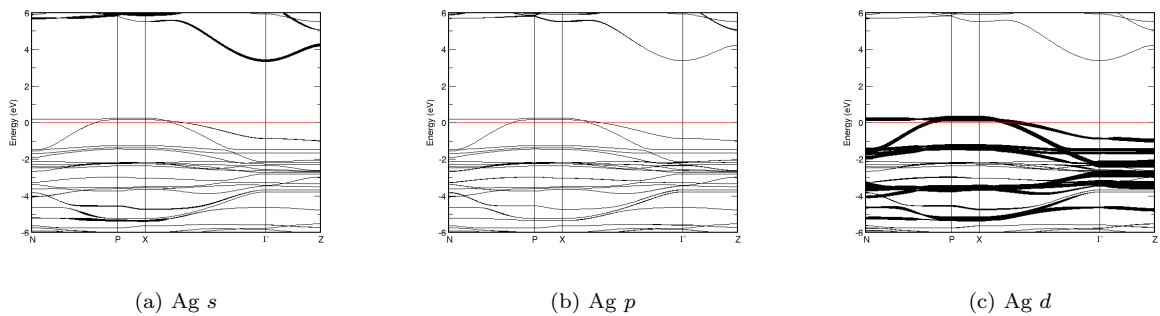

(a) Ag $s$

(b) Ag $p$

(c) Ag $d$

FIG. 388: Fat band representation of Ag in $Cs_2AgF_4$

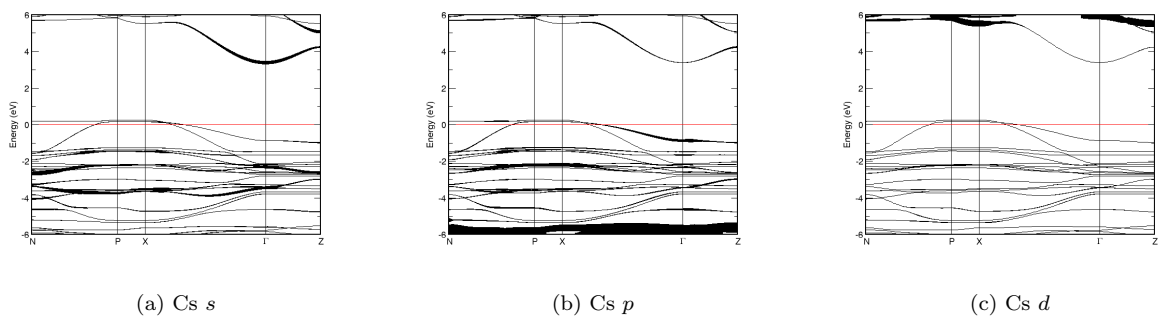

(a) Cs $s$

(b) Cs $p$

(c) Cs $d$

FIG. 389: Fat band representation of Cs in $Cs_2AgF_4$

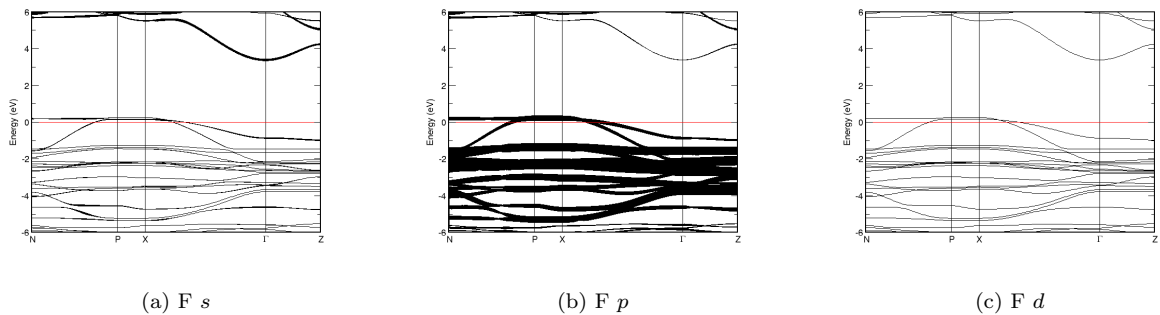

(a) F $s$

(b) F $p$

(c) F $d$

FIG. 390: Fat band representation of F in $Cs_2AgF_4$

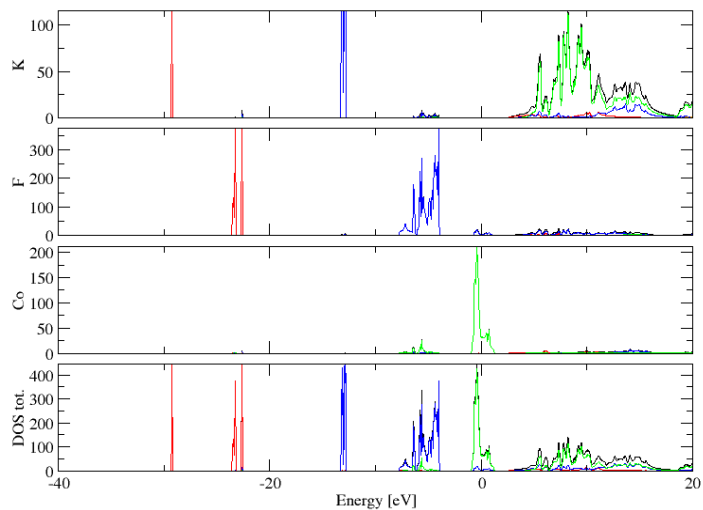

FIG. 391: (Color online) PDOS of K$_2$CoF$_4$ (ICSD #33522). The $s$-, $p$- and $d$-projected states are in red, blue and green, respectively. K$_2$CoF$_4$ crystallizes in space group I 4/m m m (#139), in a tetragonal body-centred structure.

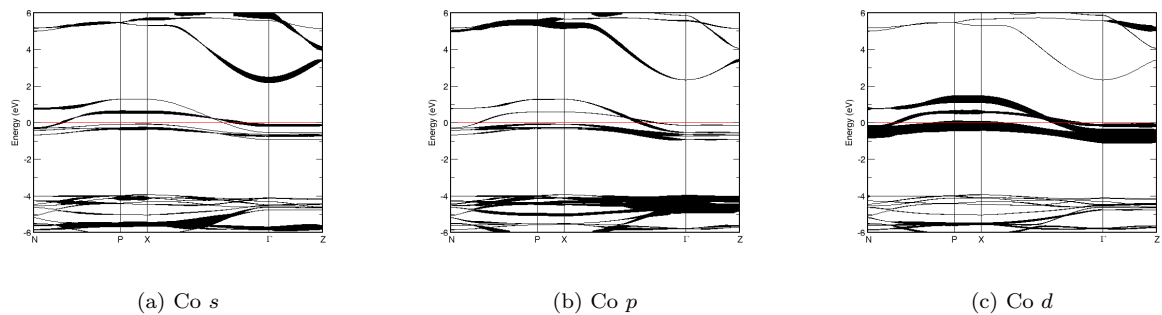

(a) Co $s$  (b) Co $p$  (c) Co $d$

FIG. 392: Fat band representation of Co in K$_2$CoF$_4$

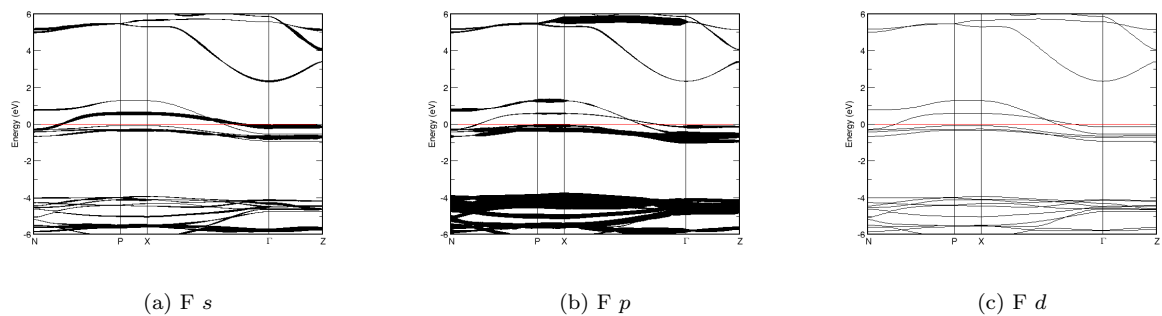

(a) F $s$  (b) F $p$  (c) F $d$

FIG. 393: Fat band representation of F in K$_2$CoF$_4$

(a) K $s$

(b) K $p$

(c) K $d$

FIG. 394: Fat band representation of K in $K_2CoF_4$

FIG. 395: (Color online) PDOS of $Rb_2CoF_4$ (ICSD #69683). The $s$-, $p$- and $d$-projected states are in red, blue and green, respectively. $Rb_2CoF_4$ crystallizes in space group I 4/m m m (#139), in a tetragonal body-centred structure.

(a) Co $s$

(b) Co $p$

(c) Co $d$

FIG. 396: Fat band representation of Co in $Rb_2CoF_4$

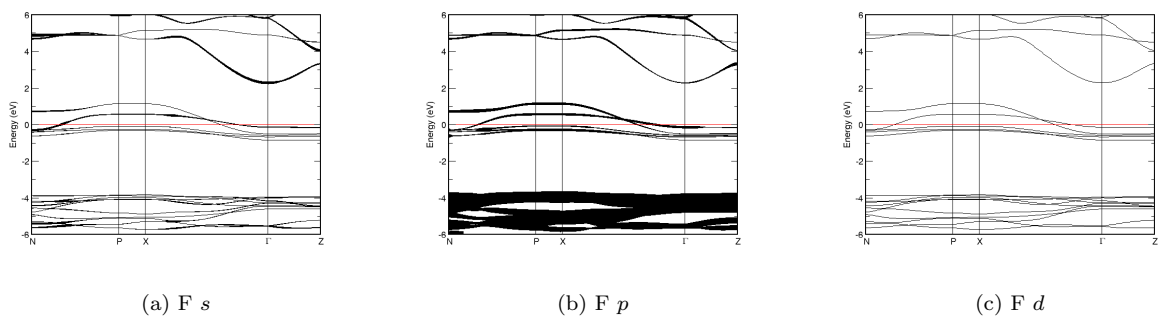

(a) F $s$                    (b) F $p$                    (c) F $d$

FIG. 397: Fat band representation of F in $Rb_2CoF_4$

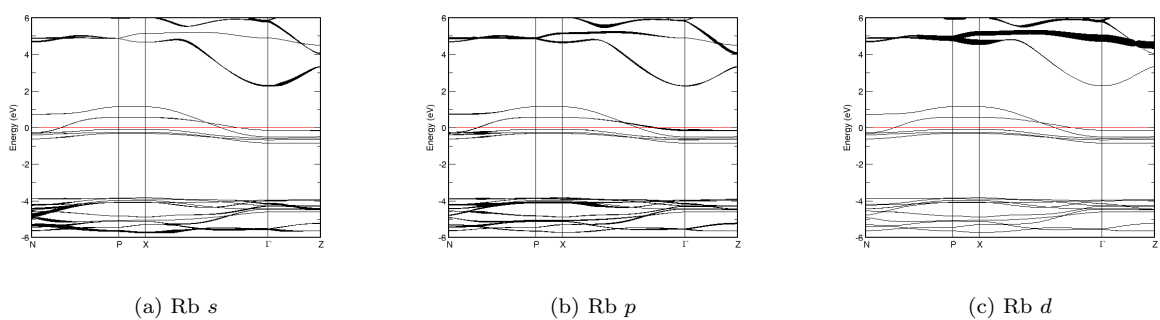

(a) Rb $s$                   (b) Rb $p$                   (c) Rb $d$

FIG. 398: Fat band representation of Rb in $Rb_2CoF_4$

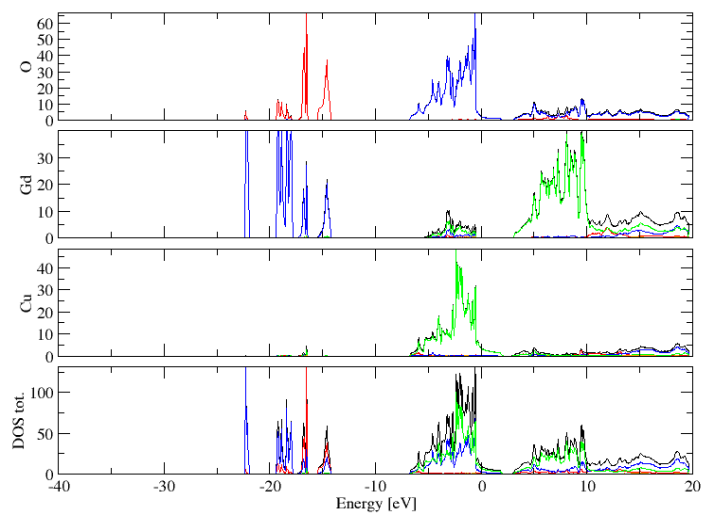

FIG. 399: (Color online) PDOS of $Gd_2(CuO_4)$ (ICSD #41844). The $s$-, $p$- and $d$-projected states are in red, blue and green, respectively. $Gd_2(CuO_4)$ crystallizes in space group I 4/m m m (#139), in a tetragonal body-centred structure.

(a) Cu $s$

(b) Cu $p$

(c) Cu $d$

FIG. 400: Fat band representation of Cu in $Gd_2(CuO_4)$

(a) Gd $s$

(b) Gd $p$

(c) Gd $d$

FIG. 401: Fat band representation of Gd in $Gd_2(CuO_4)$

(a) O $s$

(b) O $p$

(c) O $d$

FIG. 402: Fat band representation of O in $Gd_2(CuO_4)$

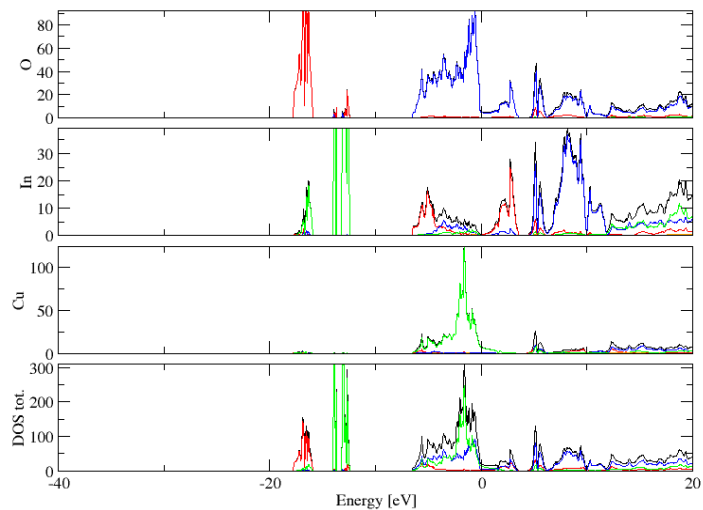

FIG. 403: (Color online) PDOS of In$_2$CuO$_4$ (ICSD #39475). The $s$-, $p$- and $d$-projected states are in red, blue and green, respectively. In$_2$CuO$_4$ crystallizes in space group I 4/m m m (#139), in a tetragonal body-centred structure.

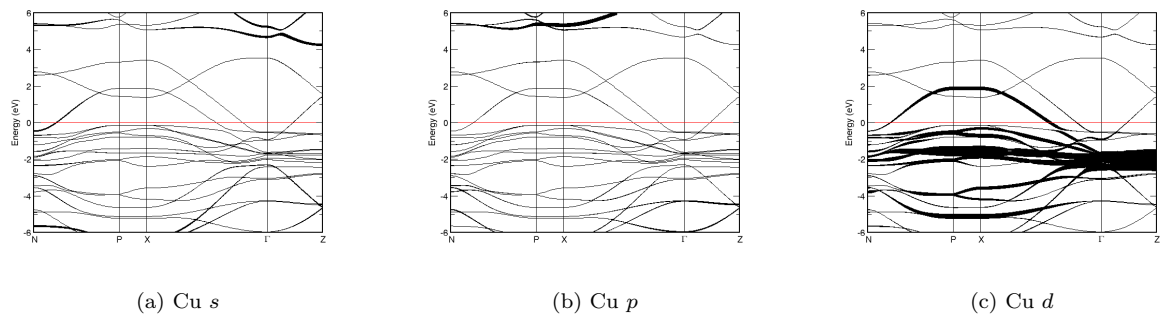

(a) Cu $s$        (b) Cu $p$        (c) Cu $d$

FIG. 404: Fat band representation of Cu in In$_2$CuO$_4$

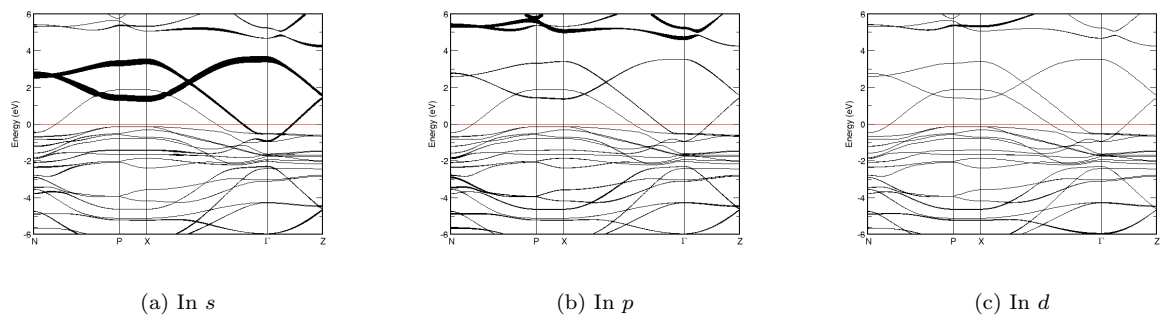

(a) In $s$        (b) In $p$        (c) In $d$

FIG. 405: Fat band representation of In in In$_2$CuO$_4$

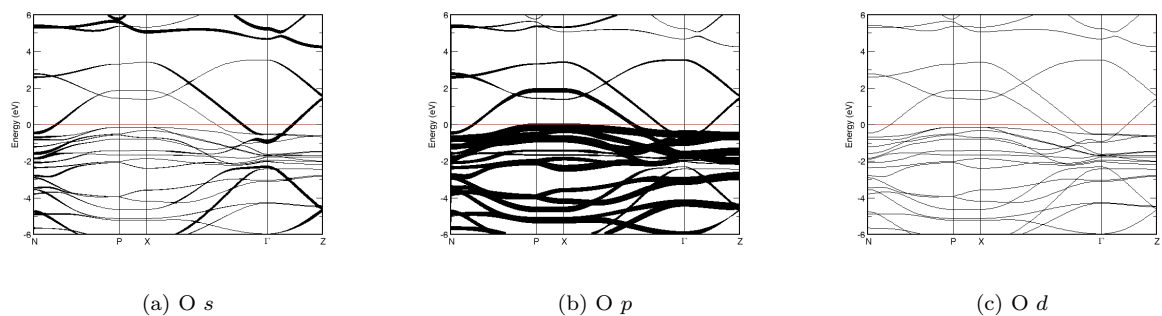

(a) O $s$      (b) O $p$      (c) O $d$

FIG. 406: Fat band representation of O in $In_2CuO_4$

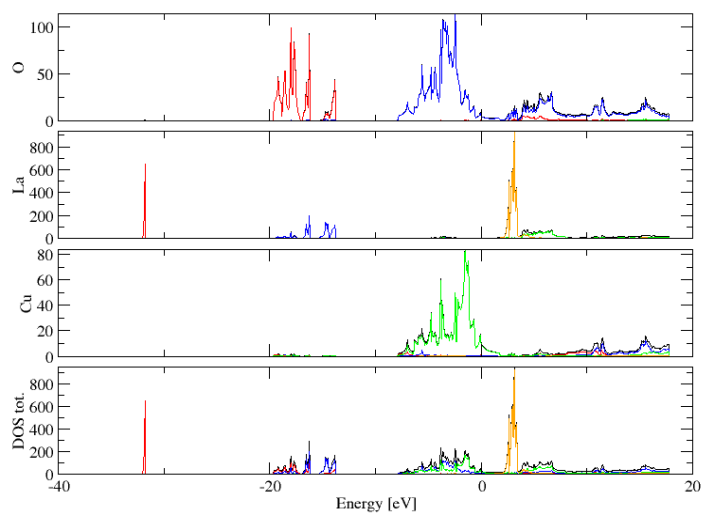

FIG. 407: (Color online) PDOS of $La_2(CuO_4)$ (ICSD #41643). The $s$-, $p$- and $d$-projected states are in red, blue and green, respectively. $La_2(CuO_4)$ crystallizes in space group I 4/m m m (#139), in a tetragonal body-centred structure.

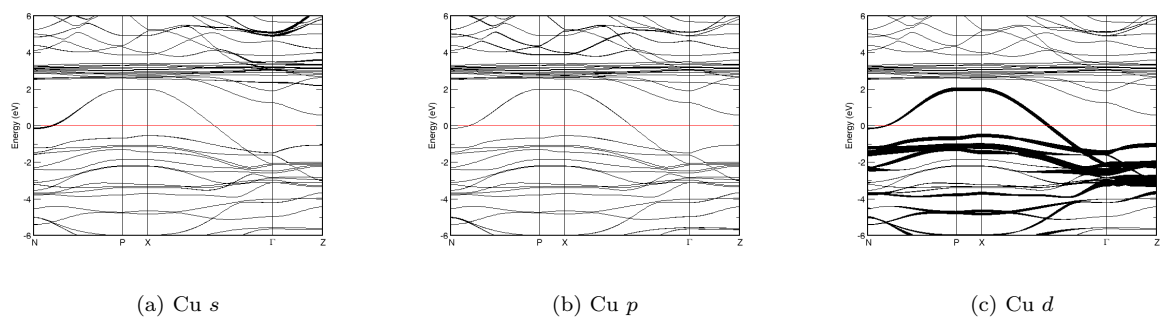

(a) Cu $s$      (b) Cu $p$      (c) Cu $d$

FIG. 408: Fat band representation of Cu in $La_2(CuO_4)$

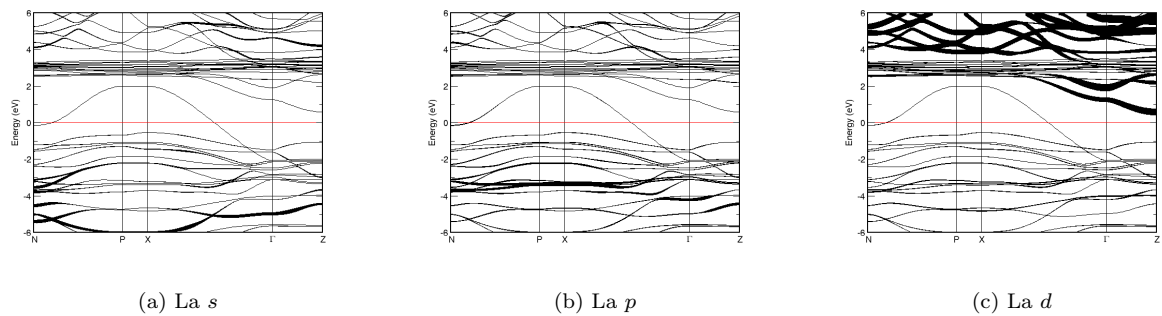

(a) La $s$          (b) La $p$          (c) La $d$

FIG. 409: Fat band representation of La in $La_2(CuO_4)$

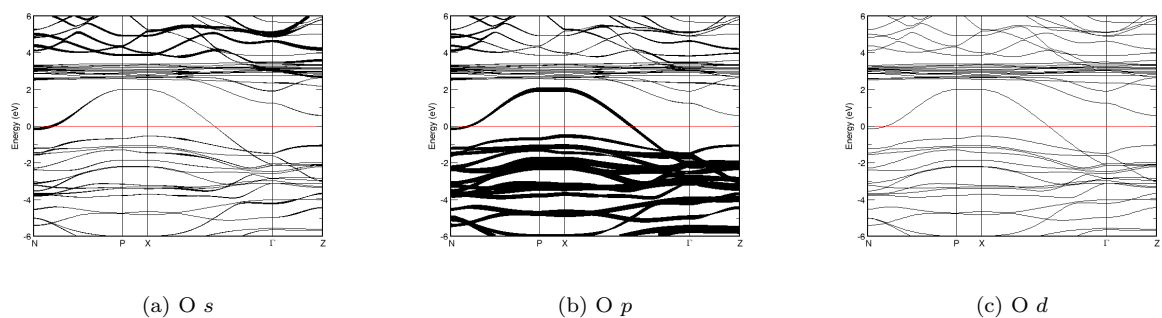

(a) O $s$          (b) O $p$          (c) O $d$

FIG. 410: Fat band representation of O in $La_2(CuO_4)$

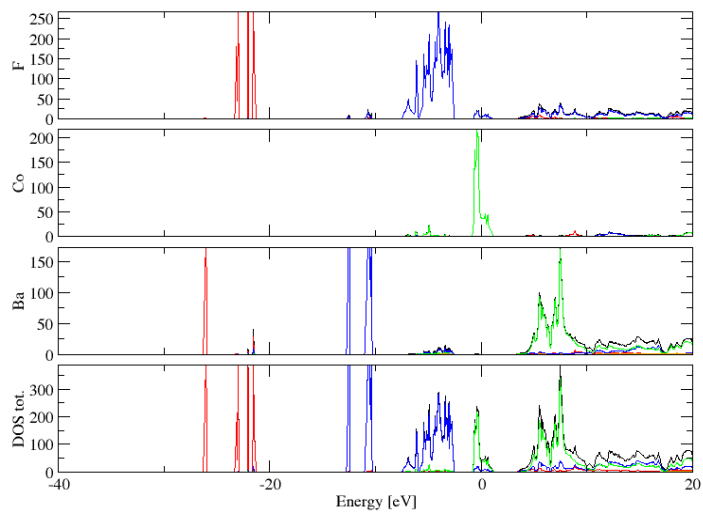

FIG. 411: (Color online) PDOS of $Ba_2CoF_6$ (ICSD #21057). The $s$-, $p$- and $d$-projected states are in red, blue and green, respectively. $Ba_2CoF_6$ crystallizes in space group I 4/m m m (#139), in a tetragonal body-centred structure.

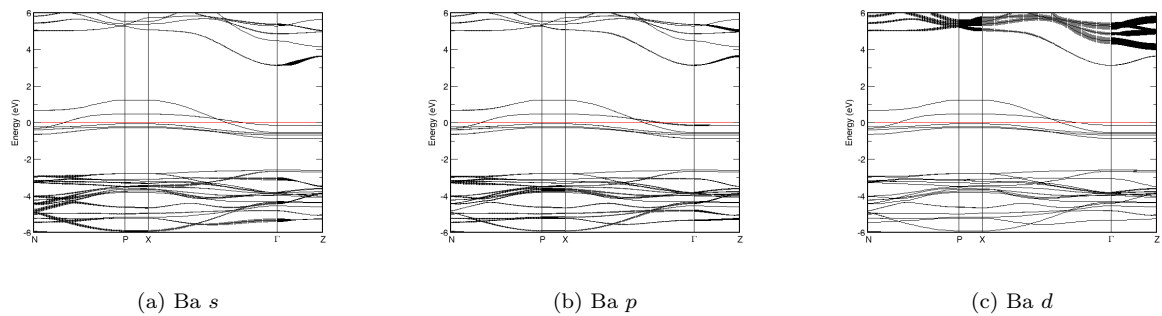

(a) Ba $s$      (b) Ba $p$      (c) Ba $d$

FIG. 412: Fat band representation of Ba in Ba$_2$CoF$_6$

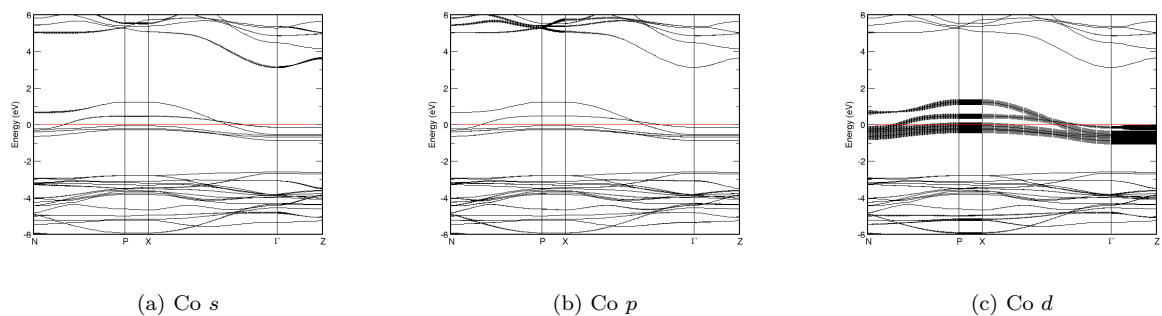

(a) Co $s$      (b) Co $p$      (c) Co $d$

FIG. 413: Fat band representation of Co in Ba$_2$CoF$_6$

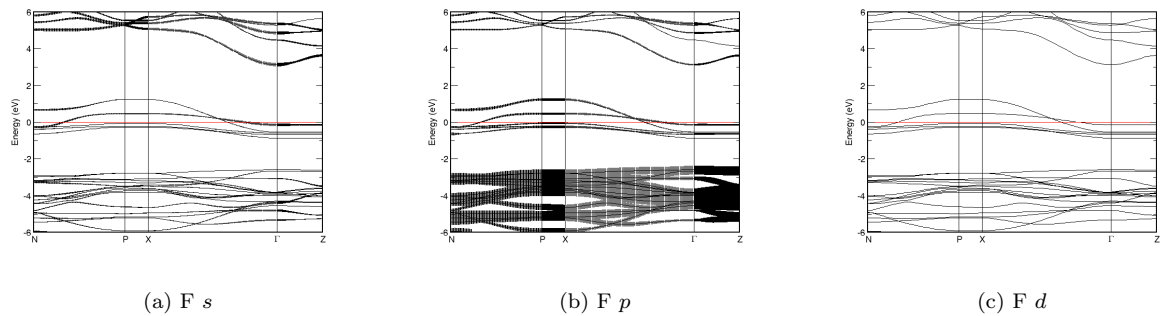

(a) F $s$      (b) F $p$      (c) F $d$

FIG. 414: Fat band representation of F in Ba$_2$CoF$_6$

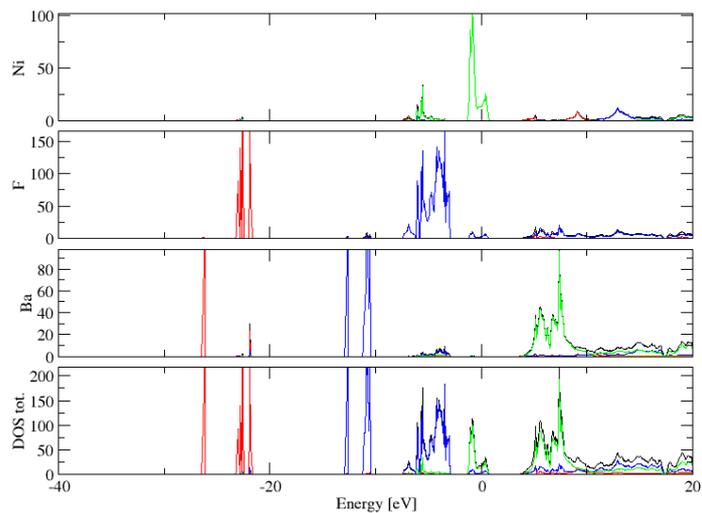

FIG. 415: (Color online) PDOS of $Ba_2NiF_6$ (ICSD #21056). The $s$-, $p$- and $d$-projected states are in red, blue and green, respectively. $Ba_2NiF_6$ crystallizes in space group I 4/m m m (#139), in a tetragonal body-centred structure.

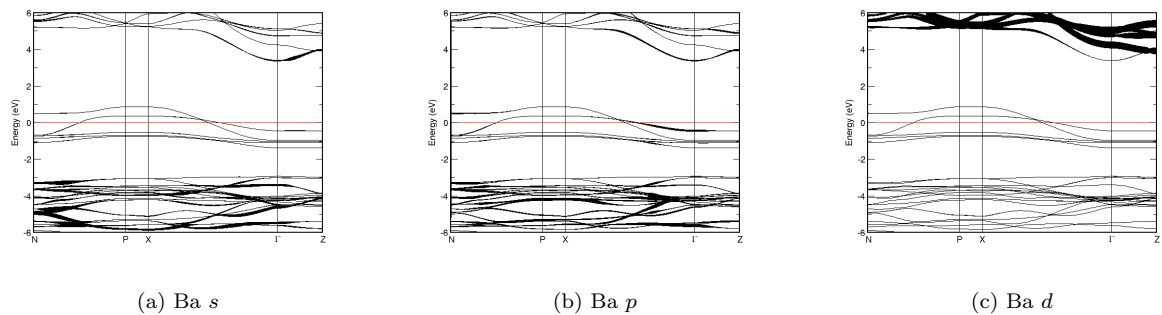

(a) Ba $s$          (b) Ba $p$          (c) Ba $d$

FIG. 416: Fat band representation of Ba in $Ba_2NiF_6$

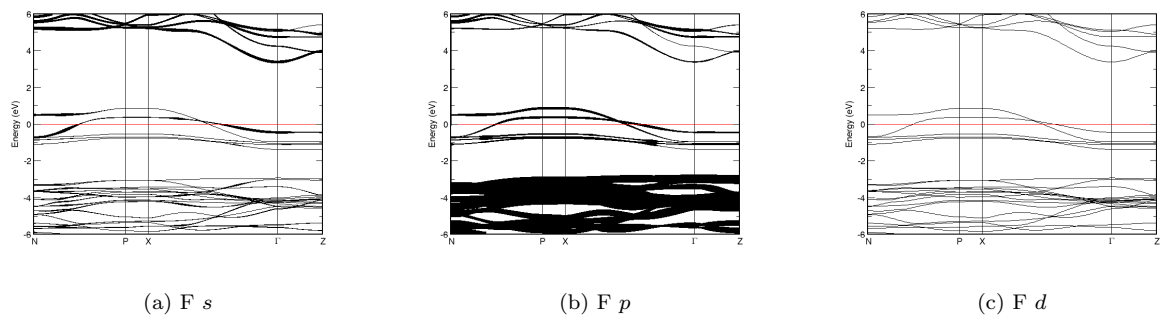

(a) F $s$          (b) F $p$          (c) F $d$

FIG. 417: Fat band representation of F in $Ba_2NiF_6$

(a) Ni $s$       (b) Ni $p$       (c) Ni $d$

FIG. 418: Fat band representation of Ni in $Ba_2NiF_6$

FIG. 419: (Color online) PDOS of $Ba_2(ZnF_6)$ (ICSD #21054). The $s$-, $p$- and $d$-projected states are in red, blue and green, respectively. $Ba_2(ZnF_6)$ crystallizes in space group I 4/m m m (#139), in a tetragonal body-centred structure.

(a) Ba $s$       (b) Ba $p$       (c) Ba $d$

FIG. 420: Fat band representation of Ba in $Ba_2(ZnF_6)$

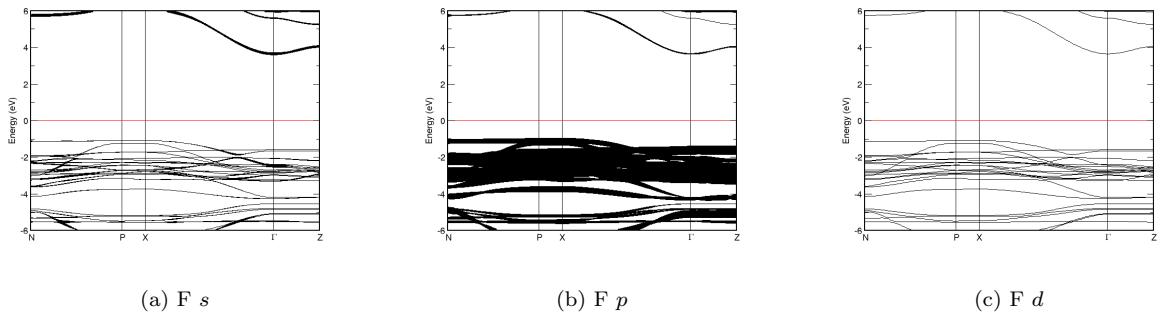

(a) F $s$      (b) F $p$      (c) F $d$

FIG. 421: Fat band representation of F in Ba$_2$(ZnF$_6$)

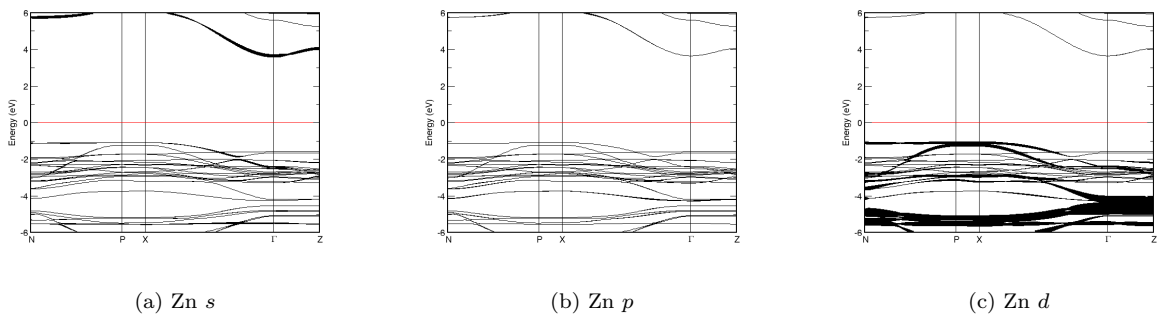

(a) Zn $s$      (b) Zn $p$      (c) Zn $d$

FIG. 422: Fat band representation of Zn in Ba$_2$(ZnF$_6$)

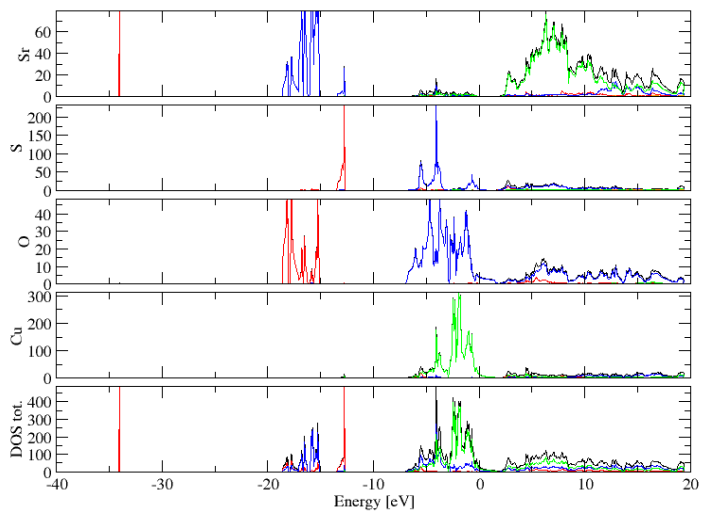

FIG. 423: (Color online) PDOS of (Cu$_2$S$_2$)(Sr$_2$CuO$_2$) (ICSD #88423). The $s$-, $p$- and $d$-projected states are in red, blue and green, respectively. (Cu$_2$S$_2$)(Sr$_2$CuO$_2$) crystallizes in space group I 4/m m m (#139), in a tetragonal body-centred structure.

(a) Cu $s$

(b) Cu $p$

(c) Cu $d$

FIG. 424: Fat band representation of Cu in $(Cu_2S_2)(Sr_2CuO_2)$

(a) O $s$

(b) O $p$

(c) O $d$

FIG. 425: Fat band representation of O in $(Cu_2S_2)(Sr_2CuO_2)$

(a) S $s$

(b) S $p$

(c) S $d$

FIG. 426: Fat band representation of S in $(Cu_2S_2)(Sr_2CuO_2)$

(a) Sr $s$

(b) Sr $p$

(c) Sr $d$

FIG. 427: Fat band representation of Sr in $(Cu_2S_2)(Sr_2CuO_2)$

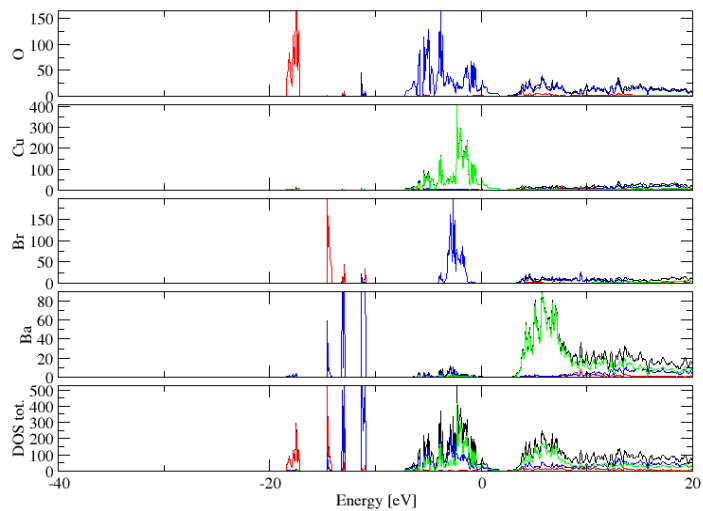

FIG. 428: (Color online) PDOS of $Ba_2Cu_3O_4Br_2$ (ICSD #36128). The $s$-, $p$- and $d$-projected states are in red, blue and green, respectively. $Ba_2Cu_3O_4Br_2$ crystallizes in space group I 4/m m m (#139), in a tetragonal body-centred structure.

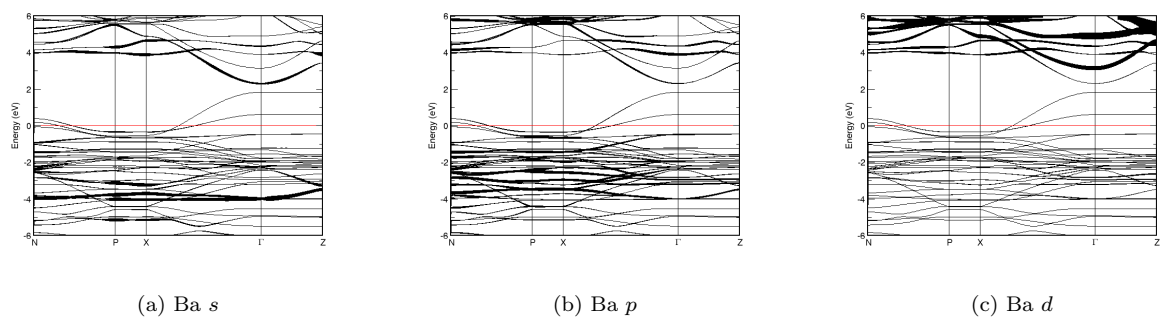

(a) Ba $s$            (b) Ba $p$            (c) Ba $d$

FIG. 429: Fat band representation of Ba in $Ba_2Cu_3O_4Br_2$

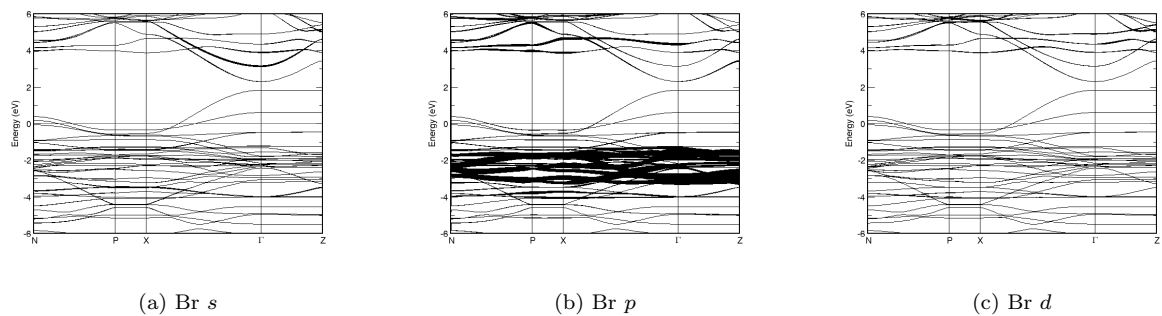

(a) Br $s$            (b) Br $p$            (c) Br $d$

FIG. 430: Fat band representation of Br in $Ba_2Cu_3O_4Br_2$

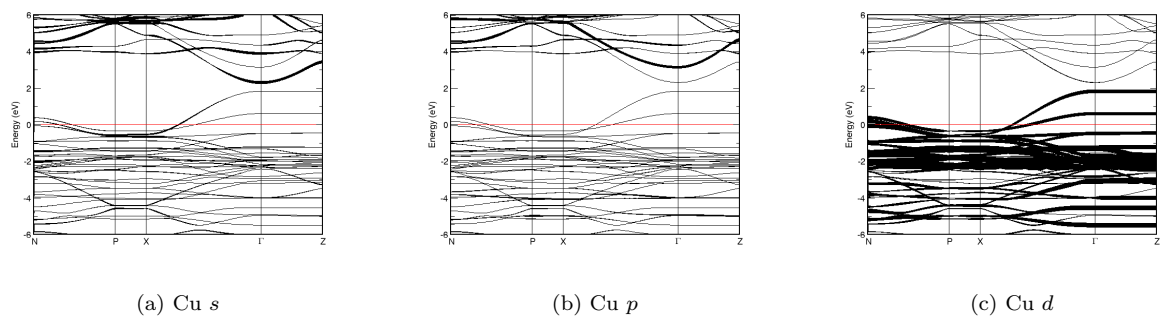

(a) Cu $s$                    (b) Cu $p$                    (c) Cu $d$

FIG. 431: Fat band representation of Cu in $Ba_2Cu_3O_4Br_2$

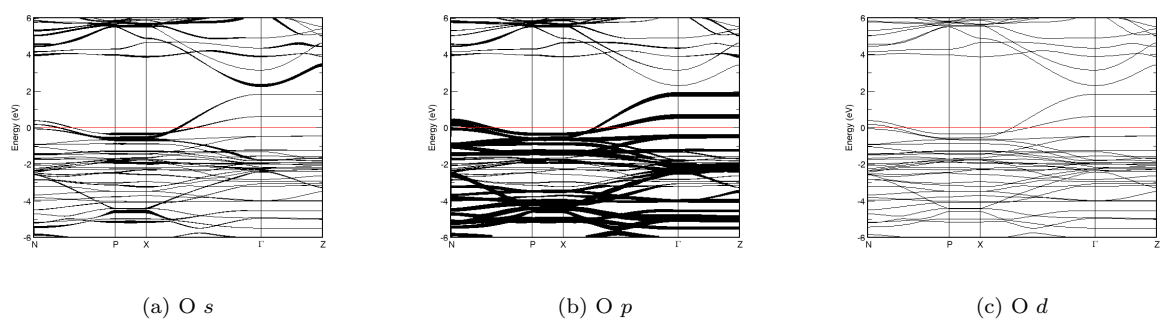

(a) O $s$                    (b) O $p$                    (c) O $d$

FIG. 432: Fat band representation of O in $Ba_2Cu_3O_4Br_2$

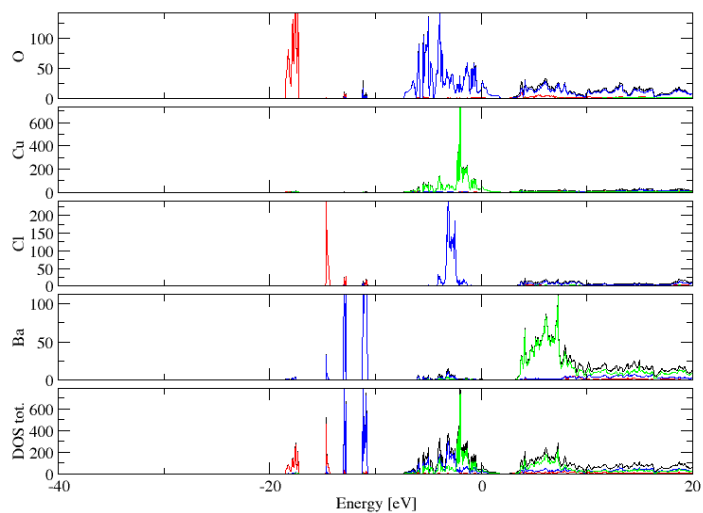

FIG. 433: (Color online) PDOS of $Ba_2Cu_3O_4Cl_2$ (ICSD #355). The $s$-, $p$- and $d$-projected states are in red, blue and green, respectively. $Ba_2Cu_3O_4Cl_2$ crystallizes in space group I 4/m m m (#139), in a tetragonal body-centred structure.

(a) Ba $s$

(b) Ba $p$

(c) Ba $d$

FIG. 434: Fat band representation of Ba in $Ba_2Cu_3O_4Cl_2$

(a) Cl $s$

(b) Cl $p$

(c) Cl $d$

FIG. 435: Fat band representation of Cl in $Ba_2Cu_3O_4Cl_2$

(a) Cu $s$

(b) Cu $p$

(c) Cu $d$

FIG. 436: Fat band representation of Cu in $Ba_2Cu_3O_4Cl_2$

(a) O $s$

(b) O $p$

(c) O $d$

FIG. 437: Fat band representation of O in $Ba_2Cu_3O_4Cl_2$

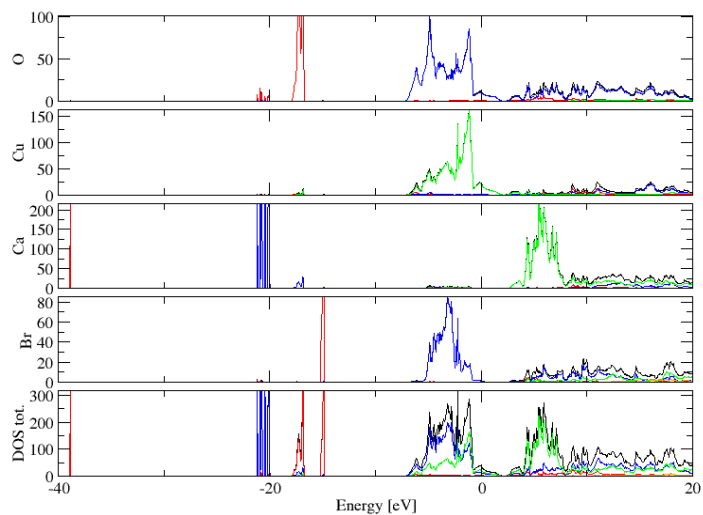

FIG. 438: (Color online) PDOS of Ca$_3$Cu$_2$O$_4$Br$_2$ (ICSD #69182). The $s$-, $p$- and $d$-projected states are in red, blue and green, respectively. Ca$_3$Cu$_2$O$_4$Br$_2$ crystallizes in space group I 4/m m m (#139), in a tetragonal body-centred structure.

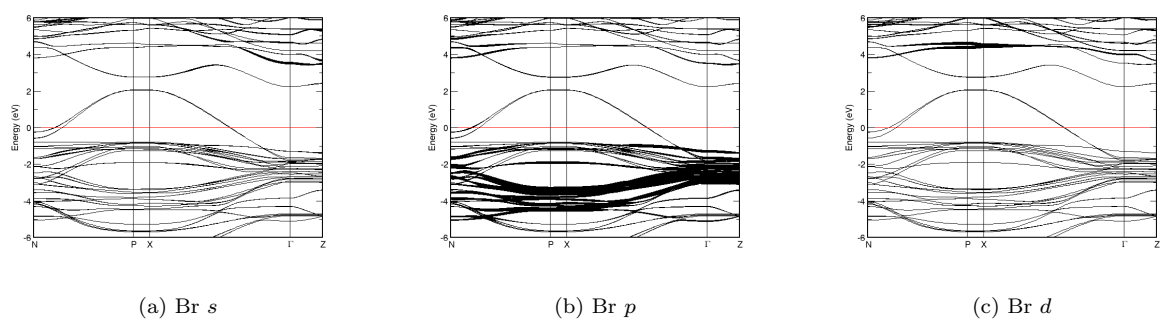

(a) Br $s$        (b) Br $p$        (c) Br $d$

FIG. 439: Fat band representation of Br in Ca$_3$Cu$_2$O$_4$Br$_2$

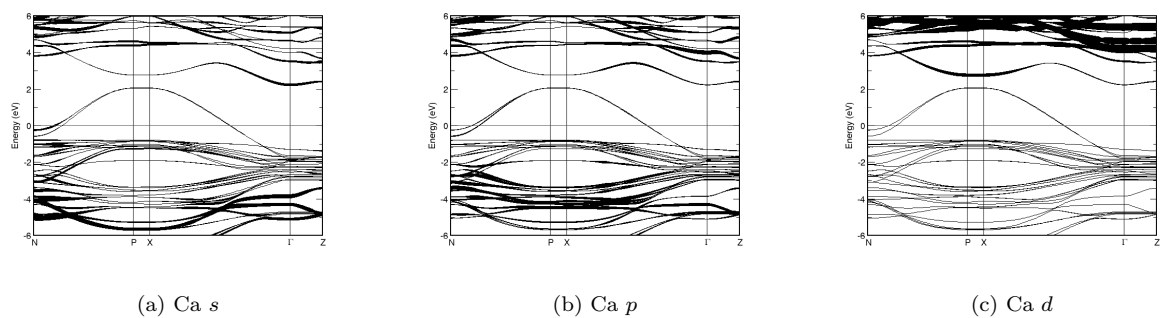

(a) Ca $s$        (b) Ca $p$        (c) Ca $d$

FIG. 440: Fat band representation of Ca in Ca$_3$Cu$_2$O$_4$Br$_2$

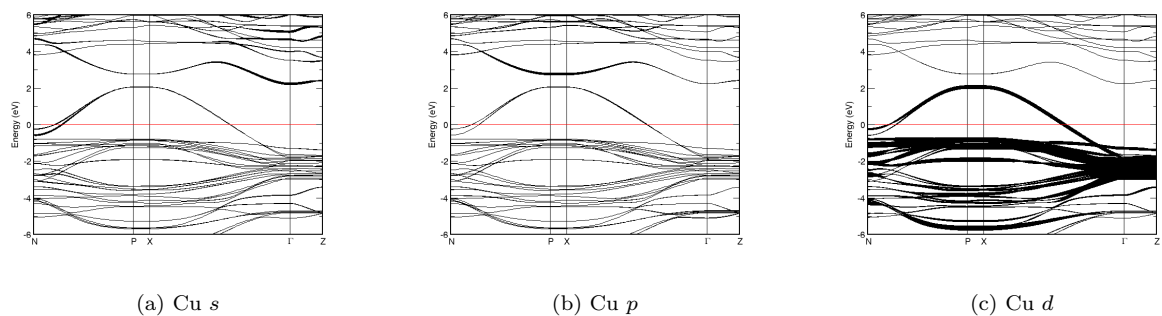

(a) Cu $s$       (b) Cu $p$       (c) Cu $d$

FIG. 441: Fat band representation of Cu in $Ca_3Cu_2O_4Br_2$

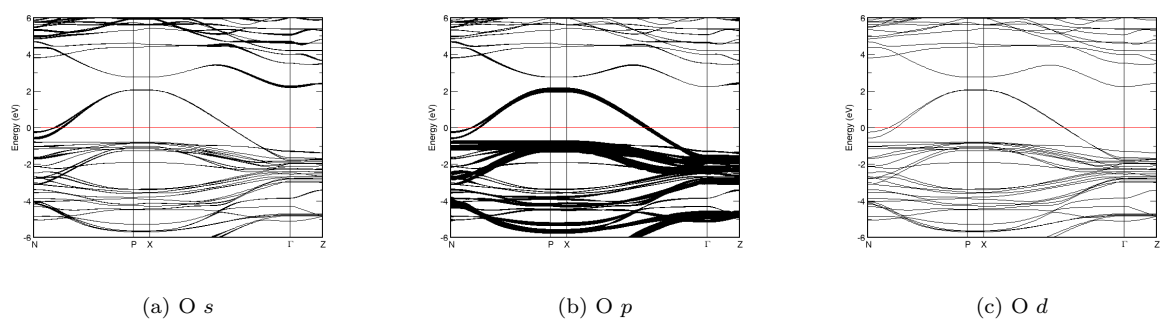

(a) O $s$       (b) O $p$       (c) O $d$

FIG. 442: Fat band representation of O in $Ca_3Cu_2O_4Br_2$

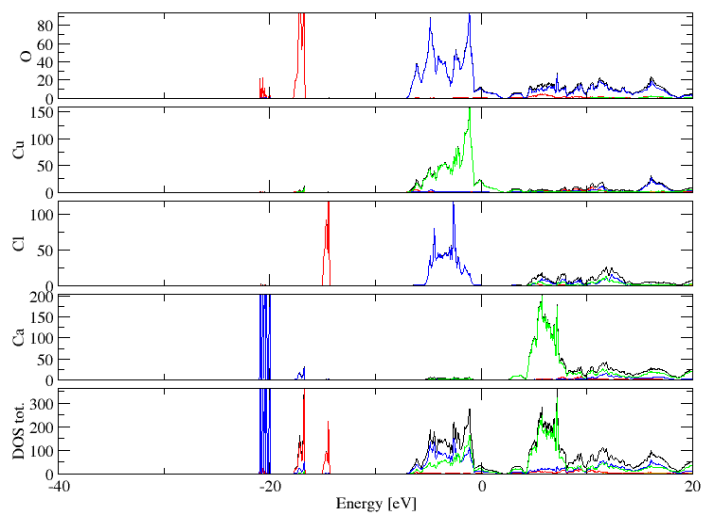

FIG. 443: (Color online) PDOS of $Ca_3Cu_2O_4Cl_2$ (ICSD #69181). The $s$-, $p$- and $d$-projected states are in red, blue and green, respectively. $Ca_3Cu_2O_4Cl_2$ crystallizes in space group I 4/m m m (#139), in a tetragonal body-centred structure.

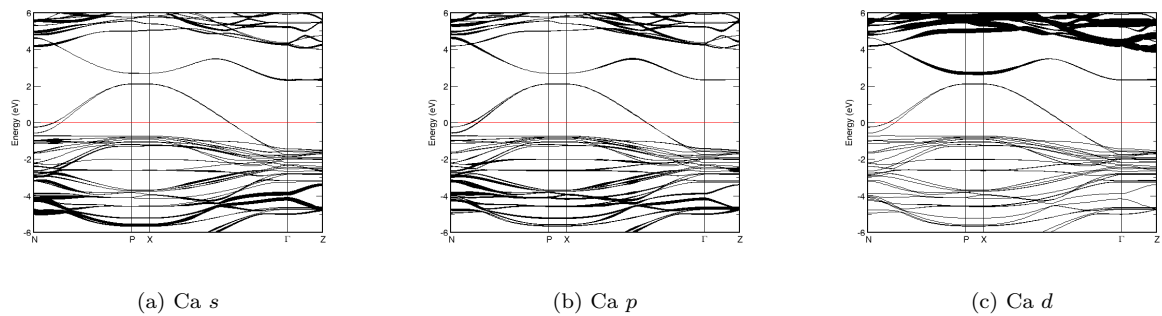

(a) Ca $s$

(b) Ca $p$

(c) Ca $d$

FIG. 444: Fat band representation of Ca in $Ca_3Cu_2O_4Cl_2$

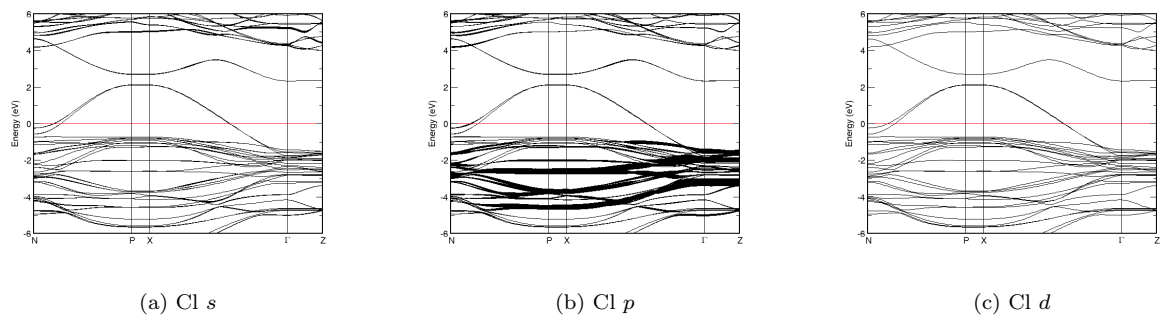

(a) Cl $s$

(b) Cl $p$

(c) Cl $d$

FIG. 445: Fat band representation of Cl in $Ca_3Cu_2O_4Cl_2$

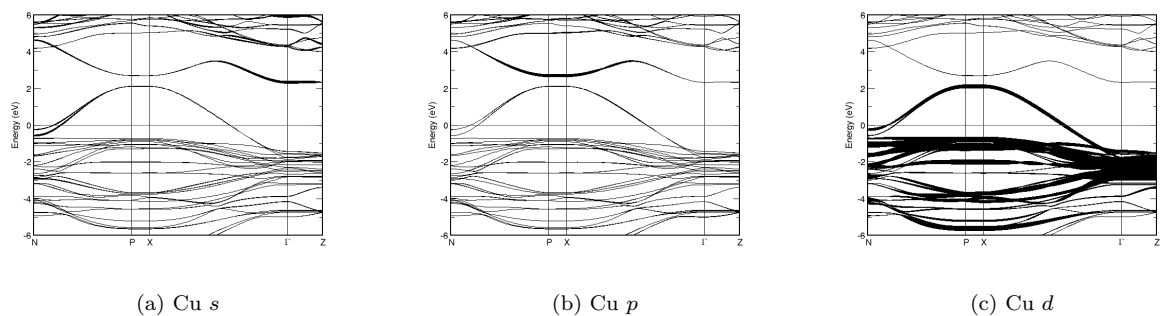

(a) Cu $s$

(b) Cu $p$

(c) Cu $d$

FIG. 446: Fat band representation of Cu in $Ca_3Cu_2O_4Cl_2$

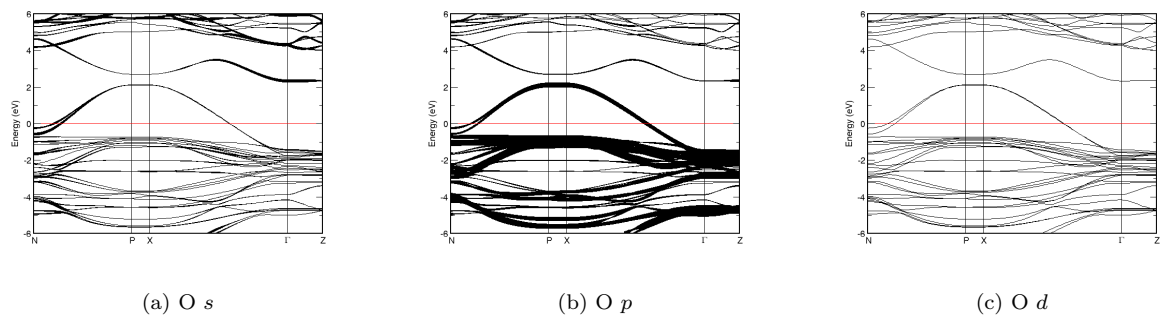

(a) O $s$

(b) O $p$

(c) O $d$

FIG. 447: Fat band representation of O in $Ca_3Cu_2O_4Cl_2$

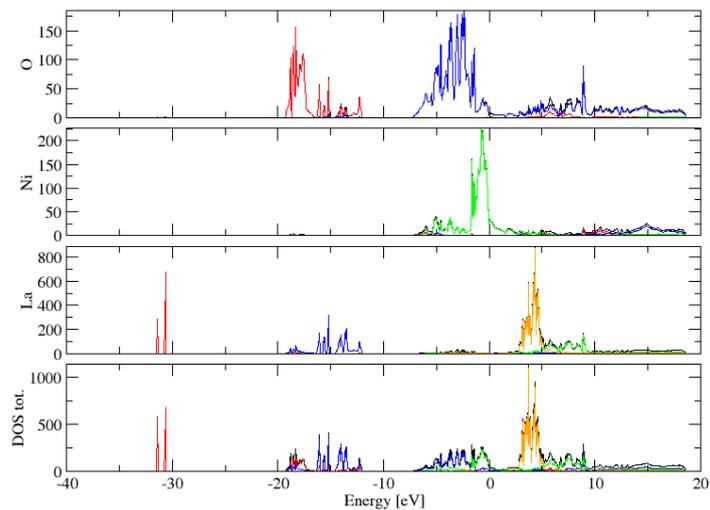

FIG. 448: (Color online) PDOS of La$_3$Ni$_2$O$_6$ (ICSD #249209). The $s$-, $p$- and $d$-projected states are in red, blue and green, respectively. La$_3$Ni$_2$O$_6$ crystallizes in space group I 4/m m m (#139), in a tetragonal body-centred structure.

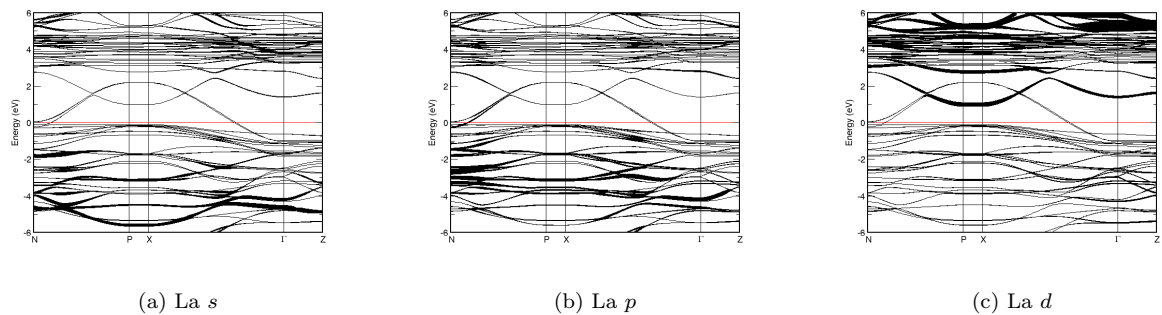

(a) La $s$                  (b) La $p$                  (c) La $d$

FIG. 449: Fat band representation of La in La$_3$Ni$_2$O$_6$

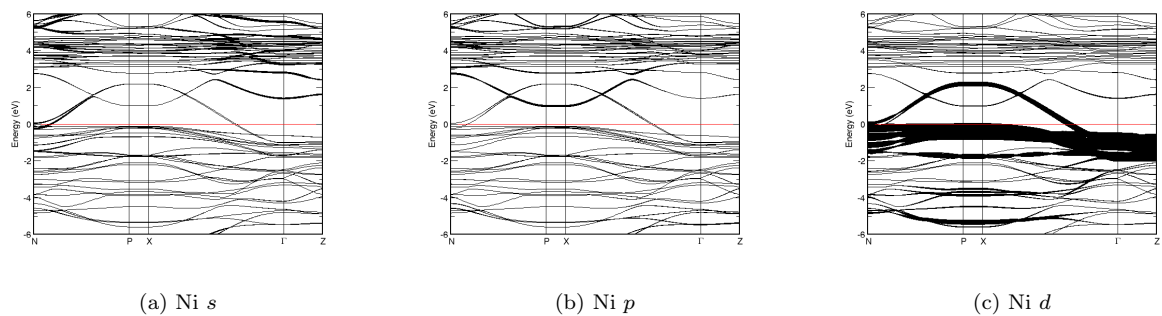

(a) Ni $s$                  (b) Ni $p$                  (c) Ni $d$

FIG. 450: Fat band representation of Ni in La$_3$Ni$_2$O$_6$

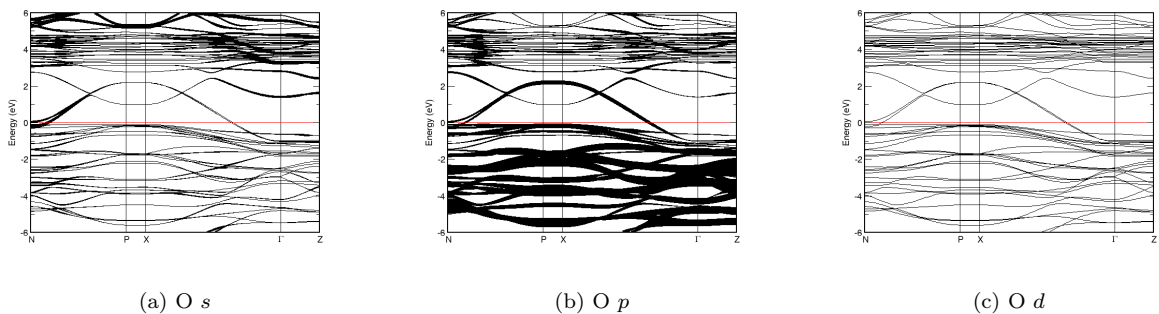

(a) O $s$        (b) O $p$        (c) O $d$

FIG. 451: Fat band representation of O in La$_3$Ni$_2$O$_6$

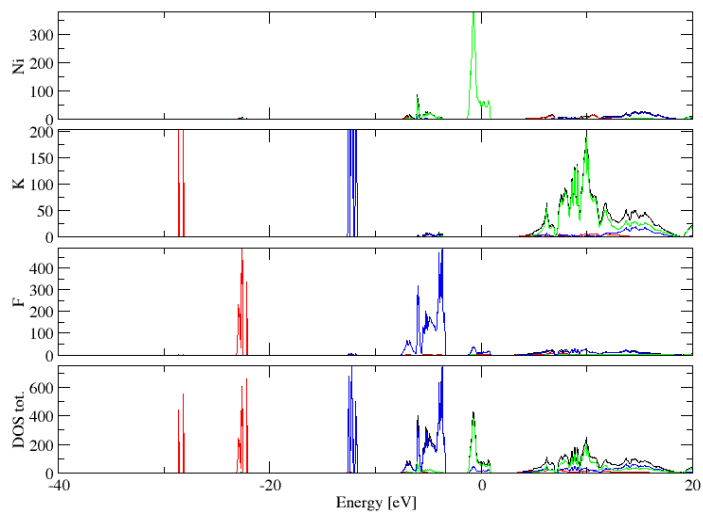

FIG. 452: (Color online) PDOS of K$_3$Ni$_2$F$_7$ (ICSD #33523). The $s$-, $p$- and $d$-projected states are in red, blue and green, respectively. K$_3$Ni$_2$F$_7$ crystallizes in space group I 4/m m m (#139), in a tetragonal body-centred structure.

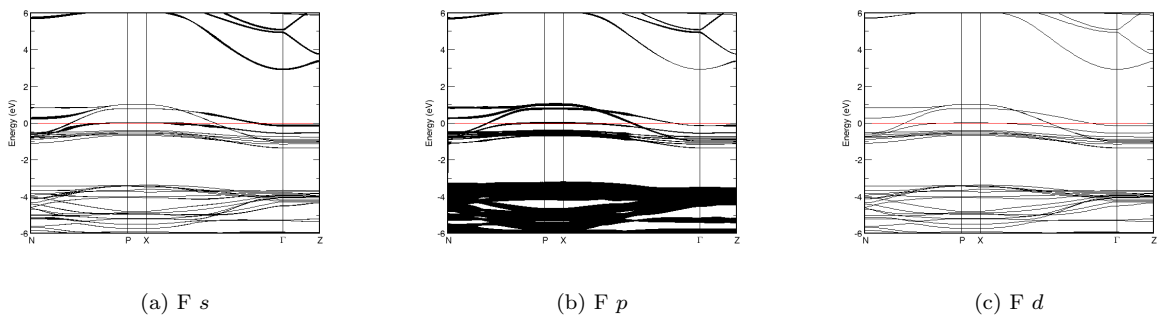

(a) F $s$        (b) F $p$        (c) F $d$

FIG. 453: Fat band representation of F in K$_3$Ni$_2$F$_7$

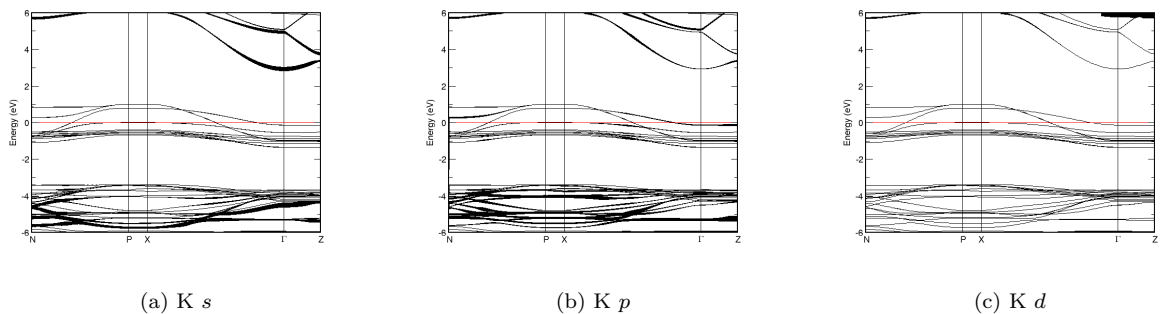

(a) K $s$          (b) K $p$          (c) K $d$

FIG. 454: Fat band representation of K in $K_3Ni_2F_7$

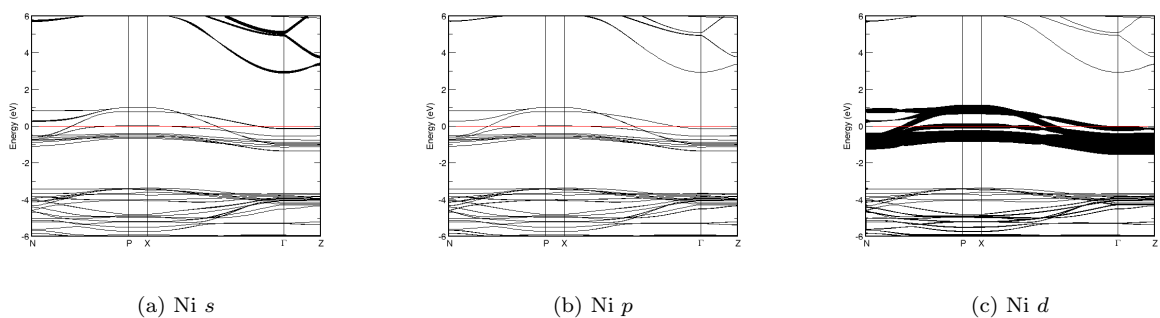

(a) Ni $s$          (b) Ni $p$          (c) Ni $d$

FIG. 455: Fat band representation of Ni in $K_3Ni_2F_7$

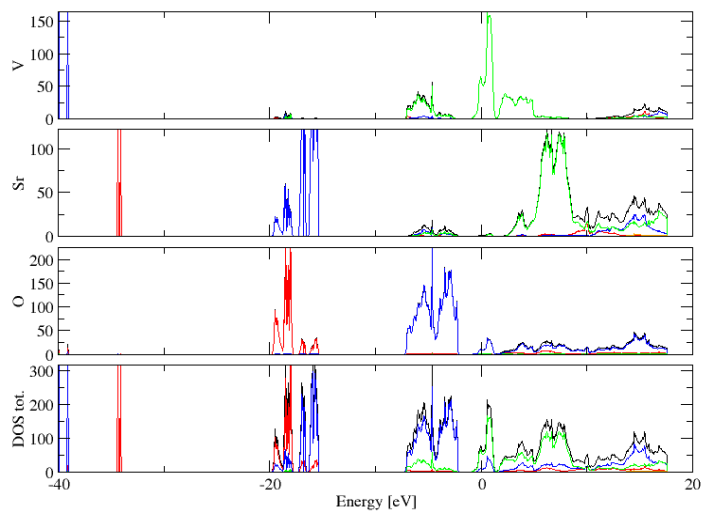

FIG. 456: (Color online) PDOS of $Sr_3V_2O_7$ (ICSD #71320). The $s$-, $p$- and $d$-projected states are in red, blue and green, respectively. $Sr_3V_2O_7$ crystallizes in space group I 4/m m m (#139), in a tetragonal body-centred structure.

(a) O $s$        (b) O $p$        (c) O $d$

FIG. 457: Fat band representation of O in $Sr_3V_2O_7$

(a) Sr $s$        (b) Sr $p$        (c) Sr $d$

FIG. 458: Fat band representation of Sr in $Sr_3V_2O_7$

(a) V $s$        (b) V $p$        (c) V $d$

FIG. 459: Fat band representation of V in $Sr_3V_2O_7$

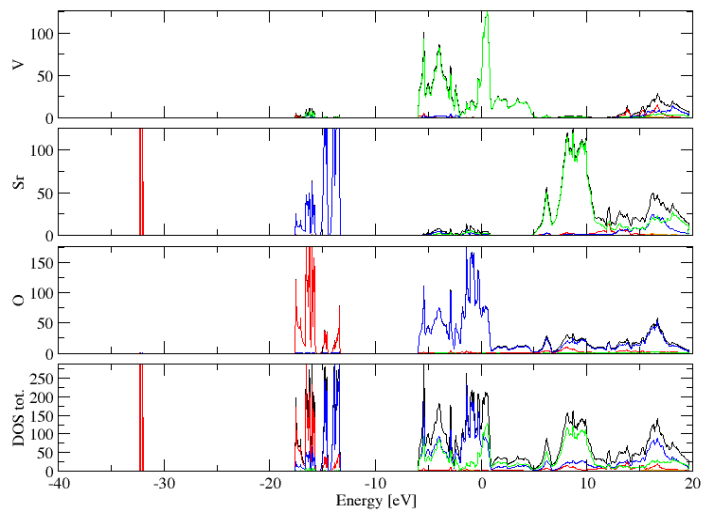

FIG. 460: (Color online) PDOS of Sr$_3$(V$_2$O$_7$) (ICSD #71451). The $s$-, $p$- and $d$-projected states are in red, blue and green, respectively. Sr$_3$(V$_2$O$_7$) crystallizes in space group I 4/m m m (#139), in a tetragonal body-centred structure.

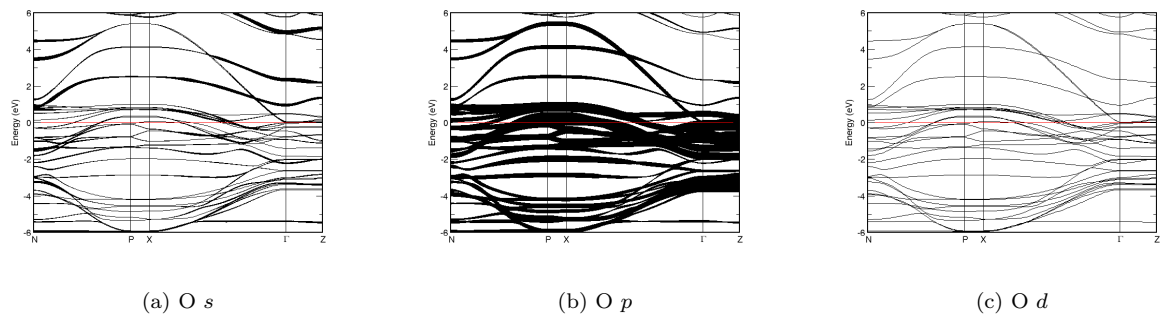

(a) O $s$         (b) O $p$         (c) O $d$

FIG. 461: Fat band representation of O in Sr$_3$(V$_2$O$_7$)

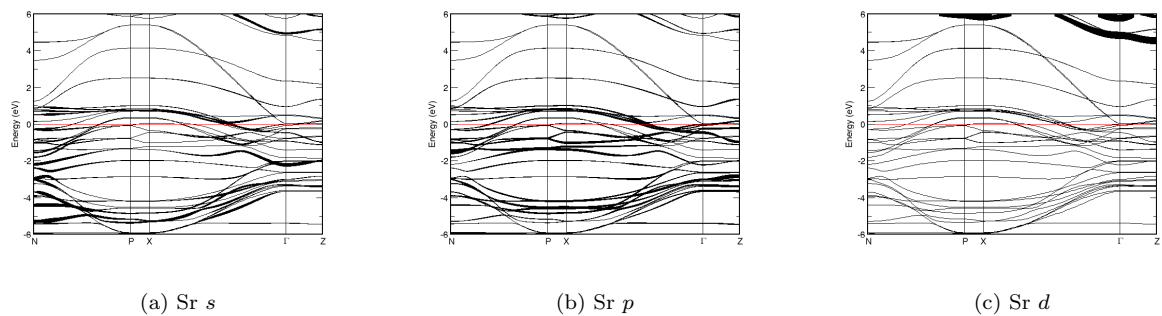

(a) Sr $s$         (b) Sr $p$         (c) Sr $d$

FIG. 462: Fat band representation of Sr in Sr$_3$(V$_2$O$_7$)

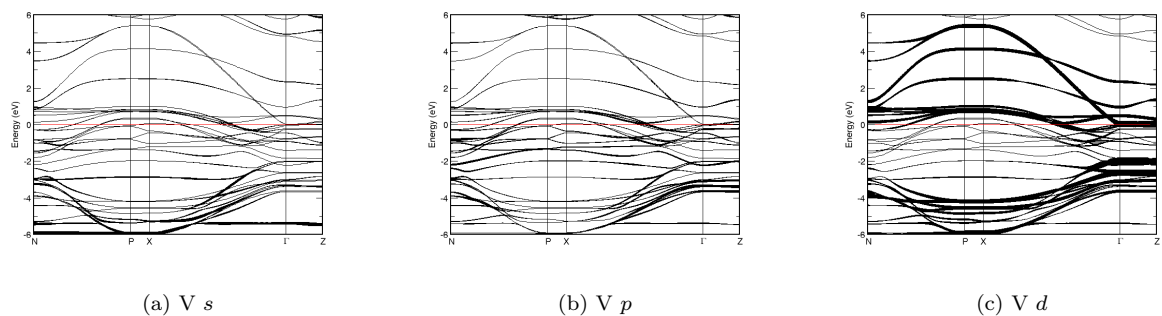

(a) V $s$      (b) V $p$      (c) V $d$

FIG. 463: Fat band representation of V in $Sr_3(V_2O_7)$

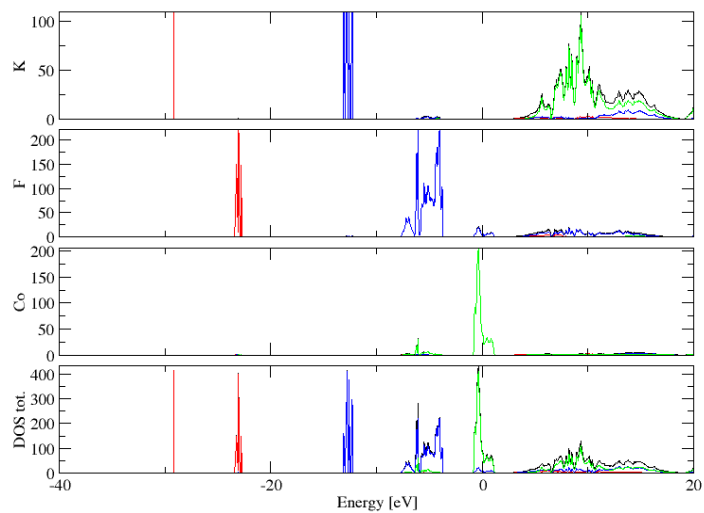

FIG. 464: (Color online) PDOS of $K_3Co_2F_7$ (ICSD #33524). The $s$-, $p$- and $d$-projected states are in red, blue and green, respectively. $K_3Co_2F_7$ crystallizes in space group I 4/m m m (#139), in a tetragonal body-centred structure.

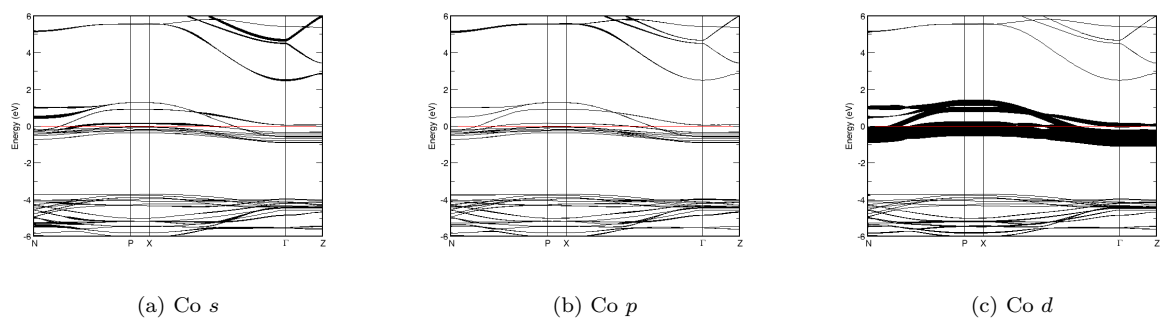

(a) Co $s$      (b) Co $p$      (c) Co $d$

FIG. 465: Fat band representation of Co in $K_3Co_2F_7$

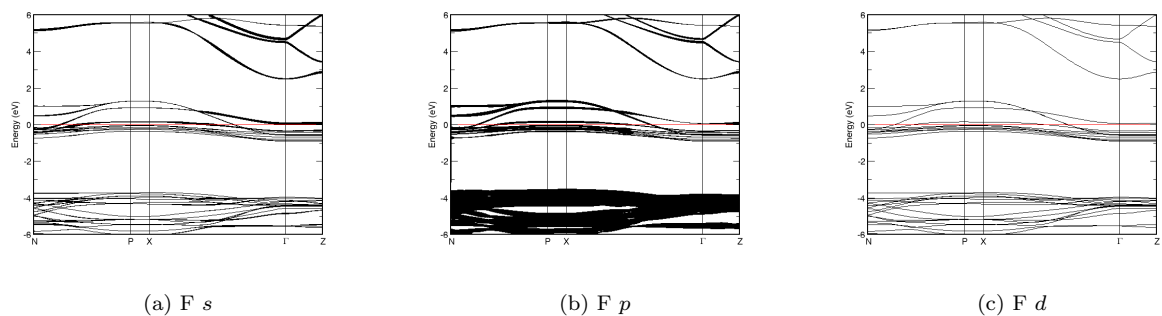

(a) F *s*     (b) F *p*     (c) F *d*

FIG. 466: Fat band representation of F in K$_3$Co$_2$F$_7$

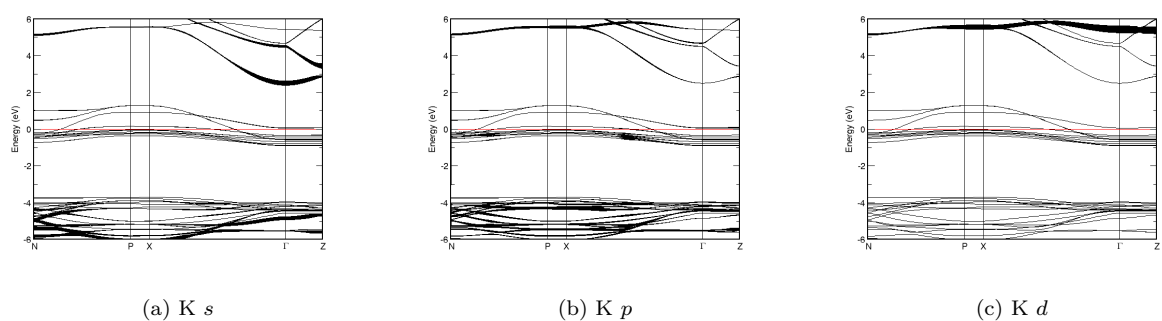

(a) K *s*     (b) K *p*     (c) K *d*

FIG. 467: Fat band representation of K in K$_3$Co$_2$F$_7$

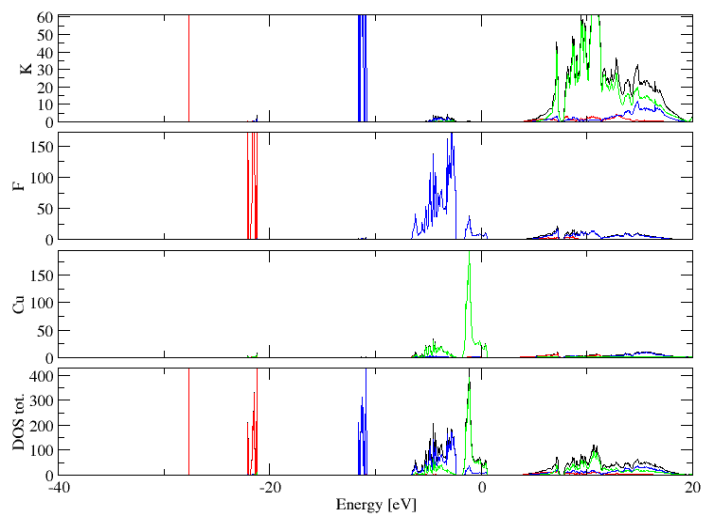

FIG. 468: (Color online) PDOS of K$_3$Cu$_2$F$_7$ (ICSD #15373). The *s*-, *p*- and *d*-projected states are in red, blue and green, respectively. K$_3$Cu$_2$F$_7$ crystallizes in space group I 4/m m m (#139), in a tetragonal body-centred structure.

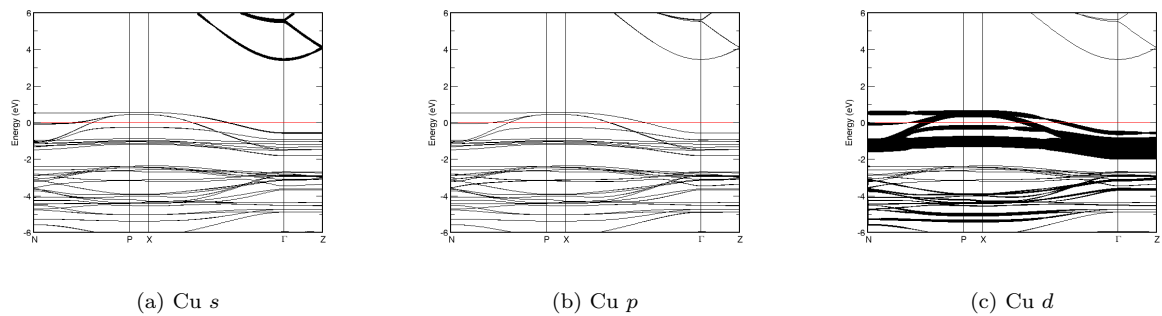

(a) Cu $s$

(b) Cu $p$

(c) Cu $d$

FIG. 469: Fat band representation of Cu in $K_3Cu_2F_7$

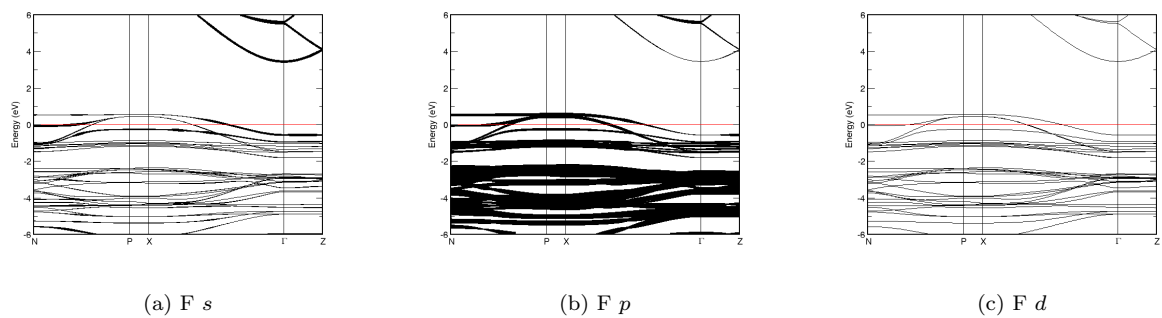

(a) F $s$

(b) F $p$

(c) F $d$

FIG. 470: Fat band representation of F in $K_3Cu_2F_7$

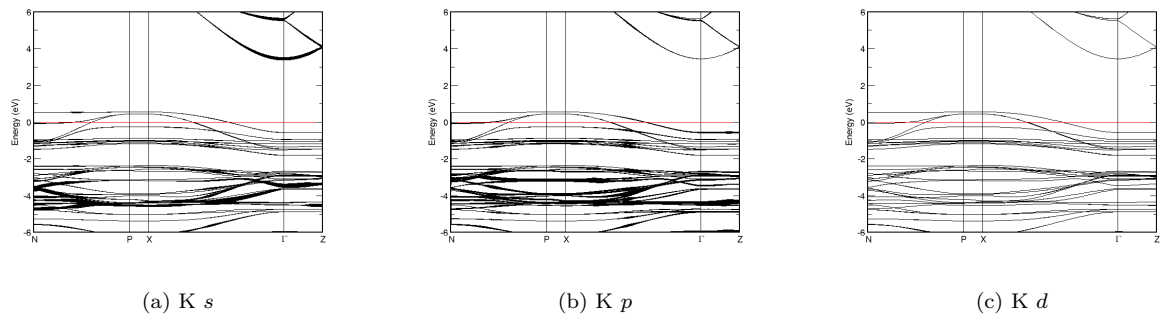

(a) K $s$

(b) K $p$

(c) K $d$

FIG. 471: Fat band representation of K in $K_3Cu_2F_7$

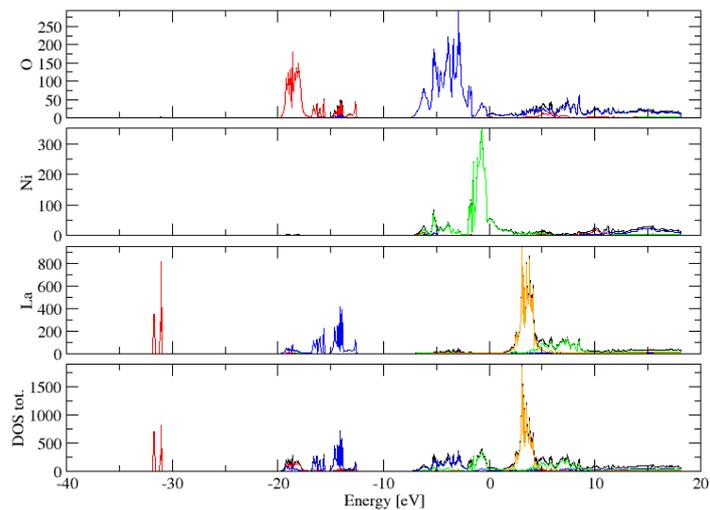

FIG. 472: (Color online) PDOS of La$_4$Ni$_3$O$_8$ (ICSD #173372). The $s$-, $p$- and $d$-projected states are in red, blue and green, respectively. La$_4$Ni$_3$O$_8$ crystallizes in space group I 4/m m m (#139), in a tetragonal body-centred structure.

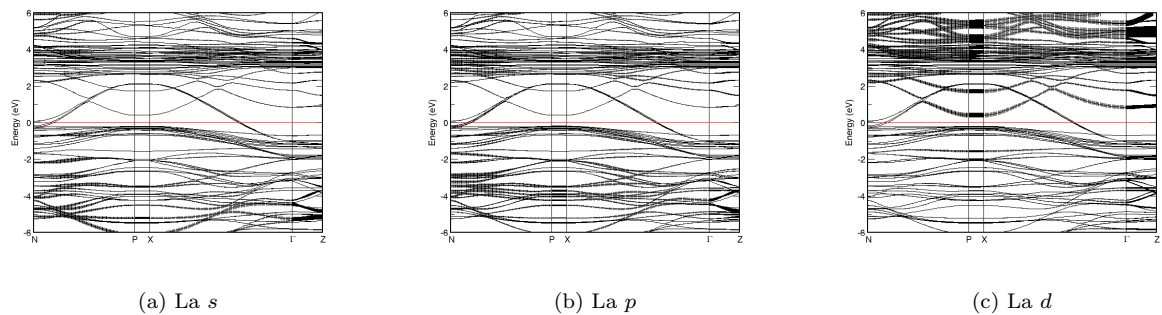

(a) La $s$         (b) La $p$         (c) La $d$

FIG. 473: Fat band representation of La in La$_4$Ni$_3$O$_8$

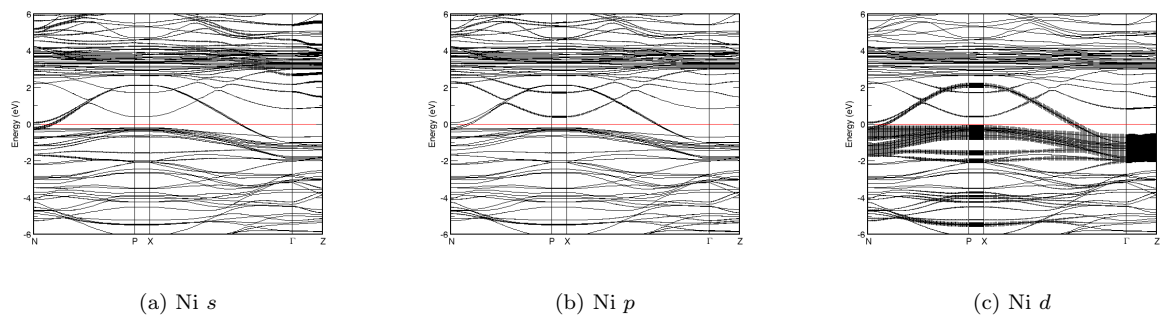

(a) Ni $s$         (b) Ni $p$         (c) Ni $d$

FIG. 474: Fat band representation of Ni in La$_4$Ni$_3$O$_8$

(a) O $s$

(b) O $p$

(c) O $d$

FIG. 475: Fat band representation of O in $La_4Ni_3O_8$

FIG. 476: (Color online) PDOS of $K_5Te_3$ (ICSD #96743). The $s$-, $p$- and $d$-projected states are in red, blue and green, respectively. $K_5Te_3$ crystallizes in space group I 4/m (#87), in a tetragonal body-centred structure.

(a) K $s$

(b) K $p$

(c) K $d$

FIG. 477: Fat band representation of K in $K_5Te_3$

(a) Te $s$

(b) Te $p$

(c) Te $d$

FIG. 478: Fat band representation of Te in $K_5Te_3$

FIG. 479: (Color online) PDOS of $CaSmCuO_3Cl$ (ICSD #86428). The $s$-, $p$- and $d$-projected states are in red, blue and green, respectively. $CaSmCuO_3Cl$ crystallizes in space group P 4/n m m Z (#129), in a tetragonal primitive structure.

(a) Ca $s$

(b) Ca $p$

(c) Ca $d$

FIG. 480: Fat band representation of Ca in $CaSmCuO_3Cl$

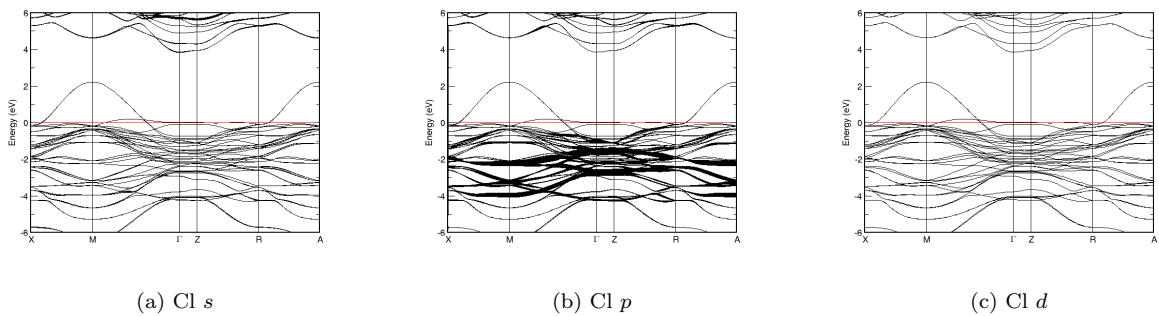

(a) Cl *s*  (b) Cl *p*  (c) Cl *d*

FIG. 481: Fat band representation of Cl in CaSmCuO₃Cl

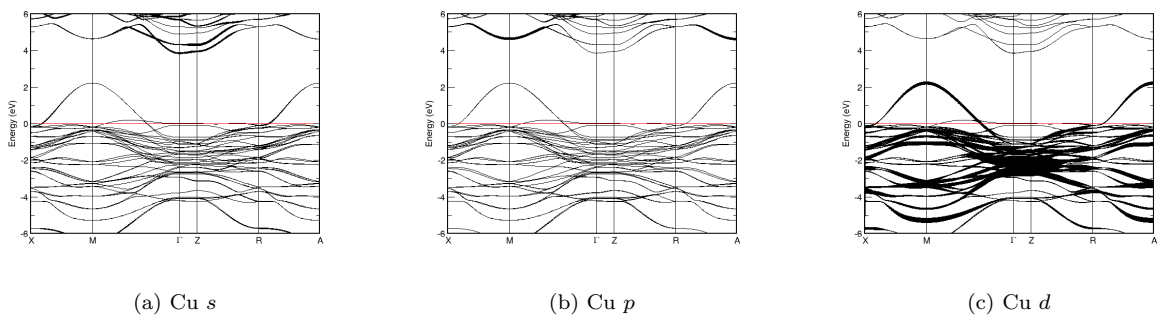

(a) Cu *s*  (b) Cu *p*  (c) Cu *d*

FIG. 482: Fat band representation of Cu in CaSmCuO₃Cl

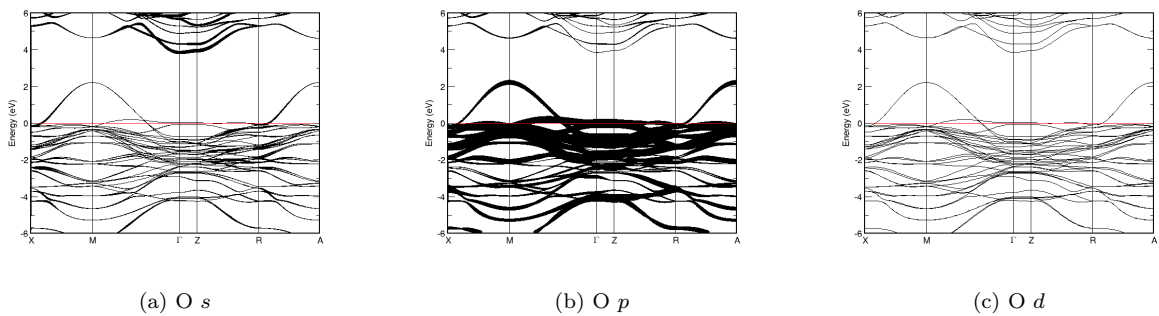

(a) O *s*  (b) O *p*  (c) O *d*

FIG. 483: Fat band representation of O in CaSmCuO₃Cl

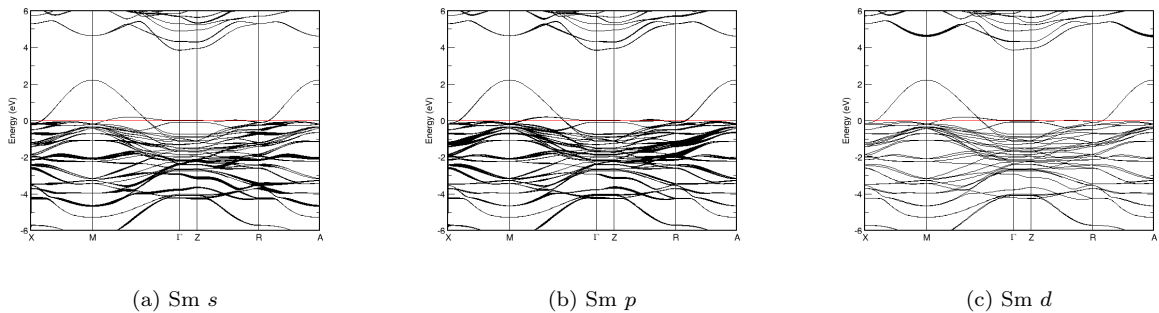

(a) Sm *s*  (b) Sm *p*  (c) Sm *d*

FIG. 484: Fat band representation of Sm in CaSmCuO₃Cl

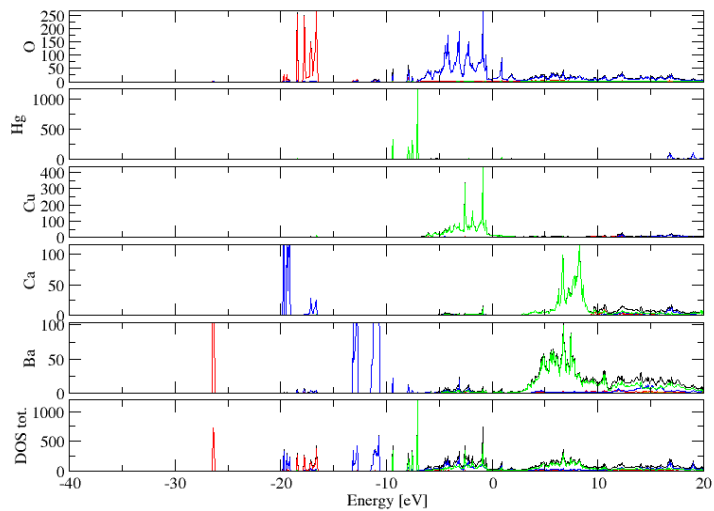

FIG. 485: (Color online) PDOS of HgBa$_2$CaCu$_2$O$_6$ (ICSD #75725). The $s$-, $p$- and $d$-projected states are in red, blue and green, respectively. HgBa$_2$CaCu$_2$O$_6$ crystallizes in space group P 4/m m m (#123), in a tetragonal primitive structure.

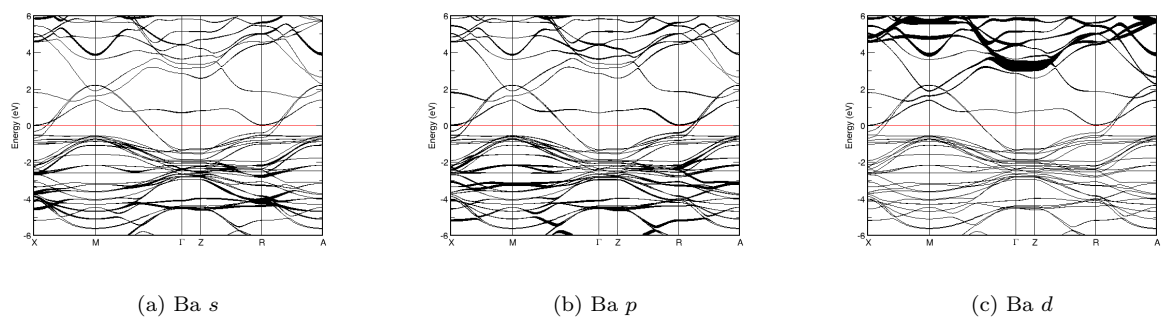

(a) Ba $s$

(b) Ba $p$

(c) Ba $d$

FIG. 486: Fat band representation of Ba in HgBa$_2$CaCu$_2$O$_6$

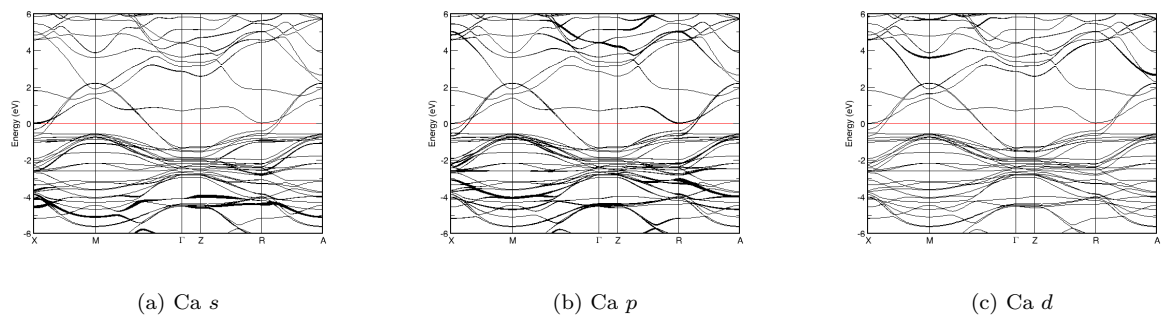

(a) Ca $s$

(b) Ca $p$

(c) Ca $d$

FIG. 487: Fat band representation of Ca in HgBa$_2$CaCu$_2$O$_6$

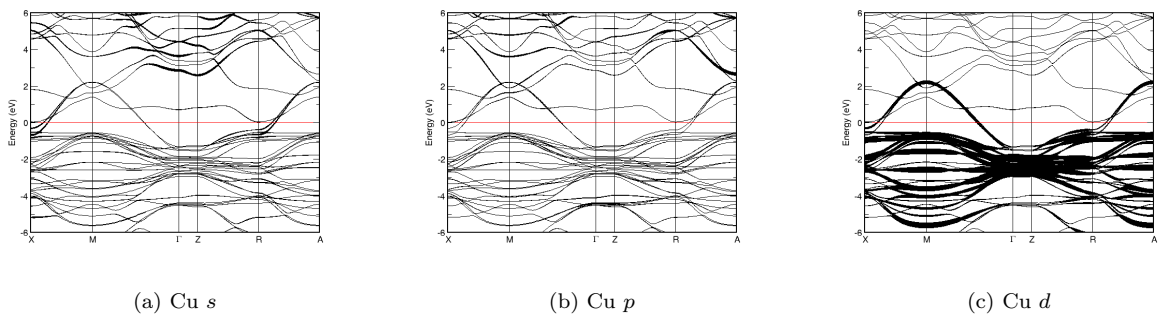

(a) Cu $s$      (b) Cu $p$      (c) Cu $d$

FIG. 488: Fat band representation of Cu in HgBa$_2$CaCu$_2$O$_6$

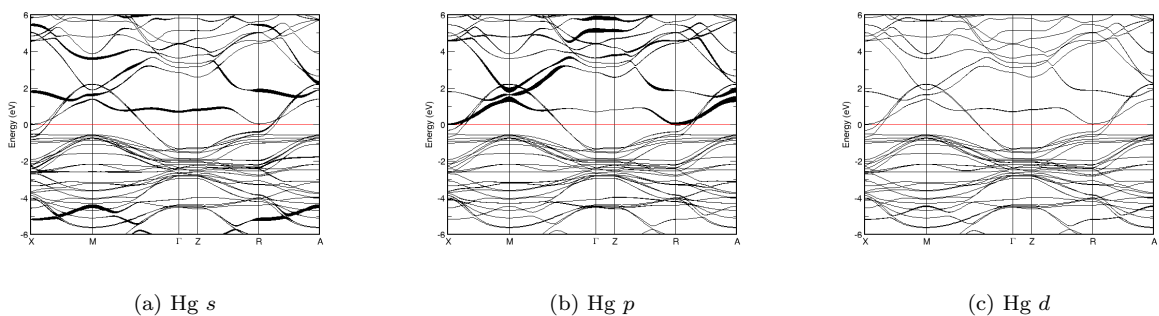

(a) Hg $s$      (b) Hg $p$      (c) Hg $d$

FIG. 489: Fat band representation of Hg in HgBa$_2$CaCu$_2$O$_6$

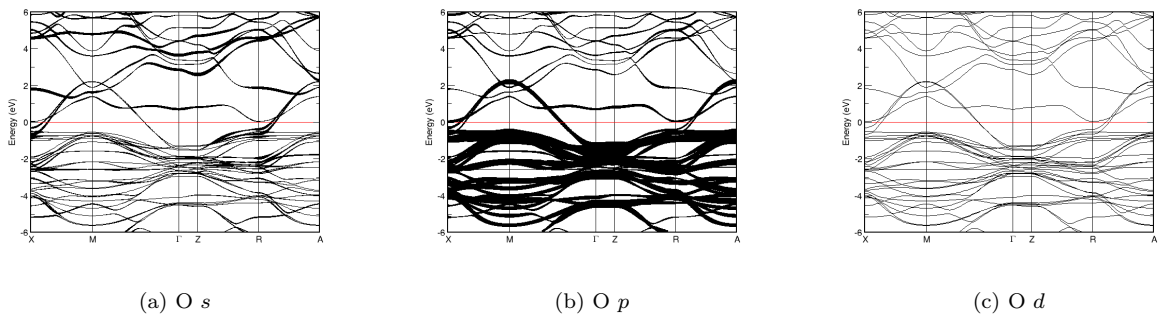

(a) O $s$      (b) O $p$      (c) O $d$

FIG. 490: Fat band representation of O in HgBa$_2$CaCu$_2$O$_6$

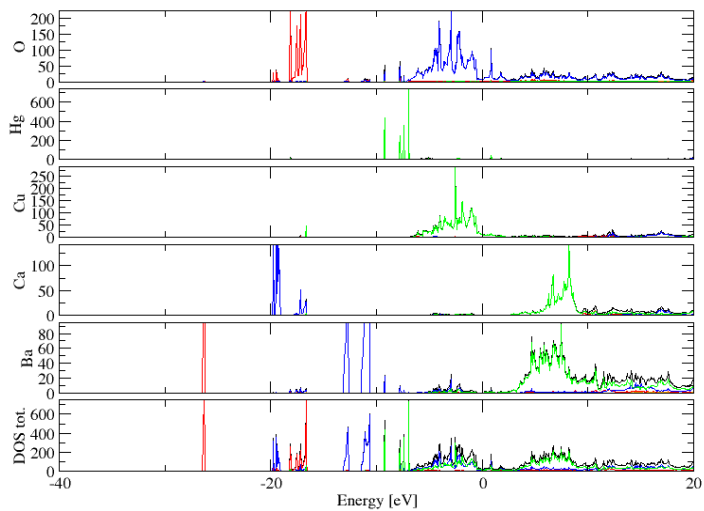

FIG. 491: (Color online) PDOS of HgBa$_2$CaCu$_2$O$_6$ (ICSD #83087). The $s$-, $p$- and $d$-projected states are in red, blue and green, respectively. HgBa$_2$CaCu$_2$O$_6$ crystallizes in space group P 4/m m m (#123), in a tetragonal primitive structure.

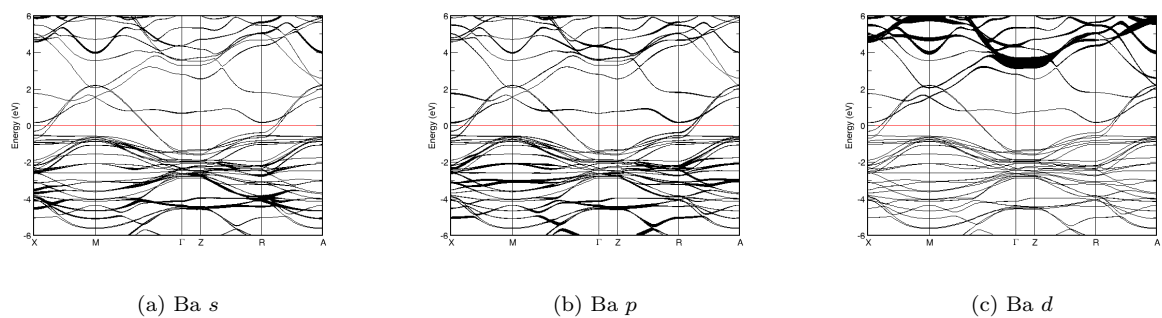

(a) Ba $s$        (b) Ba $p$        (c) Ba $d$

FIG. 492: Fat band representation of Ba in HgBa$_2$CaCu$_2$O$_6$

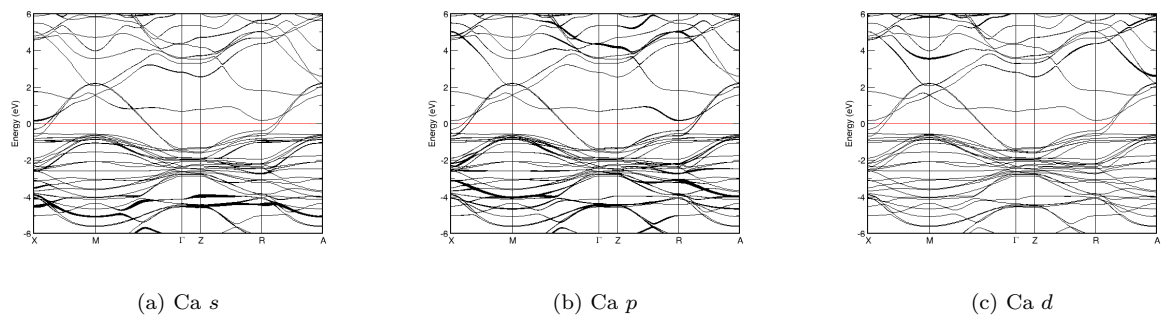

(a) Ca $s$        (b) Ca $p$        (c) Ca $d$

FIG. 493: Fat band representation of Ca in HgBa$_2$CaCu$_2$O$_6$

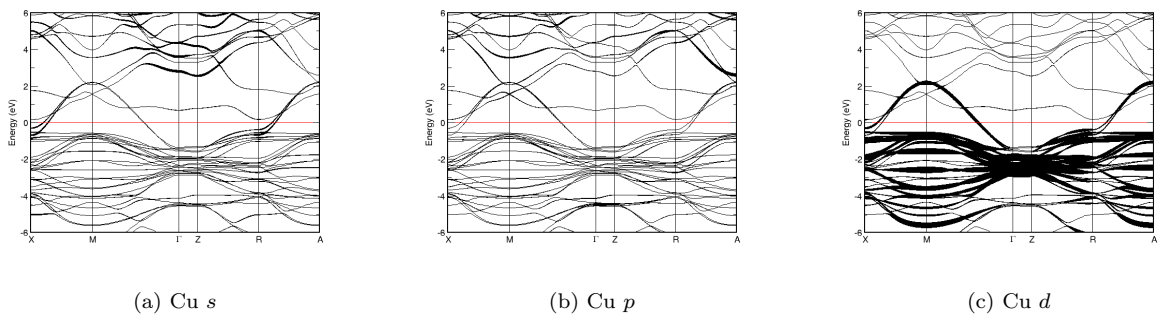

(a) Cu $s$     (b) Cu $p$     (c) Cu $d$

FIG. 494: Fat band representation of Cu in HgBa$_2$CaCu$_2$O$_6$

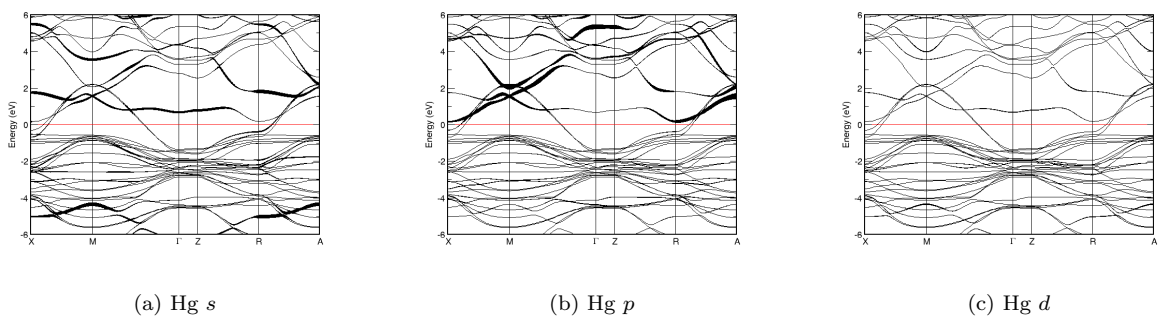

(a) Hg $s$     (b) Hg $p$     (c) Hg $d$

FIG. 495: Fat band representation of Hg in HgBa$_2$CaCu$_2$O$_6$

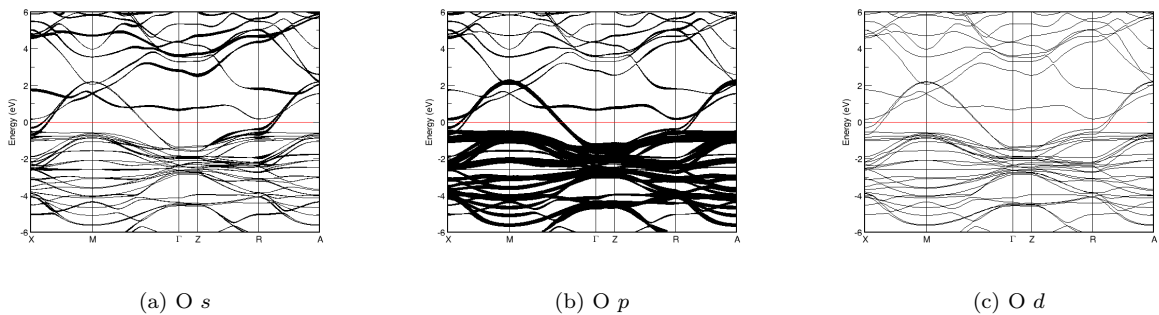

(a) O $s$     (b) O $p$     (c) O $d$

FIG. 496: Fat band representation of O in HgBa$_2$CaCu$_2$O$_6$

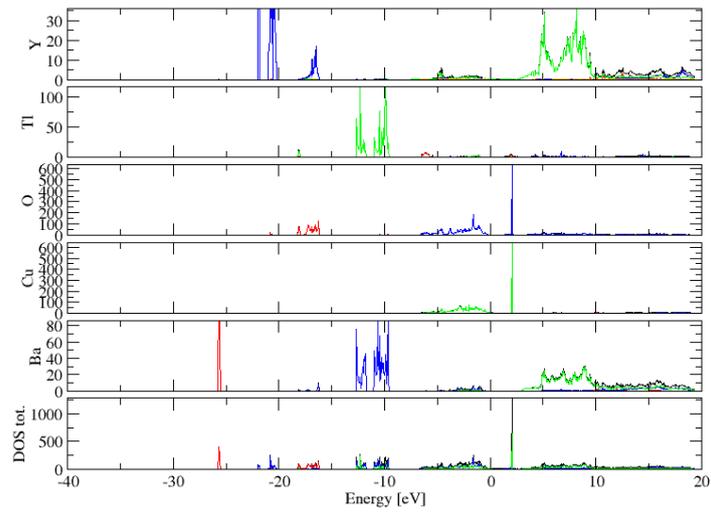

FIG. 497: (Color online) PDOS of TlYBa$_2$Cu$_2$O$_7$ (ICSD #74163). The *s*-, *p*- and *d*-projected states are in red, blue and green, respectively. TlYBa$_2$Cu$_2$O$_7$ crystallizes in space group P 4/m m m (#123), in a tetragonal primitive structure.

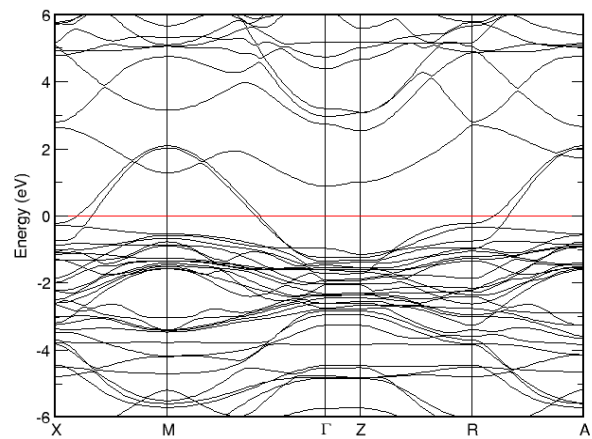

(a) E *vs.* k

FIG. 498: Band structure of TlYBa$_2$Cu$_2$O$_7$

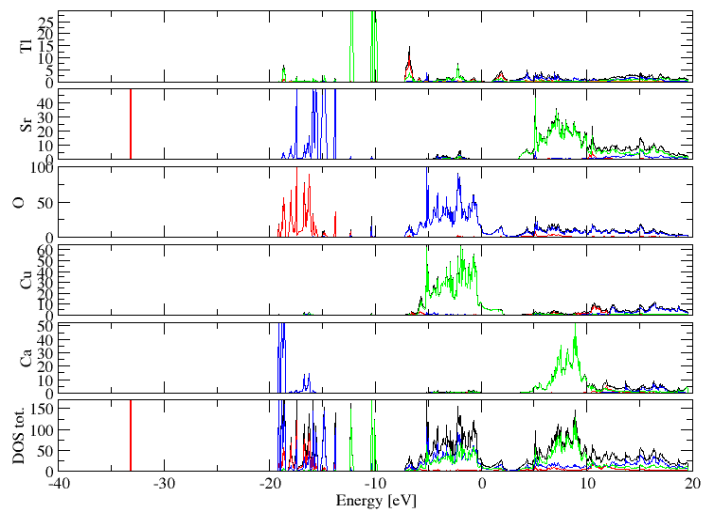

FIG. 499: (Color online) PDOS of TlCaSr$_2$Cu$_2$O$_7$ (ICSD #74165). The $s$-, $p$- and $d$-projected states are in red, blue and green, respectively. TlCaSr$_2$Cu$_2$O$_7$ crystallizes in space group P 4/m m m (#123), in a tetragonal primitive structure.

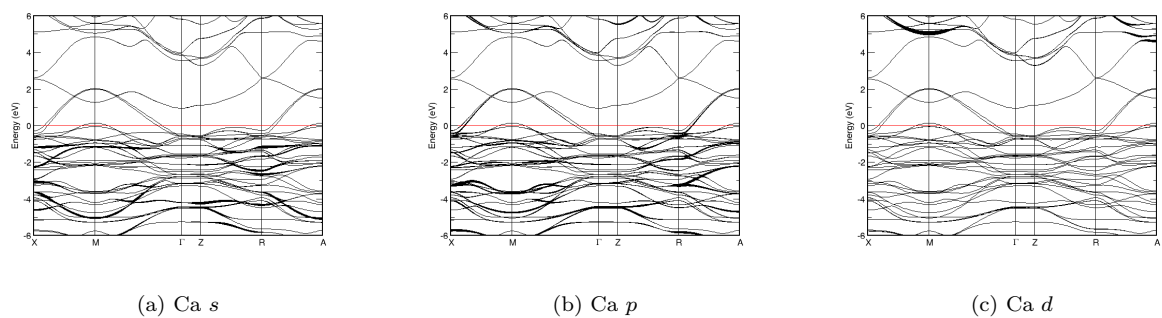

(a) Ca $s$         (b) Ca $p$         (c) Ca $d$

FIG. 500: Fat band representation of Ca in TlCaSr$_2$Cu$_2$O$_7$

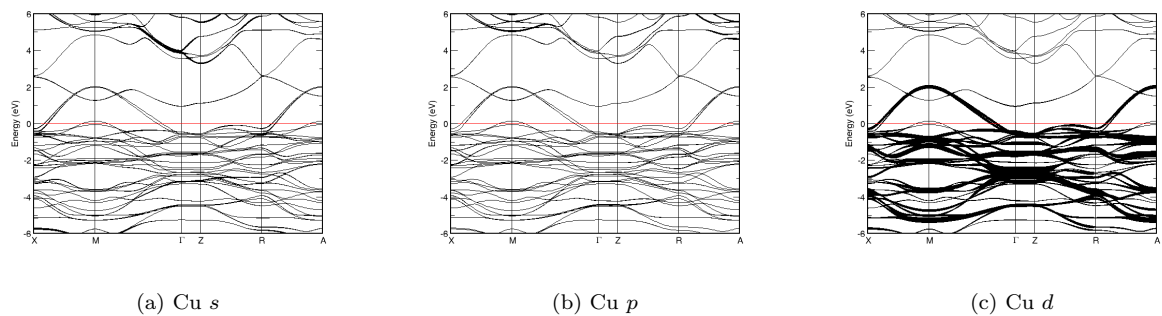

(a) Cu $s$         (b) Cu $p$         (c) Cu $d$

FIG. 501: Fat band representation of Cu in TlCaSr$_2$Cu$_2$O$_7$

(a) O $s$

(b) O $p$

(c) O $d$

FIG. 502: Fat band representation of O in TlCaSr$_2$Cu$_2$O$_7$

(a) Sr $s$

(b) Sr $p$

(c) Sr $d$

FIG. 503: Fat band representation of Sr in TlCaSr$_2$Cu$_2$O$_7$

(a) Tl $s$

(b) Tl $p$

(c) Tl $d$

FIG. 504: Fat band representation of Tl in TlCaSr$_2$Cu$_2$O$_7$

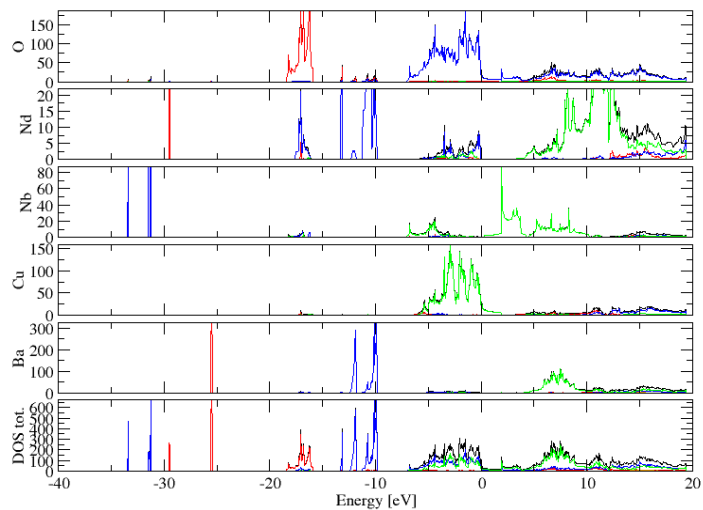

FIG. 505: (Color online) PDOS of NdBa$_2$Cu$_2$NbO$_8$ (ICSD #44255). The $s$-, $p$- and $d$-projected states are in red, blue and green, respectively. NdBa$_2$Cu$_2$NbO$_8$ crystallizes in space group P 4/m m m (#123), in a tetragonal primitive structure.

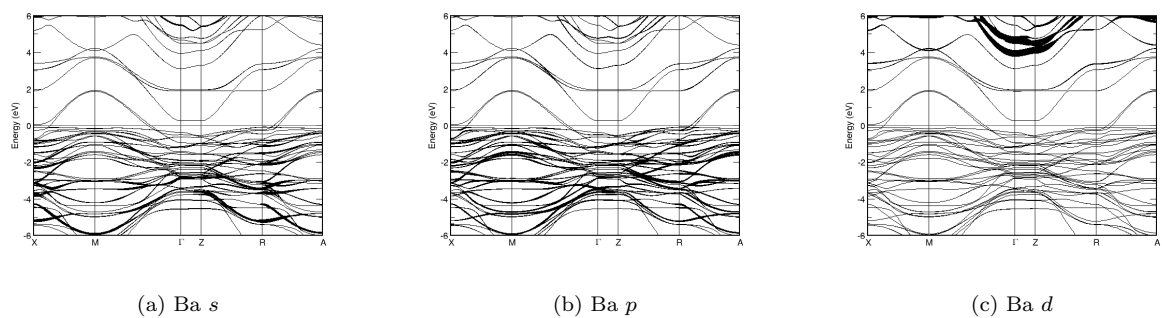

(a) Ba $s$        (b) Ba $p$        (c) Ba $d$

FIG. 506: Fat band representation of Ba in NdBa$_2$Cu$_2$NbO$_8$

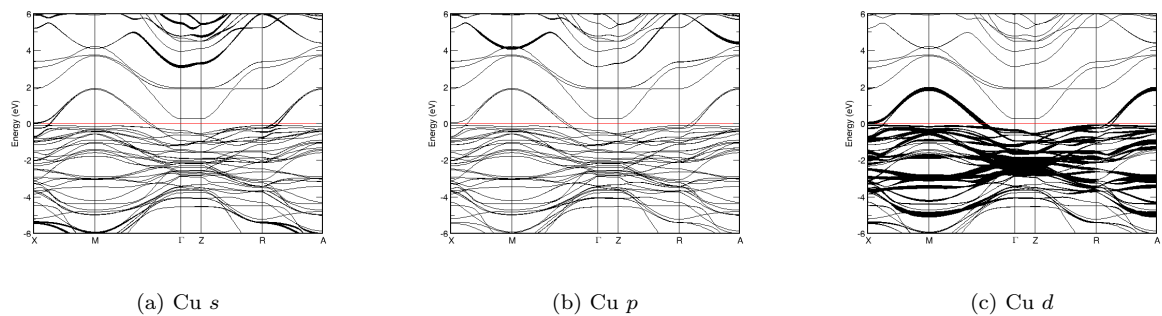

(a) Cu $s$        (b) Cu $p$        (c) Cu $d$

FIG. 507: Fat band representation of Cu in NdBa$_2$Cu$_2$NbO$_8$

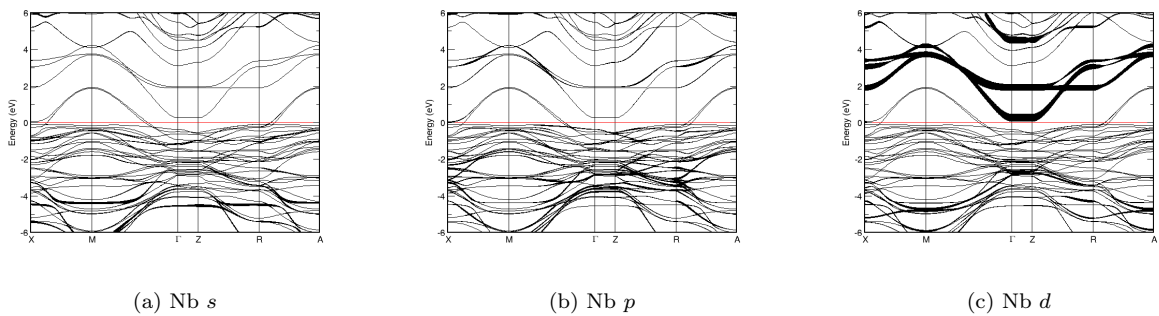

(a) Nb $s$                    (b) Nb $p$                    (c) Nb $d$

FIG. 508: Fat band representation of Nb in $NdBa_2Cu_2NbO_8$

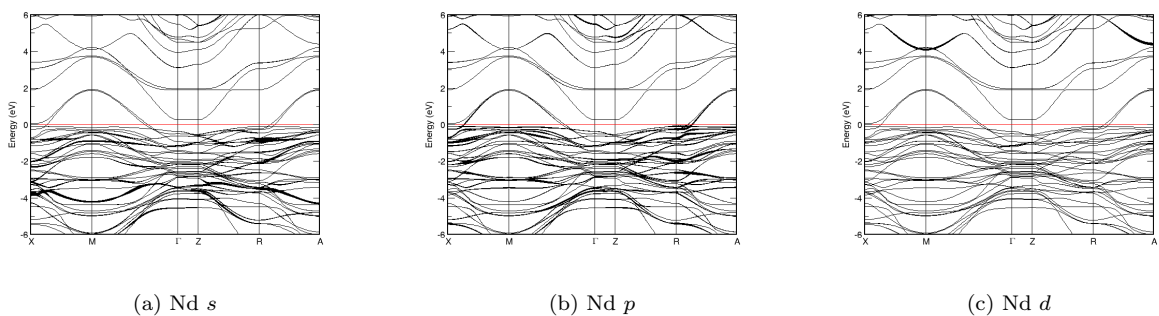

(a) Nd $s$                    (b) Nd $p$                    (c) Nd $d$

FIG. 509: Fat band representation of Nd in $NdBa_2Cu_2NbO_8$

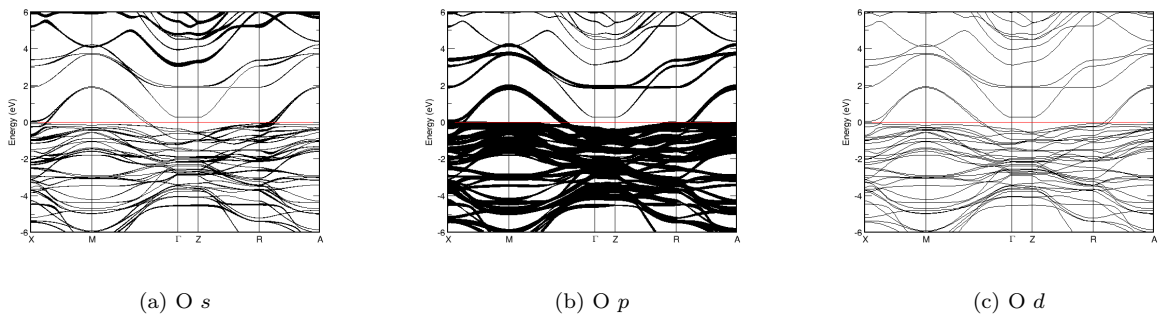

(a) O $s$                    (b) O $p$                    (c) O $d$

FIG. 510: Fat band representation of O in $NdBa_2Cu_2NbO_8$

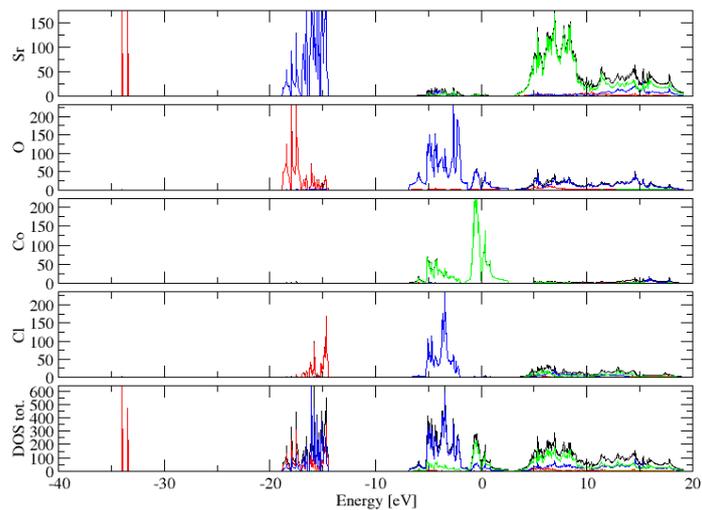

FIG. 511: (Color online) PDOS of Sr$_2$CoO$_3$Cl (ICSD #91750). The $s$-, $p$- and $d$-projected states are in red, blue and green, respectively. Sr$_2$CoO$_3$Cl crystallizes in space group P 4/n m m Z (#129), in a tetragonal primitive structure.

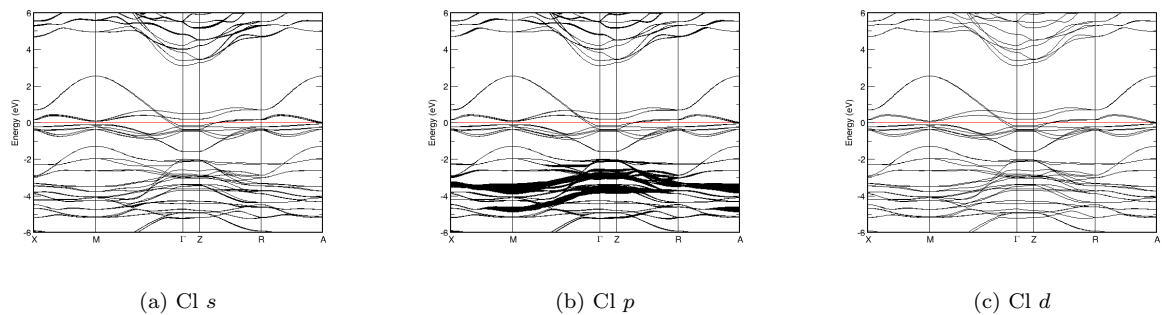

(a) Cl $s$                  (b) Cl $p$                  (c) Cl $d$

FIG. 512: Fat band representation of Cl in Sr$_2$CoO$_3$Cl

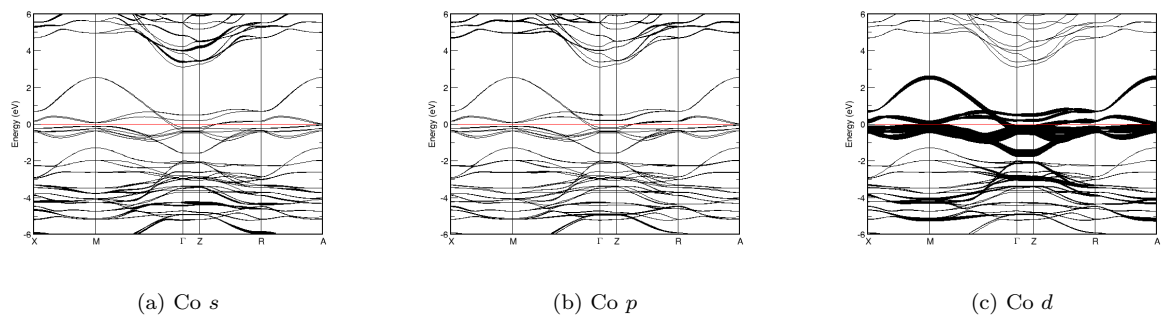

(a) Co $s$                  (b) Co $p$                  (c) Co $d$

FIG. 513: Fat band representation of Co in Sr$_2$CoO$_3$Cl

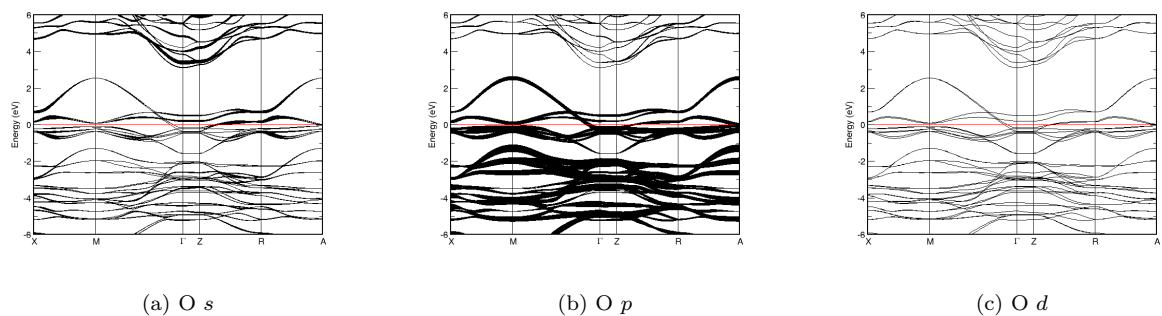

(a) O *s*  (b) O *p*  (c) O *d*

FIG. 514: Fat band representation of O in Sr$_2$CoO$_3$Cl

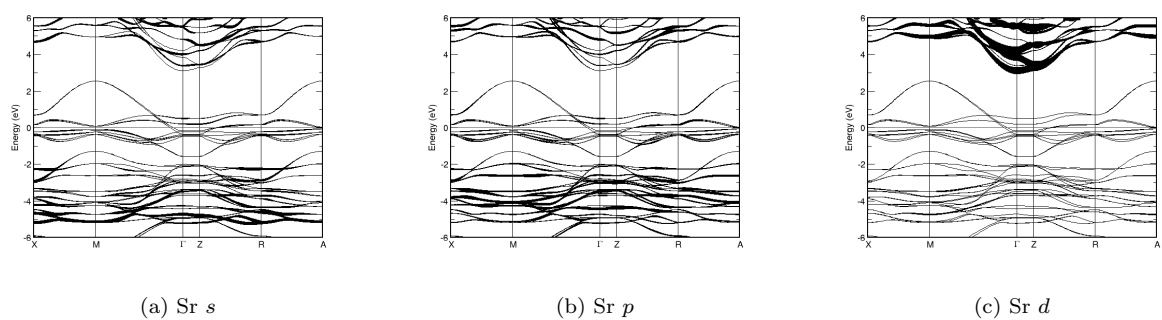

(a) Sr *s*  (b) Sr *p*  (c) Sr *d*

FIG. 515: Fat band representation of Sr in Sr$_2$CoO$_3$Cl

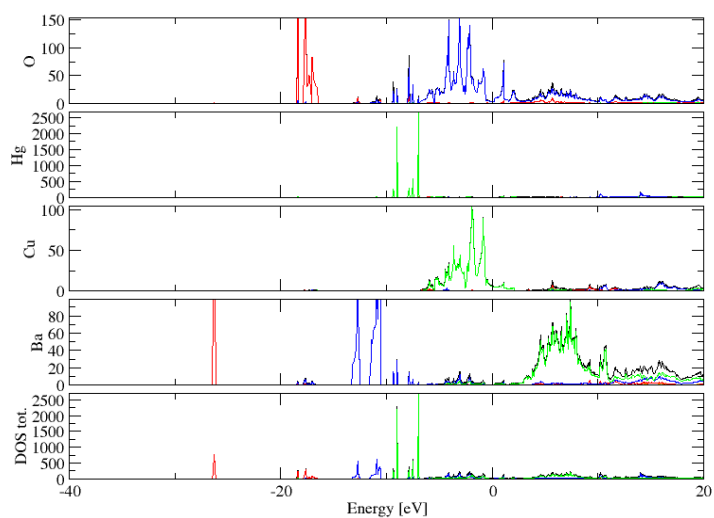

FIG. 516: (Color online) PDOS of HgBa$_2$CuO$_4$ (ICSD #75720). The *s*-, *p*- and *d*-projected states are in red, blue and green, respectively. HgBa$_2$CuO$_4$ crystallizes in space group P 4/m m m (#123), in a tetragonal primitive structure.

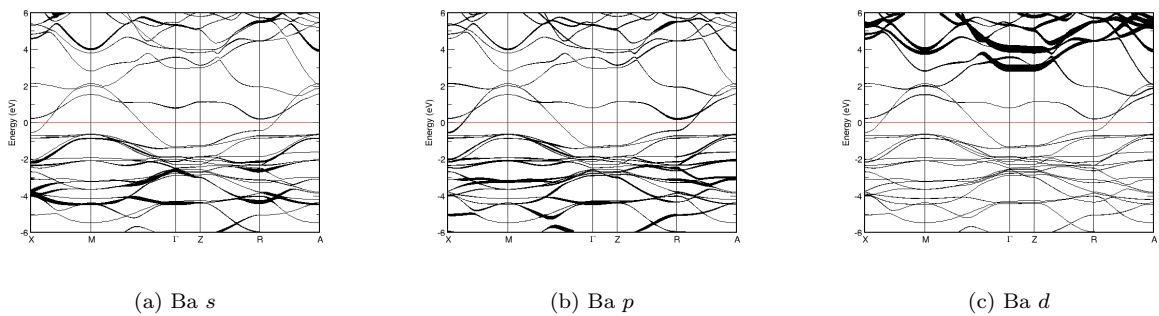

(a) Ba $s$

(b) Ba $p$

(c) Ba $d$

FIG. 517: Fat band representation of Ba in HgBa$_2$CuO$_4$

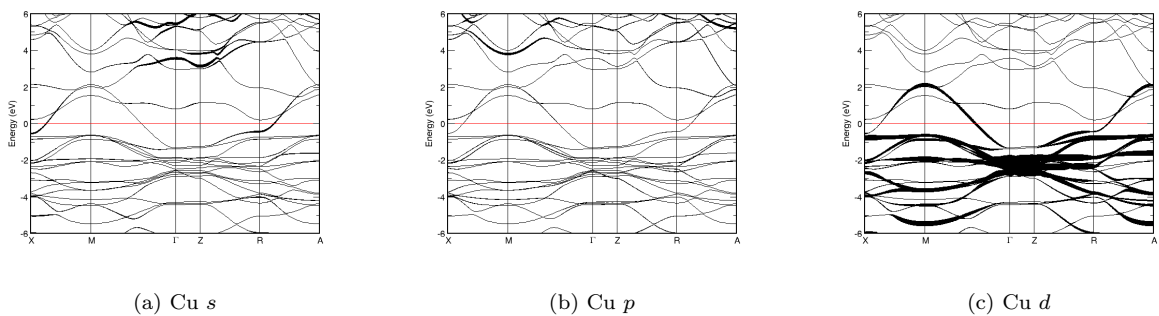

(a) Cu $s$

(b) Cu $p$

(c) Cu $d$

FIG. 518: Fat band representation of Cu in HgBa$_2$CuO$_4$

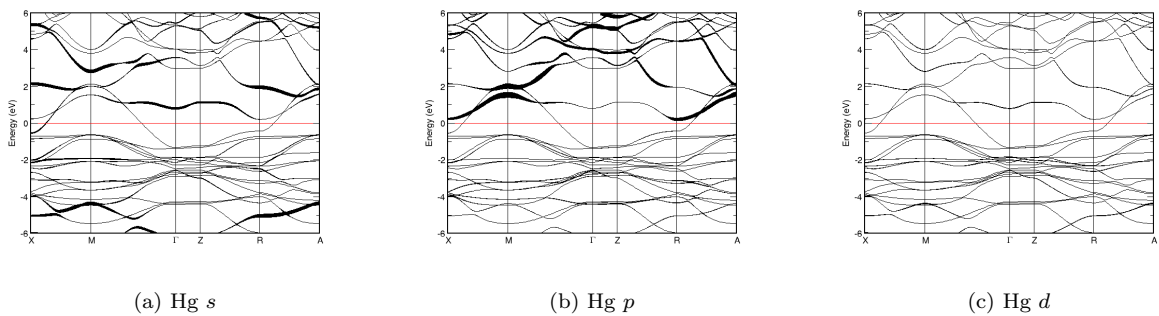

(a) Hg $s$

(b) Hg $p$

(c) Hg $d$

FIG. 519: Fat band representation of Hg in HgBa$_2$CuO$_4$

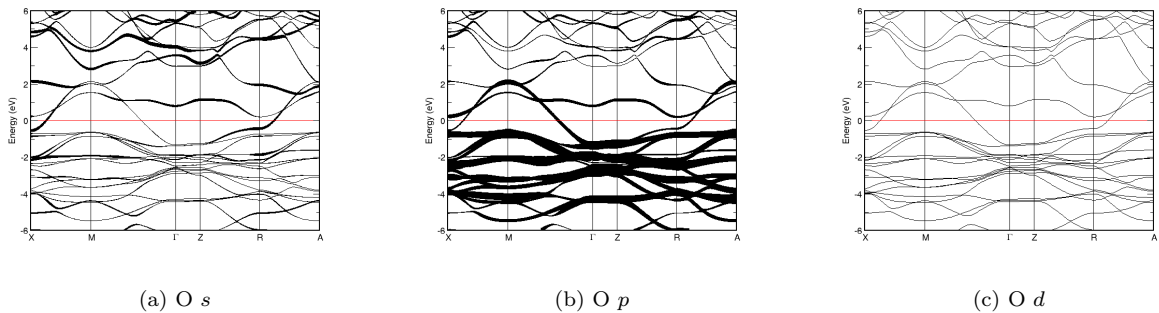

(a) O $s$

(b) O $p$

(c) O $d$

FIG. 520: Fat band representation of O in HgBa$_2$CuO$_4$

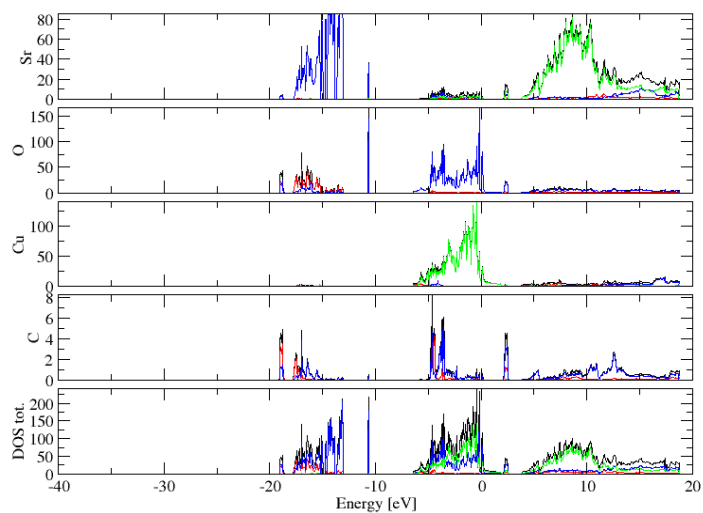

FIG. 521: (Color online) PDOS of $Sr_2CuO_2(CO_3)$ (ICSD #83096). The *s*-, *p*- and *d*-projected states are in red, blue and green, respectively. $Sr_2CuO_2(CO_3)$ crystallizes in space group P 4 21 2 (#90), in a tetragonal primitive structure.

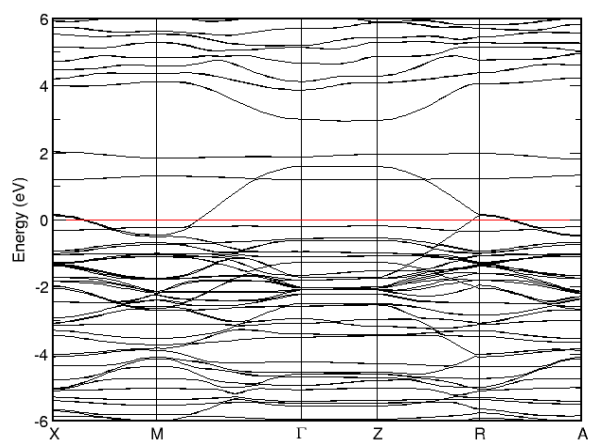

(a) E *vs.* k

FIG. 522: Band structure of $Sr_2CuO_2(CO_3)$

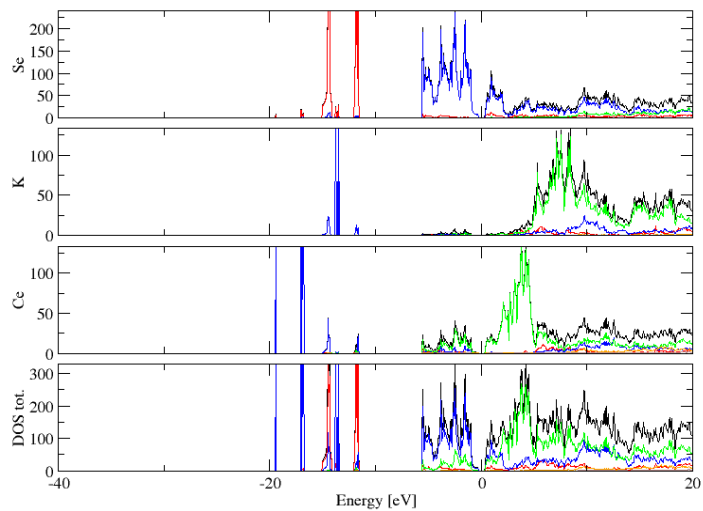

FIG. 523: (Color online) PDOS of KCeSe$_4$ (ICSD #67656). The $s$-, $p$- and $d$-projected states are in red, blue and green, respectively. KCeSe$_4$ crystallizes in space group P 4/n b m Z (#125), in a tetragonal primitive structure.

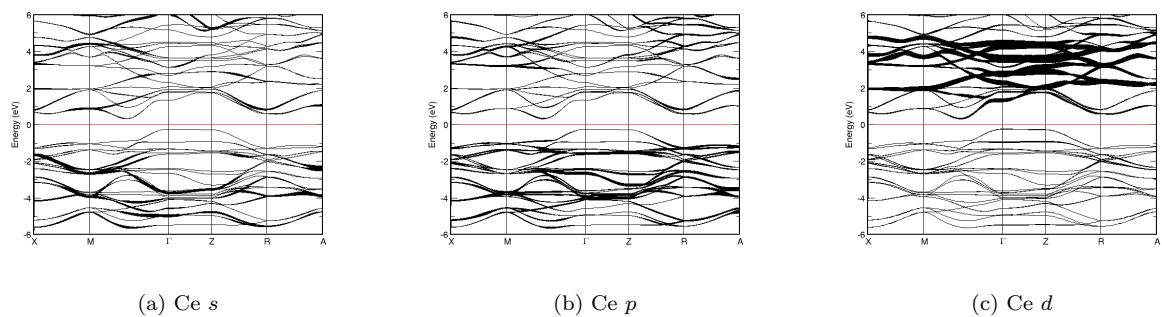

(a) Ce $s$

(b) Ce $p$

(c) Ce $d$

FIG. 524: Fat band representation of Ce in KCeSe$_4$

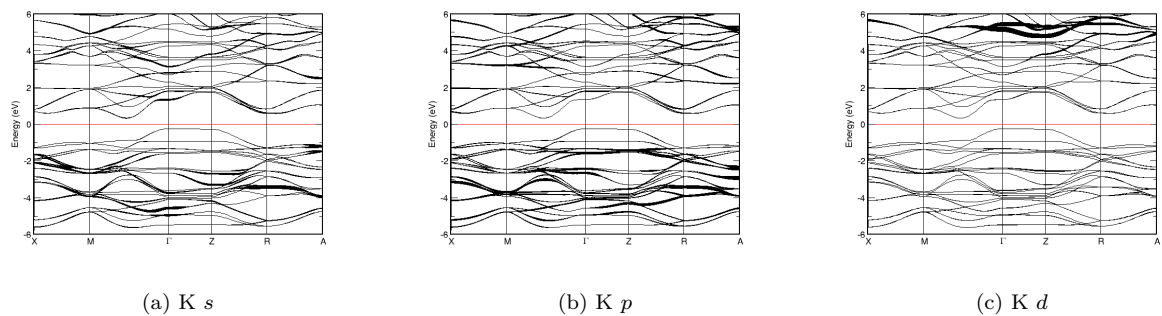

(a) K $s$

(b) K $p$

(c) K $d$

FIG. 525: Fat band representation of K in KCeSe$_4$

(a) Se $s$

(b) Se $p$

(c) Se $d$

FIG. 526: Fat band representation of Se in KCeSe$_4$

FIG. 527: (Color online) PDOS of NdLi$_2$Sb$_2$ (ICSD #36020). The $s$-, $p$- and $d$-projected states are in red, blue and green, respectively. NdLi$_2$Sb$_2$ crystallizes in space group P 4/n m m Z (#129), in a tetragonal primitive structure.

(a) Li $s$

(b) Li $p$

(c) Li $d$

FIG. 528: Fat band representation of Li in NdLi$_2$Sb$_2$

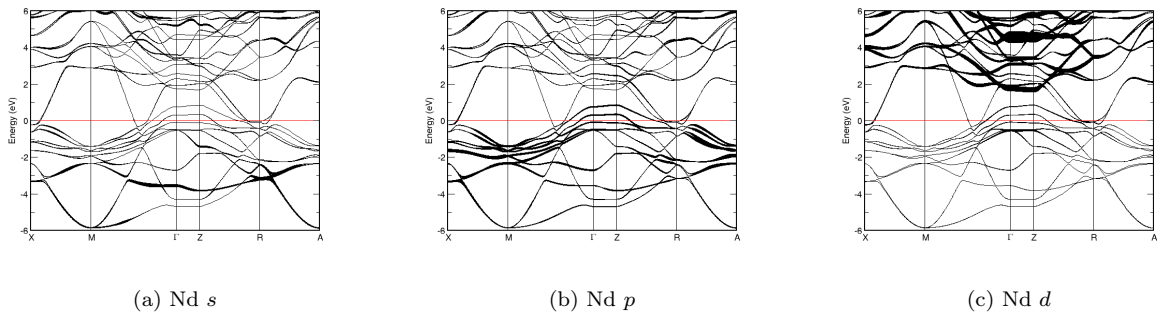

(a) Nd $s$      (b) Nd $p$      (c) Nd $d$

FIG. 529: Fat band representation of Nd in NdLi$_2$Sb$_2$

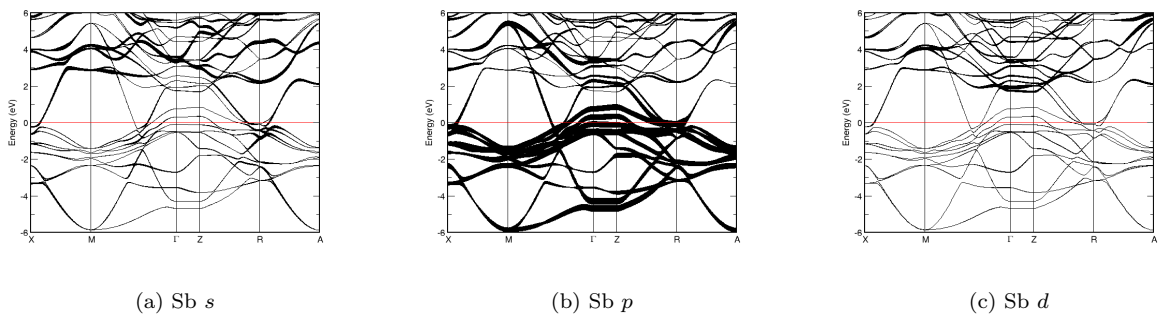

(a) Sb $s$      (b) Sb $p$      (c) Sb $d$

FIG. 530: Fat band representation of Sb in NdLi$_2$Sb$_2$

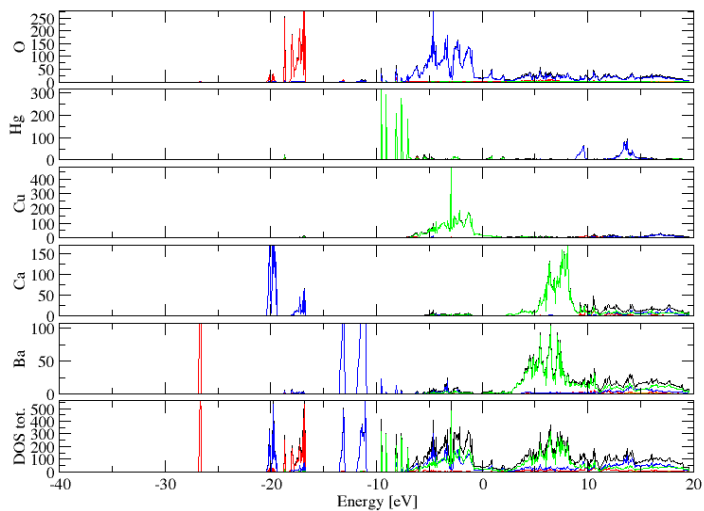

FIG. 531: (Color online) PDOS of HgBa$_2$Ca$_2$Cu$_3$O$_8$ (ICSD #75730). The $s$-, $p$- and $d$-projected states are in red, blue and green, respectively. HgBa$_2$Ca$_2$Cu$_3$O$_8$ crystallizes in space group P 4/m m m (#123), in a tetragonal primitive structure.

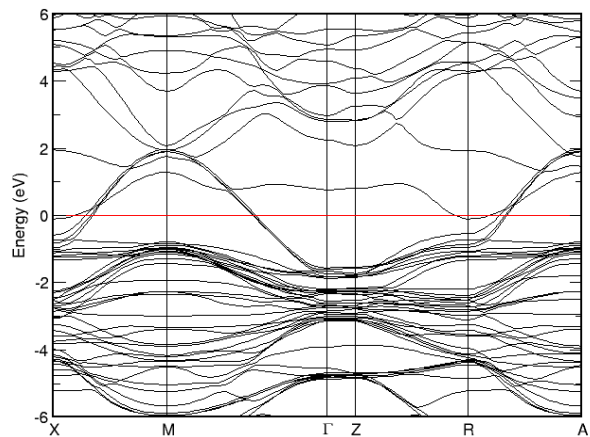

(a) E *vs.* k

FIG. 532: Band structure of HgBa$_2$Ca$_2$Cu$_3$O$_8$

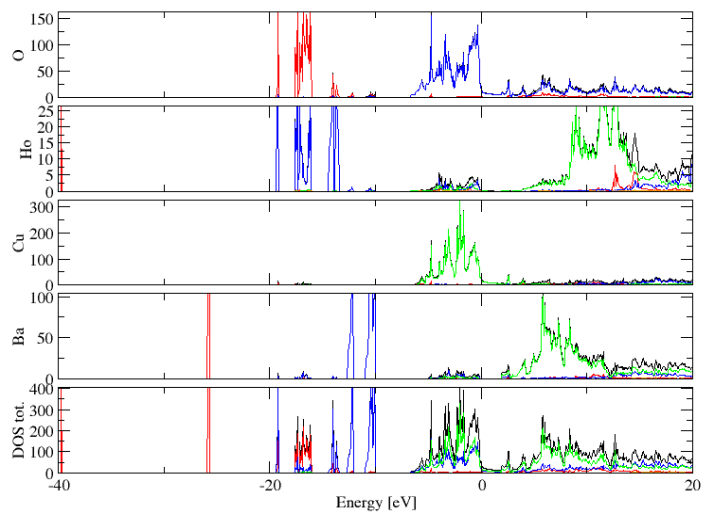

FIG. 533: (Color online) PDOS of HoBa$_2$Cu$_3$O$_6$ (ICSD #68047). The *s*-, *p*- and *d*-projected states are in red, blue and green, respectively. HoBa$_2$Cu$_3$O$_6$ crystallizes in space group P 4/m m m (#123), in a tetragonal primitive structure.

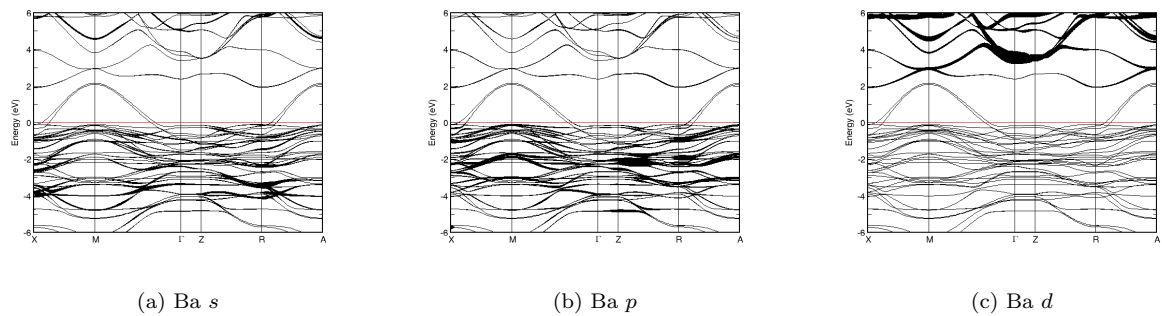

(a) Ba *s*　　　　　　(b) Ba *p*　　　　　　(c) Ba *d*

FIG. 534: Fat band representation of Ba in HoBa$_2$Cu$_3$O$_6$

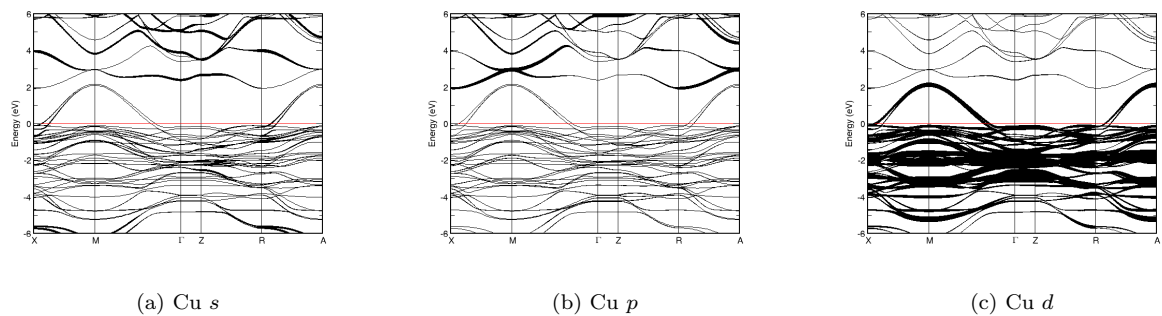

(a) Cu *s*　　　　　　(b) Cu *p*　　　　　　(c) Cu *d*

FIG. 535: Fat band representation of Cu in HoBa$_2$Cu$_3$O$_6$

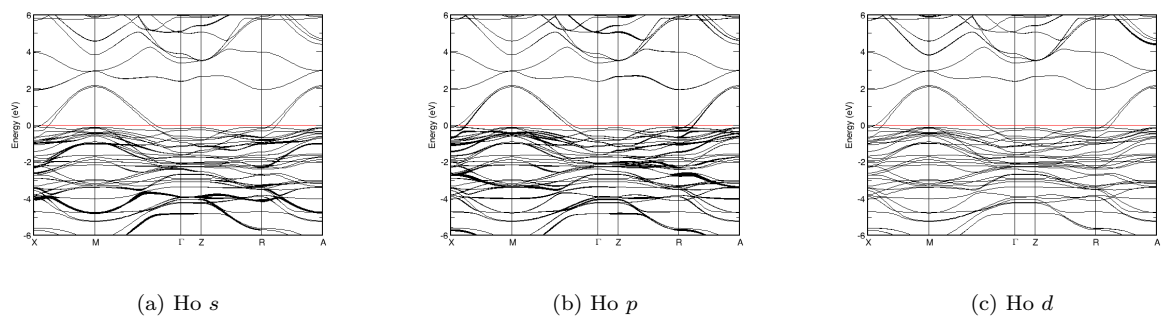

(a) Ho *s*　　　　　　(b) Ho *p*　　　　　　(c) Ho *d*

FIG. 536: Fat band representation of Ho in HoBa$_2$Cu$_3$O$_6$

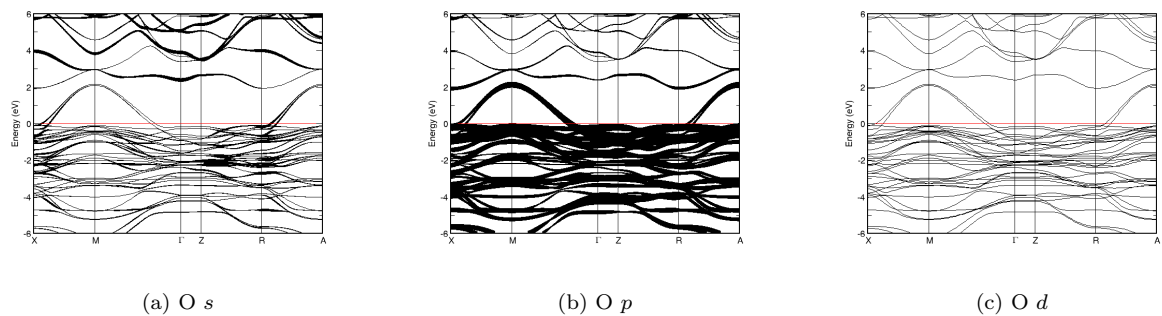

(a) O *s*　　　　　　(b) O *p*　　　　　　(c) O *d*

FIG. 537: Fat band representation of O in HoBa$_2$Cu$_3$O$_6$

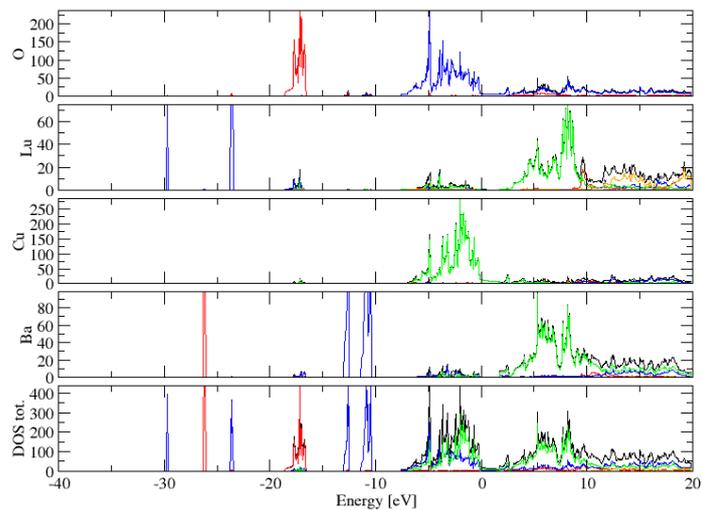

FIG. 538: (Color online) PDOS of LuBa$_2$Cu$_3$O$_6$ (ICSD #98113). The *s*-, *p*- and *d*-projected states are in red, blue and green, respectively. LuBa$_2$Cu$_3$O$_6$ crystallizes in space group P 4/m m m (#123), in a tetragonal primitive structure.

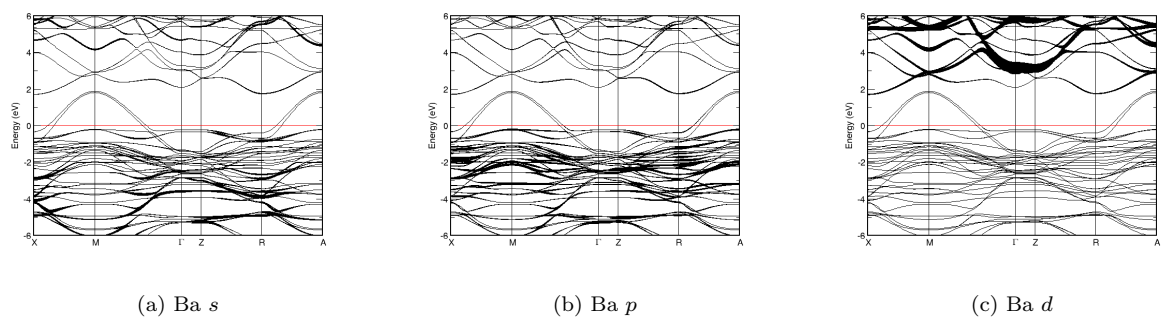

(a) Ba *s*          (b) Ba *p*          (c) Ba *d*

FIG. 539: Fat band representation of Ba in LuBa$_2$Cu$_3$O$_6$

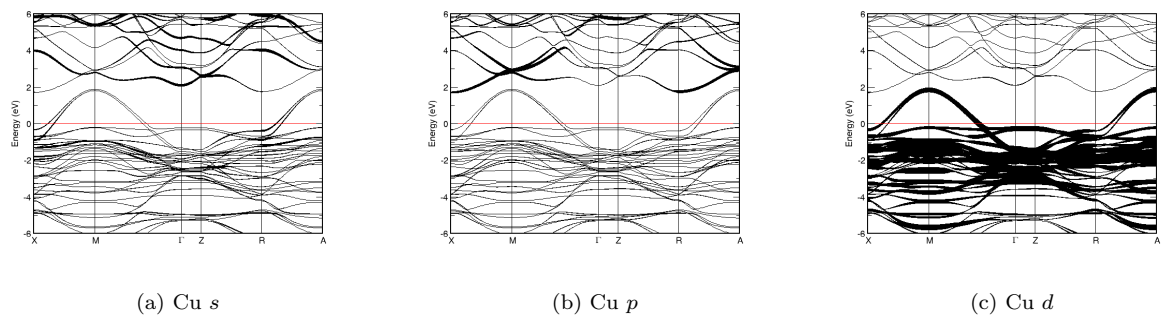

(a) Cu *s*          (b) Cu *p*          (c) Cu *d*

FIG. 540: Fat band representation of Cu in LuBa$_2$Cu$_3$O$_6$

(a) Lu *s*  (b) Lu *p*  (c) Lu *d*

FIG. 541: Fat band representation of Lu in LuBa$_2$Cu$_3$O$_6$

(a) O *s*  (b) O *p*  (c) O *d*

FIG. 542: Fat band representation of O in LuBa$_2$Cu$_3$O$_6$

FIG. 543: (Color online) PDOS of NdBa$_2$Cu$_3$O$_6$ (ICSD #83074). The *s*-, *p*- and *d*-projected states are in red, blue and green, respectively. NdBa$_2$Cu$_3$O$_6$ crystallizes in space group P 4/m m m (#123), in a tetragonal primitive structure.

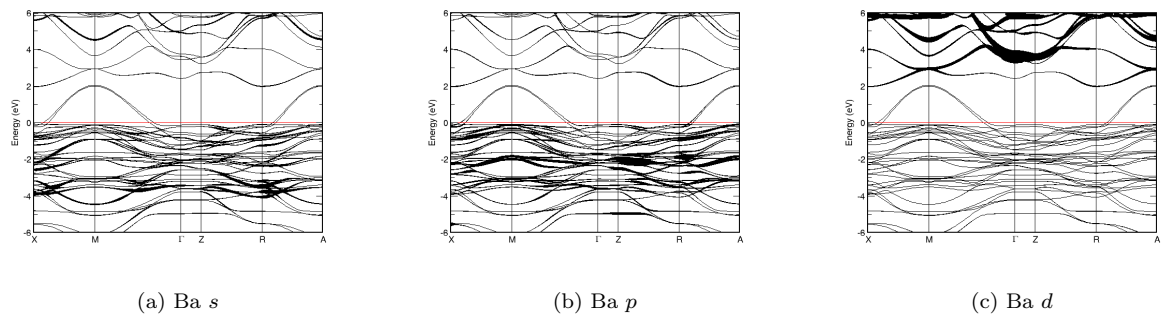

(a) Ba $s$         (b) Ba $p$         (c) Ba $d$

FIG. 544: Fat band representation of Ba in NdBa$_2$Cu$_3$O$_6$

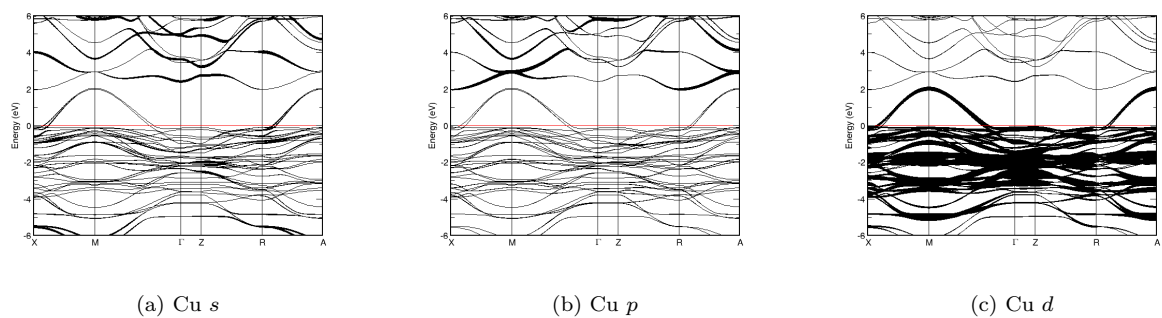

(a) Cu $s$         (b) Cu $p$         (c) Cu $d$

FIG. 545: Fat band representation of Cu in NdBa$_2$Cu$_3$O$_6$

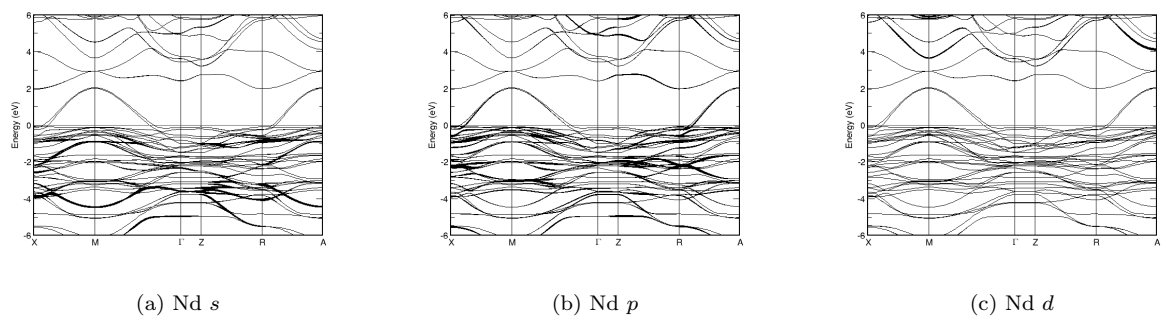

(a) Nd $s$         (b) Nd $p$         (c) Nd $d$

FIG. 546: Fat band representation of Nd in NdBa$_2$Cu$_3$O$_6$

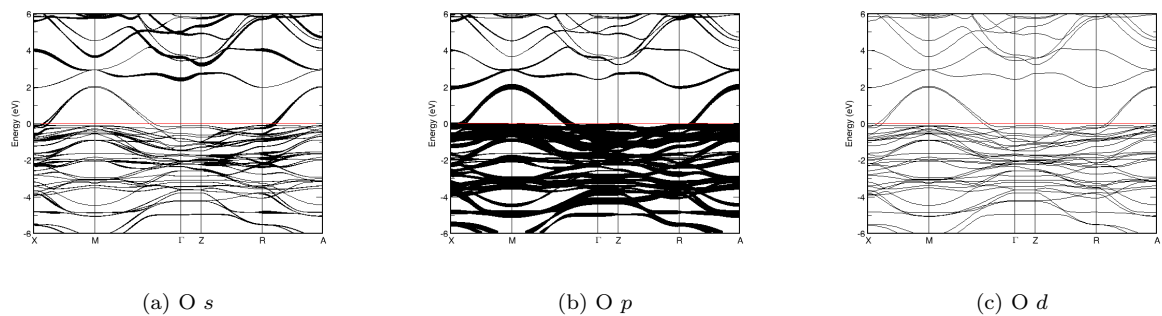

(a) O $s$         (b) O $p$         (c) O $d$

FIG. 547: Fat band representation of O in NdBa$_2$Cu$_3$O$_6$

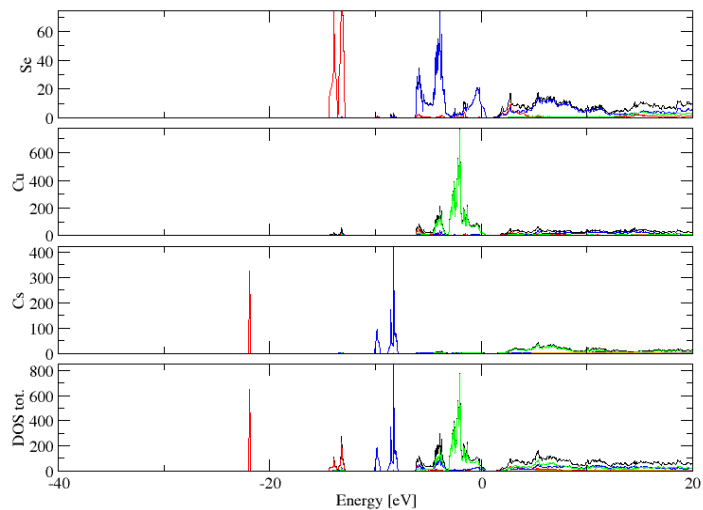

FIG. 548: (Color online) PDOS of Cs(Cu₄Se₃) (ICSD #75196). The *s*-, *p*- and *d*-projected states are in red, blue and green, respectively. Cs(Cu₄Se₃) crystallizes in space group P 4/m m m (#123), in a tetragonal primitive structure.

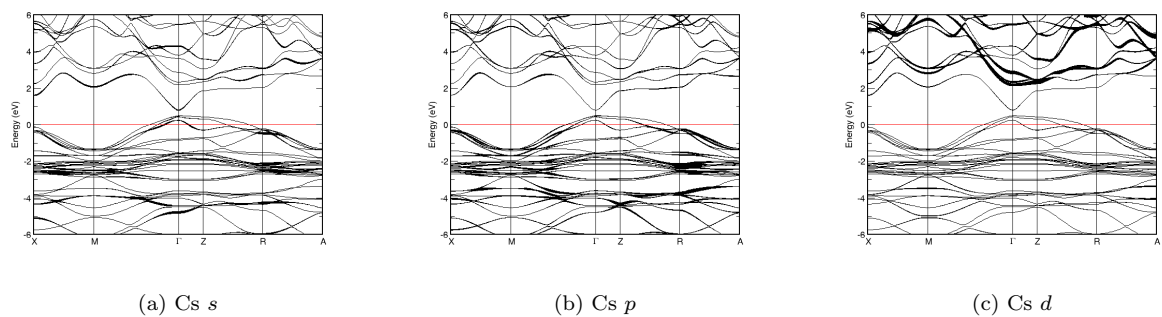

(a) Cs *s*                    (b) Cs *p*                    (c) Cs *d*

FIG. 549: Fat band representation of Cs in Cs(Cu₄Se₃)

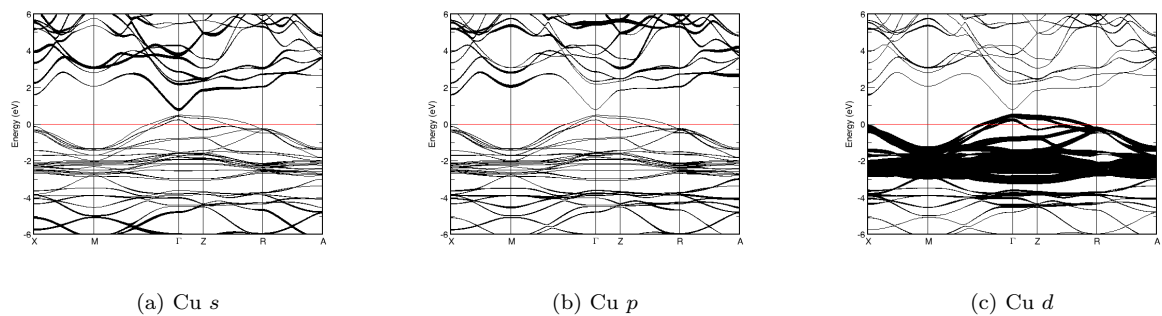

(a) Cu *s*                    (b) Cu *p*                    (c) Cu *d*

FIG. 550: Fat band representation of Cu in Cs(Cu₄Se₃)

(a) Se $s$  (b) Se $p$  (c) Se $d$

FIG. 551: Fat band representation of Se in $Cs(Cu_4Se_3)$

FIG. 552: (Color online) PDOS of $KCu_4S_3$ (ICSD #23336). The $s$-, $p$- and $d$-projected states are in red, blue and green, respectively. $KCu_4S_3$ crystallizes in space group P 4/m m m (#123), in a tetragonal primitive structure.

(a) Cu $s$  (b) Cu $p$  (c) Cu $d$

FIG. 553: Fat band representation of Cu in $KCu_4S_3$

(a) K $s$

(b) K $p$

(c) K $d$

FIG. 554: Fat band representation of K in KCu$_4$S$_3$

(a) S $s$

(b) S $p$

(c) S $d$

FIG. 555: Fat band representation of S in KCu$_4$S$_3$

FIG. 556: (Color online) PDOS of KCu$_4$Se$_3$ (ICSD #280072). The $s$-, $p$- and $d$-projected states are in red, blue and green, respectively. KCu$_4$Se$_3$ crystallizes in space group P 4/m m m (#123), in a tetragonal primitive structure.

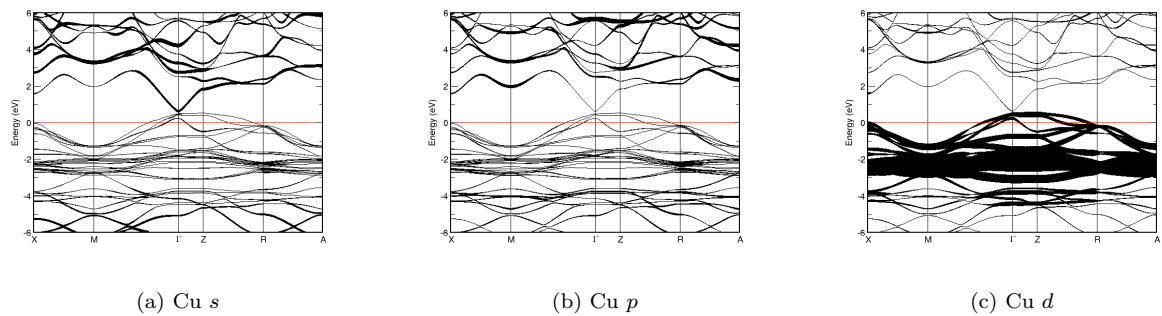

(a) Cu *s*　　　　　　(b) Cu *p*　　　　　　(c) Cu *d*

FIG. 557: Fat band representation of Cu in KCu$_4$Se$_3$

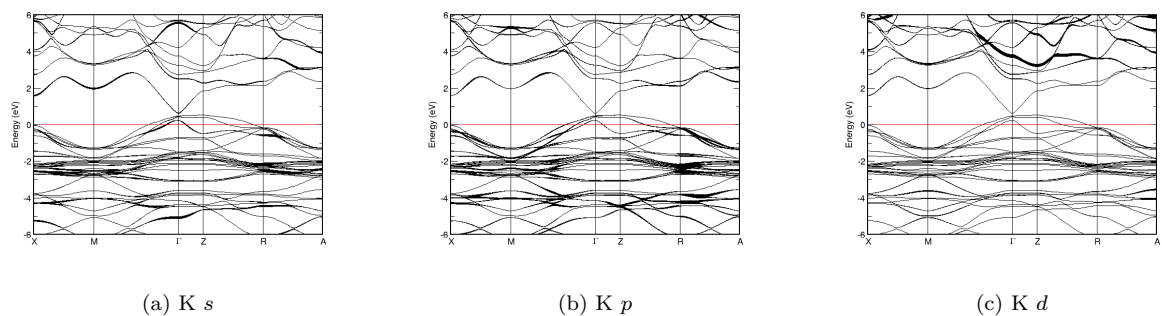

(a) K *s*　　　　　　(b) K *p*　　　　　　(c) K *d*

FIG. 558: Fat band representation of K in KCu$_4$Se$_3$

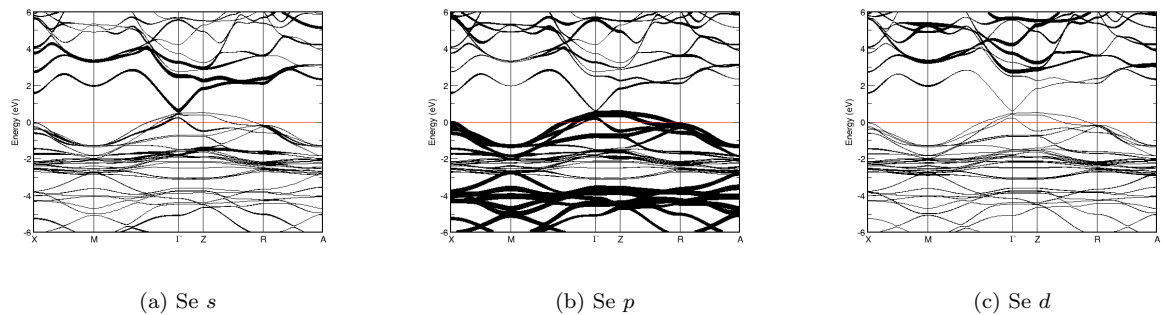

(a) Se *s*　　　　　　(b) Se *p*　　　　　　(c) Se *d*

FIG. 559: Fat band representation of Se in KCu$_4$Se$_3$